\documentclass[aps,prx,twocolumn,notitlepage,showkeys,showpacs, superscriptaddress, longbibliography, footinbib, preprintnumbers]{revtex4-2}
\usepackage{epsf}
\usepackage{bm}
\usepackage{times}
\usepackage{amsfonts}
\usepackage{amssymb}
\usepackage{amsmath,mathtools}
\usepackage{array}
\usepackage{enumerate,dsfont,here, enumitem}
\usepackage{dcolumn,multirow,bbold}
\usepackage{tikz-cd}
\usepackage{bbding}
\usepackage[colorlinks=true,citecolor=blue,linkcolor=blue]{hyperref}
\usepackage[normalem]{ulem}
\usepackage{longtable}

\usepackage{calligra}

\usepackage[whole]{bxcjkjatype}

\newcommand{\mZ}{\mathbb{Z}}
\newcommand{\R}{\mathbb{R}}
\newcommand{\bk}{{\bm{k}}} 
\newcommand{\coker}{\mathrm{Coker}~}
\newcommand{\im}{\mathrm{Im}~}
\renewcommand{\ker}{\mathrm{Ker}~}
\newcommand{\pf}{\mathrm{Pf}}
\newcommand{\calT}{\mathcal{T}}
\newcommand{\calC}{\mathcal{C}}

\newcommand{\calP}{\mathcal{P}}
\newcommand{\calJ}{\mathcal{J}}
\newcommand{\calG}{\mathcal{G}}
\newcommand{\calM}{\mathcal{M}}
\newcommand{\calGk}{\mathcal{G}_{\bk}}
\newcommand{\calA}{\mathcal{A}}

\newcommand{\calD}{\mathcal{D}}

\graphicspath{{./fig/}} 

\usepackage{braket}
\newcommand{\ii}{{\mathrm{i}}}
\newcommand{\g}{{\gamma}}
\newcommand{\G}{{\Gamma}}
\newcommand{\Z}{{\mathbb{Z}}}

\newcommand{\bx}{{\bm{x}}}

\newcommand{\tr}{{\rm tr\,}}

\usepackage{amsthm}
\newtheorem{thm}{Theorem}[section]

\newtheorem{conj}[thm]{Conjecture}

\begin{document}
\title{Towards complete characterization of topological insulators and superconductors:~A systematic construction of topological invariants based on Atiyah-Hirzebruch spectral sequence}
\author{Seishiro Ono}
\thanks{These authors contributed equally.}
\affiliation{Institute for Solid State Physics, University of Tokyo, Kashiwa 277-8581, Japan}
\affiliation{RIKEN Center for Interdisciplinary Theoretical and Mathematical Sciences (iTHEMS), RIKEN, Wako 351-0198, Japan}

\author{Ken Shiozaki}
\thanks{These authors contributed equally. \\
\href{mailto:s-ono@g.ecc.u-tokyo.ac.jp}{s-ono@g.ecc.u-tokyo.ac.jp}\\
\href{mailto:ken.shiozaki@yukawa.kyoto-u.ac.jp}{ken.shiozaki@yukawa.kyoto-u.ac.jp}}
\affiliation{Center for Gravitational Physics and Quantum Information, Yukawa Institute for Theoretical Physics, Kyoto University, Kyoto 606-8502, Japan}

\preprint{YITP-23-93}
\preprint{RIKEN-iTHEMS-Report-23}

\begin{abstract}
	The past decade has witnessed significant progress in topological materials investigation.
	Symmetry-indicator theory and topological quantum chemistry provide an efficient scheme to diagnose topological phases from only partial information of wave functions without full knowledge of topological invariants, which has resulted in a recent comprehensive materials search.
	However, not all topological phases can be captured by this framework, and topological invariants are needed for a more refined diagnosis of topological phases.
	In this study, we present a systematic framework to construct topological invariants for a large part of symmetry classes, which should be contrasted with the existing invariants discovered through one-by-one approaches.
	Our method is based on the recently developed Atiyah-Hirzebruch spectral sequence in momentum space.
	As a demonstration, we construct topological invariants for time-reversal symmetric spinful superconductors with conventional pairing symmetries of all space groups, for which symmetry indicators are silent.
	We also validate that the obtained quantities work as topological invariants by computing them for randomly generated symmetric Hamiltonians.
	Remarkably, the constructed topological invariants completely characterize $K$-groups in 159 space groups.
	Our topological invariants for normal conducting phases are defined under some gauge conditions.
	To facilitate efficient numerical simulations, we discuss how to derive gauge-independent topological invariants from the gauge-fixed topological invariants through some examples.
	Combined with first-principles calculations, our results will help us discover topological materials that could be used in next-generation devices and pave the way for a more comprehensive topological materials database.
\end{abstract}
\maketitle


\section{Introduction}
\label{sec1}
Over the past decades, topological phases of matter have attracted much attention. 
In particular, topological materials, exemplified by topological insulators~\cite{Kane-Mele1, Kane-Mele2, BHZ_Science,Fu-Kane-TRSpump,PhysRevB.75.121306, Fu-Kane-Mele_3DTI,Fu-Kane_formula,RevModPhys.82.3045} and superconductors~\cite{PhysRevLett.100.096407,PhysRevLett.102.187001,PhysRevB.81.220504,PhysRevLett.105.097001,RevModPhys.83.1057, Sato_2017}, have been intensively studied because nontrivial bulk topology gives rise to exotic surface states and fascinating response phenomena~\cite{PhysRevLett.102.146805,PhysRevB.81.205104,PhysRevB.83.035309,PhysRevB.85.165120,doi:10.1126/sciadv.1501524,Juan:2017aa,response_TSC, Luka-Ono-Watanabe2019,James_Majorana, PhysRevX.10.041041,PhysRevX.11.011001,Ahn:2022aa}.
These properties are expected to be utilized for applications in next-generation devices such as fault-tolerant quantum computers, ultrafast memories, and low-power devices~\cite{RevModPhys.80.1083,Nakatsuji-Arita_review,Bernevig:2022aa}. 
Given the diverse nature of bulk band topology~\cite{PhysRevB.88.075142,Morimoto-Furusaki2013, Shiozaki-Sato2014,Shiozaki-Sato-Gomi2016, Thorngren-Else-PRX2018, Cornfeld-Chapman, Shiozaki-CC, PhysRevResearch.3.013052}, the resulting consequences also exhibit a large variety.
This naturally raises two questions:
\vspace{-0.5\baselineskip}
\begin{enumerate}
	\setlength{\itemsep}{-2pt}
	\item[(I)] How many topological phases exist?
	\item[(II)] What quantities can distinguish materials with different topology?
\end{enumerate}
\vspace{-0.5\baselineskip}
In fact, these two questions have been central issues in condensed matter physics for the last ten years.
During this period, there have been numerous developments addressing these questions, as described below.

Symmetry is a fundamental concept in physics, which also plays a crucial role in studies of topological phases of matter. 
The existence of symmetries enriches the variety of topological phases.
Such topological phases protected by symmetries can be trivialized (continuously deformed to a trivial product state without closing the gap) once the protecting symmetries are broken.
In recent years, there has been considerable research effort to establish the full classification of topological phases with various symmetries. 
A milestone is a complete classification and characterization of stable topological phases of noninteracting systems in the presence or absence of internal symmetries, such as time-reversal and particle-hole symmetries~\cite{PhysRevB.78.195125,Kitaev_bott,Ryu_2010}.
In addition to internal symmetries, crystalline symmetries, such as rotation and spatial inversion, are also present in solids.
Recently, it has been shown that crystalline symmetries also protect topological phases~\cite{PhysRevLett.106.106802}, as exemplified by mirror Chern insulators~\cite{PhysRevB.78.045426,Tanaka:2012aa,Hsieh:2012aa, Slager:2013aa}, topological insulators and superconductors protected by nonsymmorphic symmetries~\cite{Shiozaki-Sato-Gomi2016, Wang:2016aa, Shiozaki-Sato-Gomi2017}, and higher-order topological insulators and superconductors~\cite{Benalcazar61,PhysRevLett.119.246401,PhysRevB.96.245115,PhysRevLett.119.246402,Schindler:2018aa,Schindlereaat0346,PhysRevB.97.205135,PhysRevB.97.205136,PhysRevX.9.011012,Fang2017,PhysRevB.103.184502, NM-BiBr}.
More recently, a sophisticated method to classify topological crystalline phases has been developed~\cite{TC_PRB,TC_PRX, Xiong_2018,TC_AII, R-AHSS_Song,R-AHSS_Shiozaki,defect_network,TC_bosonic}, which is based on the real-space picture.
The basic assumption behind the classification scheme is that any topological crystalline phase can be continuously deformed to a patchwork of lower-dimensional topological phases protected by onsite symmetries. 
The classification scheme has been widely applied to topological insulators~\cite{TC_AII, TC_MSG}, superconductors~\cite{PhysRevB.106.174512, Ono-Shiozaki-Watanabe2022, Shiozaki-Ono2023}, and bosonic systems~\cite{TC_PRB, R-AHSS_Song, PhysRevB.101.085137, TC_bosonic}.
This approach provides a clear insight into question (I) posed earlier.

While classifications of noninteracting fermionic systems in real space have been achieved for a large part of symmetry settings, a diagnosis of topological materials has been performed in momentum space. 
An efficient diagnostic scheme has been recently developed based on irreducible representations (irreps) of wave functions at high-symmetry points (called $0$-cells in later discussions)~\cite{PhysRevX.7.041069, SI_NC_Po,SI_SA_Watanabe, TQC,MTQC}. 
In this scheme, we can partially diagnose the topological phase of a target system by comparing irreps of wave functions at high-symmetry momenta with those of atomic insulators~\cite{SI_NC_Po,SI_SA_Watanabe, TQC,MTQC}.
Such a scheme is known as \textit{symmetry-based indicator} \cite{SI_NC_Po,SI_SA_Watanabe} or \textit{topological quantum chemistry}~\cite{TQC,MTQC}. 
The theory can be easily combined with density functional theory calculations, since the latter can give irreps at high-symmetry momenta.
In fact, comprehensive searches for topological materials have been conducted using symmetry indicators or topological quantum chemistry~\cite{catalogue0,catalogue1,catalogue2,catalogue3,catalogue4,catalogue6,catalogue_sc,doi:10.1126/science.abg9094}.
Moreover, the theory of symmetry indicators provides various topological invariants consisting of the number of irreps appearing in occupied bands and Pfaffian invariants at high-symmetry momenta~\cite{SI_PRX_Song,SI_PRX_Eslam,SI_NC_Song,SI_MSG_Fang,Corner_irrep_Hughes,Corner_Irrep_Schindler,Ono-Watanabe2018,Ono-Yanase-Watanabe2019,SI_TSC_Skurativska,SI_Shiozaki,Ono-Po-Watanabe2020,SI_TSC_Luka,Ono-Po-Shiozaki2021}. 
Although the aforementioned topological invariants are easy to compute, it is known that the representation-based diagnosis is incomplete; namely, various topological phases are undetectable only by such topological invariants consisting only of quantities defined at high-symmetry points~\cite{SI_NC_Song, nonIrrep_PRB_Cano,Naito-Takahashi-Watanabe-Murakami,PhysRevB.105.094518, Ono-Shiozaki-Watanabe2022, Li-Sun_Pfaffian, Li-Sun_1d-inv, Araya_C4T}.
This implies that the complete diagnosis of topological phases requires full knowledge of topological invariants. 
However, despite considerable advancements in classification problems, the range of available topological invariants remains limited. 
The previous constructions have typically been done on a case-by-case basis. 
To the best of our knowledge, a systematic method for identifying topological invariants has yet to be established.

\begin{figure}[t]
	\begin{center}
		\includegraphics[width=1\columnwidth]{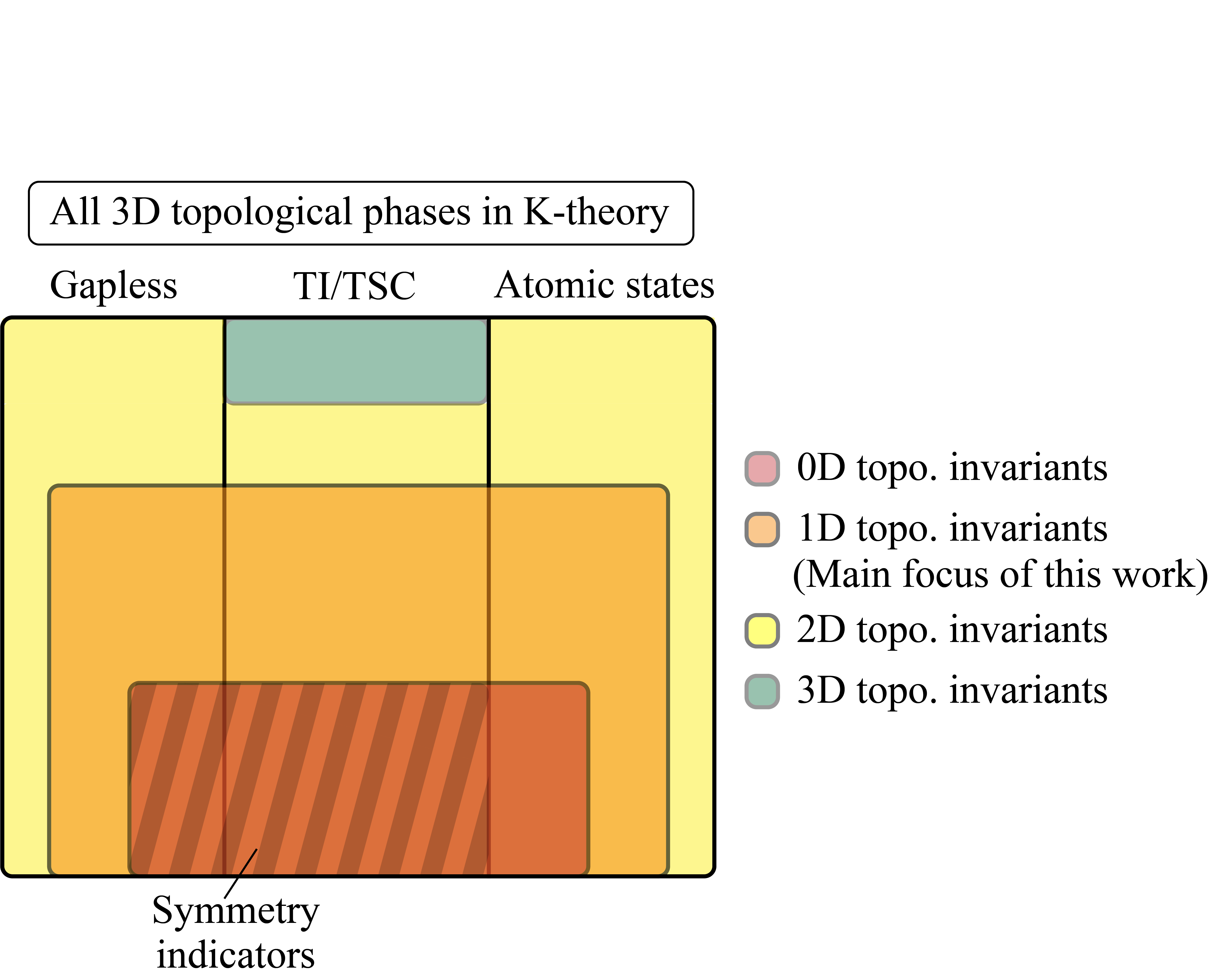}
		\caption{\label{fig:overview}
			Characterization of stable topological phases. 
			All stable gapped and gapless topological phases can be distinguished by topological invariants. 
			There are topological invariants defined on zero-, one-, two-, and three-dimensional subregions of momentum space.
			Symmetry indicators are part of zero-dimensional topological invariants (shaded region). 
			The main focus of this work is a systematic construction of topological invariants defined on one-dimensional subspaces.
			}
	\end{center}
\end{figure}

In this work, we present a systematic scheme for constructing topological invariants of normal and superconducting phases by leveraging the Atiyah-Hirzebruch spectral sequence (AHSS) in momentum space~\cite{K-AHSS}. 
See Fig.~\ref{fig:overview} for an illustration summarizing the role of our work.
The construction of topological invariants involves two pivotal steps: %
\vspace{-0.5\baselineskip}
\begin{enumerate}
	\setlength{\itemsep}{-2pt}
	\item[(i)] First, we need to identify the irreducible-representation sectors on the subregions of momentum space that enter the definition of topological invariants.
	For instance, on mirror-invariant lines and planes, the Bloch Hamiltonian can be decomposed into mirror-eigenvalue sectors, and mirror winding numbers on the former and mirror Chern numbers on the latter are then defined in each mirror-eigenvalue sector~\cite{PhysRevB.78.045426, PhysRevLett.111.056403};
	\item[(ii)] Second, we need to discover the quantities that characterize topological nature of the system. 
	For example, mirror winding and Chern numbers are computed from the winding of $q$-matrix (which is formally defined later) and the integral of Berry curvature in each mirror-eigenvalue sector. 
\end{enumerate}
\vspace{-0.5\baselineskip}
Remarkably, the momentum-space AHSS provides valuable insights into both steps.
As the first step, we discuss a general framework for extracting information about the $p$-dimensional subregions and irreps defined on them from the first differentials of AHSS.
Then, building upon the physical meaning of $E_1$-pages of AHSS---classifications of $p$-dimensional topological phases on the $p$-dimensional subregions, we define quantities for each irrep identified in the first step, which characterize topological nature of $p$-dimensional subregions in momentum space.
AHSS can also inform us how to combine the defined quantities to obtain topological invariants.
While we formulate the first step for arbitrary $p$, for the second step we focus on the case of $p=1$ and discuss how to systematically construct topological invariants defined on zero- and one-dimensional subregions.
To demonstrate our approach, we construct topological invariants for time-reversal symmetric spinful superconductors with conventional pairing symmetries in 230 space groups, for which symmetry indicators are always trivial, and numerically confirm that they are actually quantized for randomly generated symmetric Hamiltonians.
Surprisingly, $K$-groups in 159 out of the 230 space groups are completely characterized by our topological invariants.
Our construction gives topological invariants for normal conducting phases under certain gauge conditions.
For efficient numerical calculations, we discuss how to derive gauge-independent topological invariants from the gauge-dependent ones through illustrative examples where symmetry indicators do not work.

We make three remarks.
First, the present paper addresses the stable $K$-theoretic classification.
The topological invariants constructed below characterize the corresponding $K$-group, rather than the finer finite-rank classifications of fixed-size Hamiltonians or fixed-rank vector bundles.
At the same time, these invariants can also serve as diagnostic tools for topological phases appearing in such finer classifications. See App.~\ref{app:stable_vs_finite_rank} for more details.
Second, the AHSS in momentum space was originally introduced and used as a tool for classifying topological phases based on the momentum-space picture~\cite{K-AHSS, Ono-Shiozaki2022}.
On the other hand, in this work, we explore the utility of AHSS in momentum space as a versatile tool for band topology and show that the AHSS can be used to construct topological invariants.
However, we note that the AHSS data by themselves do not automatically give explicit formulas for topological invariants.
The purpose of the present work is to develop this translation systematically.
Last, many invariants based on this work can capture topological phases not diagnosed by symmetry indicators. Such phases cannot be detected from symmetry representations at high-symmetry momenta alone, and their diagnosis generally requires additional information from higher-dimensional subregions of the Brillouin zone, such as high-symmetry lines and planes. Although this may require additional computational resources compared with symmetry-indicator and topological-quantum-chemistry approaches, efficient formulas are available in some cases and are expected to be developed in others. Thus, our work would help build a comprehensive database of topological materials beyond Refs.~\cite{catalogue0,catalogue1,catalogue2,catalogue3,catalogue4,catalogue6,catalogue_sc,doi:10.1126/science.abg9094}.

\subsection{Summary of results}
\label{sec2}
Below, we present the organization of this paper with a brief summary of main results of each section.

\textbf{Section \ref{sec3}}---%
In Sec.~\ref{sec3}, we present our framework to systematically construct topological invariants of topological superconductors defined on subregions of momentum space.
Such subregions in three dimensions are defined from a fundamental domain of the Brillouin zone,
whose images under the symmetry operations cover the entire Brillouin zone
without overlap.
The open faces of the fundamental domain are two-dimensional subregions, the open edges of the faces are one-dimensional subregions, and the endpoints of the edges are zero-dimensional subregions.
Similarly, such subregions can be defined in arbitrary dimensions (see Fig.~\ref{fig:simple_cell} for an illustration in two dimensions).

\begin{enumerate}
	\item We review the basics of AHSS in momentum space.
	In the framework of AHSS, effective internal symmetry classes of irreps, so-called \textit{emergent Altland-Zirnbauer (EAZ) symmetry classes}, play an important role. 
	Roughly speaking, EAZ classes can be understood as transformation properties of irreps under time-reversal-like, particle-hole-like, and chiral-like symmetries.
	Let $\Phi^{\alpha}_{\bk}$ be a set of occupied states with irrep $\alpha$. 
	Now, consider symmetry $g$ that does not change momentum $\bk$, whose representation is denoted by $U_{\bk}(g)$.
	Then, we ask whether $U_{\bk}(g)[\Phi^{\alpha}_{\bk}]^{*} (U_{\bk}(g)\Phi^{\alpha}_{\bk} \text{ if $g$ is a chiral-like symmetry})$ possess the same irrep $\alpha$ or not.
	Depending on the transformation properties of irreps, we assign one of the tenfold symmetry classes to each irrep [cf.~Table~\ref{tab:EAZ}].
	In this section, for simplicity, we focus on the cases where EAZ classes belong to class AIII, DIII, and CI.
	\item As mentioned in Sec.~\ref{sec1}, we need to identify irreps on the subregions of momentum space, which are responsible for topological nontriviality.
	This is the first difficulty in a systematic construction of topological invariants. 
	Using the first differential of AHSS, we propose a way to overcome this difficulty.
	All information is encoded in matrices $[X^{(1)}]^{-1}, [V^{(0)}]^{-1}$, $\Sigma^{(1)}$, and $\Lambda^{(0)}$, whose implications for topological invariants are summarized below.
	The matrices $\Sigma^{(1)}$ and $\Lambda^{(0)}$ are diagonal integer-valued matrices. 
	The matrices $X^{(1)}$ and $V^{(0)}$ are $N_1$- and $N_0$-dimensional invertible matrices, respectively, which are given by
	\begin{align}
		&[X^{(1)}]^{-1} := \left(\begin{array}{cccc}
			\bm{x}_{1} &
			\bm{x}_{2} &
			\cdots &
			\bm{x}_{N_1}
		\end{array}\right)^{\top};\\
		&[V^{(0)}]^{-1} := \left(\begin{array}{cccc}
			\bm{v}_{1} &
			\bm{v}_{2} &
			\cdots &
			\bm{v}_{N_0}
		\end{array}\right)^{\top}.
	\end{align}
	\begin{itemize}
		\item $\Sigma^{(1)}$:~This matrix is obtained from the compatibility between line segments and adjacent planar subspaces in momentum space. 
		The number $r_1$ of its nonzero diagonal entries gives the number of independent $\mZ$-valued invariants detecting gapless points on the two-dimensional subspaces, namely the open faces of the fundamental domain.
		Note that $\Sigma^{(1)}$ itself does not specify the planar subspaces.
		The corresponding combinations of irreps on line segments are specified by the first $r_1$ rows of $[X^{(1)}]^{-1}$.

		\item $\Lambda^{(0)}$:
		The matrix encodes information about topological invariants for phases that remain gapped on the boundary of the chosen fundamental domain.
		The matrix $\Lambda^{(0)}$ takes the form
		\begin{equation}
			\begin{aligned}
				\quad\quad\quad\quad 
				\Lambda^{(0)}
				&=
				\text{diag}(\overbrace{1,\cdots,1}^{R_0}, \lambda_1, \cdots, \lambda_{r_0}, 0, \cdots) \\
				&\in \mathbb{Z}^{(N_1 - r_1) \times N_0}
			\end{aligned},
		\end{equation}
		where $R_0$ is the number of unity entries in the diagonal elements of $\Lambda^{(0)}$, and $\lambda_j>1\ (j=1,\cdots,r_0)$.
		There exist $\mZ_{\lambda_j}$-valued invariants $(j=1, \cdots, r_0)$ and $(\min(N_1 - r_1, N_0) - R_0 - r_0)$ independent $\mZ$-valued invariants.
		It should be noted that phases detected by these invariants are either fully gapped or gapless inside the fundamental domain.
		\item $[X^{(1)}]^{-1}$:~The rows of $[X^{(1)}]^{-1}$ specify combinations of irreps on line segments in momentum space that enter the topological invariants.
		The first $r_1$ rows of $[X^{(1)}]^{-1}$, i.e., $\{\bm{x}_j\}_{j=1}^{r_1}$, correspond to topological invariants detecting gapless points on the two-dimensional subspaces, namely the open faces of the fundamental domain.
		For $j=1,\dots,r_0$, the $(j+R_0+r_1)$-th row, namely $\bm{x}_{j+R_0+r_1}$, specifies the line-segment part of the aforementioned $\mZ_{\lambda_j}$-valued invariant.
		For $j > R_0 +r_0$ (the zero diagonal entries of $\Lambda^{(0)}$), the $(j+r_1)$-th row $\bm{x}_{j+r_1}$ specifies the line-segment parts of $\mZ$-valued invariants detecting phases without any gapless points on the boundary of the chosen fundamental domain.
		\item $[V^{(0)}]^{-1}$:~For $j=1,\dots,r_0$, the $(j+R_0)$-th row of $[V^{(0)}]^{-1}$, i.e., $\bm{v}_{j+R_0}$, specifies the irreps on zero-dimensional subregions that enter the $\mZ_{\lambda_j}$-valued invariant constructed from $\Lambda^{(0)}$.
    \end{itemize}

	All the matrices are presented in Ref.~\cite{SM}.
	In cases where EAZ classes other than AIII, DIII, and CI appear, the explanation of these matrices is presented in Appendix~\ref{app:Z2case}.
	We explain how to use them below.
	\item We construct topological invariants defined on one-dimensional subregions of momentum space. 
	Depending on EAZ classes, for each irrep, we define the quantities to characterize the topological properties in superconducting systems: 
	\begin{itemize}
		\setlength{\itemindent}{-0.5cm}
		\item EAZ class AIII and CI
		\begin{align}
			w_{\alpha} := \frac{1}{2\pi\ii \mathcal{D}_{\alpha}}\int_s d(\log \det q_{\bk}^{\alpha}-\log \det (q^{\alpha}_{\bk})^{\text{vac}});
		\end{align}
		\item EAZ class DIII
		\begin{align}
			w_{\alpha}:=\frac{1}{4\pi \ii \mathcal{D}_{\alpha}}\int_s d(\log \det q_{\bk}^{\alpha}-\log \det (q^{\alpha}_{\bk})^{\text{vac}}),
		\end{align}
	\end{itemize}
	where $q_{\bk}^{\alpha}$ is a $q$-matrix defined for irrep $\alpha$, and $\mathcal{D}_{\alpha}$ is the dimension of irrep $\alpha$.
    Here, $s$ denotes an oriented line segment in momentum space.
    The integral is taken along $s$ from one endpoint to the other according to its orientation.
    The precise definition is given in Sec.~\ref{sec:topo_invariant}.
	To make the quantity invariant under gauge transformations, we need to compute the difference of the winding between the target system and a reference system. 
	In this work, we always consider the vacuum (i.e., infinite chemical potential limit of the system in which we are interested) as the reference, and $(q^{\alpha}_{\bk})^{\text{vac}}$ denotes the $q$-matrix of the vacuum.

	\quad The combinations of these quantities yield topological invariants. As explained above, $[X^{(1)}]^{-1}$ and $[V^{(0)}]^{-1}$ tell us how to take the combinations.
	%
	%
	Then, topological invariants are given by
	\begin{itemize}[leftmargin=0.3 cm]
		\item $\mZ$-valued invariant for gapless points on planes
		\begin{align}
			&\mathcal{W}^{\text{gapless}}_{j}[H_{\bk}] = \sum_{\alpha}[\bm{x}_{j}]_{\alpha}w_{\alpha} \in \mZ
		\end{align}
		for $1\leq j \leq r_1$.
		\item $\mZ$-valued invariant for gapped phases or gapless points at generic momenta
		\begin{align}
			&\mathcal{W}^{\text{gapped}}_{j}[H_{\bk}] = \sum_{\alpha}[\bm{x}_{j}]_{\alpha}w_{\alpha} \in \mZ
		\end{align}
		for $r_1+R_0+r_0< j$.
		\item $\mZ_{\lambda_{j}}$-valued invariant for gapped phases or gapless points at generic momenta
		\begin{align}
			&\exp\hspace{-0.5mm}\left[\frac{2\pi \mathrm{i}}{\lambda_{j}^{(0)}}\mathcal{X}_{j}[H_{\bk}]\right] \nonumber \\
				& \quad\quad\quad= \frac{\exp[-2\pi \ii \sum_{\alpha}[\bm{x}_{j+r_1+R_0}]_{\alpha}w_{\alpha}/\lambda_{j}^{(0)}]}{\prod_{\beta}\left(\mathcal{Z}[q_{\bk}^{\beta}]/\mathcal{Z}[(q^{\beta}_{\bk})^{\text{vac}}]\right)^{[\bm{v}_{j}]_{\beta}}},
		\end{align}
	\end{itemize}
	where $j\leq r_0$, and $\mathcal{Z}[q_{\bk}^{\beta}]$ satisfies
	\begin{align}
		\det q^{\alpha}_{\bk} = \begin{cases}
			\left(\mathcal{Z}[q^{\alpha}_{\bk}]\right)^{\calD_\alpha} \text{ for EAZ class AIII/CI}\\
			\left(\mathcal{Z}[q^{\alpha}_{\bk}]\right)^{2\calD_\alpha} \text{ for EAZ class DIII}
		\end{cases}.
	\end{align}
	See Eq.~\eqref{eq:def_Z_q_alpha} for the precise definition of $\mathcal{Z}[q_{\bk}^{\beta}]$.
	\item To demonstrate the power of our method, we compute constructed topological invariants for randomly generated symmetric Hamiltonians of time-reversal symmetric spinful superconductors with conventional pairing symmetries in 80 layer groups and 230 space groups. 
	Importantly, symmetry indicators do not work for these symmetry classes, and our invariants completely characterize $K$-groups in 76 out of the 80 layer groups and 159 out of the 230 space groups.
\end{enumerate}

\textbf{Section \ref{sec4}}---In Sec.~\ref{sec4}, we explicitly show how to construct topological invariants from $[X^{(1)}]^{-1}$ and $[V^{(0)}]^{-1}$.
\begin{enumerate}
	\item We construct topological invariants of time-reversal symmetric spinful superconductors in layer groups $p2_1/m11, p2_1/b11$, and $p2_111$ with conventional pairing symmetries, whose $K$-groups are $\mZ$, $\mZ_2$, and $\mZ_2\times \mZ_4$, respectively.
	\item We compute our topological invariants for representative models of $K$-groups and show that our invariants fully characterize $K$-groups of them. 
\end{enumerate}

\textbf{Section \ref{sec5}}---%
In Sec.~\ref{sec5}, we discuss our construction of topological invariants for normal conducting phases. 
For normal conducting phases, the EAZ class of each irrep is always either class A, AI, or AII.
Furthermore, as far as we are interested in topological invariants defined on one-dimensional subregions, only class AI gives rise to nontrivial invariants, and the defined topological invariants are always $\mZ_2$-valued.
\begin{enumerate}
	\item As a counterpart to the winding of the $q$-matrix, we introduce the transition function $t_{\bk_a}$ given by
	\begin{align}
		t_{\bk_a} = \Phi_{\text{P}, \bk_a}^{\dagger}\Phi_{\text{Q}, \bk_a},
	\end{align}
	where $\text{P}$ and $\text{Q}$ are points, and $\bk_a$ is the midpoint of the line segment $\text{PQ}$.
	Note that the gauge choices of $\Phi_{\text{P}, \bk_a}$ and $\Phi_{\text{Q}, \bk_a}$ are generally different.
	We also carefully analyze effects of gauge transformations on the transition function.
	\item We show that the transition functions are defined for each irrep (denoted by $\tilde{t}_{\bk_a}^{\alpha}$) and that $\det \tilde{t}_{\bk_a}^{\alpha}$ is $\mZ_2$-quantized under certain gauge conditions. 
	The combinations of $\det \tilde{t}_{\bk_a}^{\alpha}$ yield the topological invariants for normal conducting phases.
	As is the case with superconducting phases, the matrices $[X^{(1)}]^{-1}, \Sigma^{(1)}$, and $\Lambda^{(0)}$ inform us about the combinations to construct topological invariants of gapless points and gapped phases. 
	See Secs.~\ref{sec:TI_z2number_Preliminary} and \ref{sec:Gauge_invariant_expression_u1} for more detailed discussions.
	\item The obtained topological invariants do not satisfy linearity for the direct sum of bands. 
	To overcome the undesirable property, we introduce the quadratic refinement of the $\mZ_2$ invariant so that the redefined invariants behave linearly with respect to the direct sum of bands.
	\item Since the topological invariants under certain gauge conditions are not useful for numerical simulations, we introduce several general ways to obtain gauge-independent expressions of topological invariants. 
	The overall strategy is as follows.
	\begin{itemize}[leftmargin=0.2cm, itemindent=0cm]
		\item[(i)] First, we define a $\text{U}(1)$-valued Wilson line by
		\begin{align}
			e^{\ii \gamma^\alpha_{\text{PQ}}} := \lim_{\mathcal{N}\rightarrow\infty} \det \prod_{j = 0}^{\mathcal{N}-1} (\Phi^\alpha_{\text{P} + (j+1)\bm{\delta}})^{\dagger}\Phi^\alpha_{\text{P} + j\bm{\delta}},
		\end{align}
		where $\bm{\delta} = (\text{Q} - \text{P})/\mathcal{N}$ and $\Phi^{\alpha}_{\bk}$ is a set of occupied states with irrep $\alpha$ on the line segment $\text{PQ}$.
		In fact, we can rewrite the determinant of the transition function in terms of the $\text{U}(1)$-valued Wilson lines.
		It should be noted that the Wilson lines are not invariant under gauge transformations at the boundary points P and Q.
		\item[(ii)] Next,  we multiply some compensation quantities to cancel out $\text{U}(1)$ values from gauge transformations.
		For instance, if we have time-reversal symmetry (TRS) $\calT$ such that $U(\calT)[U(\calT)]^{*}=-1$, the following quantity is gauge invariant:
		\begin{align}
			e^{\ii \gamma_{\text{PQ}}} \times \frac{\pf \left[ \Phi_{\text{P}}^\dag U({\cal T}) \Phi_{\text{P}}^* \right]}{\pf \left[ \Phi_{\text{Q}}^\dag U({\cal T}) \Phi_{\text{Q}}^* \right]},
		\end{align}
        where P and Q are time-reversal invariant momenta (TRIMs).
		In Sec.~\ref{sec:Symmetry-enriched Berry phase}, we discuss techniques how to find such gauge invariant quantities.
	\end{itemize}
	\item We demonstrate how to construct gauge-independent topological invariants through space groups $P222$ and $P\bar{4}$ of spinless electrons with TRS and space groups $P2$ and $P2_12_12_1$ of spinful electrons with TRS.
\end{enumerate}

\textbf{Section \ref{sec6}}---In Sec.~\ref{sec6}, we conclude the paper with outlooks for future works. 

Some technical details are discussed in appendices to avoid straying from the main subjects.

\section{General framework}
\label{sec3}
As mentioned in Sec.~\ref{sec1}, a refined diagnosis requires knowledge of topological invariants. 
In this section, we discuss a systematic framework to construct topological invariants in momentum space, for which we employ AHSS in momentum space~\cite{K-AHSS}. 
There are two crucial steps: (i) identifying irreps on $p$-dimensional subregions responsible for topological nontriviality; (ii) finding quantities to detect topological nature of systems. 
These two steps are discussed in Secs.~\ref{sec:general_frame} and \ref{sec:topo_invariant}, respectively.
We numerically verify that our topological invariants actually work for time-reversal symmetric superconductors with conventional pairing symmetries by computing the constructed topological invariants in this way for randomly generated symmetric Hamiltonians.
In Sec.~\ref{sec:numerics}, we explain how to generate random symmetric Hamiltonians and to confirm that our invariants work for these classes.

\subsection{Review on Atiyah-Hirzebruch spectral sequence in momentum space}
In this subsection, to be self-contained, we provide a brief review of AHSS in momentum space~\cite{K-AHSS}.
In particular, here we introduce basic notions of AHSS and focus on their physical meaning rather than showing how to actually compute them.
All technical details of the computation are discussed in our companion work in Ref.~\cite{Shiozaki-Ono2023}.
Readers familiar with AHSS in momentum space can skip this subsection.

Before moving on to the review of AHSS, let us briefly explain the background of AHSS.  
Let $G$ be a magnetic space group $\calM$ for normal conducting phases and a magnetic space group with particle-hole symmetry $\calP$ for superconducting phases, i.e., $G = \calM$ for normal conducting phases and $G = \calM + \calM\calP$.
An element $g = \{p_g\vert \bm{a}_g\}\in G$ transforms a point $\bm{x} \in \mathbb{R}^3$ into $p_g\bm{x}+\bm{a}_g$, where $p_g$ is an orthogonal matrix and $\bm{a}_g$ is a vector.
When $\Pi$ denotes the translation subgroup of $G$, $G/\Pi$ is isomorphic to a magnetic point group with or without particle-hole symmetry.
As a result, free fermionic topological phases on a $d$-dimensional torus $T^d$ are classified by the twisted equivariant $K$-group $^{\phi}K_{G/\Pi}^{(z,c)+n}(T^d)$~\cite{Freed2013, Shiozaki-Sato-Gomi2017} (see Appendix~\ref{app:K-theory} for a brief review of $K$-theory).
The twisted equivariant $K$-group contains symmetry data denoted by $\phi, c, z$, and $n$. 
First, $\phi$ and $c$ are defined as maps $\phi, c: G/\Pi \rightarrow \{\pm 1\}$. 
A symmetry $g \in G/\Pi$ is unitary (antiunitary) when  $\phi_g = +1\ (\phi_g = -1)$, and its symmetry representation $\mathcal{U}_{\bk}(g)$ commutes (anticommutes) with Hamiltonian $H_{\bk}$ when $c_g = +1\ (c_g = -1)$.
In other words, $\mathcal{U}_{\bk}(g)$ and $H_{\bk}$ satisfy the following relations
\begin{align}
	\mathcal{U}_{\bk}(g)H^{\phi_g}_{\bk} = c_gH_{g\bk}\mathcal{U}_{\bk}(g),
    \label{eq:Hamiltonian_G_sym}
\end{align}
where we introduce  the following notation of matrices: $A^{\phi_g} = A$ for $\phi_g = +1$ and $A^{\phi_g} = A^*$ for $\phi_g = -1$.
Next, the symbol $z$ represents the set of $\text{U}(1)$ projective factors $\{z_{\bk}(g,g')\in \text{U}(1)\}_{g,g' \in G/\Pi}$, which are defined by 
\begin{align}
	\mathcal{U}_{g'\bk}(g)\mathcal{U}^{\phi_g}_{\bk}(g') &= z_{\bk}(g,g')\mathcal{U}_{\bk}(gg').
\end{align}
Last, an integer $n$ denotes the {\it grading} in $K$-theory, on which the physical meaning of $K$-groups depends.

We also recall how the grading is implemented at the level of Hamiltonians. 
For a positive integer $n$, the degree-$(-n)$ group is represented by imposing $n$ additional Clifford generators $\Gamma_1,\ldots,\Gamma_n$ satisfying
\begin{align}
    \{\Gamma_i,\Gamma_j\}=2\delta_{ij},
\end{align}
whereas the degree-$(+n)$ group is represented by Clifford generators satisfying
\begin{align}
    \{\Gamma_i,\Gamma_j\}=-2\delta_{ij}.
\end{align}
These generators are required to transform under $G/\Pi$ as
\begin{align}
    \mathcal{U}_{\bk}(g)\Gamma_i^{\phi_g}
    =
    c_g \Gamma_i \mathcal{U}_{\bk}(g),
    \qquad
    g\in G/\Pi,\quad i=1,\ldots,n.
    \label{eq:U_and_G}
\end{align}
In addition, they are imposed as additional chiral symmetries of the Hamiltonian,
\begin{align}
    \Gamma_i H_{\bk}\Gamma_i^{-1}=-H_{\bk},
    \qquad i=1,\ldots,n.
    \label{eq:Hamiltonian_chiral_sym}
\end{align}
In this sense, the graded groups
${}^{\phi}K^{(z,c)-n}_{G/\Pi}(T^d)$ and
${}^{\phi}K^{(z,c)+n}_{G/\Pi}(T^d)$
are realized as classification groups of gapped Hamiltonians satisfying
Eqs.~\eqref{eq:Hamiltonian_G_sym} and \eqref{eq:Hamiltonian_chiral_sym}, with the above choices of Clifford generators.
In particular, $n=0$ corresponds to the classification of stable gapped Hamiltonians with no additional $\Gamma_i$'s.
See Ref.~\cite{Shiozaki-Sato-Gomi2017} for details.

The degree-$1$ $K$-group can also be interpreted as the classification of gapless Hamiltonians $\hat H_{\bk}$ with infinite-dimensional Hilbert spaces, which are realized only on the boundary of $(d+1)$-dimensional gapped systems~\cite{K-AHSS}. 
This interpretation can be formulated as follows. 
Let
\begin{align}
    V_{\bk}= i\exp[-i\chi(\hat H_{\bk})],
\end{align}
where $\chi$ is a continuous odd function satisfying $\chi(E)=\pi\,{\rm sgn}(E)$ for $|E|>E_{\rm gap}$, with $E_{\rm gap}$ the bulk energy gap of the $(d+1)$-dimensional system. 
Then the doubled Hermitian Hamiltonian
\begin{align}
    \tilde H_{\bk}
    =
    \begin{pmatrix}
        0 & V_{\bk}\\
        V_{\bk}^{\dagger} & 0
    \end{pmatrix}
\end{align}
has an intrinsic chiral symmetry generated by
\begin{align}
    \Gamma_1=i\sigma_z,
\end{align}
where $\sigma_z$ is the Pauli matrix acting on the doubled space. 
Together with the doubled symmetry operator
\begin{align}
    \tilde{\mathcal U}_{\bk}(g)
    =
    \mathcal U_{\bk}(g)\otimes
    (\sigma_x)^{(1-\phi_g c_g)/2},
\end{align}
the doubled Hamiltonian and the doubled symmetry operators satisfy the same symmetry relations as
\eqref{eq:Hamiltonian_G_sym} and \eqref{eq:U_and_G}~\cite{Shiozaki_etal_DetachableBoundayState2024}.

Unfortunately, it is not known how to directly compute twisted equivariant $K$-groups for a large part of symmetry classes, except for some simple cases such as order-two spatial symmetries and point-group symmetries~\cite{Shiozaki-Sato2014, Shiozaki-Sato-Gomi2016, Cornfeld-Chapman, Shiozaki-CC}.
Recently, it has turned out that AHSS in momentum space provides us with fruitful information about topology in momentum space. 
If we can compute AHSS completely for a given symmetry setting, it generally gives us an ``approximate $K$-group" in the sense that the obtained Abelian group is the same as the exact $K$-group as a set but not as an Abelian group~\cite{K-AHSS}. 
On the other hand, as discussed in Ref.~\cite{Shiozaki-Ono2023}, AHSS can completely determine $^{\phi}K_{G/\Pi}^{(z,c)-n}(T^d)\ (d \leq 3)$ for various symmetry settings in which we are interested.

\subsubsection{Cell decomposition}
To perform AHSS calculations, we introduce a sequence of spaces 
\begin{align}
	X_0 \subset X_1 \subset \cdots \subset T^d,
\end{align}
where $X_p$ is a $p$-dimensional subspace of $T^d$, the so-called \textit{$p$-skeleton}. 
Such a decomposition is known as \textit{cell decomposition}~\cite{K-AHSS, TC_AII, R-AHSS_Song, defect_network}. 
In the following, we explain a way to obtain $X_p$ for three dimensions.

First, we find a fundamental domain of the first Brillouin zone (BZ) that spans the entire BZ by symmetry operations without any overlap. 
Note that, by definition, any two points strictly inside the fundamental domain are not related by symmetries.
The interior of the fundamental domain is an open polyhedron, which is denoted by $D^3$.
Then, we decompose boundary objects of the fundamental domain into $p$-dimensional subregions.
The open polygons on faces of the fundamental domain are the case of $p=2$, the open edge line segments of the polygons are $p=1$, and the boundary points of the line segments are $p=0$.
They are denoted by $\{D_{i}^{p}\}_i$. 
We also assign an orientation to each of the $p$-dimensional subregions. 
For later convenience, let us define a set of all inequivalent $p$-cells in the fundamental domain by $I_{\text{orb}}^{p}$.
By acting symmetry operations on these $p$-cells, we have
\begin{align}
	C_p := \bigcup_{i \in I_{\text{orb}}^{p}} \bigcup_{g \in G/\Pi} g\left(D_{i}^{p}\right).
\end{align} 
Elements in $C_p$ are called \textit{$p$-cells}.
In this construction, each $p$-cell satisfies the following conditions: 
\begin{enumerate}
	\setlength{\itemsep}{-2pt}
	\item[(i)] The intersection of any two $p$-cells is an empty set.
	\item[(ii)] A symmetry $g \in G/\Pi$ keeps every point in $D_{i}^{p}$ invariant or transforms a point $\bk_{i}^{p}$ in $D_{i}^{p}$ into a point in another $p$-cell $g\left(D_{i}^{p}\right)$.
	More formally, for $\bk_{i}^{p} \in D_{i}^{p}$, either $g$ fixes $\bk_i^p$ in the Brillouin zone, i.e., there exists a reciprocal lattice vector $\bm G$ such that $g\bk_i^p=\bk_i^p+\bm G$, or $g\bk_i^p$ belongs to another cell $g(D_i^p)\neq D_i^p$.
	\item[(iii)] There are no isolated $p$-cells ($p \leq 2$). Every $(p-1)$-cell is in a boundary of a $p$-cell.
	\item[(iv)] The orientation of $p$-cells respects all symmetries in $G$. 
\end{enumerate}
In Supplementary Materials, we show our cell decomposition of a fundamental domain in each magnetic space group with and without particle-hole symmetry.

Using the set of $p$-cells $C_p$, we define the $p$-skeleton by
\begin{align}
	X_0 = C_0,\ X_p = X_{p-1} \bigcup C_p\ (p \geq 1). 
\end{align}
It should be noted that, in our construction, we distinguish among $p$-cells even when they are symmetry-related.

\begin{figure}[t]
	\begin{center}
		\includegraphics[width=0.99\columnwidth]{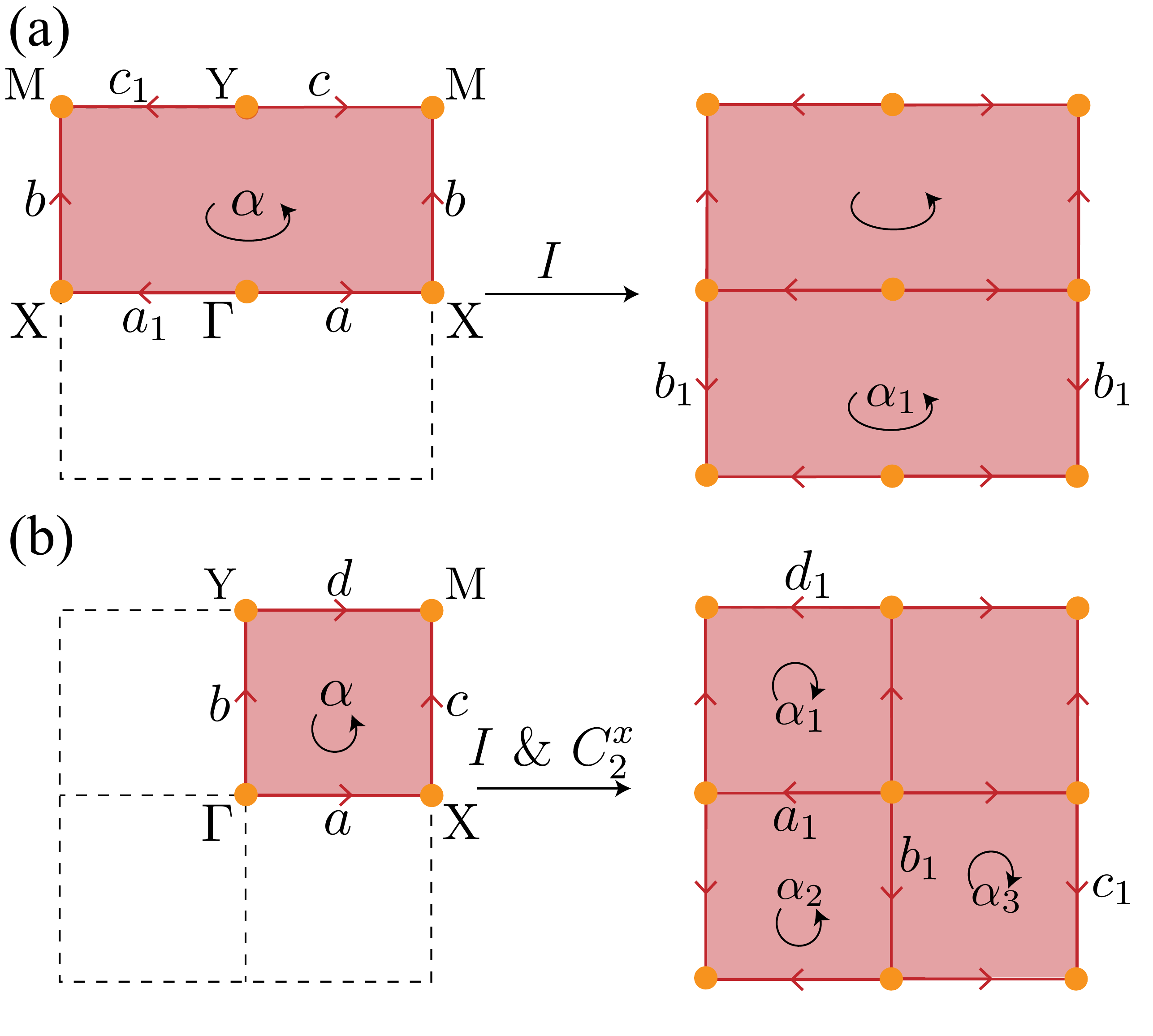}
		\caption{\label{fig:simple_cell}%
			Cell decomposition for layer groups $p\bar{1}$ (a) and $p2/m$ (b).
			First, we find a fundamental domain of BZ shown in the left panel and decompose it into points ($0$-cells), line segments ($1$-cells), and polygons ($2$-cells), which are colored in orange, red, and pink, respectively.  
			The red and black arrows represent orientations of 1-cells and 2-cells. 
			Next, we obtain a cell decomposition shown in the right panel by acting symmetry operations on the fundamental domain.}
	\end{center}
\end{figure}

Let us illustrate how to obtain the cell decomposition for layer group $p\bar{1}$ whose generators are lattice translations and spatial inversion. 
As discussed above, the first step is to find a fundamental domain of BZ. 
Here, we consider $-\pi \leq k_x \leq \pi$ and $0 \leq k_y \leq \pi$ as our fundamental domain.
Then, we decompose the fundamental domain into one $2$-cell (pink plane), five $1$-cells (solid red lines), and four $0$-cells (orange circles) in the left panel of Fig.~\ref{fig:simple_cell} (a). 
Also, $\{I_{\text{orb}}^{p}\}_{p=0}^{2}$ are given by 
\begin{align}
	&I_{\text{orb}}^{0} = \{\Gamma, \text{X}, \text{Y}, \text{M}\};\ I_{\text{orb}}^{1} = \{a, b, c\};\ I_{\text{orb}}^{2} = \{\alpha\}.
\end{align}
By acting inversion symmetry on this fundamental domain, we obtain $C_p$ as
\begin{align}
	\label{eq:C0_p-1}
	C_0 &= \{\Gamma, \text{X}, \text{Y}, \text{M}\}; \\
	\label{eq:C1_p-1}
	C_1 &= \{a, b, c, a_1, b_1, c_1\}; \\
	\label{eq:C2_p-1}
	C_2 &= \{\alpha, \alpha_1\}.
\end{align}
As mentioned above, here we assign symmetry-equivalent $p$-cells to different labels.
For example, $a = \{(k_x, 0)\ \vert\ k_x \in (0, \pi)\}$ is related to $a_1 = \{(k_x, 0)\ \vert\ k_x \in (-\pi, 0)\}$ by inversion. 

Similarly, we can construct a cell decomposition for layer group $p2/m11$ generated by translations, inversion, and twofold rotation along $x$-axis. 
As a result, we have
\begin{align}
	\label{eq:C0_p2m}
	C_0 &= \{\Gamma, \text{X}, \text{Y}, \text{M}\}; \\
	\label{eq:C1_p2m}
	C_1 &= \{a, b, c, d, a_1, b_1, c_1, d_1\}; \\
	\label{eq:C2_p2m}
	C_2 &= \{\alpha, \alpha_1, \alpha_2, \alpha_3\}.
\end{align}

\subsubsection{$E_1$-pages}
\label{sec:E1}
As a preliminary step, we find a finite subgroup of $G$ whose elements do not change $\bk$.
To do so, we first define a subgroup of $G$ by $G_{\bk} = \{g \in G\ \vert\ \phi_g p_g \bk = \bk + \bm{G} \text{ for a reciprocal lattice vector }\bm{G}\}$, which is referred to as \textit{little group}. 
It should be noted that the number of elements in $G_{\bk}$ is infinite since the translation group $\Pi$ is a subgroup of $G_{\bk}$ by definition.
Furthermore, $\Pi$ is a normal subgroup of $G_{\bk}$, i.e., $tht^{-1} \in G_{\bk}$ for $^\forall h \in G_{\bk}$ and $^\forall t \in \Pi$.
As a result, we can always obtain the desired finite group by $G_{\bk}/\Pi$.
Let us comment on projective factors of $G_{\bk}/\Pi$. 
Suppose that we have symmetry representations $U_{\bk}(g)\ (g \in G)$ such that 
\begin{align}
	U_{g'\bk}(g)U_{\bk}^{\phi_g}(g') = z^{\text{int}}(g,g')U_{\bk}(gg').
\end{align}
We can always find representations of $G_{\bk}/\Pi$ from those of $G_{\bk}$ as
\begin{align}
	\mathcal{U}_{\bk}(g) &= U_{\bk}(g)e^{i \bk \cdot \bm{a}_g},
\end{align}
where $\bm{a}_g$ is a fractional translation or zeros~\cite{PhysRevLett.117.096404}.
Correspondingly, the projective factors of $G_{\bk}/\Pi$ are given by
\begin{align}
	\mathcal{U}_{\bk}(h)\mathcal{U}^{\phi_h}_{\bk}(h') = z_{\bk}(h,h')\mathcal{U}_{\bk}(hh')\ \ (h,h' \in G_{\bk}/\Pi),
\end{align}
where $z_{\bk} (h,h')= z^{\text{int}}(h,h')e^{-i \bk\cdot(p_h \bm{a}_{h'}- \phi_h \bm{a}_{h'})}$.

Let $\phi|_{\bk},c|_{\bk}$, and $z|_{\bk}$ be symmetry data $\phi, c$, and $z$ restricted to elements in $G_{\bk}/\Pi$, respectively.
Also, since $G_{\bk}$ is common for every point $\bk \in D^p$, $G_{D^p}$ denotes the common little group. 
As discussed in Refs.~\cite{K-AHSS} and \cite{Shiozaki-Ono2023}, for a given filtration $\{X_p\}_{p=0}^{3}$, a finitely generated Abelian group $E_{1}^{p,-n}$ is defined by the relative $K$-group
\begin{align}
	E_1^{p,-n} &:= {}^\phi K^{(z,c)+p-n}_{G/\Pi}(X_p,X_{p-1}) \\
	&\cong \bigoplus_{i \in I_{\text{orb}}^{p}} {}^{\phi|_{D^p_{i}}} K^{(z|_{D^p_{i}},c|_{D^p_i})+p-n}_{G_{D^p_{i}}/\Pi}(D^p_{i}, \partial D^p_{i}) \label{eq:E1_meaning1}\\
	&\cong \bigoplus_{i\in I_{\text{orb}}^{p}} {}^{\phi|_{D^p_{i}}} \tilde K^{(z|_{D^p_{i}},c|_{D^p_i})+p-n}_{G_{D^p_{i}}/\Pi}(D^p_{i}/\partial D^p_{i})  \label{eq:E1_meaning2} \\
	&
    \cong \bigoplus_{i \in I_{\text{orb}}^{p}}  {}^{\phi|_{\bk^p_{i}}}
	  K^{(z|_{\bk^p_{i}},c|_{\bk^p_{i}})-n}_{G_{\bk^p_{i}}/\Pi}(\{\bk^p_{i}\})
    \label{eq:E1_meaning3}
    \\
	&\cong \bigoplus_{i \in I_{\text{orb}}^{p}}  {}^{\phi|_{\bk^p_{i}}}
	  \tilde{K}^{(z|_{\bk^p_{i}},c|_{\bk^p_{i}})}_{G_{\bk^p_{i}}/\Pi}(\{\bk^p_{i}\} \times \tilde{S}^n)
    \label{eq:E1_meaning4},
\end{align}
where $\bk_{i}^{p}$ is a representative point of $D_{i}^{p}$ and the definitions of the relative $K$-groups are briefly discussed in Appendix~\ref{app:K-theory}.
Here, instead of defining the $K$-groups formally, we mention physical meaning of $E_{1}^{p, -n}$, which can be interpreted in various ways since $E_{1}^{p, -n}$ is defined in terms of $K$-groups.
\begin{enumerate}
	\item[(i)] According to Eq.~\eqref{eq:E1_meaning1}, $E_{1}^{p, -n}$ represents $p$-dimensional gapped topological phases on $p$-cells but trivial on $(p-1)$-cells when $(p-n) = 0$.%
	\item[(ii)] According to Eq.~\eqref{eq:E1_meaning2}, $E_{1}^{p, -n}$ corresponds to gapped topological phases on the $p$-dimensional sphere when $(p-n) = 0$ and gapless points on the $p$-dimensional sphere when $(p-n) = 1$. %
    \item[(iii)] According to Eq.~\eqref{eq:E1_meaning3}, the same group is identified with the degree-$(-n)$ $K$-group at a representative point $\bk_i^p$ of the $p$-cell.
    This is the form used in the actual computation of the $E_1$ page: the problem reduces to irreducible representations of the little group $G_{\bk_i^p}/\Pi$ with the $n$ additional Clifford generators encoded in the grading, or equivalently to their emergent Altland-Zirnbauer symmetry classes (see below).
    \item[(iv)] According to Eq.~\eqref{eq:E1_meaning4}, the degree-$(-n)$ $K$-group at $\bk_i^p$ in Eq.~\eqref{eq:E1_meaning3} can equivalently be written as a degree-zero problem on the auxiliary sphere $\tilde S^n$. 
In this description, the $n$ Clifford generators introduced in the grading are represented by the coordinates $\tilde k_\mu$ on $\tilde S^n$ and the gamma matrices $\gamma_\mu$. 
Therefore, $E_{1}^{p,-n}$ can be understood as the direct sum of classifications of mass terms in $n$-dimensional massive Dirac Hamiltonians
\begin{align}
    \label{eq:Dirac_Sn}
    &H_{\bk^p_{i}} = \sum_{\mu = 1}^{n} \tilde{k}_{\mu}\gamma_\mu + m \gamma_0\ \ (\tilde{k}_\mu \in \tilde{S}^n),\\
    &\gamma_\mu \gamma_\nu + \gamma_\nu \gamma_\mu = 2\delta_{\mu \nu},
\end{align}
with symmetry group $G_{\bk_i^p}/\Pi$. 
Here, $\tilde S^n$ is an auxiliary sphere associated with the grading, not an additional momentum-space cell. 
Symmetries in $G_{\bk_i^p}/\Pi$ act trivially on $\tilde S^n$, and therefore serve as internal symmetries of the Dirac Hamiltonian. 
As a result, the classification is given by $\oplus_{\alpha} \pi_{n}(\mathcal{C}_{s(\alpha)}) \simeq \oplus_{\alpha} \pi_{0}(\mathcal{C}_{s(\alpha)+n})$ or $\oplus_{\alpha} \pi_{n}(\mathcal{R}_{s(\alpha)}) \simeq \oplus_{\alpha} \pi_{0}(\mathcal{R}_{s(\alpha)+n})$~\cite{Teo-Kane2010}, where $\mathcal{C}_{s(\alpha)}$ and $\mathcal{R}_{s(\alpha)}$ are classifying spaces of the Dirac Hamiltonians (see also Table~\ref{tab:EAZ}).

\end{enumerate}

To construct Abelian group $E_{1}^{p, -n}$, it is convenient to use the third or fourth interpretation.
Then, the remaining task is to obtain ${}^{\phi|_{\bk^p_{i}}}
K^{(z|_{\bk^p_{i}},c|_{\bk^p_{i}})}_{G_{\bk^p_{i}}/\Pi}(\{\bk^p_{i}\} \times \tilde{S}^n) $, i.e., classifications of the massive Dirac Hamiltonian in the presence of onsite symmetries.
In fact, this task is easily achieved by identifying an effective internal symmetry class of each irrep of the unitary part of $G_{\bk_{i}^{p}}/\Pi$~\cite{Teo-Kane2010,Cornfeld-Chapman,Shiozaki-CC}.
Such symmetry classes are known as \textit{emergent Altland-Zirnbauer symmetry class} (EAZ class).
For every $\bk$, we can always decompose $G_{\bk}/\Pi$ into the following four parts:
\begin{align}
	G_{\bk}/\Pi = \calG_{\bk} + \calA_{\bk} + \calP_{\bk} + \calJ_{\bk}, 
\end{align}
where
\begin{align}
	\calG_{\bk} &= \{g \in G_{\bk}/\Pi \ \vert\ (\phi_g, c_g) = (+1,+1)\};\\
	\calA_{\bk} &= \{g \in G_{\bk}/\Pi \ \vert\ (\phi_g, c_g) = (-1,+1)\}; \\
	\calP_{\bk} &= \{g \in G_{\bk}/\Pi \ \vert\ (\phi_g, c_g) = (-1,-1)\}; \\
	\calJ_{\bk} &= \{g \in G_{\bk}/\Pi \ \vert\ (\phi_g, c_g) = (+1,-1)\}.
\end{align}
Once we identify $\calG_{\bk}$, we can compute irreps of $\calG_{\bk}$, whose labels are denoted by $\alpha$ in the following.
For each irrep of $\calG_{\bk_{i}^{p}}$, we can compute $W^{\alpha}_{\bk_{i}^{p}}(\mathcal{P}) \in \{0, \pm1\}, W^{\alpha}_{\bk_{i}^{p}}(\calA) \in \{0, \pm 1\}$, and $W^{\alpha}_{\bk_{i}^{p}}(\mathcal{J})\in \{0, 1\}$ defined by
\begin{align}
	\label{eq:wigner_C}
	W^{\alpha}_{\bk_{i}^{p}}(\mathcal{P})\hspace{-0.5mm}&=\hspace{-0.5mm}\begin{cases}
		0 \quad  (\calP_{\bk} = \emptyset)\\
		\hspace{-0.5mm}\frac{1}{\vert \mathcal{P}_{\bk_{i}^{p}} \vert}{\displaystyle\sum_{c \in \calP_{\bk_{i}^{p}} }}\hspace{-1mm}z_{\bk_{i}^{p}}(c, c)\chi_{\bk_{i}^{p}}^{\alpha}(c^2)\in \hspace{-0.5mm}\{0, \pm 1\},
	\end{cases}\hspace{-0.5mm}\\
	W^{\alpha}_{\bk_{i}^{p}}(\calA) \hspace{-0.5mm}&=\hspace{-0.5mm}\begin{cases}
		0 \quad  (\calA_{\bk} = \emptyset)\\
		\hspace{-1mm}\frac{1}{\vert \calA_{\bk_{i}^{p}} \vert}\hspace{-0.8mm}{\displaystyle\sum_{a \in \calA_{\bk_{i}^{p}}}}\hspace{-1.33mm}z_{\bk_{i}^{p}}(a, a)\chi_{\bk_{i}^{p}}^{\alpha}(a^2)\hspace{-0.3mm} \in \hspace{-0.5mm}\{0, \pm 1\},
	\end{cases}\hspace{-0.5mm}\\
	\label{eq:wigner_G}
	W^{\alpha}_{\bk_{i}^{p}}(\mathcal{J})\hspace{-0.5mm}&=\hspace{-0.5mm} \begin{cases}
		0 \quad  (\calJ_{\bk} = \emptyset)\\
		\hspace{-0.5mm}\frac{1}{\vert \calG_{\bk_{i}^{p}} \vert}{\displaystyle \sum_{g \in \calG_{\bk_{i}^{p}} }} \frac{z_{\bk_{i}^{p}}(\gamma, \gamma^{-1}g\gamma)}{z_{\bk_{i}^{p}}(g, \gamma)} [\chi^{\alpha}_{\bk_{i}^{p}}(\gamma^{-1} g \gamma)]^{*}\chi^{\alpha}_{\bk_{i}^{p}}(g),
	\end{cases}
\end{align}
where $\chi^{\alpha}_{\bk}(g)$ is the character of an irrep of $\calGk$ for $g\in \calGk$ and $\gamma$ is a representative element of $\calJ_{\bk}$. 
These quantities inform us about how symmetries in $\calA_{\bk}, \calP_{\bk},$ and $\calJ_{\bk}$ affect irreps.
When $W_{\bk_{i}^{p}}^{\alpha}(\mathcal{V}) = 0\ (\mathcal{V}=\calP, \calA, \calJ)$ and $\mathcal{V}_{\bk_{i}^{p}}\neq\emptyset$, symmetries in $\mathcal{V}_{\bk_{i}^{p}}$ transform irrep $\alpha$ into another one.
On the other hand, when $W_{\bk_{i}^{p}}^{\alpha}(\mathcal{V}) = \pm 1$, irrep $\alpha$ is invariant under the symmetries. 
As a result, the EAZ symmetry class for $\alpha$ is identified by
\begin{align}
	\label{eq:wigner}
	W_{\bk_{i}^{p}}[\alpha] := \left(W^{\alpha}_{\bk_{i}^{p}}(\calA) , W^{\alpha}_{\bk_{i}^{p}}(\mathcal{P}), W^{\alpha}_{\bk_{i}^{p}}(\mathcal{J})\right),
\end{align}
which is called {\it Wigner criteria}~\cite{Bradley, K-AHSS}.
The classification of the $n$-dimensional massive Dirac Hamiltonians is shown in Table~\ref{tab:EAZ}.
\begin{table}[t]
	\begin{center}
		\caption{\label{tab:EAZ}The classification of mass terms in the $n$-dimensional massive Dirac Hamiltonians for each EAZ symmetry class.
		}
		\scalebox{0.95}{
		\begin{tabular}{c|c|c|c|c|c|c|c}
			\hline
			EAZ & $s(\alpha)$ & $\mathcal{C}_s$ & $W_{\bk}[\alpha]$  & $n = 0$ & $n = 1$  & $n = 2$ & $n = 3$  \\
			\hline\hline
			A & $0$ & $\frac{\text{U}(N)}{\text{U}(N-n)\times \text{U}(n)}$ & $(0,0,0)$ & $\mZ$ & $0$ & $\mZ$ & $0$ \\
			AIII & $1$& $\text{U}(N)$ & $(0,0,1)$ & $0$ & $\mZ$ & $0$ & $\mZ$ \\
			\hline\hline
			EAZ & $s(\alpha)$ & $\mathcal{R}_s$ & $W_{\bk}[\alpha]$ & $n = 0$ & $n = 1$  & $n = 2$ & $n = 3$  \\
			\hline\hline
			AI & $0$ & $\frac{\text{O}(N)}{\text{O}(N-n)\times \text{O}(n)}$&  $(1,0,0)$ & $\mZ$ & $\mZ_2$ & $\mZ_2$ & $0$ \\
			BDI &  $1$& $\text{O}(N)$ & $(1,1,1)$ & $\mZ_2$ & $\mZ_2$ & $0$ & $\mZ$\\ 
			D &  $2$ & $\text{O}(2N)/\text{U}(N)$& $(0,1,0)$ & $\mZ_2$ & $0$  & $\mZ$ & $0$\\
			DIII &  $3$& $\text{U}(N)/\text{Sp}(N)$ & $(-1,1,1)$ & $0$ & $\mZ$ & $0$ &  $0$\\
			AII &  $4$& $\frac{\text{Sp}(N)}{\text{Sp}(N-n)\times \text{Sp}(n)}$ & $(-1,0,0)$ & $\mZ$ & $0$ & $0$ & $0$\\
			CII &  $5$& $\text{Sp}(N)$ & $(-1,-1,1)$ & $0$ & $0$ & $0$ & $\mZ$\\
			C &  $6$ & $\text{Sp}(2N)/\text{U}(N)$& $(0,-1,0)$ & $0$ & $0$ & $\mZ$ & $\mZ_2$\\
			CI &  $7$& $\text{U}(N)/\text{O}(N)$ & $(1,-1,1)$ & $0$ & $\mZ$ & $\mZ_2$ & $\mZ_2$\\
			\hline
		\end{tabular}
	}
	\end{center}
\end{table}
As a result, $E_{1}^{p,-n}$ is written by
\begin{align}
	\label{eq:E1-topo}
	E^{p, -n}_{1} =\bigoplus_{i}\left(\bigoplus_{\alpha}\mZ_{2}[\bm{b}^{(p)}_{\bk^{p}_{i},\alpha}] \oplus \bigoplus_{\beta} \mZ[\bm{b}^{(p)}_{\bk^{p}_{i},\beta}]  \right),
\end{align}
where $\bm{b}^{(p)}_{\bk^{p}_{i},\alpha}$ denotes a basis of $E^{p, -n}_{1}$ that is defined for irrep $\alpha$. 
%
Also, $\mZ_2[\bm{b}^{(p)}_{\bk^{p}_{i},\alpha}]$ or $\mZ[\bm{b}^{(p)}_{\bk^{p}_{i},\alpha}]$ represents an Abelian group generated by $\bm{b}^{(p)}_{\bk^{p}_{i},\alpha}$.

\subsubsection{first differential $d_{1}^{p,-n}$ and $E_2$-pages}
It should be emphasized that, although an element of $E_{1}^{p,-n}$ corresponds to a massive Dirac Hamiltonian defined on $p$-cells, it is not necessary to be gapped on $(p+1)$-cells. 
In other words, elements of $E_{1}^{p,-n}$ are sometimes incompatible with those of $E_{1}^{p+1,-n}$. 
We implement the relation between $E_{1}^{p,-n}$ and $E_{1}^{p+1,-n}$ in a homomorphism
\begin{align}
	d_{1}^{p,-n}: E_{1}^{p,-n} \rightarrow E_{1}^{p+1,-n},
\end{align}
which is called \textit{first differential}~\cite{K-AHSS}. 
For our purpose, it is sufficient to consider the cases where $n = (p-1), p$, and $(p+1)$. 
Here the detailed calculation to construct such a homomorphism is discussed in our companion work~\cite{Shiozaki-Ono2023}.

Based on the second interpretation of $E_{1}^{p,-n}$, we present another physical meaning of $d_{1}^{p,-p}$.
Since $E_{1}^{p,-p}$ and $E_{1}^{p+1,-p}$ correspond to gapped and gapless phases, $d_{1}^{p,-p}$ can be understood as the process of generating gapless points on $(p+1)$-cells from band inversions on $p$-cells.
Then, the $\ker d_{1}^{p,-p} \subseteq E_{1}^{p,-p}$ represents the gapped Hamiltonians on $p$- and $(p+1)$-cells.
Also, $d_{1}^{p-1,-p}$ can be interpreted as the generation of trivial gapped Hamiltonians on $p$- and $(p+1)$-cells.
As a result, we define $E_2$-pages by
\begin{align}
	E_{2}^{0,0} &:= \ker d_{1}^{0,0};\\
	E_{2}^{p,-p} &:= \ker d_{1}^{p,-p}/ \im d_{1}^{p-1,-p}\ (\text{for }1\leq p \leq d-1);\\
	E_{2}^{d,-d} &:= E_{1}^{d,-d}/ \im d_{1}^{d-1,-d},
\end{align}
which represents topologically nontrivial phases gapped on $(p-1)$-, $p$-, and $(p+1)$-cells.

Furthermore, similar to $E_1$-pages, some elements of $E_{2}^{p,-p}$ might be incompatible with gapped phases on $(p+r)$-cells and contain trivial phases generated from $(p-r)$-cells.
Then, to obtain completely gapped phases, we must consider higher differential and $E_r$-pages for $r \geq 2$. 
However, the higher differentials are out of the scope of this work. 
Importantly, to the best of our knowledge, we do not have a systematic way to construct $d_{r}^{p,-n}$. 
Instead, we focus on $E_{2}$-pages and construct topological invariants for $E_{2}^{1,-1}$ in Sec.~\ref{sec:topo_invariant}, although phases corresponding to $E_{2}^{1,-1}$ are sometimes gapless on $3$-cells. 
It should be emphasized that this is still useful because it is generally difficult to detect gapless points at generic momenta.

\subsection{Step (i): identification of irreducible representations on $p$-cells for topological invariants}
\label{sec:general_frame}
As mentioned in Sec.~\ref{sec1}, we need to identify irreps on $p$-cells responsible for topological nontriviality, which is Step (i) mentioned in Introduction. 
In fact, $E^{p,-p}_{2}$ contains the information we need. 
In this subsection, we explain how to systematically extract the information from $E^{p,-p}_{2}$.
In the following, we assume that $\{E^{q, -p}_{1}\}_{q=p-1}^{p+1}$ are free Abelian groups for simplicity.
This is always true for time-reversal symmetric superconductors with conventional pairing symmetries.
It is straightforward to generalize the scheme presented here (the derivation of the matrix $X^{(p)}$ below) to the case where $\{E^{q, -p}_{1}\}_{q=p-1}^{p+1}$ contain torsion elements, as discussed in Appendix~\ref{app:Z2case}.
Let $\{\bm{b}_{i}^{(p)}\}_{i=1}^{N_p}$ be a serially numbered basis set of $E^{q,-p}_{1} = \mZ^{N_q}$, i.e., there exists $i$ such that $\bm{b}_{i}^{(q)} = \bm{b}_{\bk_{j}^{q}, \alpha}^{(q)}$.
Then, $E^{q,-p}_{1}$ is expressed by
\begin{align}
	&E^{q,-p}_{1} = \bigoplus_{i=1}^{N_q}\mZ[\bm{b}_{i}^{(q)}],
\end{align}
where $\mZ[\bm{b}_{i}^{(q)}]$ represents a free Abelian group generated by $\bm{b}_{i}^{(q)}$. 
When we use $\mathcal{B}^{(q)} = \left(\bm{b}_{1}^{(q)} \bm{b}_{2}^{(q)} \cdots \bm{b}_{N_q}^{(q)} \right)$, the first differential $d_{1}^{p,-p}: E_{1}^{p,-p} \rightarrow E_{1}^{p+1,-p}$ is represented by 
\begin{align}
	d_{1}^{p,-p}\left(\mathcal{B}^{(p)}\right) &:= \left(d_{1}^{p,-p}(\bm{b}_{1}^{(p)}) d_{1}^{p,-p}(\bm{b}_{2}^{(p)}) \cdots d_{1}^{p,-p}(\bm{b}_{N_p}^{(p)})\right) \nonumber\\
	&= \mathcal{B}^{(p+1)}M_{d_{1}^{p,-p}},
 \label{eq:def_matrix_d1p-p}
\end{align} 
where $M_{d_{1}^{p,-p}}$ is a ($N_{p+1} \times N_p$)-dimensional integer-valued matrix. 
For an integer-valued matrix, we can always find two unimodular matrices $U^{(p)}$ and $V^{(p)}$ such that 
\begin{align}
	\label{eq:Smith}
	 \Sigma^{(p)}=U^{(p)}M_{d_{1}^{q,-p}}V^{(p)},
\end{align}
where $\Sigma^{(p)}$ is an integer-valued diagonal matrix. 
The left-hand side is known as \textit{Smith normal form}.
When let $r_p$ be the matrix rank of $\Sigma^{(p)}$, each diagonal element of $\Sigma^{(p)}$ satisfies the following things:
\begin{align}
  \label{eq:Smith2}
  \begin{minipage}[l]{\linewidth}
    \begin{itemize}
      \setlength{\itemindent}{-0.5em}
      \item $[\Sigma^{(p)}]_{ii} \in \mZ_{>0}$ for $1\leq i \leq r_p$;
	 \item $[\Sigma^{(p)}]_{ii}$ can divide $[\Sigma^{(p)}]_{(i+1)(i+1)}$ for $1 \leq i \leq  r_p-1$;
	 \item $[\Sigma^{(p)}]_{ii} = 0$ for $r_p<i$ if $r_p \neq N_P$.
    \end{itemize}
  \end{minipage}
\end{align}

Then, we have
\begin{align}
	\label{eq:d_map}
	d_{1}^{p,-p}\left(\mathcal{B}'^{(p)}\right) &=d_{1}^{p,-p}\left(\mathcal{B}^{(p)} V^{(p)}\right) \nonumber \\
	&= \mathcal{B}^{(p+1)}[U^{(p)}]^{-1}U^{(p)}M_{d_{1}^{p,-p}}V^{(p)} \nonumber \\
	&= \mathcal{B}^{(p+1)}[U^{(p)}]^{-1}\Sigma^{(p)}, 
\end{align}
where we introduce $\mathcal{B}'^{(p)}=\left(\bm{b}'^{(p)}_{1} \bm{b}'^{(p)}_{2} \cdots \bm{b}'^{(p)}_{N_p} \right)=\mathcal{B}^{(p)}V^{(p)}$.
Equation~\eqref{eq:d_map} implies that we can obtain $\ker(d_{1}^{p,-p}) = \mathrm{span}\left(\bm{b}'^{(p)}_{r_p+1} \cdots \bm{b}'^{(p)}_{N_p} \right)$. 

Similarly,
\begin{align}
	\label{eq:d_map2}
	d_{1}^{p-1,-p}\left(\mathcal{B}^{(p-1)}\right)&=
	\mathcal{B}^{(p)}M_{d_{1}^{p-1,-p}}\nonumber \\
	&= \mathcal{B}'^{(p)}[V^{(p)}]^{-1}M_{d_{1}^{p-1,-p}}\nonumber\\
	&=\mathcal{B}'^{(p)}\left(\begin{array}{c}
		O_{r_p\times N_{p-1}} \\
		\hline
		\multicolumn{1}{c}{Y}
	\end{array}\right).
\end{align}
The reason why the first $r_p$ rows are zeros is that $\im d^{p-1,-p}_{1}$ should be expanded by $ \ker d^{p,-p}_{1}$.
Applying the same decomposition of Eq.~\eqref{eq:Smith} to $Y$, we rewrite the above equation as
\begin{align}
	\label{eq:d_map3}
	&d_{1}^{p-1,-p}\left(\mathcal{B}^{(p-1)}V^{(p-1)}\right)\nonumber\\
	&=\mathcal{B}'^{(p)}\left(\mathds{1}_{r_{p}}\oplus[U^{(p-1)}]^{-1}\right)
	\left(\begin{array}{c}
		O_{r_p\times N_{p-1}} \\
		\hline
		\Lambda^{(p-1)}
	\end{array}\right)\nonumber\\
	&=\mathcal{B}''^{(p)}\left(\begin{array}{c}
		O_{r_p\times N_{p-1}} \\
		\hline
		\Lambda^{(p-1)}
	\end{array}\right),
\end{align}
where $\Lambda^{(p-1)} = \text{diag}(\lambda_{1}^{(p-1)}, \lambda_{2}^{(p-1)}, \cdots, \lambda_{N_p - r_p}^{(p-1)})$ is the Smith normal form of $Y$ with the same properties in \eqref{eq:Smith2}.
Here, we also introduce $\mathcal{B}''^{(p)} = \left(\bm{b}''^{(p)}_{1} \bm{b}''^{(p)}_{2} \cdots \bm{b}''^{(p)}_{N_p} \right)=\mathcal{B}^{(p)}X^{(p)}=\mathcal{B}^{(p)}V^{(p)}\left(\mathds{1}_{r_p}\oplus[U^{(p-1)}]^{-1}\right)$.
As a result, we find 
\begin{enumerate}
	\item[(1)] $\bm{b}''^{(p)}_{i}\ (i=1,\cdots r_p)$ is not an element in $\ker d^{p,-p}_{1}$;
	
	\item[(2)] $\bm{b}''^{(p)}_{i}$ is a basis of $E_{2}^{p,-p}$ when $i > r_p$ and $\lambda_{i-r_p}^{(p-1)}\neq 1$. 
\end{enumerate}

Equations~\eqref{eq:d_map} and \eqref{eq:d_map3} inform us of topological invariants defined on $p$-cells.
Since any element of $E^{p, -p}_{1}$ can be expressed by $\mathcal{B}^{(p)}(n^{(p)}_{1}\ n^{(p)}_{2} \cdots\ n^{(p)}_{N_p})^{\top}$, 
\begin{align}
	\label{eq:X_inv}
	\bm{n}^{(p)} &= \mathcal{B}^{(p)}(n^{(p)}_{1}\ n^{(p)}_{2} \cdots\ n^{(p)}_{N_p})^{\top} \nonumber \\
	&= \mathcal{B}''^{(p)}[X^{(p)}]^{-1}(n^{(p)}_{1}\ n^{(p)}_{2} \cdots\ n^{(p)}_{N_p})^{\top}. %
\end{align}
Combining the facts (1) and (2) with Eq.~\eqref{eq:X_inv}, the row lists $\{\bm{x}_{i}^{T}\}_i$ tell us which combinations of $\{n^{(p)}_{j}\}_j$ detect topologically nontrivial nature. 
This is indeed Step (i), which we need to achieve.
Thus the remaining problem is to identify quantities that function as $\{n^{(p)}_{j}\}_j$ properly, which is discussed in the next subsection.

\subsection{Step (ii): Construction of topological invariants}
\label{sec:topo_invariant}
In preceding subsection, we find that each row list in $[X^{(p)}]^{-1}$ tells us which irreps on $p$-cells are used in topological invariants to detect nontrivial elements of $E_{2}^{p,-p}$. 
In this subsection, we discuss which quantities are assigned to them and how to take combinations of these quantities.
In the following, by focusing on the case of $p = 1$, we present topological invariants defined on the $1$-skeleton (a domain gluing all $1$-cells together with all $0$-cells).
For later convenience, $[X^{(1)}]^{-1}$ and $[V^{(0)}]^{-1}$ are denoted by
\begin{align}
	\label{eq:inv_X1}
	&[X^{(1)}]^{-1} := \left(\begin{array}{cccc}
		\bm{x}_{1} &
		\bm{x}_{2} &
		\cdots &
		\bm{x}_{N_1}
	\end{array}\right)^{\top},\\
	\label{eq:inv_V0}
	&[V^{(0)}]^{-1} := \left(\begin{array}{cccc}
		\bm{v}_{1} &
		\bm{v}_{2} &
		\cdots &
		\bm{v}_{N_0}
	\end{array}\right)^{\top}.
\end{align}

As a preparation for the following discussions, here we define \textit{q-matrix} for each irrep when an irrep is invariant under chiral-like symmetries.
In this case, we consider irreps of $\calGk + \calJ_{\bk}$ characterized by chirality, which are denoted by $\alpha\pm$. 
Each irrep of $\calGk + \calJ_{\bk}$ satisfies
\begin{align}
	\chi_{\bk}^{\alpha\pm}(g) &= \chi_{\bk}^{\alpha}(g) \quad \left(g \in \calGk\right), \\
	\chi_{\bk}^{\alpha-}(g) &= -\chi_{\bk}^{\alpha+}(g) \ \left(g \in \calJ_{\bk} \right). 
\end{align}
When $\calA_{\bk} \neq \emptyset$ and $W^{\alpha}_{\bk}(\calA)  = 0$, two irreps, denoted by $\alpha\pm$ and $\beta\pm$ here, are related by symmetries in $\calA_{\bk}$. 
In such a case, we always choose $\beta-$ such that 
\begin{align}
	\chi^{\beta-}_{\bk}(g) = \frac{z_{g, a}^{\bk}}{z_{a, a^{-1}ga}^{\bk}}[\chi^{\alpha+}_{\bk}(g)]^{*} \text{ for }g \in \calGk + \calJ_{\bk}.
\end{align}
Then, we define a projection matrix by
\begin{align}
	\label{eq:projection}
	P^{\alpha\pm} &= \frac{\mathcal{D}_{\alpha}}{\vert \calGk +  \calJ_{\bk}\vert} \sum_{g \in  \calGk +  \calJ_{\bk}} \left[\chi_{\bk}^{\alpha\pm}(g)\right]^{*}\mathcal{U}_{\bk}(g),
\end{align}
where $\mathcal{D}_{\alpha}$ is the dimension of irrep $\alpha$. 
Using the projection matrices, we also define a chiral matrix by
\begin{align}
	\Gamma^{\alpha} = P^{\alpha+} - P^{\alpha-},
\end{align}
whose eigenvalues are $0$ and $\pm 1$. 
When $\mathcal{U}^{\alpha}_{\pm}$ denotes a matrix composing of eigenvectors with eigenvalues $\pm 1$, we have 
\begin{align}
	\label{eq:qmatrix}
	q^{\alpha}_{\bk} &= [\mathcal{U}^{\alpha}_{-}]^{\dagger} H_{\bk}\mathcal{U}^{\alpha}_{+} \in \text{GL}(\mathrm{rank}\ P^{\alpha+}),
\end{align}
which is what we refer to as \textit{q-matrix}. 
It should be noted that $q^{\alpha}_{\bk}$ depends on the choice of $\mathcal{U}^{\alpha}_{\pm}$ and can be changed by gauge transformations. 
In particular, we can always choose ${\cal U}^{\alpha}_\pm$ such that the eigenvalues of $q^{\alpha}_{\bk}$ are $\calD_{\alpha}$-fold degenerate for EAZ class AIII/CI and $(2\calD_{\alpha})$-fold degenerate for EAZ class DIII [See Appendix~\ref{app:basis_q} for how to find such a basis set]. 
Then, we define ``$(\det q^{\alpha}_{\bk})^{1/\calD_{\alpha}}$'' for AIII and CI and ``$(\det q^{\alpha}_{\bk})^{1/2\calD_{\alpha}}$'' for DIII by the product of the duplicated eigenvalues of $q^{\alpha}_{\bk}$.
As a result, we introduce 
\begin{align}
	\mathcal{Z}[q^{\alpha}_{\bk}] :=\prod_{j}\pi_{j},
 \label{eq:def_Z_q_alpha}
\end{align}
where $\pi_{j}$ is the $j$-th eigenvalue of $q^{\alpha}_{\bk}$, and
\begin{align}
	\det q^{\alpha}_{\bk} = \begin{cases}
		\left(\mathcal{Z}[q^{\alpha}_{\bk}]\right)^{\calD_\alpha} \text{ for EAZ class AIII/CI}\\
		\left(\mathcal{Z}[q^{\alpha}_{\bk}]\right)^{2\calD_\alpha} \text{ for EAZ class DIII}
	\end{cases}
\end{align} 
holds. 
For EAZ class DIII with $\calD_{\alpha}=1$, when $\calA_{\bk}$ contains an order-two antiunitary symmetry denoted by $a$, one can replace $\mathcal{Z}[q^{\alpha}_{\bk}]$ by $\pf[q^{\alpha}_{\bk}]$ by choosing $\mathcal{U}^{\alpha}_{-} = U_{\bk}(a)[\mathcal{U}^{\alpha}_{+}]^{*}$.

We make a brief remark on $\Gamma^{\alpha}$.
The chiral matrix $\Gamma_\alpha$ is generally not the same as the representation of chiral symmetry.
Even when we discuss symmetry groups containing chiral symmetry with a fractional translation, the eigenvalues of $\Gamma^{\alpha}$ are $0$ and $\pm 1$.
On the other hand, ${\cal U}^{\alpha}_\pm$ depend on momenta and are not periodic. 
Correspondingly, $\mathcal{U}^{\alpha}_{+}$ and $\mathcal{U}^{\alpha}_{-}$ are interchanged up to unitary matrices of gauge transformations.
As a result, $q^{\alpha}_{\bk+\bm{G}}$ is not the same as $q^{\alpha}_{\bk}$ but rather related to $\left(q^{\alpha}_{\bk}\right)^{\dagger}$ by the unitary matrices.
Nonetheless, our construction discussed below works for such cases.
See Appendix~\ref{app:Pc3} for an example. 

\subsubsection{Revisiting physical meaning of $E_{1}^{1,-1}$}
Since we restrict ourselves to free Abelian groups $E_{1}^{1,-1}$, EAZ classes are always one of AIII, CI, or DIII.
As defined in Eq.~\eqref{eq:E1_meaning1}, $E_{1}^{1,-1}$ corresponds to the direct sum of gapped topological phases on $1$-cells with trivial gapped states on $0$-cells. 
Such topological phases are characterized by the winding of $q$-matrix, where the winding is quantized.
The simplest example is a one-dimensional class AIII system only with translation symmetry.
In this case, $C_0=\{k = \pm \pi\}$ and $C_1 =\{k\ \vert\  k \in (-\pi, \pi)\}$. 
Since the relation $H_{k=-\pi}=H_{k=\pi}$ always holds, $E_1^{1,-1} = {}^\phi K^{(z,c)+0}_{G/\Pi}(X_1 = T^1,X_{0} = C_0) = \mZ$ corresponds just ordinary one-dimensional topological phases of class AIII, which are characterized by one-dimensional winding number $w:= \frac{1}{2\pi \mathrm{i}}\int_{-\pi}^{\pi}dk \partial_k \log \det q_k \in \mZ$.
In the following, we define various topological invariants based on the relation between the windings of $q$-matrices and topological nontriviality, although the windings are not quantized for general Hamiltonians.

\subsubsection{$\mZ$ topological invariants for gapless points on $2$-cells}
First, we consider topological invariants for gapless points on $2$-cells, which corresponds to $\im d_{1}^{1,-1}$.
To define topological invariants, for each EAZ class of each irrep on a $1$-cell, we define the following quantities:
\begin{itemize}
	\item class AIII and CI
	\begin{align}
		\label{eq:winding_AIII}
		w_{\alpha} := \frac{1}{2\pi \mathrm{i}}\times \frac{1}{\mathcal{D}_{\alpha}}\int_s d(\log \det q_{s}^{\alpha}-\log \det (q^{\alpha}_{s})^{\text{vac}});
	\end{align}
\end{itemize}

\begin{itemize}
	\item class DIII
	\begin{align}
		\label{eq:winding_DIII}
		w_{\alpha}:=\frac{1}{2\pi \mathrm{i}}\times\frac{1}{2\mathcal{D}_{\alpha}}\int_s d(\log \det q_{s}^{\alpha}-\log \det (q^{\alpha}_{s})^{\text{vac}}),
	\end{align}
\end{itemize}
where the integral is from one boundary $0$-cell of the $1$-cell to another boundary $0$-cell along the orientation of the $1$-cell.
Here, we introduce $(q^{\alpha}_{\bk})^{\text{vac}}$ as the $q$-matrix of the vacuum Hamiltonian $H^{\text{vac}}_{\bk}$ defined by the infinite chemical potential
limit of $H_{\bk}$ using the same $\mathcal{U}^{\alpha}_{\pm}$.
The reason why we need the winding of $(q^{\alpha}_{\bk})^{\text{vac}}$ is that the winding of $q^{\alpha}_{\bk}$ is not invariant under global gauge transformations~\cite{Thiang_AIII} and basis transformation of $\mathcal{U}^{\alpha}_{\pm}$.
When we subtract the winding of a reference $q$-matrix from that of $q^{\alpha}_{\bk}$, the gauge dependence is resolved. 
From the fact (I) and Eq.~\eqref{eq:X_inv},  
\begin{align}
	\label{eq:gapless_inv}
	\mathcal{W}_{i}^{\text{gapless}}[H_{\bk}] := \bm{x}_{i}^{\top}(w_{1}, w_{2}, \cdots, w_{N_1})^{\top}\ (1 \leq i \leq r_1)
\end{align}
is a $\mZ$-valued topological invariant to detect gapless points on $2$-cells.

\subsubsection{$\mZ$ topological invariants for gapped phases on $2$-skeleton}
Next, we construct $\mZ$-valued topological invariants for gapped phases on $1$-skeleton, i.e., fully gapped phases or gapless only on $3$-cells. 
Indeed, we can use the same quantities as the invariants for gapless points on $2$-cells. 
From the fact (II) and Eq.~\eqref{eq:X_inv}, we find an invariant defined by
\begin{align}
	\label{eq:gapped_inv}
	\mathcal{W}_{i}^{\text{gapped}}[H_{\bk}] &:= \bm{x}_{i}^{\top}(w_{1}, w_{2}, \cdots, w_{N_1})^{\top}\ \text{for $i$ s.t. $\lambda_{i-r_{1}}^{(0)} = 0$} \nonumber \\
	&\in \mZ.
\end{align}

\subsubsection{$\mZ_k$ topological invariants for gapped phases on $2$-skeleton}
\label{sec:torsion_invs}
For $i$ such that $\lambda_{i-r_{1}}^{(0)} \notin \{0, 1\}$, we construct a $\mZ_{\lambda_{i-r_{1}}^{(0)}}$-valued topological invariant from $\bm{x}_{i}^{\top}$ and $\bm{v}_{i-r_{1}}^{\top}$ in Eqs.~\eqref{eq:inv_X1} and \eqref{eq:inv_V0}.
Similar to the case of $\mZ$-valued invariants, it is natural to characterize topological nature on $1$-cells by $w_{\alpha}$ in Eqs.~\eqref{eq:winding_AIII} and \eqref{eq:winding_DIII}.
We define the following quantities for irreps on $1$-cells:
\begin{itemize}
	\item class AIII and CI
	\begin{align}
		\label{eq:exp_winding_AIII}
		\nu_{\alpha} &:= \exp\left[-\frac{2\pi\mathrm{i}}{\lambda_{i-r_{1}}^{(0)}}w_{\alpha}\right]\nonumber\\
		&=\exp \left[\frac{-1}{\lambda_{i-r_{1}}^{(0)}\mathcal{D}_{\alpha}}\int_s \hspace{-0.5mm}d(\log \det q_{s}^{\alpha}-\log \det (q^{\alpha}_{s})^{\text{vac}}) \right];
	\end{align}
	
	\item class DIII
	\begin{align}
		\label{eq:exp_winding_DIII}
		\nu_{\alpha} &:= \exp\left[-\frac{2\pi\mathrm{i}}{\lambda_{i-r_{1}}^{(0)}}w_{\alpha}\right]\nonumber\\
		&=\exp \left[\frac{-1}{2\lambda_{i-r_{1}}^{(0)}\mathcal{D}_{\alpha}}\int_s \hspace{-0.5mm}d(\log \det q_{s}^{\alpha}-\log \det (q^{\alpha}_{s})^{\text{vac}}) \right],
	\end{align}
\end{itemize}
where the integral is from one boundary $0$-cell of the $1$-cell to another boundary $0$-cell along the orientation of the $1$-cell.
Here, $\calD_{\alpha}$ is dimension of an irrep $\alpha$ on the $1$-cell.
The necessity of $\log \det (q^{\alpha}_{s})^{\text{vac}}$ arises from the same rationale as in Eqs.~\eqref{eq:winding_AIII} and \eqref{eq:winding_DIII}.
Since $\mZ$-valued topological invariants are defined only by $\{w_\alpha\}_\alpha$ as seen above, one might expect that $\prod_{\alpha}\nu_{\alpha}^{[\bm{x}_{i}^{T}]_{\alpha}} = \exp[-2\pi \mathrm{i} \sum_{\alpha} [\bm{x}_{i}]_{\alpha}w_{\alpha}/\lambda_{i-r_{1}}^{(0)}]$ to be a $\mZ_{\lambda_{i-r_{1}}^{(0)}}$-valued topological invariant.
However, this is untrue. 
This is because
\begin{align}
	\label{eq:triv_seed}
	\left(\prod_{\alpha}\nu_{\alpha}^{[\bm{x}_{i}]_{\alpha}}\right)^{\lambda_{i-r_{1}}^{(0)}} = \exp[-2\pi \mathrm{i} \sum_{\alpha} [\bm{x}_{i}]_{\alpha}w_{\alpha}]
\end{align}
is generally not unity. 
This implies that $\prod_{\alpha}\nu_{\alpha}^{[\bm{x}_{i}^{T}]_{\alpha}}$ does not work for a $\mZ_{\lambda_{i-r_{1}}^{(0)}}$-valued topological invariant.
After integrating and summing the exponent of right-hand side of Eq.~\eqref{eq:triv_seed}, we have the product of $\left\{\mathcal{Z}[q_{\bk}^{\beta}]/\mathcal{Z}[(q^{\beta}_{\bk})^{\text{vac}}]\right\}_{\beta}$ over 0-cells. 
Then, we construct a $\mZ_{\lambda_{i-r_{1}}^{(0)}}$-valued topological invariant by combining $\prod_{\alpha}\nu_{\alpha}^{[\bm{x}_{i}^{T}]_{\alpha}}$ with the correction terms $\left\{\mathcal{Z}[q_{\bk}^{\beta}]/\mathcal{Z}[(q^{\beta}_{\bk})^{\text{vac}}]\right\}_{\beta}$.

The remaining task is to identify such correction terms. 
In fact, this is accomplished by AHSS. 
To see this, we rewrite Eq.~\eqref{eq:d_map3} as 
\begin{align}
	\label{eq:d_map4}
		&\left(d_{1}^{0,-1}(\bm{b}'^{(0)}_{1})\ d_{1}^{0,-1}(\bm{b}'^{(0)}_{2}) \cdots d_{1}^{0,-1}(\bm{b}'^{(0)}_{N_0}) \right) \nonumber \\
		&=  \left(\bm{b}''^{(1)}_{1} \bm{b}''^{(1)}_{2} \cdots \bm{b}''^{(1)}_{N_1} \right)\left(\begin{array}{c}
			O_{r_1\times N_{0}} \\
			\hline
			\Lambda^{(0)}
		\end{array}\right),
\end{align}
where $\Lambda^{(0)} = \text{diag}(\lambda_{1}^{(0)}, \lambda_{2}^{(0)}, \cdots)$ is the Smith normal form in Eq.~\eqref{eq:d_map3} for $p=1$.
For later convenience, let us suppose that the first $R_0$ diagonal elements are unity, i.e., $\lambda_{j}^{(0)} = 1$ for $1\leq j \leq R_0$.
Note that $R_0$ depends on conventions, for example, choices of the cell decomposition.
When we consider Eq.~\eqref{eq:d_map4} in terms of coordinates $\{n_{j}^{(0)}\}_{j=1}^{N_0}$ and $\{n_{j}^{(1)}\}_{j=1}^{N_1}$ of $E_{1}^{0,-1}$ and $E_{1}^{1,-1}$, this equation implies
\begin{align}
	\label{eq:d_map5}
	\lambda_{i-r_{1}}^{(0)} \bm{v}_{i-r_1}^{\top}(n_{1}^{(0)}, n_{2}^{(0)}, \cdots n_{N_0}^{(0)})\hspace{-0.5mm}^{\top} \hspace{-0.5mm}= \bm{x}_{i}^{\top}(n_{1}^{(1)}, n_{2}^{(1)}, \cdots n_{N_1}^{(1)})\hspace{-0.5mm}^{\top}.
\end{align}
Equation \eqref{eq:d_map5} hints at a relationship between quantities defined on $0$- and $1$-cells, as discussed below.
Inspired by Eq.~\eqref{eq:d_map5}, we define the following quantity
\begin{align}
	\label{eq:Z_k}
	&\exp\hspace{-0.5mm}\left[\frac{2\pi \mathrm{i}}{\lambda_{i-r_1}^{(0)}}\mathcal{X}_{i-r_1}[H_{\bk}]\right] := \frac{\prod_{\alpha}\nu_{\alpha}^{[\bm{x}_{i}]_{\alpha}}}{\prod_{\beta}\left(\mathcal{Z}[q_{\bk}^{\beta}]/\mathcal{Z}[(q^{\beta}_{\bk})^{\text{vac}}]\right)^{[\bm{v}_{i-r_1}]_{\beta}}}.
\end{align}

The $\mZ_{\lambda^{(0)}_{i-r_1}}$-quantization of $\mathcal{X}_{i-r_1}[H_{\bk}]$ is also confirmed by the following discussion. 
Let $\alpha$ and $\beta$ be an irrep on a $1$-cell and an irrep at its adjacent $0$-cell, respectively.
Compatibility relations imply that the $\mathbb{C}$-valued quantity ${\cal Z}[q_\bk^\alpha]$ defined in Eq.~\eqref{eq:def_Z_q_alpha} obeys the relation 
\begin{align}
    {\cal Z}[q_\bk^\alpha] = \prod_\beta ({\cal Z}[q_\bk^\beta])^{\Big\vert \left[M_{d_1^{0,-1}}\right]_{\alpha\beta}\Big\vert}, 
    \label{eq:compatibility_calZ}
\end{align}
where the product $\prod_\beta$ runs over the irreps on $0$-cells whose EAZ class is either AIII, DIII or CI.
The exponent $\Big\vert \left[M_{d_1^{0,-1}}\right]_{\alpha\beta}\Big\vert$ is the absolute value of a matrix element of $M_{d_1^{0,-1}}$ defined in Eq.~\eqref{eq:d_map2}.
See Appendix~\ref{sec:Compatibility relation of non-Hermitian matrix} for a derivation.
Let $s$ denote the $1$-cell going from an adjacent $0$-cell $\bk_0$ to another adjacent $0$-cell $\bk_1$.
By using Eq.~\eqref{eq:compatibility_calZ}, 
\begin{align}
    \label{eq:nu_calZ_M}
    \nu_\alpha^{\lambda^{(0)}_{i-r_1}}
    &=\frac{{\cal Z}[q_{\bk_0}^\alpha]/{\cal Z}[(q_{\bk_0}^\alpha)^{\rm vac}]}{{\cal Z}[q_{\bk_1}^\alpha]/{\cal Z}[(q_{\bk_1}^\alpha)^{\rm vac}]} \nonumber \\
    &=\prod_{\beta} \left({\cal Z}[q_{\bk}^\beta]/{\cal Z}[(q_{\bk}^\beta)^{\rm vac}] \right)^{\left[M_{d_1^{0,-1}}\right]_{\alpha\beta}}.
\end{align}
Noticing that, from Eq.~\eqref{eq:d_map3}, 
\begin{align}
    [X^{(1)}]^{-1} M_{d_1^{0,-1}} = \left(\begin{array}{c}
			O_{r_1\times N_{0}} \\
			\hline
			\Lambda^{(0)} 
		\end{array}\right) [V^{(0)}]^{-1}, 
\end{align}
we have 
\begin{align}
    &\left(\prod_{\alpha}\nu_{\alpha}^{[\bm{x}_{i}]_{\alpha}}\right)^{\lambda_{i-r_{1}}^{(0)}} \nonumber\\
    &=\prod_\beta \left({\cal Z}[q_{\bk}^\beta]/{\cal Z}[(q_{\bk}^\beta)^{\rm vac}] \right)^{\sum_\alpha [\bm{x}_i]_\alpha \big[M_{d_1^{0,-1}}\big]_{\alpha\beta}} \nonumber\\
    &= \prod_\beta \left({\cal Z}[q_{\bk}^\beta]/{\cal Z}[(q_{\bk}^\beta)^{\rm vac}] \right)^{\lambda^{(0)}_{i-r_1} [\bm{v}_{i-r_1}]_{\beta}} \nonumber\\
    &=\left( \prod_\beta \left({\cal Z}[q_{\bk}^\beta]/{\cal Z}[(q_{\bk}^\beta)^{\rm vac}] \right)^{[\bm{v}_{i-r_1}]_{\beta}} \right)^{\lambda^{(0)}_{i-r_1}}.
\end{align}
This proves $\exp\left[2\pi i {\cal X}_{i-r_1}[H_\bk]\right]=1$.
We note that the above proof is only applicable to symmetry settings where the EAZ class at 0-cells does not include BDI.
When there exists an EAZ class BDI at some 0-cell, the relation (\ref{eq:nu_calZ_M}) has a correction from irreps at 0-cells whose EAZ class is BDI, resulting in the symmetry indicator to detect gapless point on 2-cells or nontrivial extension of $E_2^{1,-1}$ by $E_2^{0,0}$.
We also numerically verify that $\mathcal{X}_{i-r_1}[H_{\bk}]\in \{0,1,\dots,\lambda_{i-r_1}^{(0)}-1\}$ is actually $\mZ_{\lambda_{i-r_1}^{(0)}}$-valued for time-reversal symmetric spinful superconductors, as discussed in Sec.~\ref{sec:numerics}.

\subsubsection{Demonstration}
\label{sec:1D_DIII}
As a demonstration, here we discuss the topological invariant of one-dimensional time-reversal symmetric topological superconductors only with translation symmetries.
It is well known that the classification of topological superconducting phases is $^{\phi}K_{G/\Pi}^{(z,c)}(T^1) = \mZ_2$ in this symmetry setting. 
In addition, the following topological invariant is also known:
\begin{align}
	\label{eq:1D_DIII}
	(-1)^{\nu} = \frac{\mathrm{Pf}[q_{k=\pi}]}{\mathrm{Pf}[q_{k=0}]}\exp\left[-\frac{1}{2}\int_{0}^{\pi} dk \partial_k \log \det q_{k}\right],
\end{align}
where we choose the basis such that $q_{k=0,\pi}^{T} = - q_{k=0,\pi}$. 
Here, we rederive the above invariant by using our method. 

\begin{figure}[t]
	\begin{center}
		\includegraphics[width=0.99\columnwidth]{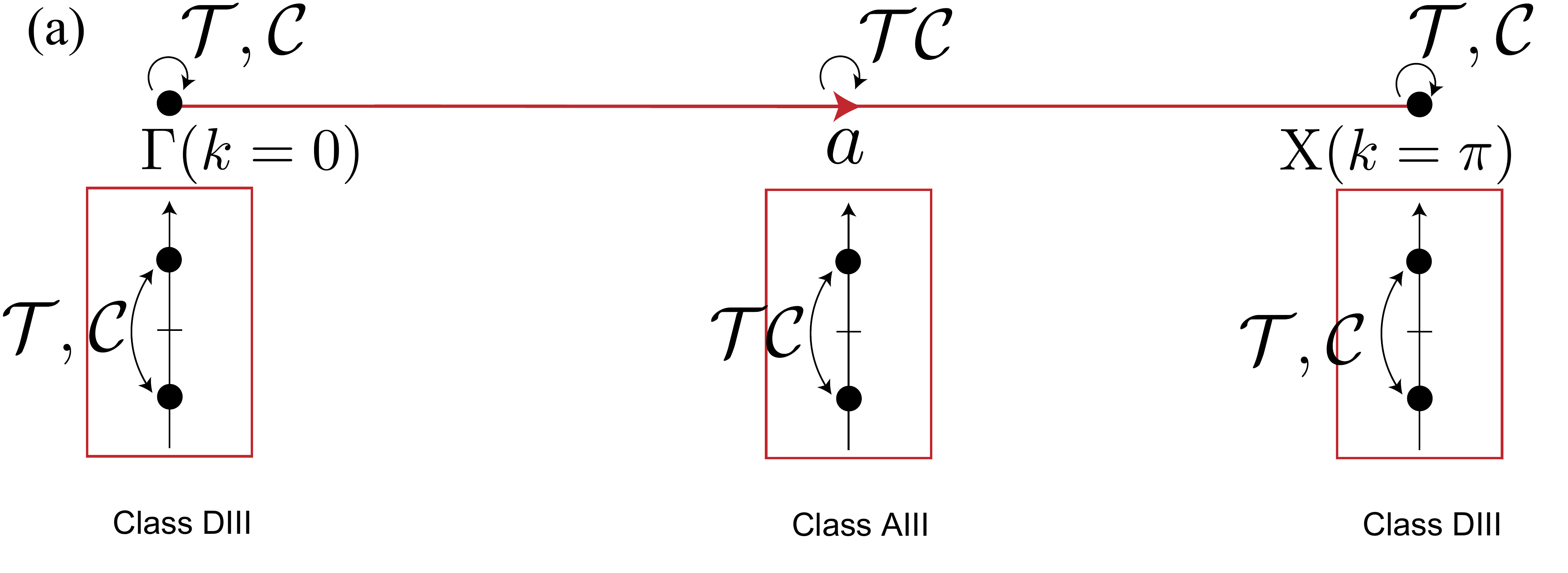}
		\caption{\label{fig:cell_1D}Illustration of the cell decomposition of the fundamental domain in rod group {\calligra p}$1$ with time-reversal symmetry $\calT$ and particle-hole symmetry $\calC$.
		The cell decomposition of BZ is given by 
		$C_0 = \{\Gamma (k = 0), \text{X}(k = \pi), \calT \text{X}\}$ and $C_1 = \{a (0<k<\pi), \calT a\}$. 
		}
	\end{center}
\end{figure}

Our cell decomposition of the fundamental domain is shown in Fig.~\ref{fig:cell_1D}.
There are two inequivalent $0$-cells denoted by $\Gamma\ (k = 0)$ and $\text{X}\ (k = \pi)$.
These $0$-cells are invariant under time-reversal and particle-hole symmetries, and thus 
\begin{align}
	E_{1}^{0, -1} = \mZ[\bm{b}^{(0)}_{\Gamma}]\oplus \mZ[\bm{b}^{(0)}_{\text{X}}].
\end{align}
On the other hand, there exists an inequivalent $1$-cell $a$, which is invariant only under chiral symmetry.
Therefore, 
\begin{align}
	E_{1}^{1, -1} = \mZ[\bm{b}^{(1)}_{a}].
\end{align}

Since $d_{1}^{1,-1}$ is not defined in one dimension, we start from Eq.~\eqref{eq:d_map2} with $r_1 = 0$ and $V^{(1)} = (1)$. 
\begin{align}
	\label{eq:d1_R1}
	d_{1}^{0,-1}\left(\bm{b}^{(0)}_{\Gamma} \bm{b}^{(0)}_{\text{X}}\right)
	&= \left(\bm{b}^{(1)}_{a}\right)\begin{pmatrix}
		2 & -2
	\end{pmatrix} \nonumber \\
	&=  \left(\bm{b}^{(1)}_{a}\right)U^{(0)}\begin{pmatrix}
		2 & 0
	\end{pmatrix}[V^{(0)}]^{-1},
\end{align}
where $X^{(1)} = U^{(0)} = (1)$ and $[V^{(0)}]^{-1} = \begin{pmatrix}
	1 & -1\\
	0 & 1
\end{pmatrix}$.

Following discussions in Sec.~\ref{sec:torsion_invs}, we construct the topological invariant defined by
\begin{align}
	&(-1)^{\mathcal{X}[H_{\bk}]} = \frac{\mathcal{Z}[q_{\text{X}}]/\mathcal{Z}[q_{\text{X}}^{\text{vac}}]}{\mathcal{Z}[q_{\Gamma}]/\mathcal{Z}[q_{\Gamma}^{\text{vac}}]}\nu_{a}[H_{\bk}]\nonumber \\
	&\quad=\frac{\mathrm{Pf}[q_{\text{X}}]}{\mathrm{Pf}[q_{\Gamma}]}\exp\left[-\frac{1}{2}\int_{\Gamma}^{X} \partial_k \log \det q_{k} dk\right]\nonumber \\
	&\quad\times\left(\frac{\mathrm{Pf}[q^{\text{vac}}_{\text{X}}]}{\mathrm{Pf}[q^{\text{vac}}_{\Gamma}]}\exp\left[-\frac{1}{2}\int_{\Gamma}^{X} \partial_k \log \det q^{\text{vac}}_{k} dk\right]\right)^{-1}.
\end{align}
It should be noted that $\Gamma$ and X are invariant under time-reversal symmetry (TRS) and that their EAZ classes are class DIII with $\calD_\alpha = 1$. 
Thus, $\mathcal{Z}[q_{k=0, \pi}]$ is replaced by $\mathrm{Pf}[q_{k=0, \pi}]$ with $q^{T}_{k=0, \pi} = -q^{T}_{k=0, \pi}$. 
As a result, our invariant is consistent with the well-known formula. 
\begin{table}[H]
	\begin{center}
		\caption{\label{tab:E1_Rp1}$E_1$- and $E_2$-pages for rod group {\calligra p}$1$.}
		\begin{tabular}{c}			
			\begin{minipage}{0.5\hsize}
				\begin{center}		
					\begin{tabular}{c|c|c}
						$n=0$ & $0$ & $0$\\
						$n=1$ & $\mZ^2$ & $\mZ$\\
						\hline
						$E^{p,-n}_{1}$ & $p=0$ & $p=1$
					\end{tabular}
				\end{center}
			\end{minipage}
			\begin{minipage}{0.5\hsize}
				\begin{center}			
					\begin{tabular}{c|c|c}
						$n=0$ & $0$ & $0$\\
						$n=1$ & $\mZ$ & $\mZ_2$\\
						\hline
						$E^{p,-n}_{2}$ & $p=0$ & $p=1$
					\end{tabular}
				\end{center}
			\end{minipage}
		\end{tabular}
	\end{center}
\end{table}

In Supplementary Materials~\cite{SM}, we present all matrices $[X^{(1)}]^{-1}$, $[V^{(0)}]^{-1}$, $\Lambda^{(0)}$, $[V^{(1)}]^{-1}$, and $\Sigma^{(1)}$ for insulators and superconductors in all magnetic space groups.
Our scheme presented in this subsection is applicable to not only time-reversal symmetric superconductors with conventional pairing symmetries but also superconductors whose $E_{1}^{1,-1}$'s are free Abelian groups. 
Also, $[X^{(1)}]^{-1}$ informs us of topological invariants for insulators and superconductors, where $E_{1}^{1,-1} = (\mZ_2)^l\ (l\in\mZ)$, under some gauge conditions, as discussed in Sec.~\ref{sec5}. 
However, although the matrices $[X^{(1)}]^{-1}$, $[V^{(0)}]^{-1}$, $\Lambda^{(0)}$, $[V^{(1)}]^{-1}$, and $\Sigma^{(1)}$ are presented, we do not have a scheme to find explicit expressions of topological invariants for superconductors such that $E_{1}^{1,-1}$'s contain both $\mZ$- and $\mZ_2$-parts.

\subsection{Numerical verification}
\label{sec:numerics}
In the preceding section, we construct $\mZ$- and $\mZ_k$-valued topological invariants. 
However, we do not provide any proof that they are actually quantized, although it is possible to confirm it one by one.
In this work, we alternatively check if they are quantized by computing invariants for randomly generated symmetric Hamiltonians defined on $1$-skeletons repeatedly. 
Here, we discuss how to numerically confirm that quantities defined in Eqs.~\eqref{eq:gapless_inv}, \eqref{eq:gapped_inv}, and \eqref{eq:Z_k} are quantized. 
For simplicity, we always consider spinful systems in nonmagnetic layer groups or nonmagnetic space groups with TRS.
In such a case, $\calM = \calG + \calG \calT$, where $\calG$ is a nonmagnetic space group or layer group.

\subsubsection{Preparation of symmetry representations and symmetric Hamiltonians}

\begin{table}[t]
	\begin{center}
		\caption{\label{tab:R-irrep}
			Irreducible representations for symmetry groups $\{e, \calT\}$ and $\{e, M_z, \calT, M_z\calT\}$ at $\bm{x}_1$.
			The latter case can happen only when we consider layer groups with mirror symmetry about $z$-direction.
		}
		\begin{tabular}{c|c|c|c|c}
			\hline
			$g\in \{e, \calT\}$ & $e$ & $\calT$ & &\\
			\hline
			$u_{\bm{x}_1}(g)$ & $\begin{pmatrix}
				1 & 0\\
				0 & 1
			\end{pmatrix}$ & $\begin{pmatrix}
			0 & 1\\
			-1 & 0
		\end{pmatrix}$ & &\\
			\hline \hline
			$g\in \{e, M_z, \calT, M_z\calT\}$ & $e$ & $M_z$ & $\calT$& $M_z\calT$\\
			\hline
			$u_{\bm{x}_1}(g)$ & $\begin{pmatrix}
				1 & 0\\
				0 & 1
			\end{pmatrix}$ & 
			$\begin{pmatrix}
				\mathrm{i} & 0\\
				0 & -\mathrm{i}
			\end{pmatrix}$
			& 
			$\begin{pmatrix}
				0 & 1\\
				-1 & 0
			\end{pmatrix}$ & $\begin{pmatrix}
			0 & \mathrm{i}\\
			\mathrm{i} & 0
		\end{pmatrix}$\\
			\hline
		\end{tabular}
	\end{center}
\end{table}

To generate symmetric Hamiltonians, we need symmetry representations $U_{\bk}(g)$. 
We can always construct $U_{\bk}(g)$ from real space as follows. 
First, we specify a generic point $\bm{x}_{1}$ in real space, where a generic point is symmetric only under nonspatial symmetries.
Once we have a generic point, we find $\{\bm{x}_l\}_{l=1}^{\vert \calG/\Pi \vert}$ and $\{g_l\}_{l=1}^{\vert \calG/\Pi \vert}$ such that $\bm{x}_l = g_{l}(\bm{x}_{1})$. 
Then, we can construct $u_{\bk}(g)\ (g\in\calM)$ by
\begin{align}
	[u_{\bk}(g)]_{\sigma \sigma';ll'} &= \frac{z^{\text{int}}_{g, g_l}}{z^{\text{int}}_{g_{l'}, g}} e^{-i g(\bk) \cdot (g(\bm{x}_l) - \bm{x}_{l'})} \delta'_{g(\bm{x}_l), \bm{x}_l} [u_{\bm{x}_1}(h)]^{\phi_g}_{\sigma\sigma'},
\end{align}
where $u_{\bm{x}_1}(h)$ is defined in Table~\ref{tab:R-irrep}, $h = g_{l'}T_{\bm{x}_{l'} - g(\bm{x_l})}gg_l\ (T_{\bm{x}_{l'} - g(\bm{x_l})} \in \Pi \text{ is a translation by }\bm{x}_{l'} - g(\bm{x_l}))$, and $\delta'_{g(\bm{x}_l), \bm{x}_{l'}}$ is $1$ only if $g(\bm{x}_l)-\bm{x}_{l'}$ is a lattice vector and $0$ otherwise.
It should be noted that $u_{\bk+\bm{G}}(g) = u_{\bk}(g)$. 
While $U_{\bk}(g) = u_{\bk}(g)$ for normal conducting phases, 
\begin{align}
	U_{\bk}(g) = \begin{cases}
		\begin{pmatrix}
			u_{\bk}(g) & O \\
			O & \chi_g u^{*}_{-\bk}(g)
		\end{pmatrix}\text{ for }g \in \mathcal{M}\\
		\begin{pmatrix}
			O & \xi u_{-\bk}(g)\\
			\chi_g u^{*}_{\bk}(g) & O
		\end{pmatrix}\text{ for }g \in \mathcal{M}\calP
	\end{cases}
\end{align}
for superconducting conducting phases and $\xi = +1 (-1)$ corresponds to the presence (absence) of $\text{SU}(2)$ symmetry.
Here, $\chi_g \in \text{U}(1)$ represents the pairing symmetry of superconducting gap function $\Delta_{\bk}$, i.e. it is defined by $u_{\bk}(g)\Delta_{\bk}^{\phi_g}u^{T}_{-\bk}(g) = \chi_g \Delta_{g\bk}\ (g \in \mathcal{M})$.

For later convenience, let us introduce the different convention of symmetry representations.
By performing a basis transformation, we can always find
\begin{align}
	\label{eq:symm_nonperiodic}
	\tilde{U}_{\bk}(g) := V^{\dagger}_{g\bk}U_{\bk}(g)V_{\bk} = e^{-i (g\bk)\cdot \bm{a}_g}D(g),
\end{align}
where $D(g)$ is a unitary matrix independent of $\bk$, and $\bm{a}_g$ is a vector that represents translation part of $g \in G$. 
The basis transformation matrix $V^{\dagger}_{\bk}$ is given by $[v_{\bk}]_{l l'} = \delta_{ll'}e^{-i \bk\cdot \bm{x}_l}$ for normal conducting phases and
\begin{align}
	\begin{pmatrix}
		v_{\bk} & O \\
		O & v_{\bk}
	\end{pmatrix}
\end{align}
for superconducting phases.

Once we have symmetry representations, we can generate a symmetric random Hamiltonian defined at a $\bk$-point $\bk_0$.
Let us suppose that we have a hermitian random matrix $h$. 
Then, we obtain a symmetric random Hamiltonian at $\bk_0$ from
\begin{align}
	\label{eq:symm_Hamiltonian}
	H_{\bk_0} = \frac{1}{\vert G_{\bk_0} \vert} \sum_{g \in G_{\bk_0}} c_g U_{\bk_0}(g) h^{\phi_g} U^{\dagger}_{\bk_0}(g).
\end{align}

\subsubsection{Adiabatic connection between two symmetric Hamiltonians}
For $0$- and $1$-cells, we can construct symmetric random Hamiltonians using Eq.~\eqref{eq:symm_Hamiltonian}. 
To compute our topological invariants obtained in Sec.~\ref{sec:topo_invariant}, we need a symmetric Hamiltonian defined on a $1$-skeleton. 
Let $\bk_0$ be a representative point of a $1$-cell, and $\bk_1$ and $\bk_2$ are its boundary $0$-cells. 
Here, employing the technique developed in Ref.~\cite{PhysRevLett.130.036601}, we connect $H_{\bk_0}$ to $H_{\bk_1}$ and $H_{\bk_2}$. 
First, we perform the basis transformation and obtain 
\begin{align}
	\tilde{H}_{\bk_i} = V^{\dagger}_{\bk_i}H_{\bk_i}V_{\bk_i}. 
\end{align}
It should be noted that $\tilde{H}_{\bk_i}$ is not periodic under shift of reciprocal lattice vector $\bm{G}$, as seen from $\tilde{H}_{\bk_i + \bm{G}} = V^{\dagger}_{\bm{G}}\tilde{H}_{\bk_i} V_{\bm{G}}$.
Then, we flatten the energy spectrum of $\tilde{H}_{\bk_0}$, $\tilde{H}_{\bk_1}$, and $\tilde{H}_{\bk_2}$.
In other words, we introduce
\begin{align}
	\tilde{Q}_{\bk_i} = \tilde{V} \begin{pmatrix}
		-\mathds{1} & O\\
		O & \mathds{1}
	\end{pmatrix} \tilde{V}^{-1},
\end{align}
where $\tilde{V}$ is a unitary matrix composed of eigenvectors of $\tilde{H}_{\bk_i}$. 
Next, we define parameterized Hamiltonians by 
\begin{align}
	\tilde{Q}_1(t_1) = (1-t_1)\tilde{Q}_{\bk_1} + t_1 \tilde{Q}_{\bk_0}, \\
	\tilde{Q}_2(t_2) = (1-t_2)\tilde{Q}_{\bk_0} + t_2 \tilde{Q}_{\bk_2},
\end{align}
where $t_1$ and $t_2$ are parameters defined by $\bk_1 + t_1 (\bk_0 - \bk_1)$ and $\bk_0 + t_2 (\bk_2 - \bk_0)$. 
Finally, by performing the inverse basis transformation, we have symmetric Hamiltonians 
\begin{align}
	Q_1(t_1) = V_{\bk_1 + t_1 (\bk_0 - \bk_1)}\tilde{Q}_1(t_1)V^{\dagger}_{\bk_1 + t_1 (\bk_0 - \bk_1)}, \\
	Q_2(t_2) = V_{\bk_0 + t_2 (\bk_2 - \bk_0)}\tilde{Q}_2(t_2)V^{\dagger}_{\bk_0 + t_2 (\bk_2 - \bk_0)},
\end{align}
which are defined on the $1$-cell and the two boundary $0$-cells.
As a result, we have a Hamiltonian given by 
\begin{align}
	H_{\bk} = \begin{cases}
		Q_1(t_1) \text{ for } \bk = \bk_1 + t_1 (\bk_0 - \bk_1) \\
		Q_2(t_2) \text{ for } \bk = \bk_0 + t_2 (\bk_2 - \bk_0).
	\end{cases}
\end{align}
Applying this scheme to all $1$-cells, we obtain a symmetric Hamiltonian defined on the $1$-skeleton.

We make the following two remarks. 
First, we should interpolate $\tilde{Q}_{\bk_i}\ (i = 0, 1, 2)$ to make interpolated Hamiltonians symmetric.
If we start from the flattened $H_{\bk_i}\ (i = 0, 1, 2)$, the interpolated Hamiltonian does not always possess all symmetries on $1$-cells. 
On the other hand, the symmetry representation $\tilde{U}_{\bk}(g)$ does not depend on momentum, as shown in Eq.~\eqref{eq:symm_nonperiodic}. 
In such a case, it is guaranteed that the interpolated Hamiltonians in the same basis of $\tilde{U}_{\bk}(g)$ respect all symmetries on the $1$-cell.
Second, gapless points sometimes appear on $1$-cells when compatibility conditions between $0$- and $1$-cells are violated.
However, our invariants are well-defined only when the Hamiltonian is gapped on the $1$-skeleton.
To eliminate gapless points on $1$-cells, we must stack some Hamiltonians so that all compatibility conditions between $0$- and $1$-cells are satisfied. 

\subsubsection{Completeness check}
To check if our invariants work, we compute topological invariants of Hamiltonians obtained in the above way. 
In our numerical calculations, we repeat this process 20 times for all layer groups and all space groups with conventional pairing symmetries ($\chi_g = +1$ for all $g \in \calM$) in the presence of TRS.
For later convenience, the Hamiltonian of the $m$-th calculation is denoted by $H^{(m)}_{\bk}$.
Let us suppose that we have the following set of topological invariants:
\begin{align}
	\label{eq:wl_list}
	&W_{l}[H_{\bk}]= (\mathcal{W}^{\text{gapless}}_{1}[H_{\bk}], \mathcal{W}^{\text{gapless}}_{2}[H_{\bk}], \cdots, \mathcal{W}^{\text{gapless}}_{r_1}[H_{\bk}])^{\top}, \\
	\label{eq:wg_list}
	&W_{\text{g}}[H_{\bk}]= (\mathcal{W}^{\text{gapped}}_{1}[H_{\bk}], \mathcal{W}^{\text{gapped}}_{2}[H_{\bk}], \cdots, \mathcal{W}^{\text{gapped}}_{N_{\text{f}}}[H_{\bk}])^{\top},\\
	\label{eq:c_list}
	&C[H_{\bk}]= (\mathcal{X}_{R_0+1}[H_{\bk}], \mathcal{X}_{R_0+2}[H_{\bk}], \cdots, \mathcal{X}_{R_0+N_{\mathrm{t}}}[H_{\bk}])^{\top},
\end{align}
where $N_{\mathrm{f}}$ and $N_{\mathrm{t}}$ are the numbers of $\mZ$- and $\mZ_k$-valued topological invariants. 
Recall that $R_0$ is the number of unity in the Smith normal form $\Lambda^{(0)}$, i.e., $\lambda_{j}^{(0)} = 1$ for $1\leq j \leq R_0$.
After computing all the topological invariants of the 20 Hamiltonians, we have three sets $\{W_{l}[H^{(m)}_{\bk}]\}_{m=1}^{20}$, $\{W_{\text{g}}[H^{(m)}_{\bk}]\}_{m=1}^{20}$, and $\{C[H^{(m)}_{\bk}]\}_{m=1}^{20}$.

Then, we check if our topological invariants can fully characterize $E_{2}^{1,-1}$. 
More precisely, we confirm that $\{W_{l}[H^{(m)}_{\bk}]\}_{m=1}^{20}$, $\{W_{\text{g}}[H^{(m)}_{\bk}]\}_{m=1}^{20}$, and $\{C[H^{(m)}_{\bk}]\}_{m=1}^{20}$ can span $\mZ^{r_1}$, $\mZ^{N_{\text{f}}}$, and $\bigoplus_{i-r_1=R_0+1}^{N_{\text{t}}+R_0}\mZ_{\lambda_{i-r_1}}$, respectively.
See Appendix~\ref{app:check} for more technical details. 

\subsection{Fermi surface formulas for topological superconductors}
\label{sec:Fermi_surfaces}
Although obtaining $q$-matrices for realistic materials is usually challenging, it is well-known that the expressions of Eq.~\eqref{eq:1D_DIII} and the winding number $w$ could be simplified when the scale of pair potentials is small enough compared with that of normal conducting phases~\cite{FS_TSC_Qi,FS_TSC_Sato}.
In this weak-pairing limit, if normal conducting phases are gapped at $k=0, \pi$ and Fermi surfaces are not degenerate, we have the following formulas
\begin{align}
	(-1)^{\nu} &= \prod_{\epsilon_{nk} = 0}\mathrm{sgn}(\delta_{nk}),\\
	w &= \frac{1}{2}\sum_{\epsilon_{nk} = 0} \mathrm{sgn}(\partial_k \epsilon_{nk})\mathrm{sgn}(\delta_{nk}),
\end{align}
where $\epsilon_{nk}$ is the $n$-th eigenenergy of the normal conducting phase and $\delta_{nk}$ is a diagonal element of the superconducting gap function in the band basis%
~\footnote{
More precisely, $\delta_{nk}$ is defined as follows. 
Let $\psi_{nk}, u(\calT),$ and $\Delta_{k}$ be an eigenvector with the eigenvalue $\epsilon_{nk}$, the unitary representation of TRS, and the superconducting gap function, respectively. 
Then, $\delta_{nk} = \psi_{nk}^{\dagger}(u(\calT)\Delta_{k}^{\dagger})\psi_{nk}$.
See Refs.~\cite{FS_TSC_Qi,FS_TSC_Sato} for the derivations.
}. 
Given that these two invariants are also defined in terms of winding of $q$-matrices and exponential function of the winding, such as Eqs.~\eqref{eq:winding_AIII}, and \eqref{eq:exp_winding_AIII}, it is natural to expect that similar formulas for our invariants also exist. 
A systematic enumeration of such formulas is beyond the scope of the present paper. We note, however, that this problem has been addressed in subsequent work~\cite{zhang2025fermisurfacediagnosistopologicalsuperconductivity}, where the invariants constructed here are further reduced to practical Fermi-surface formulas for $s$-wave-like topological superconductivity.

\section{Examples}
\label{sec4}
In this section, we derive topological invariants of topological superconductors in several symmetry settings with TRS. 
We also compute the obtained topological invariants for representative models of all possible topological phases in these symmetry settings. 

\subsection{Layer group $p2_1/m11$ with $A_g$ pairing}
\label{sec:p21m11}
Our first example is layer group $p2_1/m11$ whose generators are screw symmetry $S_x =\{2_{100}\vert (1/2, 0, 0)^{\top}\}$, mirror symmetry $M_x =\{m_{100}\vert (1/2, 0, 0)^{\top}\}$, and a translation along $y$-direction. 
According to Ref.~\cite{Ono-Shiozaki-Watanabe2022}, $^{\phi}K_{G/\Pi}^{(z,c)}(T^2) = \mZ$ in layer group $p2_1/m11$ with TRS $\calT$ and $A_g$ pairing.
However, topological invariants are not known yet. 
Here, we construct the topological invariants based on our framework. 

\begin{table}[t]
	\begin{center}
		\caption{\label{tab:irrep_LGp21/m11}
			Irreducible representations and their EAZ classes on $0$- and $1$-cells in layer group $p2_1/m11$. 
		}
		\begin{tabular}{c|c|cccc}
			\hline
			irrep. & EAZ & $e$ & $S_x$ & $I=M_xS_x$ & $M_x$\\
			\hline\hline
			$\Gamma_1, \text{Y}_1$ & AIII & $1$ & $-\mathrm{i}$ & $-1$ & $\mathrm{i}$\\
			$\Gamma_2, \text{Y}_2$ & AIII &$1$ & $-\mathrm{i}$ & $1$ & $-\mathrm{i}$\\
			$\Gamma_3, \text{Y}_3$ & AIII & $1$ & $\mathrm{i}$ & $-1$ & $-\mathrm{i}$\\
			$\Gamma_4, \text{Y}_4$ & AIII &$1$ & $\mathrm{i}$ & $1$ & $\mathrm{i}$\\
			$\text{X}_1, \text{M}_1$ & DIII &$2$ & $0$ & $0$ & $0$\\
			\hline
			irrep. & EAZ & $e$ & $S_x$ & &\\
			\hline\hline
			$a_1$, $d_1$ & AIII & $1$ & $-\mathrm{i}e^{-i k_x/2}$ & &\\
			$a_2$, $d_2$ & AIII &$1$ & $\mathrm{i}e^{-i k_x/2}$ & &\\
			\hline
			irrep. & EAZ &$e$ & $M_x$ & &\\
			\hline\hline
			$b_1$ & AIII &$1$ & $-\mathrm{i}$ & &\\
			$b_2$ & AIII &$1$ & $\mathrm{i}$ & &\\
			$c_1$ & DIII &$1$ & $\mathrm{i}$ & &\\
			$c_2$ & DIII &$1$ & $-\mathrm{i}$ & &\\
			\hline
		\end{tabular}
	\end{center}
\end{table}

Our cell decomposition of the fundamental domain is shown in Fig.~\ref{fig:simple_cell}(b), and irreps are tabulated in Table~\ref{tab:irrep_LGp21/m11}. 
Then, $E_{1}^{1,-1}=\mZ^5$ is given by
\begin{align}
	\label{eq:E111_p21m11}
	E_{1}^{1,-1} = \mZ[\bm{b}_{a,1}]\oplus\mZ[\bm{b}_{b,1}]\oplus\mZ[\bm{b}_{c,1}]\oplus\mZ[\bm{b}_{c,2}]\oplus\mZ[\bm{b}_{d,1}],
\end{align}
where we omit ``$(1)$'' from $\bm{b}_{D_{i}^{p}, \alpha}^{(1)}$ for short. 
After following procedures in Sec.~\ref{sec:topo_invariant}, we find $r_1 = 1$ and
\begin{align}
	&[X^{(1)}]^{-1} = \left(
	\begin{array}{cccccc}
		a_1 & b_1 & c_1 & c_2 & d_1\\
		\hline
		1 & -1 & 1 & 1 & -1 \\
		0 & 1 & 0 & 0 & 1 \\
		0 & 0 & 1 & 0 & 0 \\
		0 & 0 & 0 & 0 & 1 \\
		0 & 0 & -1 & 1 & 0 \\
	\end{array}
	\right);\\
	&\Lambda^{(0)} = \left(
	\begin{array}{cccc}
		1 & 0 & 0 & 0  \\
		0 & 1 & 0 & 0  \\
		0 & 0 & 1 & 0  \\
		0 & 0 & 0 & 0  \\
	\end{array}
	\right).
\end{align}
As discussed in Sec.~\ref{sec:general_frame}, the first row of $[X^{(1)}]^{-1}$ informs us about a topological invariant for gapless points on $2$-cells, which is given by
\begin{align}
	\mathcal{W}^{\text{gapless}} &= w_{a_1} - w_{b_1} + w_{c_1} + w_{c_2} - w_{d_1} \nonumber \\
	&= \frac{1}{2 \pi \mathrm{i}} \left[\int_a d \log \frac{\det q^{a_1}_{\bk}}{\det (q^{a_1}_{\bk})^{\text{vac}}}- \int_b d \log \frac{\det q^{b_1}_{\bk}}{\det (q^{b_1}_{\bk})^{\text{vac}}}\right.\nonumber\\
	&\left.+ \frac{1}{2}\sum_{i=1}^{2}\int_c d \log \frac{\det q^{c_i}_{\bk}}{\det (q^{c_i}_{\bk})^{\text{vac}}} - \int_d d \log \frac{\det q^{d_1}_{\bk}}{\det (q^{d_1}_{\bk})^{\text{vac}}} \right]\nonumber \\ 
	&= \frac{1}{2}\times\frac{1}{2 \pi \mathrm{i}} \oint_{C} (d \log \det q_{\bk} - d \log \det q_{\bk}^{\text{vac}}),
\end{align}
where the loop $C$ is defined by $\Gamma$-X-M-Y-$\Gamma$. 
In the last line, we use the relation $\det q_{\bk}^{K_2}= \det q_{\bk}^{K_1}$ for $K = a, b, d$ (See Appendix~\ref{sec:Compatibility relation of non-Hermitian matrix}).

As shown in Eq.~\eqref{eq:gapped_inv}, since only $\lambda_{4 = (5-1)} \neq 1$ and $\lambda_{4}=0$, the fifth row of $[X^{(1)}]^{-1}$ gives us a $\mZ$-valued topological invariant for gapped phases 
\begin{align}
	\label{eq:p21m_Z}
	\mathcal{W}^{\text{gapped}} &= -w_{c_1} + w_{c_2}\nonumber \\
	&= \frac{-1}{4\pi \mathrm{i}}\int_{0}^{\pi} d \log \frac{\det q_{(\pi, k_y)}^{c_1}}{\det (q_{(\pi, k_y)}^{c_1})^{\text{vac}}} \left(\frac{\det q_{(\pi, k_y)}^{c_2}}{\det (q_{(\pi, k_y)}^{c_2})^{\text{vac}}}\right)^{-1}.
\end{align}
In fact, this is the mirror winding number.

\begin{figure}[t]
	\begin{center}
		\includegraphics[width=0.99\columnwidth]{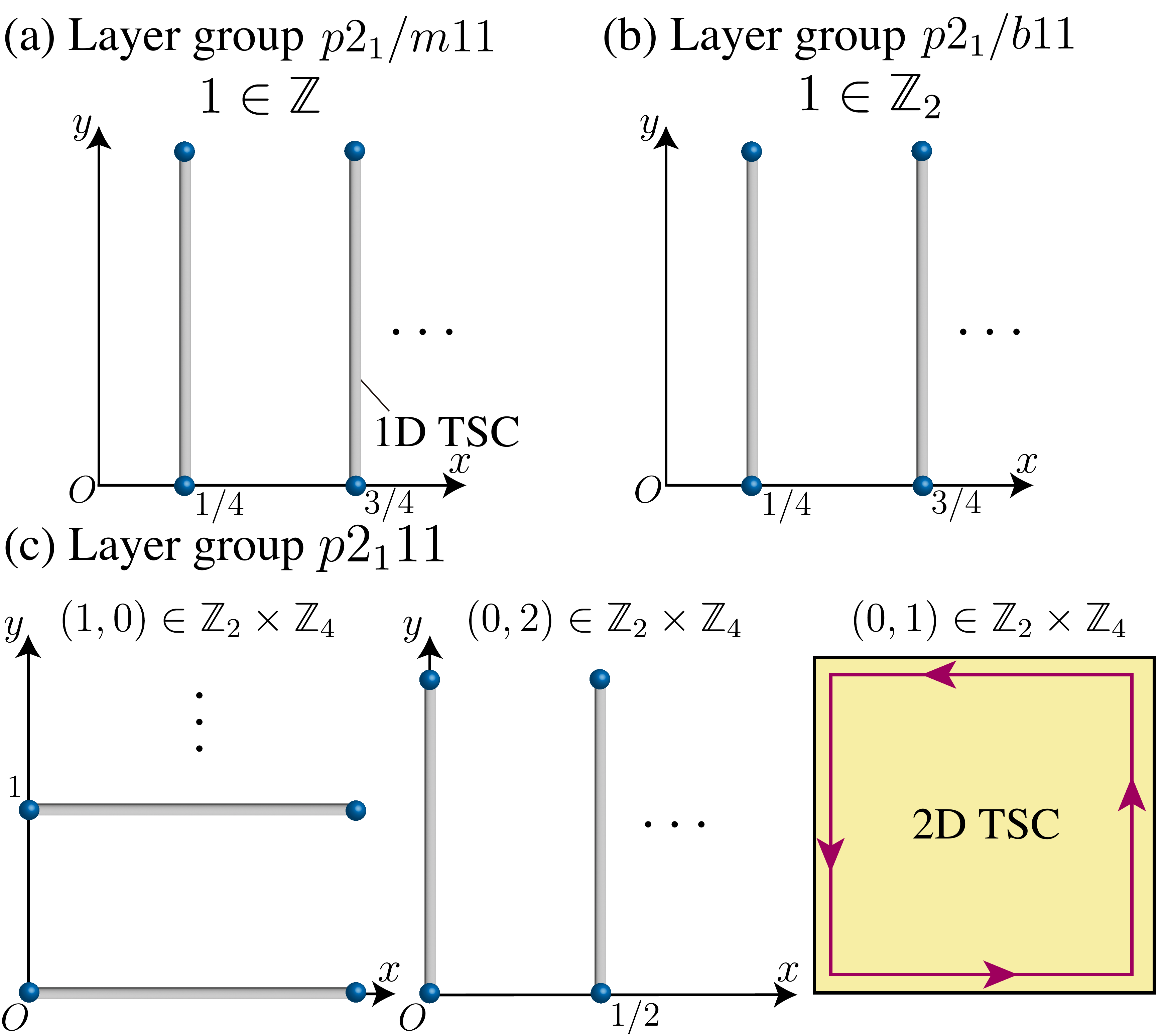}
		\caption{\label{fig:R-AHSS}
			Real-space pictures of topological phases in layer groups $p2_1/m11$ (a), $p2_1/b11$ (b), and $p2_111$ (c). 
		}
	\end{center}
\end{figure}

A generator of $^{\phi}K_{G/\Pi}^{(z,c)}(T^2) = \mZ$ is constructed from one-dimensional topological superconducting phases protected by mirror and chiral symmetries, as shown in Fig.~\ref{fig:R-AHSS}(a).
After computing $\mathcal{W}^{\text{gapped}}$ for a representative model of this generator with a random symmetric perturbation, we find that $\mathcal{W}^{\text{gapped}} =-1$.

\subsection{Layer group $p2_1/b11$ with $A_g$ pairing}
\label{sec:p21b11}
Our next example is layer group $p2_1/b11$, which is generated by screw symmetry $S_x =\{2_{100}\vert (1/2, 1/2, 0)^{\top}\}$ and glide symmetry $G_x =\{m_{100}\vert (1/2, 1/2, 0)^{\top}\}$. 
Reference~\cite{Ono-Shiozaki-Watanabe2022} shows that $^{\phi}K_{G/\Pi}^{(z,c)}(T^2) = \mZ_2$ in layer group $p2_1/b11$ with TRS $\calT$ and $A_g$ pairing.

The cell decomposition is the same as in Fig.~\ref{fig:simple_cell}(b). 
Irreducible representations and their EAZ classes are tabulated in Table~\ref{tab:irrep_LGp21/b11}.
We have $E_{1}^{1,-1} = \mZ^6$ spanned by 
\begin{align}
	E_{1}^{1,-1} = \bigoplus_{K = a,b}\mZ[\bm{b}_{K,1}]\oplus\bigoplus_{K=c,d}\left(\bigoplus_{\alpha=1,2}\mZ[\bm{b}_{K,\alpha}]\right).
\end{align}
\begin{table}[t]
	\begin{center}
		\caption{\label{tab:irrep_LGp21/b11}
			Character tables of irreps and their EAZ classes on $0$- and $1$-cells in layer group $p2_1/b11$. 
		}
		\begin{tabular}{c|c|cccc}
			\hline
			irrep. & EAZ & $e$ & $S_x$& $I$ & $G_x$\\
			\hline\hline
			$\Gamma_1$ & AIII & $1$ & $-\mathrm{i}$& $-1$ & $\mathrm{i}$\\
			$\Gamma_2$ & AIII & $1$ & $-\mathrm{i}$& $1$ & $-\mathrm{i}$\\
			$\Gamma_3$ & AIII & $1$ & $\mathrm{i}$& $-1$ & $-\mathrm{i}$\\
			$\Gamma_4$ & AIII & $1$ & $\mathrm{i}$& $1$ & $\mathrm{i}$\\
			$\text{X}_1$ & DIII & $2$ & $0$& $0$ & $0$\\
			$\text{Y}_1$ & DIII & $2$ & $0$& $0$ & $0$\\
			$\text{M}_1$ & DIII & $1$ & $1$& $-1$ & $-1$\\
			$\text{M}_2$ & DIII & $1$ & $1$& $1$ & $1$\\
			$\text{M}_3$ & DIII & $1$ & $-1$& $-1$ & $1$\\
			$\text{M}_4$ & DIII & $1$ & $-1$& $1$ & $-1$\\
			\hline
			irrep. & EAZ & $e$ & $S_x$&&\\
			\hline\hline
			$a_1$ & AIII & $1$ & $-\mathrm{i}e^{-i k_x/2}$&&\\
			$a_2$ & AIII &$1$ & $\mathrm{i}e^{-i k_x/2}$&&\\
			$d_1$ & DIII & $1$ & $\mathrm{i}e^{-i k_x/2}$&&\\
			$d_2$ & DIII &$1$ & $-\mathrm{i}e^{-i k_x/2}$&&\\
			\hline
			irrep. & EAZ &$e$ & $G_x $&&\\
			\hline\hline
			$b_1$ & AIII &$1$ & $-\mathrm{i}e^{-i k_y/2}$&&\\
			$b_2$ & AIII &$1$ & $\mathrm{i}e^{-i k_y/2}$&&\\
			$c_1$ & DIII &$1$ & $\mathrm{i}e^{-i k_y/2}$&&\\
			$c_2$ & DIII &$1$ & $-\mathrm{i}e^{-i k_y/2}$&&\\
			\hline
		\end{tabular}
	\end{center}
\end{table}
From the analyses in Sec.~\ref{sec:general_frame}, we find $r_1 = 1$ and 
\begin{align}
	&[X^{(1)}]^{-1} =\left(
	 \begin{array}{cccccc}
	 	a_1 & b_1 & c_1 & c_2 & d_1 & d_2 \\
	 	\hline
		1 & -1 & 1 & 1 & -1 & -1 \\
		0 & 1 & 0 & 0 & 0 & 2 \\
		0 & 0 & 0 & 1 & -1 & 1 \\
		0 & 0 & 0 & 0 & 0 & 1 \\
		0 & 0 & 0 & 0 & -1 & 1 \\
		0 & 0 & -1 & 1 & -1 & 1 \\
	\end{array}
	\right);
	\\
	\label{eq:lambda_p21b11}
	&\Lambda^{(0)}=\left(
	\begin{array}{ccccc}
		1 & 0 & 0 & 0 & 0  \\
		0 & 1 & 0 & 0 & 0  \\
		0 & 0 & 1 & 0 & 0  \\
		0 & 0 & 0 & 1 & 0  \\
		0 & 0 & 0 & 0 & 2  \\
	\end{array}
	\right);\\
	&[V^{(0)}]^{-1} =\left(
	 \begin{array}{cccccccc}
	 	\Gamma_1 & \Gamma_3 & \text{X}_1 & \text{Y}_1 & \text{M}_1 & \text{M}_2 & \text{M}_3 & \text{M}_4 \\
	 	\hline
		1 & 1 & 0 & 0 & 0 & 0 & -2 & -2 \\
		0 & 0 & 1 & 0 & 0 & 1 & -1 & -2 \\
		0 & 0 & 0 & 1 & 0 & 0 & -1 & -1 \\
		0 & 0 & 0 & 0 & 1 & 1 & -1 & -1 \\
		0 & 0 & 0 & 0 & 0 & 1 & 0 & -1 \\
		0 & 1 & 0 & 0 & 0 & 0 & 0 & 0 \\
		0 & 0 & 0 & 0 & 0 & 0 & 1 & 0 \\
		0 & 0 & 0 & 0 & 0 & 0 & 0 & 1 \\
	\end{array}
	\right).
\end{align}
From the first row of $[X^{(1)}]^{-1}$, we have the following $\mZ$-valued topological invariant for gapless points on $2$-cells:
\begin{align}
	\mathcal{W}^{\text{gapless}} &= w_{a_1} - w_{b_1} + w_{c_1} + w_{c_2} - w_{d_1} - w_{d_2} \nonumber \\
	&= \frac{1}{2 \pi \mathrm{i}} \left[\int_a d \log \frac{\det q^{a_1}_{\bk}}{\det (q^{a_1}_{\bk})^{\text{vac}}}- \int_b d \log \frac{\det q^{b_1}_{\bk}}{\det (q^{b_1}_{\bk})^{\text{vac}}}\right.\nonumber\\
	&\left.+ \frac{1}{2}\sum_{i=1}^{2}\left(\int_c d \log \frac{\det q^{c_i}_{\bk}}{\det (q^{c_i}_{\bk})^{\text{vac}}} - \int_d d \log \frac{\det q^{d_i}_{\bk}}{\det (q^{d_i}_{\bk})^{\text{vac}}}\right) \right]\\ 
	&= \frac{1}{2}\times\frac{1}{2 \pi \mathrm{i}} \oint_{C} (d \log \det q_{\bk} - d \log \det q_{\bk}^{\text{vac}}),
\end{align}
where the loop $C$ is defined by $\Gamma$-X-M-Y-$\Gamma$. 
We see that only $\lambda_{5}=2$ is not unity in Eq.~\eqref{eq:lambda_p21b11}, which indicates that the last row of $[X^{(1)}]^{-1}$ and the fifth row of $[V^{(0)}]^{-1}$ provide a $\mZ_2$-valued topological invariant for gapped phases.
The $\mZ_2$-valued topological invariant $\mathcal{X}$ is given by
	\begin{align}
		\label{eq:p21b_z2}
		(-1)^{\mathcal{X}} &= \frac{\mathcal{Z}[q_{\text{M}}^{\text{M}_4}]/\mathcal{Z}[(q_{\text{M}}^{\text{M}_4})^{\text{vac}}]}{\mathcal{Z}[q_{\text{M}}^{\text{M}_2}]/\mathcal{Z}[(q_{\text{M}}^{\text{M}_2})^{\text{vac}}]}\frac{\nu_{c_2}[H_{\bk}]\nu_{d_2}[H_{\bk}]}{\nu_{c_1}[H_{\bk}]\nu_{d_1}[H_{\bk}]}\nonumber \\
		&= \frac{\mathrm{Pf}[q_{\text{M}}^{\text{M}_4}]/\mathrm{Pf}[(q_{\text{M}}^{\text{M}_4})^{\text{vac}}]}{\mathrm{Pf}[q_{\text{M}}^{\text{M}_2}]/\mathrm{Pf}[(q_{\text{M}}^{\text{M}_2})^{\text{vac}}]} \nonumber \\%
		%
		&\hspace{-3mm}\times\exp\left[\frac{1}{4}\hspace{-1mm}\sum_{K=c,d}\hspace{-0.5mm}\int_K\hspace{-1.5mm} d\log \frac{\det q_{\bk}^{K_1}}{\det (q_{\bk}^{K_1})^{\text{vac}}}\hspace{-1mm}\left(\frac{\det q_{\bk}^{K_2}}{\det (q_{\bk}^{K_2})^{\text{vac}}}\right)^{-1}\right],
	\end{align}
where  we replace $\mathcal{Z}[q_{\bk}^{\alpha}]$ by $\pf [q_{\bk}^{\alpha}]$ in the second line, as mentioned in Sec.~\ref{sec:general_frame}. 

A generator of $^{\phi}K_{G/\Pi}^{(z,c)}(T^2) = \mZ_2$ is constructed from one-dimensional topological superconducting phases, as shown in Fig.~\ref{fig:R-AHSS}(b).
After computing $\mathcal{X}$ for a representative model of this generator with a random symmetric perturbation, we find that $\mathcal{X} =1$ mod $2$.

\subsection{Layer group $p2_111$ with $A$ pairing}
\label{sec:p2111}
The third example is layer group $p2_111$, whose generators are screw $S_x =\{2_{100}\vert (1/2, 0, 0)^{\top}\}$ and a translation along $y$-direction. 
In Ref.~\cite{Shiozaki-Sato-Gomi2016}, the authors show that $^{\phi}K_{G/\Pi}^{(z,c)}(T^2) = \mZ_2 \times \mZ_4$ in layer group $p2_111$ with TRS and with $A$ pairing.
However, topological invariants are not known yet. 
Here, we construct the topological invariants based on our framework. 

Again, the cell decomposition is the same as layer groups $p2_1/m11$ and $p2_1/b11$. 
Irreducible representations and their EAZ classes are tabulated in Table~\ref{tab:irrep_LGp2111}.
We have $E_{1}^{1,-1} = \mZ^6$ spanned by 
\begin{align}
	E_{1}^{1,-1} = \bigoplus_{K = b,c}\mZ[\bm{b}_{K,1}]\oplus\bigoplus_{K=a,d}\left(\bigoplus_{\alpha=1,2}\mZ[\bm{b}_{K,\alpha}]\right).
\end{align}

\begin{table}[t]
	\begin{center}
		\caption{\label{tab:irrep_LGp2111}
			Character tables of irreps and their EAZ classes on $0$- and $1$-cells in layer group $p2_111$. 
		}
		\begin{tabular}{c|c|cc}
			\hline
			irrep. & EAZ & $e$ & $S_x$\\
			\hline\hline
			$\Gamma_1, \text{Y}_1$ & AIII & $1$ & $-\mathrm{i}$\\
			$\Gamma_2, \text{Y}_2$ & AIII & $1$ & $\mathrm{i}$\\
			$\text{X}_1, \text{M}_1$ & DIII & $1$ & $-1$\\
			$\text{X}_2, \text{M}_2$ & DIII & $1$ & $1$\\
			\hline
			irrep. & EAZ & $e$ & $S_x$\\
			\hline\hline
			$a_1, d_1$ & AIII & $1$ & $-\mathrm{i}e^{-i k_x/2}$\\
			$a_2, d_2$ & AIII &$1$ & $\mathrm{i}e^{-i k_x/2}$\\
			\hline
			irrep. & EAZ &$e$ & \\
			\hline\hline
			$b$ & CI &$1$ &\\
			$c$ & DIII &$1$ &\\
			\hline
		\end{tabular}
	\end{center}
\end{table}
We perform the analyses presented in Sec.~\ref{sec:general_frame}, and then we find $r_1 = 1$ and 
\begin{align}
	&[X^{(1)}]^{-1} = \left(
	\begin{array}{cccccc}
		a_1 & a_2 & b_1 & c_1 & d_1 & d_2\\
		\hline
		1 & 1 & -1 & 2 & -1 & -1 \\
		0 & -1 & 1 & 0 & 0 & 2 \\
		0 & 0 & 0 & 1 & -1 & 1 \\
		0 & 0 & 0 & 0 & 0 & 1 \\
		0 & 2 & -1 & 0 & -1 & 1 \\
		0 & 2 & -1 & 0 & -2 & 0 \\
	\end{array}
	\right);\\
	&\Lambda^{(0)} = \left(
	\begin{array}{ccccc}
		1 & 0 & 0 & 0 & 0  \\
		0 & 1 & 0 & 0 & 0  \\
		0 & 0 & 1 & 0 & 0  \\
		0 & 0 & 0 & 2 & 0  \\
		0 & 0 & 0 & 0 & 4  \\
	\end{array}
	\right);\\
	&[V^{(0)}]^{-1} = \left(
	\begin{array}{cccccc}
		\Gamma_1 & \text{X}_1 & \text{X}_2 & \text{Y}_1 & \text{M}_1 & \text{M}_2\\
		\hline
		1 & 0 & 2 & 0 & 0 & -4 \\
		0 & 1 & 1 & 0 & 1 & -3 \\
		0 & 0 & 0 & 1 & 0 & -2 \\
		0 & 0 & -2 & 1 & 1 & -1 \\
		0 & 0 & -1 & 0 & 1 & 0 \\
		0 & 0 & 0 & 0 & 0 & 1 \\
	\end{array}
	\right).
\end{align}
In the same way as the above examples, we have a $\mZ$-valued topological invariant for gapless points on $2$-cells:
\begin{align}
	\mathcal{W}^{\text{gapless}} &= w_{a_1} + w_{a_2} - w_{b_1} + 2w_{c_1} - w_{d_1} - w_{d_2} \nonumber \\
	&= \frac{1}{2 \pi \mathrm{i}} \oint_{C} (d \log \det q_{\bk} - d \log \det q_{\bk}^{\text{vac}}).
\end{align}
We find $\lambda_4 = 2$ and $\lambda_5 = 4$. 
As a result, we construct a $\mZ_2$-valued topological invariant from the fifth row of $[X^{(1)}]^{-1}$ and the fourth row of $[V^{(0)}]^{-1}$. 
Also, a $\mZ_4$-valued topological invariant is obtained from the sixth rows of $[X^{(1)}]^{-1}$ and the fifth row of $[V^{(0)}]^{-1}$.
The $\mZ_2$- and $\mZ_4$-valued invariants, denoted by $\mathcal{X}_1$ and $\mathcal{X}_2$, are defined by 
\begin{table}[t]
	\begin{center}
		\caption{\label{tab:p2_111}
			Topological invariants for representative models of topological phases. 
		}
		\begin{tabular}{c|c}
			\hline
			$^{\phi}K_{G/\Pi}^{(z,c)}(\mathbb{R}^2) = \mZ_2 \times \mZ_4$ & $(\mathcal{X}_1 \text{ mod } 2, \mathcal{X}_2 \text{ mod } 4)$ \\
			\hline\hline
			$(1,0)$ & $(1, 2)$\\
			$(0,2)$ & $(0, 2)$\\
			$(0,1)$ & $(1, 1)$\\
			\hline
		\end{tabular}
	\end{center}
\end{table}
\begin{widetext}
	\begin{align}
		\label{eq:p21_z2}
		(-1)^{\mathcal{X}_1} &=\frac{\mathcal{Z}^2[q_{\text{X}}^{\text{X}_2}]\mathcal{Z}[q_{\text{M}}^{\text{M}_2}]}{\mathcal{Z}[q_{\text{Y}}^{\text{Y}_1}]\mathcal{Z}[q_{\text{M}}^{\text{M}_1}]}
		\left(\frac{\mathcal{Z}^2[(q_{\text{X}}^{\text{X}_2})^{\text{vac}}]\mathcal{Z}[(q_{\text{M}}^{\text{M}_2})^{\text{vac}}]}{\mathcal{Z}[(q_{\text{Y}}^{\text{Y}_1})^{\text{vac}}]\mathcal{Z}[(q_{\text{M}}^{\text{M}_1})^{\text{vac}}]}\right)^{-1}
		\frac{\nu_{a_2}^2[H_{\bk}]\nu_{d_2}[H_{\bk}]}{\nu_{b_1}[H_{\bk}]\nu_{d_1}[H_{\bk}]} \nonumber \\
		&= \frac{\det q_{\text{X}}^{\text{X}_2}\pf[q_{\text{M}}^{\text{M}_2}]}{\det q_{\text{Y}}^{\text{Y}_1}\pf[q_{\text{M}}^{\text{M}_1}]}\left(\frac{\det (q_{\text{X}}^{\text{X}_2})^{\text{vac}}\pf[(q_{\text{M}}^{\text{M}_2})^{\text{vac}}]}{\det (q_{\text{Y}}^{\text{Y}_1})^{\text{vac}}\pf[(q_{\text{M}}^{\text{M}_1})^{\text{vac}}]}\right)^{-1} \nonumber \\
		&\quad\times
		\exp\left[-\frac{1}{2}\int d\left(2\log \frac{\det q_{(k_x, 0)}^{a_2}}{\det (q_{(k_x, 0)}^{a_2})^{\text{vac}}} - \log \frac{\det q_{(0, k_y)}^{b_1}}{\det (q_{(0, k_y)}^{b_1})^{\text{vac}}}  - \log  \frac{\det q_{(k_x, \pi)}^{d_1}}{\det (q_{(k_x, \pi)}^{d_1})^{\text{vac}}} + \log \frac{\det q_{(k_x, \pi)}^{d_2}}{\det (q_{(k_x, \pi)}^{d_2})^{\text{vac}}}\right) \right],\\
		\label{eq:p21_z4}
		\mathrm{i}^{\mathcal{X}_2} &= \frac{\mathcal{Z}[q_{\text{X}}^{\text{X}_2}]/\mathcal{Z}[(q_{\text{X}}^{\text{X}_2})^{\text{vac}}]}{\mathcal{Z}[q_{\text{M}}^{\text{M}_1}]/\mathcal{Z}[(q_{\text{M}}^{\text{M}_1})^{\text{vac}}]}\frac{\nu_{a_2}^2[H_{\bk}]}{\nu_{b_1}[H_{\bk}]\nu_{d_1}^{2}[H_{\bk}]}\nonumber\\
		&= \frac{\pf[q_{\text{X}}^{\text{X}_2}]}{\pf[q_{\text{M}}^{\text{M}_1}]}\frac{\pf[(q_{\text{M}}^{\text{M}_1})^{\text{vac}}]}{\pf[(q_{\text{X}}^{\text{X}_2})^{\text{vac}}]}
		\exp\left[-\frac{1}{4}\int d\left(2\log \frac{\det q_{(k_x, 0)}^{a_2}}{\det (q_{(k_x, 0)}^{a_2})^{\text{vac}}} - \log \frac{\det q_{(0, k_y)}^{b_1}}{\det (q_{(0, k_y)}^{b_1})^{\text{vac}}}  - 2\log  \frac{\det q_{(k_x, \pi)}^{d_1}}{\det (q_{(k_x, \pi)}^{d_1})^{\text{vac}}}\right) \right].
	\end{align}
\end{widetext}
Real-space pictures of topological phases are shown in Fig.~\ref{fig:R-AHSS}(c), as discussed in Ref.~\cite{Ono-Shiozaki-Watanabe2022}. 
From $(\mathcal{X}_1, \mathcal{X}_2)$ for a representative model of each phase, we find that $(\mathcal{X}_1, \mathcal{X}_2)$ can fully characterize $^{\phi}K_{G/\Pi}^{(z,c)}(T^2) = \mZ_2 \times \mZ_4$, as shown in Table~\ref{tab:p2_111}.


\section{Topological invariants for normal conducting phases}
\label{sec5}


So far, we have discussed topological invariants when $E_{1}^{p,-1}\ (p = 0, 1, 2)$ are free Abelian groups. 
Typically, this happens to time-reversal symmetric superconductors with conventional pairing symmetries.
On the other hand, this is not always the case for superconductors with unconventional pairing symmetries and normal conducting systems.
In particular, when we are interested in normal conducting systems, $E_{1}^{p,-1}$ always takes the form of $(\mZ_2)^{l}\ (l \in \mZ_{\geq 0})$. 
This is because an EAZ class of each irrep on a $p$-cell is always any one of A, AI, and AII, and only AI gives $\mZ_2$ as shown in Table~\ref{tab:EAZ}.
Here, let us discuss how to construct topological invariants for insulators and semimetals.

The construction of topological invariants for normal conducting phases also requires the similar two steps discussed in Secs.~\ref{sec:general_frame} and \ref{sec:topo_invariant}.
The only difference in the first step is the existence of $\mZ_2$-parts in $E_{1}^{p, -1}$. 
The discussions in Sec.~\ref{sec:general_frame} are easily generalized, as shown in Sec.~\ref{sec:X_TI}. 
On the other hand, due to the absence of any chiral symmetry, we need to develop a different characterization of $E_{1}^{1, -1}$ for normal conducting phases.
In Sec.~\ref{sec:z2inv_1skeleton}, we introduce the transition function as $\mZ_2$-valued quantity under some gauge conditions. 
As a result, we first construct topological invariants under some gauge conditions from AHSS.
Since the formulas with fixed gauges are generally not useful for actual computations, we then rewrite the formulas in terms of gauge-independent quantities. 
In the following, we explain our strategy through some examples. 
We leave a systematic implementation of gauge-invariant formulas as a future work. 

\subsection{Some generalities}

Before moving on to the construction of invariants on the 1-skeleton, we summarize in this section general remarks not limited to the 1-skeleton.

\subsubsection{Types of topological invariants}
In insulating systems, $E_2^{0,0}$ is a free abelian group, and $E_2^{1,-1}$ consists solely of $\mZ_2$. 
Since free abelian groups do not undergo nontrivial extensions, the topological invariants defined on the 1-skeleton take values in the group $\mZ_2$. 
On the other hand, the group $E_3^{2,-2}$ may be nontrivially extended by invariants on the 1-skeleton, so there can be, for example, $\mZ_4$ invariants.

\subsubsection{Notations}
We summarize the notations used in this section, which are different from those in Sec.~\ref{sec3}. 
Let $G$ be a magnetic point group, and $\phi: G \to \{\pm 1\}$ be a homomorphism specifying whether an element $g \in G$ is unitary or antiunitary. 
Introduce the notation for a matrix $X$, $X^{\phi_g} = X$ when $\phi_g=1$ and $X^{\phi_g}=X^*$ when $\phi_g=-1$. Let $p_g \in \text{O}(d)$ be the point group action of $g$, and $\bm{a}_g$ be a (fractional) translation.
The action on real space is $g:\bm{x}\mapsto p_g\bm{x} + \bm{a}_g$. 
Denote the group action on the momentum space $\bk \in T^d$ as $g:\bm{k} \mapsto g\bm{k} = \phi_g p_g \bm{k}$. Write $z^{\rm int}(g,h)$ as the factor system for internal degrees of freedom, and the factor system in the momentum space as
\begin{align}
    z_\bk(g,h) = z^{\rm int}(g,h) e^{-i\bm{k}\cdot (p_g \bm{a}_h+\bm{a}_h-\bm{a}_{gh})} 
\end{align}
for $g,h \in G$, where $z_\bk(g,h)$ satisfies the following cocycle condition 
\begin{align}
    &z_{g^{-1}\bk}(h,l)^{\phi_g} z_{\bk}(gh,l)^{-1} z_{\bk}(g,hl) z_\bk(g,h)^{-1} = 1  
\end{align}
for $g,h,l \in G$~ \cite{Shiozaki-Sato-Gomi2017}. 
Let $\{U_\bk(g)\}_{g \in G}$ be symmetry operators globally defined in the momentum space $T^d$, satisfying the following $\bk$-dependent cocycle condition:
\begin{align}
    U_{h\bk}(g)U_\bk(h)^{\phi_g}=z_{gh\bk}(g,h)U_\bk(gh),\quad g,h \in G.
    \label{eq:Ukg_group_str}
\end{align}
A periodic Hamiltonian $H_\bk$ in the momentum space $T^d$ satisfies the following symmetry 
\begin{align}
    U_\bk(g)H_\bk^{\phi_g}U_\bk(g)^\dag = H_{g\bk},\quad g \in G.
    \label{eq:G_sym}
\end{align}

\subsubsection{Classification of equivariant vector bundle}
\label{sec:Classification of equivariant vector bundle}
In general, the classification of isomorphism classes of vector bundles is to classify transition functions. 
Let $\{U_i\}_{i \in I}$ be a covering of the momentum space $T^d$ compatible with the symmetry, i.e., 
\begin{align}
	T^d = \bigcup_i U_i,
\end{align}
where $U_i$ is called \textit{patch} in the following.
Each patch $U_i$ corresponds to the Poincar\'e dual of a 0-cell $\bk_i$ in the cell decomposition of the AHSS.
By design, the action of $g$ on each patch $g(U_i) = \{g\bk \in T^d\ \vert\ \bk \in U_i\}$ is some patch, and the group action on patch labels is denoted as $i \mapsto g(i)$. 
Namely, $g(U_i) = U_{g(i)}$. 
Each patch $U_i$ is contractible, and the Bloch states $\Phi_{i,\bk}$ on each patch $U_i$ can be chosen continuously. 
Let $N$ be the number of occupied states. 
On a two-patch intersection $U_{ij} = U_i \cap U_j$, the transition function is defined as 
\begin{align}
    t_{ij,\bk} = \Phi_{i,\bk}^\dag \Phi_{j,\bk} \in \text{U}(N),\quad \bk \in U_{ij}.
\end{align}
The transition function $t_{ij,\bk}$ satisfies the following cocycle condition over a three-patch intersection $U_{ijl} = U_i\cap U_j \cap U_l$:
\begin{align}
    t_{ij,\bk}t_{jl,\bk}t_{li,\bk} = \mathds{1}_{N_{\rm occ}}, \quad \bk \in U_{ijl}.
\end{align}
The intersection $U_{ij}$ is generally not contractible and may have multiple connected components. 

Under certain symmetry constraints (gauge fixing conditions) on the Bloch states of each patch, the desired topological classification is obtained from the homotopy equivalence class of transition functions $t_{ij,\bk}$ invariant against residual gauge transformations.
In each patch, due to the symmetry relation in Eq.~\eqref{eq:G_sym}, the Bloch state $\Phi_{i,\bk}$ defines a unitary matrix, called \textit{sewing matrix}, $w_{i,\bk}(g \in G) \in \text{U}(N)$ by
\begin{align}
	\label{eq:gauge_cond_Phi}
    U_\bk(g) \Phi_{i,\bk}^{\phi_g} = \Phi_{g(i),g\bk} w_{i,\bk}(g),\quad \bk \in U_i,\quad g \in G.
\end{align}
Here, note that the matrix $w_{i,\bk}$, unlike $U_\bk(g)$, is defined only on the patch $\bk \in U_i$ and satisfies the same product structure (\ref{eq:Ukg_group_str}) as $U_\bk(g)$, namely,
\begin{align}
    w_{h(i),h\bk}(g) w_{i,\bk}(h)^{\phi_g} = z_{gh\bk}(g,h) w_{i,\bk}(gh), \quad g,h \in G. 
    \label{eq:cocycle_cond_w}
\end{align}
The following fact is the starting point of construction: 
\begin{thm}
	There exists a continuous unitary matrices $V_{i,\bk \in U_i}$ on patches $U_i$ such that 
	\begin{align}
		&w_{i,\bk}(g) V_{i,\bk}^{\phi_g} = V_{g(i),g\bk} e^{-i g (\bk-\bk_i)\cdot \bm{a}_g} w_{i,\bk_i}(g), \nonumber\\
		&\bk \in U_i, \quad g \in G, 
		\label{eq:theorem_canonical_gauge}
	\end{align}
	hold true. 
\end{thm}
See Appendix~\ref{app:proof_canonical_gauge} for a proof.
This implies that through a gauge transformation $\Phi_{i,\bk} \mapsto \Phi_{i,\bk} V_{i,\bk}$, one can always fix the gauge in such a way that the symmetry operator in each patch can be the symmetry operator $w_{i,\bk_i}(g)$ at the 0-cell $\bk_i$ up to the factor $e^{-i g (\bk-\bk_i)\cdot \bm{a}_g}$ determined solely by the magnetic space group data.
It is noted that the unitary equivalence class of the representation matrix $w_{i,\bk_i}(g)$ at the 0-cell $\bk_i$ is determined by the numbers of irreps of the little group $G_{\bk_i} = \{g \in G|g\bk_i=\bk_i\}$ at $\bk_i$. 
Consequently, the gauge fixing condition for each patch depends only on the representation at 0-cell $\bk_i$.

Let us choose a gauge fixing condition $\{w_{i,\bk}(g)\}_{i \in I, g\in G}$ that satisfies Eq.~\eqref{eq:cocycle_cond_w}. 
Under the gauge fixing condition \eqref{eq:gauge_cond_Phi}, the transition functions satisfy the following symmetry:
\begin{align}
    &w_{i,\bk}(g) t_{ij,\bk}^{\phi_g} w_{j,\bk}(g)^\dag = t_{g(i)g(j),g\bk}, \nonumber\\
    &\bk \in U_{ij}, \quad g \in G.
    \label{eq:sym_transition_func}
\end{align}
The residual gauge transformations, which leave the gauge fixing condition (\ref{eq:gauge_cond_Phi}) invariant, are given by 
\begin{align}
    \Phi_{i,\bk} \mapsto \Phi_{i,\bk} W_{i,\bk}, \quad W_{i,\bk} \in \text{U}(N), 
    \label{eq:residual_gauge_tr}
\end{align}
with the matrices $W_{i,\bk}$ on $U_i$ satisfying the condition 
\begin{align}
    &w_{i,\bk}(g) W_{i,\bk}^{\phi_g} w_{i,\bk}(g)^\dag
    =W_{g(i),g\bk},\nonumber\\
    &\bk \in U_i, \quad g \in G. 
    \label{eq:symm_gauge_tr}
\end{align}
Under the residual gauge transformation in \eqref{eq:residual_gauge_tr}, the transition functions change as 
\begin{align}
    t_{ij,\bk} \mapsto W_{i,\bk}t_{ij,\bk}W_{j,\bk}^\dag.
\end{align}
The set of homotopy equivalence classes of the transition functions $t_{ij,\bk}$, when divided by the equivalence relation generated by the residual gauge transformation in Eq.~\eqref{eq:residual_gauge_tr}, yields the data for topological invariants.

The classification of topological invariants defined by the transition functions $t_{ij,\bk}$ is generally difficult. 
However, when limiting the Bloch states to the 1-skeleton of momentum space, the topological invariants are classified by $E_2^{1,-1}$ and constructed explicitly, as discussed in the next section.

\subsection{The construction of $\mZ_2$ topological invariants over 1-skeleton}
\label{sec:z2inv_1skeleton}

In this section, we discuss the construction of $\mZ_2$ topological invariants defined from the occupied states over the 1-skeleton in momentum space. 
In the following discussions, we assume that any set of representations of occupied states at 0-cells satisfies the compatibility relations to ensure the existence of the gap over the 1-skeleton, resulting in that the set of representations at 0-cells is an element of $E_2^{0,0}$. 

\subsubsection{Introducing a gauge constraint over 1-cells}
\begin{figure}[t]
	\begin{center}
		\includegraphics[width=0.99\columnwidth]{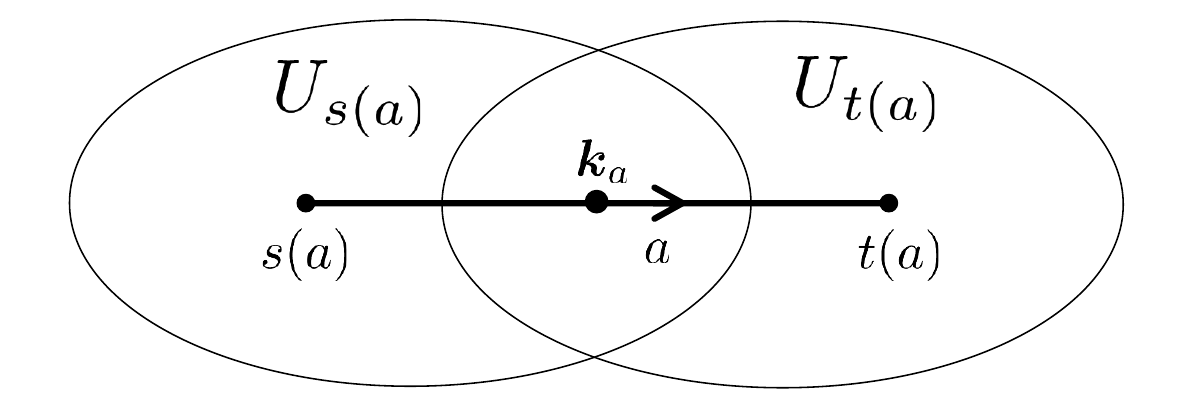}
		\caption{\label{fig:pathc_1cell}
			Illustration of 1-cell $a$ and its start point $s(a)$ and terminal point $t(a)$.
			The transition function is defined on the overlap of patches $U_{s(a)}$ and $U_{t(a)}$, which are the neighborhoods of 0-cells $s(a)$ and $t(a)$, respectively.
			The point $\bk_a$ is a representative point in the intersection $U_{s(a)} \cap U_{t(a)} \cap a$. 
   		}
	\end{center}
\end{figure}

Let $s(a)$ and $t(a)$ be the start and terminal point of a 1-cell $a$, respectively [see Fig.~\ref{fig:pathc_1cell}].  
We choose a representative point $\bk_a \in a$ and examine the symmetry constraint of the transition function at the point $\bk_a$:
\begin{align}
	\label{eq:transition_function_raw}
    t_{\bk_a} = \Phi_{s(a),\bk_a}^\dag \Phi_{t(a),\bk_a} \in \text{U}(N).
\end{align}

Let us define the subgroup $G_a = \{g \in G\ \vert \ g\bk = \bk {\rm\ for\ } ^\forall\bk \in a\}$, whose elements keep points on 1-cell $a$ invariant. 
The transition function then satisfies the symmetry constraint in Eq.~\eqref{eq:sym_transition_func}:
\begin{align}
    &w_{s(a),\bk_a}(g) t_{\bk_a}^{\phi_g} w_{t(a),\bk_a}(g)^\dag = t_{\bk_a}, \quad g \in G_a.
    \label{eq:sym_t}
\end{align}
Although $w_{s(a),\bk_a}(g)$ and $w_{t(a),\bk_a}(g)$ are unitary equivalent as representations of $G_a$ with the common factor system $z_{\bk_a}(g,h)$, $w_{s(a),\bk_a}(g)$ and $w_{t(a),\bk_a}(g)$ are generally different.
As a result, no apparent constraint arises in eigenvalues of the transition function $t_{\bk_a}$

To make restrictions on the eigenvalues explicit, we fix a representation matrix $w_{a,\bk}(g)\ (g\in G)$ for 1-cell $a$, which obeys the globally defined factor system $z_\bk(g,h)$: 
\begin{align}
    &w_{h(a),h\bk}(g) w_{a,\bk}(h)^{\phi_g} = z_{gh\bk}(g,h) w_{a,\bk}(gh),\nonumber \\
    &g,h \in G.
\end{align}
Here, $g(a)$ denotes a 1-cell mapped from 1-cell $a$ by $g \in G$.
The matrix $w_{a,\bk}(g)$ can be considered as a gauge fixing condition on 1-cell: 
\begin{align}
    U_{\bk}(g) \Phi_{a,\bk}^{\phi_g} = \Phi_{g(a),g\bk} w_{a,\bk}(g), \quad g \in G.
\end{align}

Let $V_{i\to a,\bk}$ denote the unitary matrix for a basis transformation from the 0-cell $i$ to an adjacent 1-cell $a$:
\begin{align}
    w_{i,\bk}(g) V_{i\to a,\bk}^{\phi_g} = V_{i\to a,\bk} w_{a,\bk}(g), \quad 
    g \in G_a.
    \label{eq:Via_def}
\end{align}
(As will be detailed later, $V_{i\to a,\bk}$ is not unique.)
Using this basis transformation, we redefine 
\begin{align}
    \tilde t_{\bk_a}:= V_{s(a) \to a,\bk_a}^\dag t_{\bk_a} V_{t(a)\to a,\bk_a}.
    \label{eq:tilde_t_def}
\end{align}
The transition function $\tilde t_{\bk_a}$ then satisfies the symmetry 
\begin{align}
    w_{a,\bk_a}(g) \tilde t_{\bk_a}^{\phi_g} w_{a,\bk_a}(g)^\dag = \tilde t_{\bk_a},\quad g \in G_a,
    \label{eq:sym_tilde_t}
\end{align}
imposing a constraint on the eigenvalues of $\tilde t_{\bk_a}$.

\subsubsection{Preliminary discussion: For block diagonalized $w_{a,\mathbf{k}_a}(g)$}
\label{sec:TI_z2number_Preliminary}
To examine the eigenvalue structure of $\tilde t_{\bk_a}$, we first consider the case where the representation matrices $w_{a,\bk_a}(g)$ are block diagonalized for each irrep. 

Let $G^0_{a} = \{g \in G_a | \phi_g = 1\}$ be the subgroup of $G_a$ consisting of unitary elements. 
Denote the irreducible character of $\alpha$-irrep of $G_a^0$ as $\chi^\alpha_\bk(g \in G_a^0)$. 
When $G_a \backslash G^0_a \neq \emptyset$, choose a representative ${\sf t} \in G_a \backslash G^0_a$, and ${\sf t}\alpha$ denotes the irrep related to $\alpha$ by ${\sf t}$. 
Note that ${\sf t}^2 \neq e$ in general.
The character of ${\sf t}\alpha$ is given by $\chi_{\bk}^{{\sf t}\alpha}(g) = \frac{z_{\bk_a}(g,{\sf t})}{z_{\bk_a}({\sf t},{\sf t}^{-1}g{\sf t})} \chi_{\bk_a}^\alpha({\sf t}^{-1}g{\sf t})^*$ for $^\forall g \in G_a^0$. 
Choose a set of representation matrices for each irrep and denote them as $\rho^\alpha_\bk(g)$. 
Let ${\cal D}_\alpha$ be the matrix dimension (representation dimension) of $\rho^\alpha_\bk(g)$.

We consider the following block diagonalized form for the representation matrix
\begin{align}
w_{a,\bk_a}(g) = \bigoplus_\alpha \rho^{\alpha}_{\bk_a}(g) \otimes \mathds{1}_{n_\alpha}\ (g\in G_a^0), 
\label{eq:rep_mat_w}
\end{align}
where $n_\alpha$ is the number of $\alpha$-irrep contained in $w_{\bk_a}(g)$.
With this choice, $\tilde t_{\bk_a}$ is block diagonalized as
\begin{align}
	\label{eq:tilde_t_alpha_def}
    \tilde t_{\bk_a} = \bigoplus_\alpha \mathds{1}_{{\cal D}_\alpha} \otimes \tilde t^\alpha_{\bk_a}, \quad 
    \tilde t^\alpha_{\bk_a} \in \text{U}(n_\alpha). 
\end{align}
Furthermore, due to ${\sf t}$-symmetry, the unitary matrix $\tilde t^\alpha_{\bk_a}$ of each sector is subjected to further constraint depending on the EAZ class of $\alpha$-irrep, as described below.

\paragraph{EAZ class A, i.e., no antiunitary symmetry present.}
In this case, there are no further restrictions on $\tilde t^\alpha_{\bk_a}$.
Since $\pi_0[\text{U}(n_\alpha)]=0$, no topologically nontrivial classification arises.
In particular, $\det \tilde t^\alpha_{\bk_a}$ can take any $\text{U}(1)$ value.

\paragraph{EAZ class A$_T$, i.e., antiunitary symmetry ${\sf t}$ exists but $\alpha$ and ${\sf t}\alpha$ are different.}
In this case, there are no further restrictions on $\tilde t^\alpha_{\bk_a}$. 
Since $\pi_0[\text{U}(n_\alpha)]=0$, no topologically nontrivial classification arises.
For $\rho^{\alpha}_\bk(g)$, the representation matrix of the irrep ${\sf t}\alpha$ can be fixed as $\rho^{{\sf t}\alpha}_\bk(g) = \frac{z_{\bk_a}(g,{\sf t})}{z_{\bk_a}({\sf t},{\sf t}^{-1}g{\sf t})} \rho_{\bk_a}^\alpha({\sf t}^{-1}g{\sf t})^*$ for $g\in G_a^0$. 
In doing so, since ${\sf t}({\sf t} \alpha) \sim \alpha$, there exists a unitary matrix $U\in \text{U}(\calD_\alpha)$ such that $\rho^{\alpha}_\bk(g)= \frac{z_\bk(g,{\sf t})}{z_\bk({\sf t},{\sf t}^{-1}g{\sf t})} U \rho^{{\sf t}\alpha}_\bk(g)^* U^\dag$ and $U$ is given by $z_\bk({\sf t},{\sf t})\rho^{\alpha}_\bk({\sf t}^2)$. 
As a result, the representation matrix of ${\sf t}$ in the $\alpha\oplus {\sf t}\alpha$ sector becomes 
\begin{align}
    w_{a,\bk}({\sf t})|_{(\alpha \oplus {\sf t}\alpha)^{\oplus n_\alpha}} = \begin{pmatrix}
        O & z_\bk({\sf t},{\sf t})\rho^{\alpha}_\bk({\sf t}^2) \\
        \mathds{1}_{\calD_\alpha} & O \\
    \end{pmatrix} \otimes \mathds{1}_{n_\alpha}.
\end{align}
Note that $\calD_\alpha=\calD_{{\sf t}\alpha}$ and $n_\alpha=n_{{\sf t}\alpha}$.
Under this representation, due to ${\sf t}$-symmetry, $\tilde t^{{\sf t}\alpha}_{\bk_a} = (\tilde t^{\alpha}_{\bk_a})^*$ holds. 
In particular,
\begin{align}
    \det \tilde t^\alpha_{\bk_a} \det \tilde t^{{\sf t}\alpha}_{\bk_a} = 1.
\end{align}

\paragraph{EAZ class AI.}
There exists a unitary matrix $\rho^\alpha_{\bk}({\sf t})\in \text{U}({\cal D}_\alpha)$ such that $\rho^\alpha_{\bk}({\sf t}) \rho^\alpha_{\bk}({\sf t})^* = z_{\bk}({\sf t},{\sf t}) \rho^\alpha_{\bk}({\sf t}^2)$~\cite{K-AHSS}.
The representation matrix of ${\sf t}$ of $\alpha$-irrep is given by $w_{a,\bk}|_{\alpha^{\oplus n_\alpha}} = \rho^\alpha_{\bk}(\sf t) \otimes \mathds{1}_{n_\alpha}$. 
Consequently, $(\tilde t^\alpha_{\bk_a})^*=\tilde t^\alpha_{\bk_a}$ due to ${\sf t}$-symmetry, meaning that $\tilde t^\alpha_{\bk_a} \in \text{O}(n_\alpha)$. 
Since $\pi_0[\text{O}(n_\alpha)]=\mZ_2$, a $\mZ_2$ classification arises from the transition functions on 1-cell, which is determined by the sign 
\begin{align}
    \det \tilde t^\alpha_{\bk_a} \in \{\pm 1\}.
    \label{eq:z2_inv_1cell}
\end{align}

It turns out that the sign $\det \tilde t^\alpha_{\bk_a}$ depends on the basis transformation $V_{i\to a,\bk}$. 
For a triple of representation matrices $w_{i,\bk}(g), w_{a,\bk}(g)$ and the basis transformation $V_{i\to a,\bk}$, let us consider a substitution of $V_{i\to a,\bk}$ as 
\begin{align}
    &V_{i\to a,\bk} \mapsto V_{i\to a,\bk} \delta V_{i \to a,\bk}, \\
    &w_{a,\bk}(g) (\delta V_{i\to a,\bk})^{\phi_g} w_{a,\bk}(g)^\dag = \delta V_{i\to a,\bk}. 
    \label{eq:deltaV_sym}
\end{align}
This substitution does not alter the relation (\ref{eq:Via_def}).
Considering the form of $w_{a,\bk}(g)$ in Eq.~\eqref{eq:rep_mat_w} and the symmetry relation in Eq.~\eqref{eq:deltaV_sym}, we have
\begin{align}
    &\delta V_{i\to a,\bk} = \bigoplus_\alpha \mathds{1}_{{\cal D}_\alpha} \otimes \delta V_{i\to a,\bk}^\alpha. 
\end{align}
In particular, for the block of $\alpha$-irrep whose EAZ class is AI, $\delta V_{i\to a,\bk}^\alpha \in \text{O}(n_\alpha)$. 
This substitution leads to a change in the unitary matrix $\tilde t^\alpha_{\bk_a}$ defined in Eq.~\eqref{eq:tilde_t_alpha_def}:
\begin{align}
    \tilde t^\alpha_{\bk_a} \mapsto (\delta V^\alpha_{s(a)\to a,\bk_a})^\dag \tilde t^\alpha_{\bk_a} \delta V^\alpha_{t(a)\to a,\bk_a}. 
\end{align}
Consequently, the sign $\det \tilde t^\alpha_{\bk_a}$ changes as well:
\begin{align}
    \det \tilde t^\alpha_{\bk_a}\mapsto 
    \det (\delta V^\alpha_{s(a)\to a,\bk_a})^\dag 
    \det \tilde t^\alpha_{\bk_a}
    \det \delta V^\alpha_{t(a)\to a,\bk_a}. 
\end{align}

\paragraph{EAZ class AII.}
Due to Kramers' degeneracy, $n_\alpha$ is an even number. 
There exists a unitary matrix $\rho^\alpha_{\bk}({\sf t})\in \text{U}({\cal D}_\alpha)$ such that $\rho^\alpha_{\bk}({\sf t}) \rho^\alpha_{\bk}({\sf t})^* = -z_{\bk}({\sf t},{\sf t}) \rho^\alpha_{\bk_a}({\sf t}^2)$. 
Using this matrix $\rho^\alpha_{\bk_a}({\sf t})$, the representation matrix of ${\sf t}$ in the $\alpha$-block is given by $w_{a,\bk}|_{\alpha^{\oplus n_\alpha}} = \rho^\alpha_{\bk}(\sf t) \otimes (\ii \sigma_y) \otimes \mathds{1}_{n_\alpha/2}$. 
Consequently, due to ${\sf t}$-symmetry, $(\ii \sigma_y)(\tilde t^\alpha_{\bk_a})^* (\ii \sigma_y)^\dag=\tilde t^\alpha_{\bk_a}$, meaning that the $\tilde t^\alpha_{\bk_a}$ matrix belongs to the symplectic group $\text{Sp}(n_\alpha/2) = \{X \in \text{U}(n_\alpha)\ \vert \ X^\top (\ii\sigma_y) X = \ii \sigma_y\}$. 
Since $\pi_0[\text{Sp}(n_\alpha/2)]=0$, no nontrivial classification arises. 
Furthermore, as the eigenvalues of the matrix $\tilde t^\alpha_{\bk_a}$ appear in complex conjugate pairs $(\lambda,\lambda^*)$, $\det \tilde t^\alpha_{\bk}=1$.

\subsubsection{Gauge invariant expression of $\det \tilde t^\alpha_{\mathbf{k}_a}$}
\label{sec:Gauge_invariant_expression_u1}
The $\text{U}(1)$-valued quantity $\det \tilde t^\alpha_{\bk_a}$ is definable even if the representation matrix $w_{a,\bk}(g)$ is not block diagonalized as in the expression \eqref{eq:rep_mat_w}.
By introducing the orthogonal projection onto the $\alpha$-irrep defined by 
\begin{align}
    P^\alpha_{a,\bk}=\frac{{\cal D}_\alpha}{|G_a^0|} \sum_{g \in G_a^0} \chi_{\bk}^{\alpha}(g)^* w_{a,\bk}(g), 
\end{align}
the projection of the transition function $\tilde t_{\bk_a}$ onto the $\alpha$-sector is given by
\begin{align}
    P^\alpha_{a,\bk_a} \tilde t_{\bk_a} P^\alpha_{a,\bk_a} = P^\alpha_{a,\bk_a} \tilde t_{\bk_a} = \tilde t_{\bk_a} P^\alpha_{a,\bk_a}.
\end{align}
The non-zero eigenvalues of the projection $P^\alpha_{a,\bk_a} \tilde t_{\bk_a} P^\alpha_{a,\bk_a}$ consist of $n_\alpha$ eigenvalues $\lambda_1,\dots \lambda_{n_\alpha}$, and each of them are ${\cal D}_\alpha$-fold degenerated. 
We define
\begin{align}
    \label{eq:xi_alpha_def}
    \xi^\alpha_{\bk_a}(\tilde t_{\bk_a}):= \prod_{i=1}^{n_\alpha} \lambda_i.
\end{align}
The discussion in the previous section can be summarized for each EAZ class as:
\begin{align}
\begin{array}{ll}
    {\rm A}: & \xi_{\bk_a}^\alpha(\tilde t_{\bk_a}) \in \text{U}(1), \\
    {\rm A}_T: & \xi_{\bk_a}^\alpha(\tilde t_{\bk_a})=\xi_{\bk_a}^{{\sf t}\alpha}(\tilde t_{\bk_a})^* \in \text{U}(1),  \\
    {\rm AI}: & \xi_{\bk_a}^\alpha(\tilde t_{\bk_a}) \in \{\pm 1\}, \\
    {\rm AII}: & \xi_{\bk_a}^\alpha(\tilde t_{\bk_a})=1.  \\
\end{array}
\label{eq:xi_and_EAZ}
\end{align}

Some remarks are in order. 
\begin{itemize}
	\setlength{\itemsep}{-0.5pt}
	\item[---] The $\text{U}(1)$-valued quantity $\xi_{\bk_a}^\alpha(\tilde t_{\bk_a})$ behaves as a product with respect to direct sums:
	\begin{align}
		\xi_{\bk_a}^\alpha(\tilde t_{\bk_a} \oplus \tilde t'_{\bk_a})
		=\xi_{\bk_a}^\alpha(\tilde t_{\bk_a}) \xi_{\bk_a}^\alpha(\tilde t'_{\bk_a}).
	\end{align}
	\item[---] As can be seen from the expression
	\begin{align}
		\xi_{\bk_a}^\alpha(\tilde t_{\bk_a}) 
		= 
		\xi_{\bk_a}^\alpha\left(V_{s(a)\to a,\bk_a}^\dag \Phi_{s(a),\bk_a}^\dag \Phi_{t(a),\bk_a} V_{t(a)\to a,\bk_a}\right),
	\end{align}
	it is emphasized again that the $\text{U}(1)$-valued quantity $\xi_{\bk_a}^\alpha(\tilde t_{\bk_a})$ depends on the basis transformation matrix $V_{i\to a,\bk}$.
	\item[---] For irreps with EAZ class AI, a set of signs $\xi_{\bk_a}^\alpha(\tilde t_{\bk_a})$ represents an entry of $E_1^{1,-1}$ in the AHSS.
\end{itemize}

\subsubsection{Residual patch gauge transformation}
We see how the residual patch gauge transformation (\ref{eq:residual_gauge_tr}) changes the $\text{U}(1)$-valued quantity $\xi^\alpha_{\bk_a}(\tilde t_{\mathbf{k}_a})$. 
The transition function $\tilde t_{\bk_a}$ defined in Eq.~\eqref{eq:tilde_t_def} changes under a gauge transformation as 
\begin{align}
    &\tilde t_{\bk_a} \mapsto \tilde W_{s(a),\bk_a} \tilde t_{\bk_a} \tilde W_{t(a),\bk_a}^\dag, \\
    &\tilde W_{i,\bk}:= V_{i\to a,\bk}^\dag W_{i,\bk} V_{i\to a,\bk},
\end{align}
where the matrices $\tilde W_{s(a),\bk}$ and $\tilde W_{t(a),\bk}$ satisfy the same symmetry \eqref{eq:deltaV_sym} as $\delta V_{i\to a,\bk}$ with the representation matrix $w_{a,\bk}(g)$;~%
namely, 
\begin{align}
	w_{a,\bk}(g) \tilde W_{i,\bk}^{\phi_g}  w_{a,\bk}(g)^\dag = \tilde W_{i,\bk} \ \text{ for $g\in G_a$}.
\end{align}
Thus, all of $\tilde W_{s(a),\bk_a}, \tilde t_{\bk_a},$ and $\tilde W_{t(a),\bk_a}^\dag$ are block diagonalized in $\alpha$-sector.
Consequently,
\begin{align}
    &\xi^\alpha_{\bk_a}(\tilde W_{s(a),\bk_a} \tilde t_{\bk_a} \tilde W_{t(a),\bk_a}^\dag)\nonumber\\
    &=\xi^\alpha_{\bk_a}(\tilde W_{s(a),\bk_a}) \xi^\alpha_{\bk_a}(\tilde t_{\bk_a}) \xi^\alpha_{\bk_a}(\tilde W_{t(a),\bk_a}^\dag)
\end{align}
holds, meaning that it becomes the product of each $\text{U}(1)$ value. 
Moreover, the $\text{U}(1)$ value $\xi^\alpha_{\bk_a}(\tilde t_{\bk_a})$ in Eq.~\eqref{eq:xi_alpha_def} is defined by eigenvalues, and thus it does not depend on the choice of the representation matrix $w_{a,\bk}(g \in G_a)$.
Then, it follows that
\begin{align}
    &\xi^\alpha_{\bk_a}(\tilde W_{s(a),\bk_a} \tilde t_{\bk_a} \tilde W_{t(a),\bk_a}^\dag)\nonumber\\
    &=\xi^\alpha_{\bk_a}(W_{s(a),\bk_a}) \xi^\alpha_{\bk_a}(\tilde t_{\bk_a}) \xi^\alpha_{\bk_a}(W_{t(a),\bk_a}^\dag).
\end{align}

In particular, when the EAZ class of $\alpha$-irrep is AI, the value $\xi^\alpha_{\bk_a}(W_{s(a),\bk_a})$ is quantized to $\mZ_2$, thus it does not depend on the momentum $\bk \in a$ on the 1-cell $a$. 
Therefore, it coincides with the sign at the 0-cell $s(a)$:
\begin{align}
    \xi^\alpha_{\bk_a}(W_{s(a),\bk_a}) 
    = \left.\xi^\alpha_{\bk_a}(W_{s(a),\bk_{a}}) \right|_{\bk_a \to \bk_{s(a)}}\quad 
    \mbox{for AI}. 
\end{align}
Recall that momentum at 0-cell $i$ is denoted as $\bk_i$.
The same applies to $\xi^\alpha_{\bk_a}(W^\dag_{t(a),\bk_a})$.

It should be noted that, although the above $\left. \xi^\alpha_{\bk_a}(W_{i,\bk_a})\right|_{\bk_a \to \bk_i}$ is defined at $0$-cell $\bk_i$, it is computed for irreps on the adjacent $1$-cell.
On the other hand, with the stabilizer group $G_{\bk_i} = \{g \in G \ |\ g \bk_i = \bk_i\}$, a $\text{U}(1)$-valued quantity $\xi^{\beta(\bk_i)}_{\bk_i}(W_{i, \bk_i})$ is similarly defined for each irrep $\beta(\bk_i)$ of the unitary subgroup $G_{\bk_i}^0 = \ker \phi \cap G_{\bk_i}$.
The irreducible decomposition of irrep $\beta(\bk_i)$ of $G_{\bk_i}^0$ into the irreps $\{\alpha\}_\alpha$ of $G^0_a$ is given by
\begin{align}
    \beta(\bk_i)|_{G_a^0} = \bigoplus_\alpha \alpha^{\oplus n^{\beta(\bk_i)}_\alpha}.
\end{align}
Then, the following relation holds:
\begin{align}
	\label{eq:eq:xi_1cell_decom}
    \left. \xi^\alpha_{\bk_a}(W_{i,\bk_a})\right|_{\bk_a \to \bk_i}
    = \prod_{\beta(\bk_i)} \left[\xi^{\beta(\bk_i)}_{\bk_i}(W_{i, \bk_i})\right]^{n^{\beta(\bk_i)}_{\alpha}}.
\end{align}
For proof, see Appendix~\ref{sec:Compatibility relation of non-Hermitian matrix}.

Note that if $G_a$ includes an antiunitary element ${\sf t}$, then $G_{\bk_i}$ also includes ${\sf t} \in G_{\bk_i}$.
For an $\alpha$-irrep of $G_a$ with EAZ class AI, it follows that the EAZ of the irrep $\beta(\bk_i)$ on the right-hand-side of Eq.~\eqref{eq:eq:xi_1cell_decom} is either AI, AII, or A$_T$. 
Furthermore, considering Eq.~\eqref{eq:xi_and_EAZ}, the contribution from AII is $\xi^{\beta(\bk_i)}_{\bk_i}(W_{i, \bk_i})=1$, and from A$_T$ is $\xi^{\beta(\bk_i)}_{\bk_i}(W_{i, \bk_i}) \xi^{{\sf t}\beta(\bk_i)}_{\bk_i}(W_{i, \bk_i})=1$, leaving only contributions from AI.
Therefore, we obtain
\begin{align}
    \left.\xi^{\alpha \in {\rm AI}}_{\bk_a}(W_{i, \bk_a})\right|_{\bk_a \to \bk_i} = 
    \prod_{\beta(\bk_i) \in {\rm AI}} 
    \left[\xi^{\beta(\bk_i)}_{\bk_i}(W_{i, \bk_i})\right]^{n^{\beta(\bk_i)}_\alpha}.
\end{align}
Here, $\prod_{\beta(\bk_i) \in {\rm AI}}$ denotes the product over irreps of $G^0_{\bk_i}$ whose EAZ is given by AI.
Eventually, the change in the sign $\xi^{\alpha \in {\rm AI}}_{\bk_a}(\tilde t_{\mathbf{k}_a}) \in \{\pm 1\}$ under the residual patch gauge transformation (\ref{eq:residual_gauge_tr}) is given by
\begin{align}
    \xi^{\alpha \in {\rm AI}}_{\bk_a}(\tilde t_{\mathbf{k}_a}) 
    &\mapsto \xi^{\alpha \in {\rm AI}}_{\bk_a}(\tilde t_{\mathbf{k}_a}) \nonumber\\
    &\times \prod_{\beta(\bk_{s(a)}) \in {\rm AI}} \left[\xi^{\beta(\bk_{s(a)})}_{\bk_{s(a)}}(W_{s(a),\bk_{s(a)}})\right]^{n^{\beta(\bk_{s(a)})}_\alpha} \nonumber\\
    &\times \prod_{\beta(\bk_{t(a)}) \in {\rm AI}} \left[\xi^{\beta(\bk_{t(a)})}_{\bk_{t(a)}}(W_{t(a),\bk_{t(a)}})\right]^{n^{\beta(\bk_{t(a)})}_\alpha}.
    \label{eq:eq:xi_1cell_decom_AI}
\end{align}
This relation (\ref{eq:eq:xi_1cell_decom_AI}) is nothing but the differential $d_1^{0,-1}:E_1^{0,-1} \to E_1^{1,-1}$ in the AHSS, and therefore, the $\mZ_2$ invariants over the 1-skeleton are classified by the group $E_{1}^{1,-1}/\im d_1^{0,-1}$. 

The group $E_{1}^{1,-1}/\im d_1^{0,-1}$ includes $\mZ_2$ invariants that detect gapless points at points inside 2-cells, which are called ``$\pi$-Berry phase'' in literature.
Such $\mZ_2$ invariants are given by the coimage ${\rm Coim}\ \hspace{-0.5mm}d_1^{1,-1}$ of the differential $d_1^{1,-1}:E_1^{1,-1} \to E_1^{2,-1}$ and depend on a choice of cell decomposition.
In contrast, $E_2^{1,-1}=\ker d_1^{1,-1}/\im d_1^{0,-1}$ is independent of cell decomposition.
Therefore, $E_2^{1,-1}$ gives the classification of $\mZ_2$ invariants that characterize gapped phases over $2$-skeleton, i.e., insulators and gapless points at a general point inside the 3-cell, independent of the cell decomposition.

\subsubsection{Gauge-invariant product for $\mZ_2$ invariant}
\label{sec:X_TI}
The gauge invariant combination of the signs $\xi^{\alpha \in {\rm AI}}_{\bk_a}(\tilde t_{\mathbf{k}_a})$ constitutes a $\mZ_2$ invariant, which can be calculated from the first differential in AHSS. 
The construction is parallel to that discussed in Section \ref{sec:general_frame}, except that it is $\mZ_2$-valued rather than $\mZ$-valued. 
Hence, only the results are briefly stated here.

Let $E_1^{p,-1} = \bigoplus_{i=1}^{N_p} \mZ_2[\bm{b}_i^{(p)}]$. We denote the matrix composed of basis vectors as ${\cal B}^{(p)} = (\bm{b}_1^{(p)},\dots,\bm{b}_{N_p}^{(p)})$, and write the first differential as
\begin{align}
    d_1^{p,-1}\left( {\cal B}^{(p)}\right)
    ={\cal B}^{(p+1)} M_{d_1^{p,-1}},
\end{align}
where $M_{d_1^{p,-1}}$ is a $\mZ_2$-valued $N_{p+1} \times N_p$ matrix.
Let the Smith normal form of $M_{d_1^{1,-1}}$ be
\begin{align}
    U^{(1)} M_{d_1^{1,-1}} V^{(1)}=\begin{pmatrix}
      \mathds{1}_{r_1}&O\\  
      O&O\\
    \end{pmatrix},
\end{align}
and define the matrix $Y$ by
\begin{align}
[V^{(1)}]^{-1} M_{d_1^{0,-1}}
    = \begin{pmatrix}
        O_{r_1 \times N_0}\\
        Y
    \end{pmatrix}.
\end{align}
The Smith normal form of $Y$ is denoted by
\begin{align}
    U^{(0)}YV^{(0)}= 
    \begin{pmatrix}
      \mathds{1}_{R_0}&O\\  
      O&O\\
    \end{pmatrix},
\end{align}
and the corresponding basis transformation matrix is defined by
\begin{align}
    X^{(1)} = V^{(1)}(\mathds{1}_{r_1} \oplus [U^{(0)}]^{-1}).
\end{align}
We introduce the vectors $\bm{x}_1,\dots,\bm{x}_{N_1}$ by
\begin{align}
    [X^{(1)}]^{-1} := \left(\begin{array}{cccc}
		\bm{x}_{1} &
		\bm{x}_{2} &
		\cdots &
		\bm{x}_{N_1}
	\end{array}\right)^{\top}.
\end{align}
For a gapped band structure $E$ on the 1-skeleton, let us define a $\mZ_2$-valued $\nu_i(E)\in \{0,1\}$ as
\begin{align}
    (-1)^{\nu_i(E)} := \prod_{a,\alpha} [\xi^{\alpha}_{\bk_a}(\tilde t_{\bk_a})]^{[x_i]_{(a,\alpha)}},\quad i=1,\dots,N_1, 
\end{align}
where the components of the vector $\bm{x}_i$ are indexed by the pairs $(a,\alpha)$ representing the basis of $E_1^{1,-1}$, consisting of a 1-cell $a$ and an $\alpha$-irrep on $a$ whose EAZ class is AI. 
Thus, $\nu_1(E),\dots,\nu_{r_1}(E)$ correspond to $\mZ_2$ invariants that detect gapless points in the 2-cell protected by the $\pi$-Berry phase within the given cell decomposition, while $\nu_{r_1+R_0+1}(E),\dots,\nu_{N_1}(E)$ correspond to $\mZ_2$ invariants characterizing the gapped Bloch wave functions on the 1-skeleton.

On the other hand, $\nu_{r_1+1}(E),\dots,\nu_{r_1+R_0}(E)$ are not invariants as they change under residual patch gauge transformations.

\subsubsection{Incompatibility of $\mZ_2$ invariants with band sum}
\label{sec:Incompatibility of Z2 invariants with band sum}
As noted in Secs.~\ref{sec:TI_z2number_Preliminary} and \ref{sec:Gauge_invariant_expression_u1}, the sign $\xi^{\alpha \in {\rm AI}}_{\bk_a}(\tilde t_{\mathbf{k}_a}) \in \{\pm 1\}$ depends on the basis transformation $V_{i \to a,\bk}$.
Therefore, it is not immediately clear whether the $\mZ_2$ invariant $\nu(E)$ maintains an additive structure for the direct sum of bands $E \oplus F$, i.e., whether $\nu(E \oplus F) \equiv \nu(E) + \nu(F)$ holds, where ``$\equiv$'' indicates that the values of the right-hand and left-hand sides are equal modulo two in the following.
As discussed below, we discover that the relation does not hold and that a correction term is involved.

As we are interested in the direct sum $E \oplus F$ of bands, we introduce an index $B \in \{E,F\}$ specifying the bands $E$ and $F$.
Now, we order the irreps at each 0-cell $i$, which is labeled by $\beta(\bk_i) = 1(\bk_i),\dots,m_i(\bk_i)$.
Similarly, for each 1-cell $a$, order the irreps as $\alpha(\bk_a)=1(\bk_a),\dots,m_a(\bk_a)$ and denote the representation matrix of irrep $\alpha(\bk_a)$ as $u^{\alpha(\bk_a)}_{\bk_a}(g)$.
The irreducible decompositions of band $B$ at 0-cell $i$ and 1-cell $a$ can be written as $\bigoplus_{\beta=1}^{m_i} n^B_{\beta(\bk_i)} \beta(\bk_i)$ and $\bigoplus_{\alpha=1}^{m_a} n^B_{\alpha(\bk_a)} \alpha(\bk_a)$ with $n^B_{\beta(\bk_i)} \in \mZ_{\geq 0}$ and $n^B_{\alpha(\bk_a)} \in \mZ_{\geq 0}$, respectively.

After some calculations discussed in Appendix~\ref{app:sum_E_F}, we find that 
\begin{widetext}
	\begin{align}
		\label{eq:sum_E_F}
		\nu_i(E \oplus F)
		\equiv \nu_i(E)+\nu_i(F)+\delta \nu_i(E|_{E_2^{0,0}},F|_{E_2^{0,0}}), 
	\end{align}
	where
	\begin{align}
		(-1)^{\delta \nu_i(E|_{E_2^{0,0}},F|_{E_2^{0,0}})} 
		:=
		&\prod_{(a,\alpha)} \left[\delta\xi^{\alpha(\bk_a)}_{\bk_a,s(a) \to a}(E|_{E_2^{0,0}},F|_{E_2^{0,0}}) \delta\xi^{\alpha(\bk_a)}_{\bk_a,t(a) \to a}(E|_{E_2^{0,0}},F|_{E_2^{0,0}})\right]^{[x_i]_{(a,\alpha)}}; \\
		\delta\xi^{\alpha(\bk_a)}_{\bk_a,i \to a}(E|_{E_2^{0,0}},F|_{E_2^{0,0}})
		&=(-1)^{\sum_{1\leq \beta_F<m_i}\sum_{\beta_F<\beta_E\leq m_i}
			n^F_{\beta_F(\bk_i)}n^{\beta_F(\bk_i)}_{\alpha(\bk_a)}
			n^E_{\beta_E(\bk_i)}n^{\beta_E(\bk_i)}_{\alpha(\bk_a)}     
		}\in \{\pm 1\},
		\label{eq:delta_xi_formula}
	\end{align}
where the notation $B|_{E_2^{0,0}}$ represents the restriction to the 0-cell, and the correction term indeed depends only on elements of $E_2^{0,0}$ in the 0-cell.
See Appendix~\ref{app:sum_E_F} for a derivation of Eqs.~\eqref{eq:sum_E_F}-\eqref{eq:delta_xi_formula}.
It is important to note that $\nu_i(E \oplus F) \equiv \nu_i(F \oplus E)$ may not always hold true.

\subsubsection{Quadratic refinement and redefinition of $\mZ_2$ invariants}
\label{sec:redef_inv}
As seen in Eq.~\eqref{eq:sum_E_F}, the introduced $\mZ_2$ invariants do not follow the sum rule. 
Then, it is natural to ask if it is possible to redefine $\mZ_2$ invariants that satisfy linearity for the direct sum of bands. 
Here we introduce a technique, known as \textit{quadratic refinement}, to construct the desired $\mZ_2$ invariants from $\nu$.

From Eq.~\eqref{eq:delta_xi_formula}, we have
\begin{align}
    \delta \nu_i(E|_{E_2^{0,0}},F|_{E_2^{0,0}})
    &\equiv 
    \sum_{(a,\alpha)} [x_i]_{(a,\alpha)} \Bigg[
    \sum_{\substack{\beta_E(\bk_{s(a)}),\beta_F(\bk_{s(a)})\\1\leq \beta_F(\bk_{s(a)})<\beta_E(\bk_{s(a)})\leq m_{s(a)}}}
    n^F_{\beta_F(\bk_{s(a)})}n^{\beta_F(\bk_{s(a)})}_{\alpha(\bk_a)}
    n^E_{\beta_E(\bk_{s(a)})}n^{\beta_E(\bk_{s(a)})}_{\alpha(\bk_a)} \nonumber\\
    &\qquad +
    \sum_{\substack{\beta_E(\bk_{t(a)}),\beta_F(\bk_{t(a)})\\1\leq \beta_F(\bk_{t(a)})<\beta_E(\bk_{t(a)})\leq m_{t(a)}}}
    n^F_{\beta_F(\bk_{t(a)})}n^{\beta_F(\bk_{t(a)})}_{\alpha(\bk_a)}
    n^E_{\beta_E(\bk_{t(a)})}n^{\beta_E(\bk_{t(a)})}_{\alpha(\bk_a)}    
    \Bigg] \quad \mod 2. 
\end{align}
\end{widetext}
As the number of irreps behaves linearly with respect to the direct sum of bands, $n^{E\oplus F}_{\beta(\bk_i)} = n^{E}_{\beta(\bk_i)}+n^F_{\beta(\bk_i)}$, the correction $\delta \nu_i(E|_{E_2^{0,0}},F|_{E_2^{0,0}})$ is a bilinear form
\begin{align}
    \delta \nu_i: E_2^{0,0} \times E_2^{0,0} \to \mZ_2. 
\end{align}

We aim to redefine the invariant $\nu_i(E)$ using $\delta \nu_i(E|_{E_2^{0,0}},F|_{E_2^{0,0}})$ so that the redefined $\mZ_2$ invariant satisfies linearity for the direct sum of bands. 
As will be discussed later, such a redefinition is possible if $\delta \nu_i$ is symmetric.
Based on an empirical rule, we conjecture the following:
\begin{conj}
	The bilinear form $\delta \nu_i$ is symmetric for the invariants of gapless points $i \in \{1,\dots,r_1\}$ and for the invariants of gapped insulators $i \in \{r_1+R_0+1,\dots,N_1\}$. 
	In other words,
	\begin{align}
   		&\delta \nu_i(\bm{n},\bm{n}') \equiv \delta \nu_i(\bm{n}',\bm{n}), \quad \bm{n},\bm{n}' \in E_2^{0,0} \nonumber \\
    	&\mbox{for }i \in \{ 1,\dots,r_1, r_1+R_0+1,\dots,N_1\}
    	\label{eq:delta_nu_sym}
	\end{align}
	holds true.
\end{conj}
We numerically verified conjecture (\ref{eq:delta_nu_sym}) to be correct for the 528 types of 2D magnetic layer and 1651 types of 3D magnetic space groups in spinless and spinful electronic systems.
We leave an analytical proof applicable to arbitrary spatial dimensions and factor systems for a future problem.
Note that for indices $i \in \{r_1,\dots,r_1+R_0\}$ for gauge-dependent ones, and for representations $\bm{n},\bm{n}' \in E_1^{0,0}$ that do not satisfy the compatibility relations, Eq.~\eqref{eq:delta_nu_sym} does not hold.

For a given symmetric bilinear form $\delta \nu_i$, there exists a function $q_i: E_2^{0,0} \to \mZ_2$ (note that it is not linear) satisfying:
\begin{align}
    &\delta \nu_i(\bm{n},\bm{n}') \equiv q_i(\bm{n}+\bm{n}') + q_i(\bm{n}) +q_i(\bm{n}'), \nonumber\\
    &i \in \{ 1,\dots,r_1, r_1+R_0+1,\dots,N_1\}.
\end{align}
The function $q_i$ is referred to as the quadratic refinement of $\delta \nu_i$.
This quadratic refinement $q_i$ is not unique and has the redundancy of ${\rm Hom}(E_2^{0,0},\mZ_2)$.
For the existence proof, see Appendix~\ref{sec:quadratic_refinement}.
We redefine $\nu_i(E)$ using this $q_i$ as follows:
\begin{align}
    &\tilde \nu_i(E) :\equiv \nu_i(E) + q_i(E|_{E_2^{0,0}}), \nonumber\\
    &i \in \{ 1,\dots,r_1, r_1+R_0+1,\dots,N_1\}.
    \label{eq:z2inv_redefined}
\end{align}
Then, 
\begin{align}
    &\tilde \nu_i(E \oplus F) \equiv \tilde \nu_i(E) + \tilde \nu_i(F) \quad \mod 2, \nonumber\\
    &i \in \{ 1,\dots,r_1, r_1+R_0+1,\dots,N_1\}, 
\end{align}
holds true, constituting a $\mZ_2$ invariant that behaves linearly with respect to the direct sum of bands.

A generic form of the quadratic refinement $q_i$ is given below.
Choose a basis set of $E_2^{0,0}$ such that $E_2^{0,0} = \bigoplus_{\rho=1}^d \mZ[\bm{b}_\rho]$ and write $[\delta \nu_i]_{\rho\sigma} = \delta \nu_i(\bm{b}_\rho,\bm{b}_\sigma)$.
Using the floor function $\lfloor x \rfloor = \max \{n \in \mZ \ \vert \ n \leq x\}$, a generic form of $q_i$ from (\ref{eq:qr_formula}) is
\begin{align}
    q_i\left(\sum_{\rho=1}^d n_\rho \bm{b}_\rho\right)
    &\equiv \sum_{\rho=1}^d \left\lfloor \frac{n_\rho}{2} \right\rfloor [\delta \nu_i]_{\rho\rho} \nonumber\\
    &+ \sum_{1\leq \rho<\sigma \leq d} [\delta \nu_i]_{\rho\sigma} n_\rho n_\sigma
    + \sum_{\rho=1}^d a_\rho n_\rho
    \label{eq:quadratic_refinement_result}
\end{align}
where $(a_1,\dots,a_d) \in \{0,1\}^{\times d}$ represent the redundancy in quadratic refinement and can be freely chosen.

\subsubsection{Comments on the approach from symmetry operators}
Without imposing symmetry, the transition functions on 1-cells is an element of the unitary group $\text{U}(N)$ which is path connected, allowing for the existence of global and continuous Bloch states $\Phi_\bk$ on the 1-skeleton of the momentum space.
Then, by defining $w_\bk(g) = \Phi_{g\bk}^\dag U_{\bk}(g) \Phi_\bk^{\phi_g}$ for $g\in G$, a continuous matrix of symmetry transformations $w_\bk(g)$ on the 1-skeleton is obtained.
It is expected that the set of the homotopy equivalence classes of symmetry matrices $\{w_\bk(g)\}_{g \in G}$ up to gauge transformations of the Bloch state $\Phi_\bk$ may provide $\mZ_2$ invariants.
This paper does not discuss this approach further and leaves it as a future direction.

\subsection{Symmetry-enriched Berry phase}
\label{sec:Symmetry-enriched Berry phase}
In the previous section \ref{sec:z2inv_1skeleton}, we constructed a $\mZ_2$ invariant \eqref{eq:z2inv_redefined} defined on the 1-skeleton of the momentum space under the gauge fixing condition (\ref{eq:gauge_fixing_patch_irrep}).
However, requirements of continuous Bloch states and symmetry constraints are not practical for numerical calculations.
In this section, we show how to construct an invariant from Bloch states which are independently given at each point in the mesh-approximated momentum space.
Note that this paper only outlines the approach and some specific examples, leaving the numerical implementation for arbitrary symmetry classes as a future problem. 

For a 1-cell $\bk_0\bk_1 = \{(1-t)\bk_0 + t \bk_1\ \vert \ t \in [0,1]\}$, let $\Phi^\alpha_{\bk} = (\phi^\alpha_{1,\bk},\dots, \phi^\alpha_{N_{\rm occ}^\alpha,\bk})$ denote an orthonormal set of occupied states of the Hamiltonian $H_\bk$ belonging to $\alpha$-irrep. 
Here, $N_{\rm occ}^\alpha$ is the number of occupied states with $\alpha$-irrep.
Recall that ${\cal D}_\alpha$ is the dimension of $\alpha$-irrep.
When $\calD_\alpha=1$, the $\text{U}(1)$-valued Wilson line is defined as 
\begin{align}
	\label{eq:irrep_Berry}
    e^{\ii \gamma^\alpha_{\bk_0 \bk_1}} := \lim_{\mathcal{N}\rightarrow\infty} \det \prod_{j = 0}^{\mathcal{N}-1} (\Phi^\alpha_{\bk_0 + (j+1)\bm{\delta}})^{\dagger}\Phi^\alpha_{\bk_0 + j\bm{\delta}}. 
\end{align}
Here, $\bm{\delta} = (\bk_1 - \bk_0)/\mathcal{N}$.
With a smooth gauge of $\Phi_\bk$, it can be rewritten as 
\begin{align}
    e^{\ii \gamma^\alpha_{\bk_0 \bk_1}} = \exp\left[ - \int_{\bk_0}^{\bk_1} \tr {\cal A}^\alpha_\bk \right]
\end{align}
with ${\cal A}^\alpha_\bk = (\Phi^\alpha_\bk)^\dag d \Phi^\alpha_\bk$ the Berry connection. 
On the other hand, when $\calD_\alpha>1$, we must impose a gauge constraint on the occupied states $\Phi_\bk^\alpha$ to extract the $\mZ_2$ invariant developed in Sec.~\ref{sec:z2inv_1skeleton}. 
We briefly outline a construction of the Wilson line for $\calD_\alpha>1$ in Sec.~\ref{sec:higher_dim_Wilson_line}.

The Wilson line $e^{\ii \gamma^\alpha_{\bk_0 \bk_1}}$ is not gauge invariant as it changes under gauge transformations at the endpoints $\Phi^\alpha_{\bk} \mapsto \Phi^\alpha_{\bk} W^\alpha_{\bk}, W^\alpha_{\bk} \in \text{U}(N_{\rm occ}^\alpha)$, as seen in 
\begin{align}
    e^{\ii \gamma^\alpha_{\bk_0 \bk_1}} \mapsto 
    e^{\ii \gamma^\alpha_{\bk_0 \bk_1}} \frac{\det W^\alpha_{\bk_0}}{\det W^\alpha_{\bk_1}}. 
    \label{eq:gauge_tr_Wilson_line}
\end{align}
When the start and endpoints coincide $\bk_0=\bk_1$, i.e., for a loop, the Wilson line becomes a gauge invariant and is referred to as the Berry phase. 
However, in constructing invariants from the AHSS, unlike the usual Berry phase, the start and end points do not always coincide. 
Thus, it is unclear whether the group $E_2^{1,-1}$ can be translated into a gauge-invariant Berry phase expression. 
Nevertheless, we find several cases for which the gauge dependence of the Wilson line at the endpoints (\ref{eq:gauge_tr_Wilson_line}) can be canceled using TRS and band degeneracy at high-symmetry points. 

Before moving to specific examples in Secs.~\ref{sec:Example: spinless systems in space group P2221'} and ~\ref{sec:TI_exs}, we summarize some patterns for acquiring gauge invariance for ${\cal D}_\alpha=1$ below. 
It is currently unclear whether the scenarios described here are exhaustive.
The general theory of symmetry-enriched Wilson lines including irreps with dimensions ${\cal D}_\alpha>1$ is left as future work.

\subsubsection{Class AII TRS and Pfaffian}
\label{sec:AII_Pfaffian}
In class AII, we have TRS with $U(\calT) U(\calT)^* = -1$.
At time-reversal invariant momentum (TRIM) $\bk= -\bk+\bm{G}$ for a reciprocal lattice vector $\bm{G}$, the Pfaffian can be defined:
\begin{align}
    \pf \left[ \Phi_\bk^\dag U({\cal T}) \Phi_\bk^* \right] \in \text{U}(1).
\end{align}
The Pfaffian changes under a gauge transformation $\Phi_\bk \mapsto \Phi_\bk W_\bk$ as follows:
\begin{align}
    \pf \left[ \Phi_\bk^\dag U({\cal T}) \Phi_\bk^* \right]
    \mapsto 
    \pf \left[ \Phi_\bk^\dag U({\cal T}) \Phi_\bk^* \right] \det W_\bk^*. 
\end{align}
Using this property, the Wilson line along a line segment connecting two TRIMs $\bk_0$ and $\bk_1$ can be corrected to be gauge invariant:
\begin{align}
    e^{\ii \gamma^{\rm T}_{\bk_0\bk_1}}= 
    e^{\ii \gamma_{\bk_0\bk_1}} \times \frac{\pf \left[ \Phi_{\bk_0}^\dag U({\cal T}) \Phi_{\bk_0}^* \right]}{\pf \left[ \Phi_{\bk_1}^\dag U({\cal T}) \Phi_{\bk_1}^* \right]}. 
\end{align}
Here, $\gamma^{\rm T}_{\bk_0\bk_1}/(2\pi) \mod 1$ corresponds to the partial polarization \cite{Fu-Kane-TRSpump}. 
In this way, a gauge invariant quantity is defined on a line segment $\bk_0\bk_1$, not on a loop. 
The partial polarization $e^{\ii \gamma^{\rm T}_{\bk_0\bk_1}}$ may give various $\mZ_2$ invariant in spinful systems in 1-skeleton if it is quantized and also constitute invariants defined on 2-skeleton like the Kane-Mele invariant~\cite{Fu-Kane-TRSpump}. 

\subsubsection{Class ${\rm A}_T$ TRS and Pfaffian of modified magnetic operator}
We consider the case where a magnetic symmetry $a \in G_{\bk_0}$ transforms an irrep $\beta$ of the unitary subgroup of $G_{\bk_0}$ into another inequivalent irrep $a(\beta)$. 
In such a case, the EAZ class is denoted by ${\rm A}_T$, and the Wilson line over a 1-cell $\bk_0\bk_1$ can be gauge-invariant at $\bk_0$. 
Let $\Phi^{\beta}_{\bk_0}$ and $\Phi^{a(\beta)}_{\bk_0}$ be the Bloch states in the $\beta$- and $a (\beta)$-irreps at $\bk_0$, respectively.
The sewing matrix of them, $(\Phi_{\bk_0}^{a(\beta)})^\dag U_{\bk_0}(a) (\Phi_{\bk_0}^{\beta})^*$, is unitary and its determinant is a ${\rm U}(1)$ value. 
Suppose that both the irreps $\beta$ and $a(\beta)$ are identical to an irrep $\alpha$ of the unitary subgroup $G^0_{\bk_0\bk_1}$ of the 1-cell $\bk_0\bk_1$. 
Then, the product 
\begin{align}
	e^{\ii\gamma^\alpha_{\bk_0\bk_1}} \times 
	\det \left[ (\Phi_{\bk_0}^{a(\beta)})^\dag U_{\bk_0}(a) (\Phi_{\bk_0}^{\beta})^* \right]
\end{align}
is gauge-invariant if the irreps $\beta$ and $a(\beta)$ are ordered at $\bk_0$ as in 
\begin{align}
	\Phi^{\alpha}_{\bk_0}
	= (\Phi^\beta_{\bk_0},\Phi^{a(\beta)}_{\bk_0}). 
	\label{eq:AT_gauge_fixing_cond}
\end{align}
The ordering condition \eqref{eq:AT_gauge_fixing_cond} generally breaks the additivity of the Wilson line under the band sum, as discussed in Sec.~\ref{sec:Incompatibility of Z2 invariants with band sum}. 
While it is possible to restore additivity by incorporating a factor of the quadratic refinement discussed in Sec.~\ref{sec:redef_inv}, there is an alternative approach to circumvent the ordering condition \eqref{eq:AT_gauge_fixing_cond} and still maintain the gauge invariance of the Wilson line~\cite{Li-Sun_Pfaffian, Araya_C4T}. 
Let $P^\beta_{\bk_0}$ and $P^{a(\beta)}_{\bk_0}$ be the orthogonal projectors onto $\beta$ and $a(\beta)$-irreps, respectively. 
Using the equality $U_{\bk_0}(a)(P^\beta_{\bk_0})^*=P^{a(\beta)}_{\bk_0}U_{\bk_0}(a)$ and the fact that the character $\chi^\beta_{\bk_0}(g)$ for $g \in G^0_{\bk_0}$ is ${\rm U}(1)$-valued (because we are assuming $\beta$ is a one-dimensional irrep), it is found that the matrix 
\begin{align}
	U_\Theta = \left(P^\beta_{\bk_0}-z_{\bk_0}(a,a) [\chi^\beta_{\bk_0}(a^2)]^* P^{a(\beta)}_{\bk_0}\right) U_{\bk_0}(a)
\end{align}
is unitary and satisfies that $U_\Theta U_\Theta^* = P^\beta_{\bk_0}+P^{a\beta}_{\bk_0} = P^\alpha_{\bk_0}$. 
(The matrix $U_\Theta$ with the complex conjugation is the same as $\Theta$ in Ref.~\cite{Li-Sun_Pfaffian} up to a ${\rm U}(1)$ phase.) 
Therefore, we can define the Pfaffian $\pf [ \Phi^\alpha_{\bk_0} U_{\Theta} (\Phi^\alpha_{\bk_0})^*] \in {\rm U}(1)$, and the product 
\begin{align}
	e^{\ii\gamma^\alpha_{\bk_0\bk_1}} \times \pf [ \Phi^\alpha_{\bk_0} U_{\Theta} (\Phi^\alpha_{\bk_0})^*]
\end{align}
is gauge-invariant at $\bk_0$ without the ordering condition \eqref{eq:AT_gauge_fixing_cond}. 

\subsubsection{Class AI TRS and source/sink of Wilson lines}
\label{sec:source_sink_wilson_line}
In class AI, we have TRS with $U(\calT) U(\calT)^* = 1$. 
The determinant of the sewing matrix of the TRS operator for the occupied states at TRIM, $\det \left[ \Phi_\bk^\dag U({\cal T}) \Phi_\bk^* \right]$, takes a value in $\text{U}(1)$. 
Under gauge transformations $\Phi_\bk \mapsto \Phi_\bk W_\bk$, it transforms as follows:
\begin{align}
    \det \left[ \Phi_\bk^\dag U({\cal T}) \Phi_\bk^* \right]
    \mapsto 
    \det \left[ \Phi_\bk^\dag U({\cal T}) \Phi_\bk^* \right] (\det W_\bk^*)^2. 
\end{align}
Using this gauge transformation property, the product of Wilson lines along two effective line segments $\bk_0\bk_1, \bk_0\bk_2$ emanating from a TRIM $\bk_0$ can be corrected to acquire gauge invariance at $\bk_0$:
\begin{align}
    e^{\ii \gamma_{\bk_0\bk_1}} \times e^{\ii \gamma_{\bk_0\bk_2}} \times \det \left[ \Phi_{\bk_0}^\dag U({\cal T}) \Phi_{\bk_0}^* \right].
\end{align}
This expression is not gauge invariant at the points $\bk_1, \bk_2$, but it may be made gauge invariant through contributions from other line segments. 
For detailed examples, see Sec.~\ref{sec:Example: spinless systems in space group P2221'}.

\subsubsection{Band degeneracy at high-symmetry point}
\label{sec:degeneracy}
There is a method to connect Wilson lines in a gauge invariant way by utilizing band degeneracy at high-symmetry points.
Let us illustrate this through the following simple example.

We consider a spinful electronic system with the space group $P222$. 
Let us write the elements of the point group as $D_2 = \{e,C_2^x,C_2^y,C_2^x C_2^y\}$. 
At the point $\Gamma = (0,0,0)$, due to the non-commutativity of the representation matrices $U(C_2^x)U(C_2^y)=-U(C_2^y)U(C_2^x)$, the irrep is given by a single two-dimensional representation.
On the other hand, along the symmetric line $\Sigma = \Gamma \text{X}\ (\text{X}=(\pi,0,0))$, there exist two one-dimensional irreps $\Sigma_3, \Sigma_4$, specified by the irreducible characters $\chi_\bk^{\Sigma_3}(C_2^x)=-i$ and $\chi_\bk^{\Sigma_4}(C_2^x)=i$. 
Similarly, along the symmetric line $\Delta = \Gamma \text{Y}$ ($\text{Y}=(0,\pi,0)$), there are two one-dimensional irreps $\Delta_3, \Delta_4$ with irreducible characters $\chi_\bk^{\Delta_3}(C_2^y)=-i$ and $\chi_\bk^{\Delta_4}(C_2^y)=i$. 
At the Gamma point, the two irreps along the $\Sigma$ and $\Delta$ lines become degenerate and form a single two-dimensional irrep. 

The $\Sigma_4$-irrep has a finite overlap with $\Delta_4$-irrep at the Gamma point, allowing the Wilson lines to be connected in a gauge invariant way. 
In fact, if the orthonormal set of occupied states with $\Sigma_4$-irrep at the Gamma point is denoted by $\Phi^{\Sigma_4}_\Gamma$, then the orthonormal set with $\Delta_4$-irrep can be given, apart from a $\text{U}(N_{\rm occ})$ gauge degree of freedom (where $N_{\rm occ}$ is the number of occupied states for $\Sigma_4$-irrep), as 
\begin{align}
    \Phi^{\Delta_4}_\Gamma = \frac{1}{\sqrt{2}}\left(U(e)+ \chi^{\Delta_4}_\Gamma(C_2^y)^* U(C_2^y)\right) \Phi^{\Sigma_4}_\Gamma.
\end{align}
Then, $(\Phi^{\Sigma_4}_\Gamma)^\dag \Phi^{\Delta_4}_\Gamma = \frac{1}{\sqrt{2}} \mathds{1}_{N_{\rm occ}}$. 
Therefore, when the orthonormal sets $\Phi^{\Sigma_4}_\bk$ and $\Phi^{\Delta_4}_\bk$ are given on the $\Sigma$ and $\Delta$ lines independently, the following product takes values in $\text{U}(1)$ and does not have gauge ambiguity at the Gamma point:
\begin{align}
    e^{\ii \gamma^{\Sigma_3}_{\text{X} \Gamma}} \times e^{\ii \gamma^{\Delta_3}_{\Gamma \text{Y}}} \times \det \left[ (\Phi^{\Delta_3}_\Gamma)^\dag \Phi^{\Sigma_3}_\Gamma \right] \times 2^{N_{\rm occ}/2}.
\end{align}
Again, the above quantity is not invariant under gauge transformations at $\text{X}$ and $\text{Y}$.
However, we could obtain gauge-independent invariants by combining other techniques.
See Sec.~\ref{sec:spinful_P2121211'} for an example of this scenario.

\subsubsection{EAZ class AI and $\mZ_2$-quantization of Berry phase}
When EAZ class on a 1-cell is AI, the Berry phase on the 1-cell is essentially $\mZ_2$-quantized, as described below.  
For simplicity, let us consider an antiunitary symmetry constraint $U_\bk(a)H_\bk^* U_\bk(a)^\dag=H_\bk$ with $U_\bk(a) U_\bk(a)^*=\mathds{1}$ at $\bk$ inside of a 1-cell PQ. 
The orthonormal set of occupied states $\Phi_\bk$ defines unitary matrix $w_\bk(a) = \Phi_\bk^\dag U_\bk(a) \Phi_\bk^* \in \text{U}(N_{\rm occ})$.
Note that $w_\bk(a)^\top=w_\bk(a)$. 
Then, the $\text{U}(1)$-valued Wilson line $e^{\ii\gamma_{\text{PQ}}}$ has the following symmetry constraint:
\begin{align}
    e^{2\ii\gamma_{\text{PQ}}}
    &=e^{\int_{\text{P}\to \text{Q}}{\rm tr} [\Phi_\bk^\dag U_\bk(a)^T dU_\bk(a)^* \Phi_\bk]}\times\frac{\det w_\text{Q}(a)}{\det w_\text{P}(a)}.
\end{align}
If the sign ambiguity of the square root is denoted as $(-1)^\nu$, then 
\begin{align}
    e^{\ii\gamma_{\text{PQ}}}
    &=e^{\frac{1}{2} \int_{\text{P}\to \text{Q}}{\rm tr} [\Phi_\bk^\dag U_\bk(a)^T dU_\bk(a)^* \Phi_\bk]}\times (-1)^\nu\times \sqrt{\frac{\det w_\text{Q}(a)}{\det w_\text{P}(a)}}. 
    \label{eq:TRS_constraint_AI_BP}
\end{align}
Note that the first factor on the right-hand side is independent of the $\text{U}(N_{\rm occ})$ gauge of $\Phi_\bk$. 
Equation \eqref{eq:TRS_constraint_AI_BP} implies that the $\text{U}(1)$ Wilson line, when fixing the gauge of the occupied states $\Phi_\text{P}$ and $\Phi_\text{Q}$ at the endpoints, is $\mZ_2$-quantized apart from the first factor on the right-hand side.
Particularly, when the start and end points are the same $\text{P}=\text{Q}$ (namely, a loop), the factor $(-1)^\nu$ provides a $\mZ_2$ invariant given by
\begin{align}
	\label{eq:Z2-berry}
    (-1)^\nu 
    &=e^{-\ii\gamma_{\text{P}\to \text{P}}}\times e^{\frac{1}{2} \int_{\text{P}\to \text{P}}{\rm tr} [\Phi_\bk^\dag U_\bk(a)^T dU_\bk(a)^* \Phi_\bk]},
\end{align}
This point is elaborated on in Sec.~\ref{sec:Revisiting physical meaning of E11-1}.

\subsubsection{EAZ class AII and triviality of Berry phase}
\label{sec:Class AII EAZ and triviality of Berry phase}
On the other hand, when EAZ class in a 1-cell is AII, the $\text{U}(1)$ Wilson line always takes a trivial value in the following sense.
For simplicity, consider an antiunitary symmetry with $U_\bk(a)U_\bk(a)^*=-1$. 
In this case, since $w_\bk(a)^\top=-w_\bk(a)$, the Pfaffian of $w_\bk(a)$ is defined, and the $\text{U}(1)$ Wilson line is determined by the occupied states at the endpoints, except for a gauge-invariant factor:
\begin{align}
    e^{\ii\gamma_{\text{P}\to \text{Q}}}
    &=e^{\frac{1}{2} \int_{\text{P}\to \text{Q}}{\rm tr} [\Phi_\bk^\dag U_\bk(a)^T dU_\bk(a)^* \Phi_\bk]}\times \frac{\pf[\Phi_\text{Q}^\dag U_\text{Q}(a) \Phi_\text{Q}^*]}{\pf[\Phi_\text{P}^\dag U_\text{P}(a) \Phi_\text{P}^*]}.
\end{align}
When we consider a loop, the $\mZ_2$ invariant in Eq.~\eqref{eq:Z2-berry} is always trivial.
Therefore, in the case of EAZ Class AII, there is no nontrivial classification in a 1-cell.

\subsubsection{Wilson line for high dimensional irreducible representation}
\label{sec:higher_dim_Wilson_line}
We outline a construction of the Wilson line $e^{\ii \gamma^\alpha_{\bk_0\bk_1}}$ for irreps whose dimension ${\cal D}_\alpha$ is larger than $1$.
To do so, it is useful to employ the {\it non-periodic} basis, which is used in numerical verification of topological invariants of superconductors discussed in Sec.~\ref{sec:numerics}. 
The difference between periodic and non-periodic bases originates from the definition of the Fourier transformation.
Let $\hat{c}_{\bm{R}, \mathrm{a}}^{\dagger}$ be a fermionic creation operator at a Wyckoff position $\bm{x}_{\mathrm{a}}$ in a unit cell at $\bm{R}$. %
Then, there are two ways to define the Fourier transformation: (i) $\hat{c}_{\bm{k}, \mathrm{a}}^{\dagger} = \frac{1}{\sqrt{V}}\sum_{\bm{R}}\hat{c}_{\bm{R}, \mathrm{a}}^{\dagger}e^{\ii \bk \cdot \bm{R}}$; (ii) $\hat{\tilde{c}}_{\bm{k}, \mathrm{a}}^{\dagger} = \frac{1}{\sqrt{V}}\sum_{\bm{R}}\hat{c}_{\bm{R}, \mathrm{a}}^{\dagger}e^{\ii \bk \cdot (\bm{R}+\bm{x}_{\mathrm{a}})}$. 
The former definition gives rise to the periodic basis, and the latter one leads to the non-periodic one. 
These two bases are related by unitary matrix $V_{\bk}$ introduced in Sec.~\ref{sec:numerics}.
In the non-periodic basis, the symmetry representation takes the form of $\tilde{U}_\bk(g) = e^{-ig\bk\cdot \bm{a}_g} D(g)$, i.e., the product of $e^{-ig\bk\cdot \bm{a}_g}$ with (fractional) translation $\bm{a}_g$ and a $\bk$-independent unitary matrix $D(g)$.

Let $G_{\bk_0\bk_1}$ be the little group of a 1-cell $\bk_0\bk_1$ and $G_{\bk_0\bk_1}^0$ be its unitary part. 
Let $\Phi_\bk^\alpha$ denote the Bloch states in the $\alpha$-irrep of $G_{\bk_0\bk_1}^0$ over the 1-cell $\bk_0\bk_1$. 
At each $\bk\in \bk_0\bk_1$, the representation matrix for $g \in G_{\bk_0\bk_1}^0$ can be $\bk$-independent as it satisfies 
\begin{align}
    D(g) \Phi^\alpha_\bk = \Phi^\alpha_\bk w^\alpha(g).
    \label{eq:D_alpha>1_gauge_fixing}
\end{align} 
In such a gauge, the overlap matrix $(\Phi^{\alpha}_{\bk+\bm{\delta}})^\dag \Phi^\alpha_\bk$ for adjacent two momenta obeys the symmetry constraint $w^\alpha(g) [(\Phi^{\alpha}_{\bk+\bm{\delta}})^\dag \Phi^\alpha_\bk] w^\alpha(g)^\dag = (\Phi^{\alpha}_{\bk+\bm{\delta}})^\dag \Phi^\alpha_\bk$ for $g \in G^0_{\bk_0\bk_1}$, meaning that the eigenvalues of $(\Phi^{\alpha}_{\bk+\bm{\delta}})^\dag \Phi^\alpha_\bk$ consists of $n_\alpha$ eigenvalues of $\lambda_1,\dots,\lambda_{n_\alpha}$, and each of them are ${\cal D}_\alpha$-fold degenerated.
As in Eq.~\eqref{eq:xi_alpha_def}, we introduce 
\begin{align}
    \xi^\alpha_{\bk}\left( (\Phi^{\alpha}_{\bk+\bm{\delta}})^\dag \Phi^\alpha_{\bk} \right)
    = \prod_{i=1}^{n_\alpha} \lambda_i, 
\end{align}
and the Wilson line for ${\cal D}_\alpha>1$ is defined as 
\begin{align}
    e^{\ii\gamma^\alpha_{\bk_0\bk_1}}
    := \lim_{{\cal N}\to \infty}
    \prod_{j=0}^{{\cal N}-1} \xi^\alpha_{\bk_0+j\bm{\delta}} \left( (\Phi^{\alpha}_{\bk_0+(j+1)\bm{\delta}})^\dag \Phi^\alpha_{\bk_0+j\bm{\delta}} \right).
\end{align}
The gauge transformation of the Wilson line $e^{\ii\gamma^\alpha_{\bk_0\bk_1}}$ for ${\cal D}_\alpha>1$ is similar to Eq.~\eqref{eq:gauge_tr_Wilson_line} under the gauge constraint \eqref{eq:D_alpha>1_gauge_fixing}.

We are not going into detail about the cases where ${\cal D}_\alpha>1$, since in this paper we employ the periodic basis for $H_\bk$ in which the Hamiltonian $H_\bk$ is invariant under the shift by reciprocal lattice vectors. 
We leave the definition and numerical implementation of the symmetry-enriched Berry phase in the non-periodic basis as a future problem.

\subsection{Example: one-dimensional $P\calT$-symmetric systems}
\label{sec:Revisiting physical meaning of E11-1}

\begin{figure}[t]
	\begin{center}
		\includegraphics[width=0.99\columnwidth]{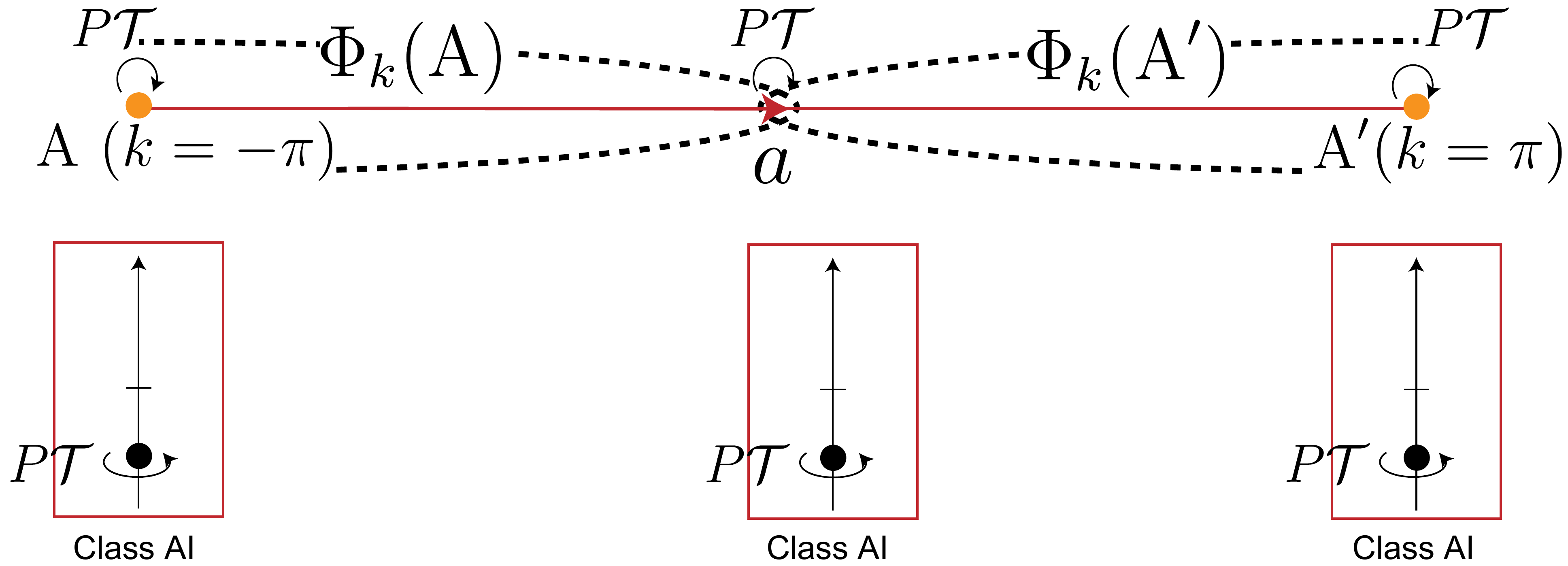}
		\caption{\label{fig:cell_PT_1D}Illustration of the cell decomposition of the fundamental domain in $P\calT$-symmetric one-dimensional systems.
			The cell decomposition of BZ is given by $C_0 = \{\text{A} (k = -\pi), \text{A}'(k = \pi)\}$ and $C_1 = \{a (-\pi<k<\pi)\}$. 
		}
	\end{center}
\end{figure}
We begin by discussing a well-studied example is $P\calT$-symmetric one-dimensional systems with $U_{k}(P\calT)[U_{k}(P\calT)]^{*} = +1$~\cite{Fang_PT_monopole,Ahn_PT,Ahn_C2T}.
Our cell decomposition is shown in Fig.~\ref{fig:cell_PT_1D}, where $C_0=\{\text{A}, \text{A}'\} = \{k = -\pi, k = \pi\}$ and $C_1 = \{a\} = \{k\ \vert\  k \in (-\pi, \pi)\}$. 
Since the relation $H_{k=-\pi}=H_{k=\pi}$ always holds, $E_{1}^{1,-1} = {}^\phi K^{(z,c)+0}_{G/\Pi}(X_1 = T^1,X_{0} = C_0) \simeq \pi_0(\text{O}(N)) = \mZ_2$.
Furthermore, since $d_{1}^{0,-1}$ is trivial, $E_{2}^{1,-1} = E_{1}^{1,-1}=\mZ_2$.

As discussed in Sec.~\ref{sec:z2inv_1skeleton}, when we fix the gauge degrees of freedom, the determinant of transition function is a $\mZ_2$ invariant, which is given by
\begin{align}
	\tilde{t}_{\text{A}\text{A}'} := \Phi^{\dagger}_{\text{A}, k=0}\Phi_{\text{A}', k=0} \in \text{O}(N_{\text{occ}}), 
\end{align}
where $\Phi_{\text{A}, k}$ and $\Phi_{\text{A}', k}$ are wave functions around A and $\text{A}'$ that satisfy $U_{k}(P\calT)\Phi_{\text{A}, k}^{*}= \Phi_{\text{A}, k}$ and $U_{k}(P\calT)\Phi_{\text{A}', k}^{*} = \Phi_{\text{A}', k}$. 
Also, $N_{\text{occ}}$ is the number of occupied bands.
It should be noted that, although $k = \pm \pi$ are the same aside from a reciprocal lattice vector, it is not always the case that the gauge choices around A and $\text{A}'$ are the same.

Although we have a concrete expression of the $\mZ_2$ topological invariant, it is not practical since we need to fix the gauge degrees of freedom. 
For numerical computations, it is essential to obtain expressions of topological invariants that do not depend on gauge choices.
To achieve this, we first define the following gauge-independent quantity
\begin{align}
	e^{\mathrm{i}\gamma} := \lim_{\mathcal{N}\rightarrow\infty} \det \prod_{j = 0}^{\mathcal{N}-1} \Phi^{\dagger}_{- \pi + (j+1) \delta}\Phi_{-\pi + j\delta}, 
\end{align}
where $\delta = 2\pi/\mathcal{N}$ and $\Phi_{k = - \pi} = \Phi_{k = \pi}$.
Next, we discuss the relation between $\det t_{\text{A}\text{A}'}$ and $e^{\mathrm{i}\gamma}$.
To relate $e^{\mathrm{i}\gamma}$ to $\det t_{\text{A}\text{A}'}$, we consider the two patches shown in Fig.~\ref{fig:cell_PT_1D}. 
Then, 
\begin{align}
	\label{eq:wilson_loop_PT}
	&e^{\mathrm{i}\gamma}
	= \lim_{\mathcal{N}\rightarrow\infty} \left[\det \prod_{j = 0}^{\mathcal{N}-1} \Phi^{\dagger}_{\text{A}', (j+1) \delta/2}\Phi_{\text{A}', j\delta/2}\right] \nonumber\\
	&\quad\quad\times\det t_{\text{A}\text{A}'}^{\dagger} \left[\det\prod_{j = 0}^{\mathcal{N}-1} \Phi^{\dagger}_{\text{A}, - \pi + (j+1) \delta/2}\Phi_{\text{A}, -\pi + j\delta/2}\right] \nonumber\\
	&= \exp\left[-\int\mathrm{tr}\Phi^{\dagger}_{\text{A}', k}d\Phi_{\text{A}', k}-\int\mathrm{tr}\Phi^{\dagger}_{\text{A}, k}d\Phi_{\text{A}, k}\right]\det t_{\text{A}\text{A}'}^{\dagger}.
\end{align}
Furthermore, we impose the gauge condition $U_{k}(P\calT)\Phi^{*}_{k} = \Phi_{k}$ on the right-hand side. 
For this gauge condition, since $\int dk\ \mathrm{tr} \Phi^{\dagger}_{k}\partial_k\Phi_{k}  = - \int dk (\mathrm{tr} \Phi^{\dagger}_{k}\partial_k\Phi_{k})^*$, we have
\begin{align}
	\label{eq:PT_rep}
	\int \mathrm{tr} \Phi^{\dagger}_{k}d\Phi_{k} 
	= -\frac{1}{2}\int \mathrm{tr} \Phi^{\dagger}_{k}\left(U_{k}(P\calT) dU^{\dagger}_{k}(P\calT)\right)\Phi_{k},
\end{align}
where the right-hand side is gauge-invariant. 
Finally, combining Eqs.~\eqref{eq:wilson_loop_PT} and \eqref{eq:PT_rep}, we arrive at the gauge-independent $\mZ_2$-valued topological invariant
\begin{align}
	(-1)^{\nu} &= e^{-\mathrm{i}\gamma}\exp\left[\frac{1}{2}\int_{-\pi}^{\pi}\mathrm{tr} \Phi_{k}^{\dagger}\left(U_{k}(P\calT)d U^{\dagger}_{k}(P\calT)\right)\Phi_{k} \right].
\end{align}

Let us comment on what this topological invariant indicates. 
In this symmetry setting, there are two inequivalent atomic insulators: One has electrons at $x = R\ (R\in \mZ)$ and the other has electrons at $x = R+1/2$.
This topological invariant is trivial for the former case and nontrivial for the latter case.

\subsection{Example: spinless systems in space group $P222$ with TRS}
\label{sec:Example: spinless systems in space group P2221'}
Consider a spinless electronic system with space group symmetry $P222$ and TRS. 
The $K$-group is given by $\mZ^{13}+\mZ_2$, and its generators are represented by atomic insulators. 
Moreover, comparing the $E_2$ pages of the momentum-space and real-space AHSS, it is found that the $\mZ_2$ part can be detected by the $\mZ_2$ invariant on the 1-skeleton~\cite{Shiozaki-Ono2023}.
This section illustrates constructing the $\mZ_2$ invariant based on the gauge fixing condition in Sec.~\ref{sec:z2inv_1skeleton} and its Berry phase expression without a gauge fixing condition using the method described in Sec.~\ref{sec:source_sink_wilson_line}.

Let $H_\bk$ be a Hamiltonian periodic in the BZ. 
Symmetries and factor systems are summarized in 
\begin{align}
&U_\bk(\calT) H_\bk^* U_\bk(\calT)^\dag = H_{-\bk}, \\
&U_\bk(C_2^\mu) H_\bk U_\bk(C_2^\mu)^\dag = H_{C_2^\mu\bk},\label{eq:sym_P222_C2mu}\\ 
&U_{-\bk}(\calT)U_\bk(\calT)^* = 1, \\
&U_{C_2^\mu\bk}(C_2^\mu)U_\bk(C_2^\mu) = 1,\\
&U_{C_2^\mu\bk}(\calT)U_\bk(C_2^\mu)^*=U_{-\bk}(C_2^\mu)U_\bk(\calT),\\
&U_{C_2^\nu\bk}(C_2^\mu)U_\bk(C_2^\nu)=U_{C_2^\mu\bk}(C_2^\nu)U_\bk(C_2^\mu). 
\end{align}
Here, $\mu,\nu\in \{x,y,z\}$, and $C_2^\mu$ denotes a twofold rotation about the $\mu$-axis. 
The cell decomposition is shown in Fig.~\ref{fig:TI_P212121} (d). 
At TRIMs $\text{P} \in \{\Gamma,\text{X},\text{Y},\text{S},\text{Z},\text{U},\text{T},\text{R}\}$, four one-dimensional irreps $\beta \in \{A,B_1,B_2,B_3\}$ exist, and their irreducible characters are given below, independent of TRIMs:
\begin{align}
\begin{array}{l|llll}
\chi^\beta(g)&e&C_2^x&C_2^y&C_2^z\\
\hline
A&1&1&1&1\\
B_1&1&-1&-1&1\\
B_2&1&-1&1&-1\\
B_3&1&1&-1&-1\\
\end{array}
\end{align}
On the other hand, two one-dimensional irreps $\alpha \in \{A,B\}$ exist along the twofold rotation axes (1-cells), with irreducible characters given in the following table:
\begin{align}
\begin{array}{l|llll}
\chi^\alpha(g)&e&C_2^\mu\\
\hline
A&1&1\\
B&1&-1\\
\end{array}    
\end{align}
The EAZ classes for them are all class AI.

The compatibility relations along each rotation axis are given as:
\begin{align}
    &\mbox{$C_2^x$-axis: }\quad A = A \oplus B_3,\quad B = B_1 \oplus B_2, \\
    &\mbox{$C_2^y$-axis: }\quad A = A \oplus B_2,\quad B = B_1 \oplus B_3, \\
    &\mbox{$C_2^z$-axis: }\quad A = A \oplus B_1,\quad B = B_2 \oplus B_3. 
\end{align}
As we are interested in invariants of insulators gapped on 1-skeleton, we impose the following compatibility conditions on every 1-cell. 
Let $n_{\beta(\text{P})}$ the number of the $\beta$-irreps at TRIM $\text{P}$.
Depending on which twofold rotation axis the 1-cell $\text{P}\text{Q}$ lies on, the compatibility conditions are represented as follows:
\begin{align}
&\mbox{$C_2^x$-axis}: \nonumber\\
&\left\{ \begin{array}{ll}
n_{A(\text{P})}+n_{B_3(\text{P})}=n_{A(\text{Q})}+n_{B_3(\text{Q})},\\
n_{B_1(\text{P})}+n_{B_2(\text{P})}=n_{B_1(\text{Q})}+n_{B_2(\text{Q})},\\
\end{array}\right.\label{eq:comp1}\\
&\mbox{$C_2^y$-axis}: \nonumber\\
&\left\{ \begin{array}{ll}
n_{A(\text{P})}+n_{B_2(\text{P})}=n_{A(\text{Q})}+n_{B_2(\text{Q})},\\
n_{B_1(\text{P})}+n_{B_3(\text{P})}=n_{B_1(\text{Q})}+n_{B_3(\text{Q})},\\
\end{array}\right.\label{eq:comp2}\\
&\mbox{$C_2^z$-axis}: \nonumber\\
&\left\{ \begin{array}{ll}
n_{A(\text{P})}+n_{B_1(\text{P})}=n_{A(\text{Q})}+n_{B_1(\text{Q})},\\
n_{B_2(\text{P})}+n_{B_3(\text{P})}=n_{B_2(\text{Q})}+n_{B_3(\text{Q})}.\\
\end{array}\right.\label{eq:comp3}
\end{align}
The sublattice $\left\{ \{n_{\beta(\text{P})}\}_{i,\beta} \in E_1^{0,0} \cong \mZ^{32} \Big| (\ref{eq:comp1}), (\ref{eq:comp2}), (\ref{eq:comp3})\right\}$ that satisfies the above compatibility conditions is $E_2^{0,0}$.

\subsubsection{Constructing the $\mZ_2$ invariant using gauge fixing condition}
In the neighborhood $U_\text{P}$ of TRIM $\text{P}$, the set of occupied states $\Phi_{\text{P},\bk \in U_\text{P}}$ of $H_\bk$ can be chosen as follows, based on section \ref{sec:Classification of equivariant vector bundle}. 
First, it is assumed that $\Phi_{\text{P},\bk}$ is block-diagonalized by irreps at 0-cell $\text{P}$ as in 
\begin{align}
&\Phi_{\text{P},\bk} = \Phi^A_{\text{P},\bk} \oplus \Phi^{B_1}_{\text{P},\bk} \oplus \Phi^{B_2}_{\text{P},\bk} \oplus \Phi^{B_3}_{\text{P},\bk}. 
\label{eq:Bloch_with_gauge_cond_P2221'}
\end{align}
Furthermore, for each block $\beta \in \{A,B_1,B_2,B_3\}$, the occupied states $\Phi_{\text{P},\bk}^\beta$ satisfy the following symmetries: 
\begin{align}
&U_\bk(C_2^\mu) \Phi^\beta_{\text{P},\bk} = \chi^\beta(C_2^\mu) \Phi^\beta_{\text{P}, C_2^\mu(\bk-\text{P})+\text{P}},\label{eq:g2}\\
&U_\bk(\calT) \left[\Phi_{\text{P},\bk}^\beta\right]^* = \Phi_{\text{P},-(\bk-\text{P})+\text{P}}^\beta.\label{eq:g1}
\end{align}
In particular, at points $\bk$ on the $C_2^\mu$-axis where $C_2^\mu (\bk-\text{P})=\bk-\text{P}$, it holds that 
\begin{align}
    &U_\bk(C_2^\mu) \Phi^\beta_{\text{P},\bk} = \chi^\beta(C_2^\mu) \Phi^\beta_{\text{P},\bk}.
\end{align}

For each of the 12 independent 1-cells $a$ shown in Fig.~\ref{fig:TI_P212121} (d), we define transition functions $t_a^{\alpha}$ for each irrep $\alpha = A,B$. 
Let us write the 1-cell going from TRIM $\text{P}$ to $\text{Q}$ by $\text{P}\text{Q}$ and its midpoint as $\bk_{\text{P}\text{Q}} = (\text{P}+\text{Q})/2$. 
Here we focus on the 1-cell $\G \text{X}$. 
The Bloch states around 0-cells $\G$ and $\text{X}$ are arranged in the order $A \oplus B_3$ and $B_1 \oplus B_2$ for $A$ and $B$ irreps over 1-cell, respectively, and then we define the transition function at the midpoint $\bk_{\G \text{X}}$ as 
\begin{align}
&t^A_{\G \text{X}} := 
\left(\Phi^A_{\G,\bk_{\G \text{X}}}\oplus \Phi^{B_3}_{\G,\bk_{\G \text{X}}}\right)^\dag 
\left(\Phi^A_{\text{X},\bk_{\G \text{X}}}\oplus \Phi^{B_3}_{\text{X},\bk_{\G \text{X}}}\right), \label{eq:tr_func_GX_A}\\
&t^B_{\G \text{X}} := 
\left(\Phi^{B_1}_{\G,\bk_{\G \text{X}}}\oplus \Phi^{B_2}_{\G,\bk_{\G \text{X}}}\right)^\dag 
\left(\Phi^{B_1}_{\text{X},\bk_{\G \text{X}}}\oplus \Phi^{B_2}_{\text{X},\bk_{\G \text{X}}}\right). \label{eq:tr_func_GX_B}
\end{align}
Due to $C_2^z \calT$ symmetry, the following holds:
\begin{align}
&(t^A_{\G \text{X}})^*
=
(\mathds{1}_{n_{A(\G)}}\oplus-\mathds{1}_{n_{B_3(\G)}})
t^A_{\G \text{X}}
(\mathds{1}_{n_{A(\text{X})}}\oplus -\mathds{1}_{n_{B_3(\text{X})}}), \\
&(t^B_{\G \text{X}})^* = 
(\mathds{1}_{n_{B_1(\G)}}\oplus -\mathds{1}_{n_{B_2(\G)}})
t^B_{\G \text{X}}
(\mathds{1}_{n_{B_1(\text{X})}}\oplus -\mathds{1}_{n_{B_2(\text{X})}}). 
\end{align}
Under the gauge fixing conditions \eqref{eq:g2} and \eqref{eq:g1}, the determinants of the transition functions of each sector are quantized as in 
\begin{align}
&(\det t^A_{\G \text{X}})^* 
= (-1)^{n_{B_3(\G)}+n_{B_3(\text{X})}} \det t^A_{\G \text{X}}, \\
&(\det t^B_{\G \text{X}})^* 
= (-1)^{n_{B_2(\G)}+n_{B_2(\text{X})}} \det t^B_{\G \text{X}}.
\end{align}
Since the transition functions are unitary matrices, we can also write them as
\begin{align}
&(\det t^A_{\G \text{X}})^2
= (-1)^{n_{B_3(\G)}+n_{B_3(\text{X})}}, \\
&(\det t^B_{\G \text{X}})^2
= (-1)^{n_{B_2(\G)}+n_{B_2(\text{X})}}. 
\end{align}

For other 1-cells on the $C_2^x$-axis, define the transition function in the same way as Eqs.~\eqref{eq:tr_func_GX_A} and \eqref{eq:tr_func_GX_B}, and for the $C_2^y,C_2^z$ axes,
\begin{align}
&\mbox{$C_2^y$-axis}: \nonumber\\
&t^A_{\text{P} \text{Q}} := 
\left(\Phi^A_{\text{P},\bk_{\text{P}\text{Q}}}\oplus \Phi^{B_2}_{\text{P},\bk_{\text{P}\text{Q}}}\right)^\dag 
\left(\Phi^A_{\text{Q},\bk_{\text{P}\text{Q}}}\oplus \Phi^{B_2}_{\text{Q},\bk_{\text{P}\text{Q}}}\right), \\
&t^B_{\text{P} \text{Q}} := 
\left(\Phi^{B_1}_{\text{P},\bk_{\text{P}\text{Q}}}\oplus \Phi^{B_3}_{\text{P},\bk_{\text{P}\text{Q}}}\right)^\dag 
\left(\Phi^{B_1}_{\text{Q},\bk_{\text{P}\text{Q}}}\oplus \Phi^{B_3}_{\text{Q},\bk_{\text{P}\text{Q}}}\right), 
\end{align}
\begin{align}
&\mbox{$C_2^z$-axis}: \nonumber\\
&t^A_{\text{P} \text{Q}} := 
\left(\Phi^A_{\text{P},\bk_{\text{P}\text{Q}}}\oplus \Phi^{B_1}_{\text{P},\bk_{\text{P}\text{Q}}}\right)^\dag 
\left(\Phi^A_{\text{Q},\bk_{\text{P}\text{Q}}}\oplus \Phi^{B_1}_{\text{Q},\bk_{\text{P}\text{Q}}}\right), \\
&t^B_{\text{P} \text{Q}} := 
\left(\Phi^{B_2}_{\text{P},\bk_{\text{P}\text{Q}}}\oplus \Phi^{B_3}_{\text{P},\bk_{\text{P}\text{Q}}}\right)^\dag 
\left(\Phi^{B_2}_{\text{Q},\bk_{\text{P}\text{Q}}}\oplus \Phi^{B_3}_{\text{Q},\bk_{\text{P}\text{Q}}}\right). 
\end{align}
Note that by reversing the orientation of a 1-cell, we have 
\begin{align}
    t^\alpha_{\text{Q}\text{P}}=(t^\alpha_{\text{P}\text{Q}})^\dag.
\end{align}
The constraints are summarized as
\begin{align}
&\mbox{$C_2^x$-axis}:\quad \left\{\begin{array}{ll}
(\det t^A_{\text{P} \text{Q}})^2
= (-1)^{n_{B_3(\text{P})}+n_{B_3(\text{Q})}}, \\
(\det t^B_{\text{P} \text{Q}})^2
= (-1)^{n_{B_2(\text{P})}+n_{B_2(\text{Q})}}, \\
\end{array}\right. \label{eq:s1}\\
&\mbox{$C_2^y$-axis}:\quad \left\{\begin{array}{ll}
(\det t^A_{\text{P} \text{Q}})^2
= (-1)^{n_{B_2(\text{P})}+n_{B_2(\text{Q})}}, \\
(\det t^B_{\text{P} \text{Q}})^2
= (-1)^{n_{B_3(\text{P})}+n_{B_3(\text{Q})}}, \\
\end{array}\right.\label{eq:s2}\\
&\mbox{$C_z$-axis}:\quad \left\{\begin{array}{ll}
(\det t^A_{\text{P} \text{Q}})^2
= (-1)^{n_{B_1(\text{P})}+n_{B_1(\text{Q})}}, \\
(\det t^B_{\text{P} \text{Q}})^2
= (-1)^{n_{B_3(\text{P})}+n_{B_3(\text{Q})}}. \\
\end{array}\right.\label{eq:s3}
\end{align}
This defines 24 $\mZ_2$-valued quantities
\begin{align}
    \zeta^\alpha_{\text{P}\text{Q}}:= \det t^\alpha_{\text{P}\text{Q}},\quad \text{P}\text{Q} \in \{\mbox{1-cells}\}, \alpha \in \{A,B\}. 
\end{align}
Note that values of $\zeta^\alpha_{\text{P}\text{Q}}$ depend on representations at 0-cells. 
In particular, if $\zeta^\alpha_{\text{P}\text{Q}} \in \{\pm i\}$, then $\zeta^\alpha_{\text{Q}\text{P}} = - \zeta^\alpha_{\text{P}\text{Q}}$~\footnote{
The definition of the $\mZ_2$ value $\zeta^\alpha_{\text{P}\text{Q}}$ here is made without introducing the basis transformation matrix $V_{i\to a,\bk}$ as in Sec.~\ref{sec:Gauge_invariant_expression_u1}, so it differs slightly from the $\mZ_2$ value $\xi^\alpha_{\bk_a}$ in Sec.~\ref{sec:Gauge_invariant_expression_u1}. 
Therefore, $\zeta^\alpha_{\text{P}\text{Q}}$ can take quantized values other than $\pm 1$.}.

The combinations of $\zeta^\alpha_{\text{P}\text{Q}}$ that are invariant under residual gauge transformations, which do not change the gauge fixing conditions \eqref{eq:g2} and \eqref{eq:g1}, become the desired topological invariants.
Consider a residual gauge transformation in the neighborhood of the 0-cell $\text{P}$, 
\begin{align}
\Phi_{\text{P},\bk}\mapsto 
\Phi_{\text{P},\bk} W_{\text{P},\bk},\quad W_{\text{P},\bk} \in \text{U}(N),
\end{align}
where $N$ is the number of occupied states. 
The gauge transformation $W_{\text{P},\bk}$ should satisfy the symmetry \eqref{eq:symm_gauge_tr}.
That is, for the symmetry transformation of the occupied states 
\begin{align}
    &w_{\text{P},\bk}(C_2^\mu) 
    = \left(\chi^A(C_2^\mu) \mathds{1}_{n_{A(\text{P})}} \right)\oplus 
    \left(\chi^{B_1}(C_2^\mu) \mathds{1}_{n_{B_1(\text{P})}} \right) \nonumber\\
    &\quad \oplus \left( \chi^{B_2}(C_2^\mu) \mathds{1}_{n_{B_2(\text{P})}} \right)\oplus 
    \left(\chi^{B_3}(C_2^\mu) \mathds{1}_{n_{B_3(\text{P})}}\right), 
\end{align}
$W_{\text{P},\bk}$ must satisfy
\begin{align}
    &w_{\text{P},\bk}(C_2^\mu) W_{\text{P},\bk} w_{\text{P},\bk}(C_2^\mu)^\dag = W_{\text{P},C_2^\mu(\bk-\text{P})+\text{P}}
    \label{eq:sym_gauge_tr_P2221'_1}
\end{align}
for $\mu \in \{x,y,z\}$ and
\begin{align}
   &W_{\text{P},\bk}^* = W_{\text{P},-(\bk-\text{P})+\text{P}}. 
   \label{eq:sym_gauge_tr_P2221'_2}
\end{align}
For example, over the $C_2^x$-axis satisfying $C_2^x (\bk-\text{P}) =\bk-\text{P}$, due to $C_2^x$ symmetry, the gauge transformation takes a form as 
\begin{align}
    W_{\text{P},\bk} = \begin{pmatrix}
        a&0&0&b\\
        0&c&d&0\\
        0&e&f&0\\
        g&0&0&h\\
    \end{pmatrix},\quad \mbox{$\bk \in C_2^x$-axis}. 
\end{align}
Let us denote
\begin{align}
    W^{A\oplus B_3}_{\text{P},\bk} = \begin{pmatrix}
        a&b\\
        g&h\\
    \end{pmatrix}, \quad 
    W^{B_1\oplus B_2}_{\text{P},\bk} = \begin{pmatrix}
        c&d\\
        e&f\\
    \end{pmatrix}.  
\end{align}
Due to $C_2^z\cal T$ symmetry, they satisfy
\begin{align}
    &(\mathds{1}_{n_{A(\text{P})}}\oplus -\mathds{1}_{n_{B_3(\text{P})}})
    [W^{A \oplus B_3}_{\text{P},\bk}]^* 
    \nonumber \\
    &= W^{A\oplus B_3}_{\text{P},\bk} (\mathds{1}_{n_{A(\text{P})}}\oplus -\mathds{1}_{n_{B_3(\text{P})}}), \\
    &(\mathds{1}_{n_{B_1(\text{P})}}\oplus -\mathds{1}_{n_{B_2(\text{P})}})
    [W^{B_1\oplus B_2}_{\text{P},\bk}]^* 
    \nonumber\\
    &= W^{B_1\oplus B_2}_{\text{P},\bk} (\mathds{1}_{n_{B_1(\text{P})}}\oplus -\mathds{1}_{n_{B_2(\text{P})}}). 
\end{align}
In particular, the determinant of each sector is quantized as 
\begin{align}
    \det W^{A\oplus B_3}_{\text{P},\bk},  
    \det W^{B_1\oplus B_2}_{\text{P},\bk}\in \{\pm 1\}, 
    \quad \mbox{$\bk \in C_2^x$-axis}, 
\end{align}
and they are constant along the $C_2^x$ axis. 
In particular, at TRIM $\bk = \text{P}$, due to symmetry (\ref{eq:sym_gauge_tr_P2221'_1}), we have a block-diagonalized form 
\begin{align}
W_{\text{P},\text{P}} = W^A_{\text{P},\text{P}} \oplus W^{B_1}_{\text{P},\text{P}} \oplus W^{B_2}_{\text{P},\text{P}} \oplus W^{B_3}_{\text{P},\text{P}},
\end{align}
and due to (\ref{eq:sym_gauge_tr_P2221'_2}), the determinant of each sector is also quantized:
\begin{align}
    \eta^{\beta \in \{A,B_1,B_2,B_3\}}_\text{P}:= \det W^{\beta}_{\text{P},\text{P}} \in \{\pm 1\}.
\end{align}
Therefore,
\begin{align}
    \det W^{A\oplus B_3}_{\text{P},\bk} 
    &= \det W^{A\oplus B_3}_{\text{P},\text{P}} \nonumber\\
    &= \det (W^{A}_{\text{P},\text{P}} \oplus W^{B_3}_{\text{P},\text{P}} )
    =\eta^A_\text{P} \eta^{B_3}_\text{P}. 
\end{align}
In this way, for a 1-cell $\text{P}\text{Q}$ parallel to $C_2^x$-axis, the transition function $t^\alpha_{\text{P}\text{Q}}$ transforms under gauge transformation as 
\begin{align}
    &t^{A}_{\text{P}\text{Q}} \mapsto (W^{A \oplus B_3}_{\text{P},\bk_{\text{P}\text{Q}}})^\dag t^A_{\text{P}\text{Q}} W^{A\oplus B_3}_{\text{Q},\bk_{\text{P}\text{Q}}}, \\
    &t^{B}_{\text{P}\text{Q}} \mapsto (W^{B_1 \oplus B_2}_{\text{P},\bk_{\text{P}\text{Q}}})^\dag t^B_{\text{P}\text{Q}} W^{B_1\oplus B_2}_{\text{Q},\bk_{\text{P}\text{Q}}}, 
\end{align}
so the change of $\mZ_2$ values is given by 
\begin{align}
&\mbox{$C_2^x$-axis: }\nonumber\\
    &\zeta^A_{\text{P}\text{Q}}\mapsto
    \zeta^A_{\text{P}\text{Q}} \eta^A_\text{P} \eta^{B_3}_\text{P} \eta^A_\text{Q} \eta^{B_3}_\text{Q}, \label{eq:gauge_tr_P2221'_1}\\
    &\zeta^B_{\text{P}\text{Q}}\mapsto
    \zeta^{B}_{\text{P}\text{Q}} \eta^{B_1}_\text{P} \eta^{B_2}_\text{P} \eta^{B_1}_\text{Q} \eta^{B_2}_\text{Q},
\end{align}
Similarly, for the $C_2^y,C_2^z$-axes, the gauge transformation for $\mZ_2$ values $\zeta^\alpha_{\text{P}\text{Q}}$ can be obtained: 
\begin{align}
&\mbox{$C_2^y$-axis: }\nonumber\\
    &\zeta^A_{\text{P}\text{Q}}\mapsto
    \zeta^A_{\text{P}\text{Q}} \eta^{A}_\text{P} \eta^{B_2}_\text{P} \eta^{A}_\text{Q} \eta^{B_2}_\text{Q}, \\
    &\zeta^B_{\text{P}\text{Q}}\mapsto
    \zeta^{B}_{\text{P}\text{Q}} \eta^{B_1}_\text{P} \eta^{B_3}_\text{P} \eta^{B_1}_\text{Q} \eta^{B_3}_\text{Q}, \\
&\mbox{$C_2^z$-axis: }\nonumber\\
    &\zeta^A_{\text{P}\text{Q}}\mapsto
    \zeta^A_{\text{P}\text{Q}} \eta^{A}_\text{P} \eta^{B_1}_\text{P} \eta^{A}_\text{Q} \eta^{B_1}_\text{Q}, \\
    &\zeta^B_{\text{P}\text{Q}}\mapsto
    \zeta^{B}_{\text{P}\text{Q}} \eta^{B_2}_\text{P} \eta^{B_3}_\text{P} \eta^{B_2}_\text{Q} \eta^{B_3}_\text{Q}.    \label{eq:gauge_tr_P2221'_-1}
\end{align}

The gauge transformations \eqref{eq:gauge_tr_P2221'_1}-\eqref{eq:gauge_tr_P2221'_-1} correspond to the differential $d_1^{0,-1}: \mZ_2^{32} \to \mZ_2^{24}$ in AHSS. 
We find that $\coker d_1^{0,-1} \cong \mZ_2^6$, hence there exist six gauge-invariant and independent combinations.
Since $E_2^{1,-1} \cong \mZ_2$, five of them are $\mZ_2$ invariants detecting gapless points in 2-cells. 
Indeed, the following six ``$\pi$ Berry phases'' along the boundaries of the six 2-cells shown in Fig.~\ref{fig:TI_P212121} (d) are gauge-invariant and detect gapless points in 2-cells:
\begin{align}
&\tau_{\G \text{X}\text{S}\text{Y}} := \zeta^A_{\G \text{X}}\zeta^B_{\G \text{X}}\zeta^A_{\text{X}\text{S}}\zeta^B_{\text{X}\text{S}}\zeta^A_{\text{S}\text{Y}}\zeta^B_{\text{S}\text{Y}}\zeta^A_{\text{Y}\G}\zeta^B_{\text{Y}\G}, \\
&\tau_{\text{Z}\text{U}\text{R}\text{T}} = \zeta^A_{\text{Z}\text{U}}\zeta^B_{\text{Z}\text{U}}\zeta^A_{\text{U}\text{R}}\zeta^B_{\text{U}\text{R}}\zeta^A_{\text{R}\text{T}}\zeta^B_{\text{R}\text{T}}\zeta^A_{\text{T}\text{Z}}\zeta^B_{\text{T}\text{Z}}, \\
&\tau_{\G \text{X}\text{U}\text{Z}} = \zeta^A_{\G \text{X}}\zeta^B_{\G \text{X}}\zeta^A_{\text{X}\text{U}}\zeta^B_{\text{X}\text{U}}\zeta^A_{\text{U}\text{Z}}\zeta^B_{\text{U}\text{Z}}\zeta^A_{\text{Z}\G}\zeta^B_{\text{Z}\G}, \\
&\tau_{\text{Y}\text{S}\text{R}\text{T}} = \zeta^A_{\text{Y}\text{S}}\zeta^B_{\text{Y}\text{S}}\zeta^A_{\text{S}\text{R}}\zeta^B_{\text{S}\text{R}}\zeta^A_{\text{R}\text{T}}\zeta^B_{\text{R}\text{T}}\zeta^A_{\text{T}\text{Y}}\zeta^B_{\text{T}\text{Y}}, \\
&\tau_{\G \text{Y}\text{T}\text{Z}} = \zeta^A_{\G \text{Y}}\zeta^B_{\G \text{Y}}\zeta^A_{\text{Y}\text{T}}\zeta^B_{\text{Y}\text{T}}\zeta^A_{\text{T}\text{Z}}\zeta^B_{\text{T}\text{Z}}\zeta^A_{\text{Z}\G}\zeta^B_{\text{Z}\G}, \\
&\tau_{\text{X}\text{S}\text{R}\text{U}} = \zeta^A_{\text{X}\text{S}}\zeta^B_{\text{X}\text{S}}\zeta^A_{\text{S}\text{R}}\zeta^B_{\text{S}\text{R}}\zeta^A_{\text{R}\text{U}}\zeta^B_{\text{R}\text{U}}\zeta^A_{\text{U}\text{X}}\zeta^B_{\text{U}\text{X}}.
\end{align}
These are not independent; using Eqs.~\eqref{eq:s1}-\eqref{eq:s3}, the following relation exists 
\begin{align}
&\tau_{\G \text{X}\text{S}\text{Y}} \tau_{\text{Z}\text{U}\text{R}\text{T}} \tau_{\G \text{X}\text{U}\text{Z}} \tau_{\text{Y}\text{S}\text{R}\text{T}} \tau_{\G \text{Y}\text{T}\text{Z}} \tau_{\text{X}\text{S}\text{R}\text{U}} \nonumber\\
&= \prod_{\text{P}\text{Q} \in \{\rm 1\text{-}cells\}} (\zeta^A_{\text{P}\text{Q}}\zeta^B_{\text{P}\text{Q}})^2 =1. 
\end{align}

The remaining one $\mZ_2$ number is given by the product of the transition functions for the $B$ representation in all 1-cells.
\begin{align}
&(-1)^{\nu} := 
\prod_{\text{P}\text{Q} \in \{\rm 1\text{-}cells\}} \zeta^B_{\text{P}\text{Q}} \nonumber\\
&=
\zeta^B_{\G \text{X}}\zeta^B_{\text{X}\text{S}}\zeta^B_{\text{S}\text{Y}}\zeta^B_{\text{Y}\G}
\zeta^B_{\text{Z}\text{T}}\zeta^B_{\text{T}\text{R}}\zeta^B_{\text{R}\text{U}}\zeta^B_{\text{U}\text{Z}}
\zeta^B_{\text{Z}\G}\zeta^B_{\text{X}\text{U}}\zeta^B_{\text{Y}\text{T}}\zeta^B_{\text{R}\text{S}}.
\label{eq:z2inv_p2221'}
\end{align}
A choice of orientations of 1-cells is arbitrary, but here it was introduced as a natural choice when removing the gauge fixing condition later.
Using Eqs.~\eqref{eq:s1}-\eqref{eq:s3}, we find that 
\begin{align}
\prod_{\text{P}\text{Q} \in \{\rm 1\text{-}cells\}} (\zeta^B_{\text{P}\text{Q}})^2
= \prod_{\text{P} \in \{\rm 0\text{-}cells\}}(-1)^{n_{B_2(\text{P})}}, 
\end{align}
and further using the compatibility conditions \eqref{eq:comp1}-\eqref{eq:comp3}, $\sum_{\text{P} \in {\rm 0\text{-}cells}}n_{B_2(\text{P})} \equiv 0$. 
Therefore, $\nu$ is quantized as 
\begin{align}
(-1)^{\nu} \in \{\pm 1\}. 
\end{align}

It should be shown that there indeed exist band structures whose $\nu$ is $0$ and $1$, to claim that $\nu$ is meaningful.
We pose this to Sec.~\ref{sec:P2221'_z2inv_ai_model}.

\subsubsection{The Berry phase formula}
\label{sec:P2221'_Z2_Berry_phase_formula}
The $\mZ_2$ invariant formula $\nu$ \eqref{eq:z2inv_p2221'} needs the gauge fixing conditions \eqref{eq:g2} and \eqref{eq:g1}, hence it lacks practicality.
In this section, we seek an expression that does not require any gauge fixing conditions.
Looking at the expression \eqref{eq:z2inv_p2221'}, it is reasonable to consider the product of Berry phases for the $B$-irrep in all 1-cells.

Introduce a mesh of the 1-skeleton $X_1$, and for each 0-cell, we compute the Bloch states of each irrep: 
\begin{align}
(\Phi^A_{\text{P}},\Phi^{B_1}_{\text{P}},\Phi^{B_2}_{\text{P}},\Phi^{B_3}_{\text{P}}),\quad 
\text{P} \in \{\mbox{0-cells}\}. 
\end{align}
For 1-cells, we compute the Bloch states for $A$ and $B$ irreps:
\begin{align}
(\Phi^A_\bk,\Phi^B_\bk),\quad \bk \in \{\mbox{1-cells}\}.
\end{align}
For each 1-cell $\text{P}\text{Q}$, depending on the twofold rotation axis, we compute the $\text{U}(1)$ Wilson line associated with the $B$ representation as defined below:
\begin{widetext}
\begin{align}
    &\mbox{$C_2^x$-axis:} \quad 
    e^{\ii\g^B_{\text{P}\text{Q}}}
    := \lim_{{\cal N}\to \infty}
    \det\left[ 
    (\Phi^{B_1}_\text{Q},\Phi^{B_2}_\text{Q})^\dag 
    \left( \prod_{j=1}^{{\cal N}-1} \Phi^B_{\text{P}+j\bm{\delta}} (\Phi^B_{\text{P}+j \bm{\delta}})^\dag \right)
    (\Phi^{B_1}_\text{P},\Phi^{B_2}_\text{P})
    \right], \label{eq:P2221'_Berry_Phase_c2x}\\
    &\mbox{$C_2^y$-axis:} \quad 
    e^{\ii\g^B_{\text{P}\text{Q}}}
    := \lim_{{\cal N}\to \infty}
    \det\left[ 
    (\Phi^{B_1}_\text{Q},\Phi^{B_3}_\text{Q})^\dag 
    \left( \prod_{j=1}^{{\cal N}-1} \Phi^B_{\text{P}+j\bm{\delta}} (\Phi^B_{\text{P}+j \bm{\delta}})^\dag \right)
    (\Phi^{B_1}_\text{P},\Phi^{B_3}_\text{P})
    \right], \\
    &\mbox{$C_2^z$-axis:} \quad 
    e^{\ii\g^B_{\text{P}\text{Q}}}
    := \lim_{{\cal N}\to \infty}
    \det\left[ 
    (\Phi^{B_2}_\text{Q},\Phi^{B_3}_\text{Q})^\dag 
    \left( \prod_{j=1}^{{\cal N}-1} \Phi^B_{\text{P}+j\bm{\delta}} (\Phi^B_{\text{P}+j \bm{\delta}})^\dag \right)
    (\Phi^{B_2}_\text{P},\Phi^{B_3}_\text{P})
    \right].
\end{align}
Here, $\bm{\delta} = (\text{Q}-\text{P})/{\cal N}$.
The Wilson line is not gauge invariant as it depends on the gauge at endpoints. 
Under the gauge transformation at 0-cell by 
\begin{align}
    \Phi^\beta_P \mapsto \Phi^\beta_P W^\beta_P, \quad W^\beta_P \in \text{U}(n_{\beta(P)}), 
\end{align}
the Wilson line transforms as 
\begin{align}
    &\mbox{$C_2^x$-axis:} \quad 
    e^{\ii\g^B_{\text{P}\text{Q}}}
    \mapsto (\det W^{B_1}_\text{Q})^* (\det W^{B_2}_\text{Q})^* e^{\ii\g^B_{\text{P}\text{Q}}} \det W^{B_1}_\text{P} \det W^{B_2}_\text{P}, \\
    &\mbox{$C_2^y$-axis:} \quad 
    e^{\ii\g^B_{\text{P}\text{Q}}}
    \mapsto (\det W^{B_1}_\text{Q})^* (\det W^{B_3}_\text{Q})^* e^{\ii\g^B_{\text{P}\text{Q}}} \det W^{B_1}_\text{P} \det W^{B_3}_\text{P}, \\
    &\mbox{$C_2^z$-axis:} \quad 
    e^{\ii\g^B_{\text{P}\text{Q}}}
    \mapsto (\det W^{B_2}_\text{Q})^* (\det W^{B_3}_\text{Q})^* e^{\ii\g^B_{\text{P}\text{Q}}} \det W^{B_2}_\text{P} \det W^{B_3}_\text{P}. 
\end{align}
Now consider the product of the following Wilson lines 
\begin{align}
z:=
e^{\ii\g^B_{\G \text{X}}} e^{\ii\g^B_{\text{X}\text{S}}} e^{\ii\g^B_{\text{S}\text{Y}}} e^{\ii\g^B_{\text{Y}\G}} 
\times 
e^{\ii\g^B_{\text{Z}\text{T}}} e^{\ii\g^B_{\text{T}\text{R}}} e^{\ii\g^B_{\text{R}\text{U}}} e^{\ii\g^B_{\text{U}\text{Z}}}
\times
e^{\ii\g^B_{\text{Z}\G}} e^{\ii\g^B_{\text{X}\text{U}}} e^{\ii\g^B_{\text{Y}\text{T}}} e^{\ii\g^B_{\text{R}\text{S}}}. 
\end{align}
The $\text{U}(1)$ value $z$ is still not gauge invariant and transforms into
\begin{align}
z \mapsto z \times 
\prod_{\text{P} \in \{\G,\text{S},\text{U},\text{T}\}} (\det W^{B_3}_{\text{P}})^{-2} \times 
\prod_{\text{P} \in \{\text{X},\text{Y},\text{Z},\text{R}\}} (\det W^{B_3}_{\text{P}})^{2}. 
\label{eq:gauge_dep_P222}
\end{align}
However, using the prescription described in \ref{sec:source_sink_wilson_line}, $z$ can be corrected to be gauge invariant using $\calT$ symmetry. 
In the present case,
\begin{align}
e^{\ii \g} := z \times 
\prod_{\text{P} \in \{\G,\text{S},\text{U},\text{T}\}} \det \left[(\Phi^{B_3}_{\text{P}})^\dag U_\text{P}(\calT) (\Phi^{B_3}_{\text{P}})^*\right]^{-1} \times 
\prod_{\text{P} \in \{\text{X},\text{Y},\text{Z},\text{R}\}} \det \left[(\Phi^{B_3}_{\text{P}})^\dag U_\text{P}(\calT) (\Phi^{B_3}_{\text{P}})^*\right]
\end{align}
is gauge-invariant.

To explore the relationship between the $\mZ_2$ invariant $\nu$ and $e^{i\g}$, we deform $e^{i\g}$ by taking a smooth gauge that satisfies the gauge fixing conditions \eqref{eq:g2} and \eqref{eq:g1} near 0-cells.
First, from the TRS gauge \eqref{eq:g1}, it follows that $\det [(\Phi^{B_3}_{\text{P}})^\dag U_\text{P}(\calT) (\Phi^{B_3}_{\text{P}})^*]=1$.
For Wilson lines, let us consider $e^{\ii\g_{\G \text{X}}}$ concretely. 
Taking the basis \eqref{eq:Bloch_with_gauge_cond_P2221'} that satisfies gauge fixing conditions \eqref{eq:g2} and \eqref{eq:g1} near the $\G$ and $\text{X}$ points, the Wilson line becomes 
\begin{align}
e^{\ii\g^B_{\G \text{X}}}
&=
\exp \left[-\int_{\bk_{\G \text{X}}}^\text{X} \tr \left[ (\Phi^{B_1}_{\G,\bk}, \Phi^{B_2}_{\G,\bk})^\dag d (\Phi^{B_1}_{\G,\bk}, \Phi^{B_2}_{\G,\bk}) \right] \right] \nonumber \\
&\times 
\det \left[(\Phi^{B_1}_{\text{X},\bk_{\G \text{X}}},\Phi^{B_2}_{\text{X},\bk_{\G \text{X}}})^\dag (\Phi^{B_1}_{\G,\bk_{\G \text{X}}}, \Phi^{B_2}_{\G,\bk_{\G \text{X}}})\right] \times 
\exp \left[ -\int_\G^{\bk_{\G \text{X}}} \tr \left[ (\Phi^{B_1}_{\G,\bk}, \Phi^{B_2}_{\G,\bk})^\dag d (\Phi^{B_1}_{\G,\bk}, \Phi^{B_2}_{\G,\bk}) \right]\right] \\
&=
\exp \left[-\int_{\bk_{\G \text{X}}}^\text{X} \tr \left[ (\Phi^{B_1}_{\G,\bk}, \Phi^{B_2}_{\G,\bk})^\dag d (\Phi^{B_1}_{\G,\bk}, \Phi^{B_2}_{\G,\bk}) \right] \right] \times \zeta^B_{\text{X}\G} \times 
\exp \left[-\int_\G^{\bk_{\G \text{X}}} \tr \left[ (\Phi^{B_1}_{\G,\bk}, \Phi^{B_2}_{\G,\bk})^\dag d (\Phi^{B_1}_{\G,\bk}, \Phi^{B_2}_{\G,\bk}) \right] \right]. 
\end{align}
For the line integrals of the exponents, using $C_2^z \calT$ symmetry, 
\begin{align}
&\int_\G^{\bk_{\G \text{X}}} \tr \left[ (\Phi^{B_1}_{\G,\bk}, \Phi^{B_2}_{\G,\bk})^\dag d (\Phi^{B_1}_{\G,\bk}, \Phi^{B_2}_{\G,\bk}) \right]^*\nonumber \\
&=
\int_\G^{\bk_{\G \text{X}}}\tr \Bigg[
\begin{pmatrix}
\mathds{1}_{n_{B_1(\G)}}\\
&-\mathds{1}_{n_{B_2(\G)}}
\end{pmatrix}
(\Phi^{B_1}_{\G,\bk}, \Phi^{B_2}_{\G,\bk})^\dag U_\bk(C_2^z\calT) 
d \left\{ U_\bk(C_2^z\calT)^\dag (\Phi^{B_1}_{\G,\bk}, \Phi^{B_2}_{\G,\bk}) 
\begin{pmatrix}
\mathds{1}_{n_{B_1(\G)}}\\
&-\mathds{1}_{n_{B_2(\G)}}
\end{pmatrix}
\right\} \Bigg]\\
&=
\int_\G^{\bk_{\G \text{X}}}\tr \left[ (\Phi^{B_1}_{\G,\bk}, \Phi^{B_2}_{\G,\bk})^\dag d (\Phi^{B_1}_{\G,\bk}, \Phi^{B_2}_{\G,\bk}) \right]+
\int_\G^{\bk_{\G \text{X}}}\tr \left[ (\Phi^{B_1}_{\G,\bk}, \Phi^{B_2}_{\G,\bk})^\dag U_\bk(C_2^z\calT) dU_\bk(C_2^z\calT)^\dag (\Phi^{B_1}_{\G,\bk}, \Phi^{B_2}_{\G,\bk}) \right].
\end{align}
Since $\tr {\cal A}^* = - \tr {\cal A}$, we obtain 
\begin{align}
&\int_\G^{\bk_{\G \text{X}}} \tr \left[ (\Phi^{B_1}_{\G,\bk}, \Phi^{B_2}_{\G,\bk})^\dag d (\Phi^{B_1}_{\G,\bk}, \Phi^{B_2}_{\G,\bk}) \right]
=-\frac{1}{2} \int_\G^{\bk_{\G \text{X}}}\tr \left[ (\Phi^{B_1}_{\G,\bk}, \Phi^{B_2}_{\G,\bk})^\dag U_\bk(C_2^z\calT) dU_\bk(C_2^z\calT)^\dag (\Phi^{B_1}_{\G,\bk}, \Phi^{B_2}_{\G,\bk}) \right]. 
\end{align}
This expression is gauge invariant, independent of gauge transformations $\Phi^B_\bk = (\Phi^{B_1}_{\G,\bk}, \Phi^{B_2}_{\G,\bk}) \mapsto \Phi^B_\bk W^B_\bk$ with $W^B_\bk \in \text{U}(n_{B_1(\G)}+n_{B_2(\G)})$. 
Similarly, for the Wilson line $e^{\ii\g^B_{\G \text{X}}}$ we find the relation:
\begin{align}
e^{\ii\g^B_{\G \text{X}}}
=\zeta^B_{\text{X} \G} \times 
\exp \frac{1}{2}\int_{\G\to \text{X}} \tr \left[
(\Phi^B_\bk)^\dag U_\bk(C_2^z\calT) d U_\bk(C_2^z\calT)^\dag \Phi^B_\bk \right].
\end{align}
The same result can be obtained by replacing $C_2^z\calT$ with $C_2^y\calT$. 
In fact, the ratio of contributions from the correction terms of $C_2^z\calT$ and $C_2^y\calT$ is $\exp \frac{1}{2}\int_{\G\to \text{X}} \tr [(\Phi^B_\bk)^\dag U_\bk(C_2^x) d U_\bk(C_2^x)^\dag \Phi^B_{\bk}]$. 
However, considering $U_\bk(C_2^x)^\dag =U_\bk(C_2^x), U_\bk(C_2^x) \Phi^B_\bk = - \Phi^B_\bk$ on the 1-cell $\bk \in \G \text{X}$, we can show that $\tr [\Phi_\bk^{B\dag} U_\bk(C_2^x) d U_\bk(C_2^x)^\dag \Phi^B_\bk] = \frac{1}{2} \tr [\Phi^{B\dag}_\bk d U_\bk(C_2^x)^2 \Phi^B_\bk] = 0$.

Applying the same to other Wilson lines, we derive the following relation between the $\mZ_2$ invariant $\nu$ and the gauge-invariant Berry phase $e^{\ii\g}$:
\begin{align}
(-1)^\nu 
&= e^{-\ii\g} \times 
\exp \left[\frac{1}{2} \left(\oint_{\G\to \text{X}\to \text{S} \to \text{Y} \to \G}+\oint_{\text{Z}\to \text{T} \to \text{R}\to \text{U}\to \text{Z}}\right) 
\tr \left[(\Phi^B_\bk)^\dag U_\bk(C_2^z\calT) d U_\bk(C_2^z\calT)^\dag \Phi^B_\bk \right]\right]\nonumber\\
&\times 
\exp
\left[ \frac{1}{2} \left(\int_{\text{Z}\to \G}+\int_{\text{X}\to \text{U}}+\int_{\text{Y}\to \text{T}}+\int_{\text{R} \to \text{S}}\right) \tr \left[(\Phi^B_\bk)^\dag U_\bk(C_2^x\calT) d U_\bk(C_2^x\calT)^\dag \Phi^B_\bk\right] \right].
\label{eq:z2inv_P2221'_berry}
\end{align}
\end{widetext}
Here, we used $\zeta^B_{\text{Q}\text{P}} = (\zeta^B_{\text{P}\text{Q}})^{-1}$. 
This provides us a gauge-invariant expression of the $\mZ_2$ invariant $\nu$.
However, there is a subtle ambiguity in $\nu$ from the ordering of irreps, as described below.

\subsubsection{Band sum and quadratic refinement}
It is important to note that the $\mZ_2$ invariant \eqref{eq:z2inv_p2221'} depends on the ordering of irreps in the Bloch states at the 0-cells.
For instance, in the expression of the transition function $t^B_{\G \text{X}}$ \eqref{eq:tr_func_GX_B}, if we swap the Bloch states near the $\G$ point, redefining $t^B_{\G \text{X}} \to \left(\Phi^{B_2}_{\G,\bk_{\G \text{X}}}\oplus \Phi^{B_1}_{\G,\bk_{\G \text{X}}}\right)^\dag 
\left(\Phi^{B_1}_{\text{X},\bk_{\G \text{X}}}\oplus \Phi^{B_2}_{\text{X},\bk_{\G \text{X}}}\right)$, a sign $(-1)^{n_{B_1(\G)}n_{B_2(\G)}}$ arises due to the swapping of irreps.
This issue of ordering dependency of irreps also occurs in the Berry phase formula \eqref{eq:z2inv_P2221'_berry} of the $\mZ_2$ invariant. 
For example, in the expression of the Wilson line \eqref{eq:P2221'_Berry_Phase_c2x} parallel to the $C_2^x$ axis, if we swap the order of Bloch states at the starting point $\text{P}$ from $(\Phi^{B_1}_\text{P},\Phi^{B_2}_\text{P})$ to $(\Phi^{B_2}_\text{P},\Phi^{B_1}_\text{P})$, a sign $(-1)^{n_{B_1(\text{P})} n_{B_2(\text{P})}}$ appears.
This ordering dependency leads to the non-additive nature of the $\mZ_2$ invariant $\nu$ for the direct sum of bands. 
However, as discussed in Sec.~\ref{sec:redef_inv}, it is possible to redefine $\nu$ to preserve the additive structure.

Recall that the order of irreps in the expression of the $\mZ_2$ invariant $\nu$ for a 0-cell $\text{P}$ was
\begin{align}
&\text{$C_2^x$-axis}: \quad (\Phi^{B_1}_\text{P},\Phi^{B_2}_\text{P}), \\
&\text{$C_2^y$-axis}: \quad (\Phi^{B_1}_\text{P},\Phi^{B_3}_\text{P}), \\
&\text{$C_2^z$-axis}: \quad (\Phi^{B_2}_\text{P},\Phi^{B_3}_\text{P}). 
\end{align}
Let $\Phi^{B,\beta}_{\text{P}}$ and $n^B_{\beta(\text{P})}$ be the Bloch states of $\beta$-irrep at 0-cell $\text{P}$ for band $B \in \{E,F\}$ and the number of $\beta$-irrep, respectively. 
Here we focus on the $C_2^x$-axis. 
Considering the direct sum $E \oplus F$ as a single band structure, the above ordering rule implies that 
\begin{align}
(\Phi^{E,B_1}_\text{P}\oplus \Phi^{F,B_1}_\text{P}) \oplus (\Phi^{E,B_2}_\text{P}\oplus \Phi^{F,B_2}_\text{P}). 
\end{align}
On the one hand, firstly ordering each band $E$ and $F$ as above and considering the direct sum, we have 
\begin{align}
(\Phi^{E,B_1}_\text{P}\oplus \Phi^{E,B_2}_\text{P}) \oplus (\Phi^{F,B_1}_\text{P}\oplus \Phi^{F,B_2}_\text{P}). 
\end{align}
They are related with a permutation matrix $S$ from the right whose determinant is 
\begin{align}
\det S = (-1)^{n^E_{B_2(\text{P})} n^F_{B_1(\text{P})}}. 
\end{align}
Similar contributions arise from the Wilson lines on the $C_2^y$ and $C_2^z$ axes. 
The non-additive nature of the invariant $\nu$ is summarized as 
\begin{align}
&\nu(E \oplus F)
\equiv \nu(E)+\nu(F)+\delta \nu(E|_{E_2^{0,0}},F|_{E_2^{0,0}})
\end{align}
with 
\begin{align}
&\delta \nu(E|_{E_2^{0,0}},F|_{E_2^{0,0}}) \equiv 
\sum_{\text{P} \in \{\rm 0\text{-}cells\}} 
\Big( n^E_{B_2(\text{P})} n^F_{B_1(\text{P})}+\nonumber \\
&\qquad \qquad n^E_{B_3(\text{P})} n^F_{B_1(\text{P})}+n^E_{B_3(\text{P})} n^F_{B_2(\text{P})}
\Big). 
\label{eq:inter}
\end{align}
Note that $\delta \nu$ depends only on elements of $E_2^{0,0}$.

Since the number of irreps behaves additively with respect to the sum of bands, $\delta \nu$ is a bilinear form.
Moreover, although $\delta \nu(x,y)$ is not symmetric for elements of $E_1^{0,0}$, we find that it is symmetric when restricted to $E_2^{0,0}$, 
\begin{align}
    \delta \nu (x,y) \equiv \delta \nu(y,x), \quad x,y \in E_2^{0,0}. 
\end{align}
Furthermore, we find that the diagonal terms are zero, $\delta (x,x) \equiv 0$ for $x \in E_2^{0,0}$. 
(This is not necessary for the existence of quadratic refinement below.) 
Therefore, according to Sec.~\ref{sec:redef_inv}, there exists a quadratic refinement $q: E_2^{0,0} \to \{0,1\}$ of $\delta \nu$, which satisfies
\begin{align}
    \delta \nu(x,y)\equiv q(x+y)+q(x)+q(y), \quad x,y \in E_2^{0,0}.
\end{align}
By explicitly constructing the basis of $E_2^{0,0}$ and calculating expression \eqref{eq:quadratic_refinement_result}, we obtain the following formula:
\begin{widetext}
\begin{align}
q\left(\{n_{\beta(\text{P})}\}_{\text{P},\beta}\right)
&\equiv
n_{B_3(\text{R})} \left(n_{B_3(\text{S})}+n_{B_2(\text{T})}+n_{B_3(\text{T})}+n_{B_2(\text{U})}+n_{B_3(\text{U})}+n_{B_3(\text{Z})}\right) \nonumber \\
&+n_{B_2(\text{R})} \left(n_{B_3(\text{T})}+n_{B_3(\text{U})}+n_{B_3(\text{X})}+n_{B_3(\text{Y})}\right) \nonumber \\
&+n_{B_2(\text{T})} \left(n_{B_3(\text{S})}+n_{B_3(\text{U})}+n_{B_3(\text{X})}\right)+n_{B_3(\text{Y})} \left(n_{B_3(\text{S})}+n_{B_2(\text{U})}+n_{B_3(\text{X})}+n_{B_3(\text{Z})}\right) \nonumber \\
&+n_{B_3(\text{S})} n_{B_2(\text{U})}+n_{B_3(\text{X})} \left(n_{B_3(\text{S})}+n_{B_3(\text{Z})}\right)+n_{B_3(\text{S})} n_{B_3(\text{Z})} \nonumber \\
&+n_{B_3(\text{T})} \left(n_{B_2(\text{U})}+n_{B_3(\text{U})}+n_{B_3(\text{Y})}+n_{B_3(\text{Z})}\right)+n_{B_3(\text{U})} \left(n_{B_3(\text{X})}+n_{B_3(\text{Z})}\right)\nonumber \\
&+n_{B_3(\text{S})}+n_{B_3(\text{T})}+n_{B_3(\text{U})}.
\label{eq:qr_P2221'}
\end{align}
\end{widetext}
See Appendix~\ref{app:derivation_of_qr_P2221'} for a derivation. 
Using this quadratic refinement $q$, we redefine the $\mZ_2$ invariant as
\begin{align}
    \tilde \nu(E) := \nu(E) + q(E|_{E_2^{0,0}})
\end{align}
to recover linearity
\begin{align}
    \tilde \nu(E \oplus F) \equiv \tilde \nu(E) + \tilde \nu(F).
\end{align}

In the formula \eqref{eq:qr_P2221'}, the linear term $n_{B_3(\text{S})}+n_{B_3(\text{T})}+n_{B_3(\text{U})}$ were chosen so that for any atomic insulator with a single band, i.e., an atomic insulator at Wyckoff position $\bm{x}_0$ with $\beta$-irrep, denoted as $a_{\bx_0}^\beta$, the following holds 
\begin{align}
\nu(a_{\bx_0}^\beta) \equiv \tilde \nu(a_{\bx_0}^\beta).
\label{eq:constraint_tilde_nu}
\end{align}
Namely, we impose the constraint 
\begin{align}
    q(a_{\bx_0}^\beta) \equiv 0
    \label{eq:condition_q_P2221'}
\end{align}
on the quadratic refinement $q$.
See the next section. 

\subsubsection{$\mZ_2$ invariant for real-space models}
\label{sec:P2221'_z2inv_ai_model}
According to the real-space AHSS~\cite{Shiozaki-Ono2023}, the $K$-group is given by $\mZ^{13}+\mZ_2$, and all are generated by atomic insulators (or formal differences of their direct sums).

First, we calculate the $\mZ_2$ invariant $\nu$ for atomic insulators with a single band.
We denote the atomic insulator obtained by placing an irrep $\beta \in \{A,B_1,B_2,B_3\}$ of the magnetic point group $2221'$ at a position shifted by the displacement vector $\bm{x}_0$ from the unit cell center as $a^\beta_{\bm{x}_0}$. 
The symmetry matrices are given by $U_\bk(\calT)=1$ and 
\begin{align}
U_\bk(C_2^\mu) =\chi^\beta (C_2^\mu) e^{-i\bk \cdot (\bm{x}_0-C_2^\mu \bm{x}_0)},\quad \mu \in \{x,y,z\}. 
\end{align}
The Bloch wave function can be $\Phi_\bk \equiv 1$ independent of the momentum space, and no contribution arises from the Wilson line, making $e^{\ii\g}=1$. 
Calculating the factor arising from the symmetry transformation in Eq.~\eqref{eq:z2inv_P2221'_berry}, we get 
\begin{align}
\tilde \nu(a^{\beta}_{\bm{x}_0}) \equiv \nu(a^{\beta}_{\bm{x}_0}) \equiv \left\{
\begin{array}{ll}
   1 & \bm{x}_0 = (\frac{1}{2},\frac{1}{2},\frac{1}{2}), \\
   0 & \mbox{else}.
\end{array}\right.
\end{align}
Here, the redefined $\tilde \nu$ is chosen to satisfy Eq.~\eqref{eq:constraint_tilde_nu}.

From the real-space AHSS, the $\mZ_2$ part of the $K$-group is given by the formal difference $[E]-[F]$ of the following two insulating states $E$ and $F$: 
\begin{align}
&E= a^A_{(0,0,0)}\oplus a^A_{(\frac{1}{2},\frac{1}{2},0)}\oplus a^A_{(0,\frac{1}{2},\frac{1}{2})}\oplus a^A_{(\frac{1}{2},0,\frac{1}{2})},\\
&F= a^A_{(\frac{1}{2},0,0)}\oplus a^A_{(0,\frac{1}{2},0)}\oplus a^A_{(0,0,\frac{1}{2})}\oplus a^A_{(\frac{1}{2},\frac{1}{2},\frac{1}{2})}.
\end{align}
Note that insulators $E$ and $F$ show identical representations at TRIMs and thus cannot be distinguished by representations at 0-cells. 
Due to the linearity of the $\mZ_2$ invariant $\tilde \nu$, we have
\begin{align}
\tilde \nu(E) = 0,\quad 
\tilde \nu(F) = 1, 
\end{align}
The $\mZ_2$ invariant $\tilde \nu$ successfully distinguishes between the two insulators $E$ and $F$.

Let us comment on an interesting physical consequence of nontrivial $\mZ_2$ nature.
The space group $P222$ is a subgroup of many other space groups, which allows us to define the $\mZ_2$ invariant for the supergroups of $P222$.
Recently, it has shown that, although atomic insulators do not exhibit any protected gapless surface states, certain types of atomic insulators still possess mid-gap surface states, known as fractional corner charges~\cite{Benalcazar61, Corner_irrep_Hughes,Corner_Irrep_Schindler}. 
Furthermore, there exist fractionally quantized corner charges that cannot be captured only by irreps at $0$-cells~\cite{Naito-Takahashi-Watanabe-Murakami}.
Indeed, the $\mathbb{Z}_2$ invariant constructed here plays a key role in the corner-charge formulas of Ref.~\cite{Wada-Naito-Ono-Shiozaki-Murakami_2023}, where it is combined with irreps at 0-cells to detect fractional corner charges in various tetrahedral and cubic space groups beyond high-symmetry-point data.

\subsubsection{$\Z_2$ invariant for adiabatic cycle without TRS}
In the absence of TRS, although the gauge dependency (\ref{eq:gauge_dep_P222}) of $z$ implies that $z$ itself is not physically observable, $z$ serves as a $\mZ_2$ invariant of adiabatic cycles. 
To see this, consider a one-parameter family of Hamiltonians $H_{\bk}(\theta \in [0,2\pi])$ satisfying $C_2^\mu$-symmetry constraints (\ref{eq:sym_P222_C2mu}) for $\mu=x,y,z$. 
We also assume the periodicity $H_\bk(2\pi) = H_\bk(0)$. 
At TRIMs $\text{P}$, the Bloch states $\Phi_\text{P}^{B_3}(\theta)$ in the $B_3$-irrep can be continuous and periodic in $\theta$. 
In doing so, $z(\theta)$ is also continuous and periodic in $\theta$, and we can define the winding number 
\begin{align}
    N = \frac{1}{2\pi \ii} \oint d\theta \frac{\partial}{\partial\theta} z(\theta) \in \mZ. 
\end{align}
Not every integer of $N$ is gauge invariant: under a gauge transformation $\Phi^{B_3}_\text{P}(\theta) \mapsto \Phi_\text{P}^{B_3}(\theta) W_\text{P}^{B_3}(\theta)$ where $W_\text{P}^{B_3}(\theta)$ is continuous and periodic in $\theta$, $N$ changes as 
\begin{align}
    N \mapsto N + 2 \left(-\sum_{\text{P} \in \{\Gamma,\text{S},\text{U},\text{T}\}} M_P + \sum_{\text{P} \in \{\text{X},\text{Y},\text{Z},\text{R}\}} M_P\right) 
\end{align}
with $M_P$ the winding number of gauge transformation 
\begin{align}
    M_P = \frac{1}{2\pi \ii} \oint d\theta \frac{\partial}{\partial\theta} \log \det W_\text{P}^{B_3}(\theta) \in \mZ.
\end{align}
Therefore, only the parity of $N$ is gauge-invariant.

We note that the existence of $\mZ_2$ invariant of adiabatic cycles for the space group $P222$ is consistent with the $E_2$-page $E_2^{1,-2} = \Z_2$ of the momentum-space AHSS~\cite{K-AHSS, Shiozaki-Ono2023}.

\subsection{Other examples}
\label{sec:TI_exs}
In this section, we present illustrative examples demonstrating the definition of $\Z_2$ invariants using the scenarios listed in Sec.~\ref{sec:Symmetry-enriched Berry phase}, comparison with the real-space model (Sec.~\ref{sec:Example: spinful systems in space group P2 with TRS}), and $\mZ_2$ invariant detecting a Weyl point in generic point (Sec.~\ref{Example: spinless systems with space group P-4 with TRS}).

\begin{figure}[t]
	\centering
	\includegraphics[width=\columnwidth]{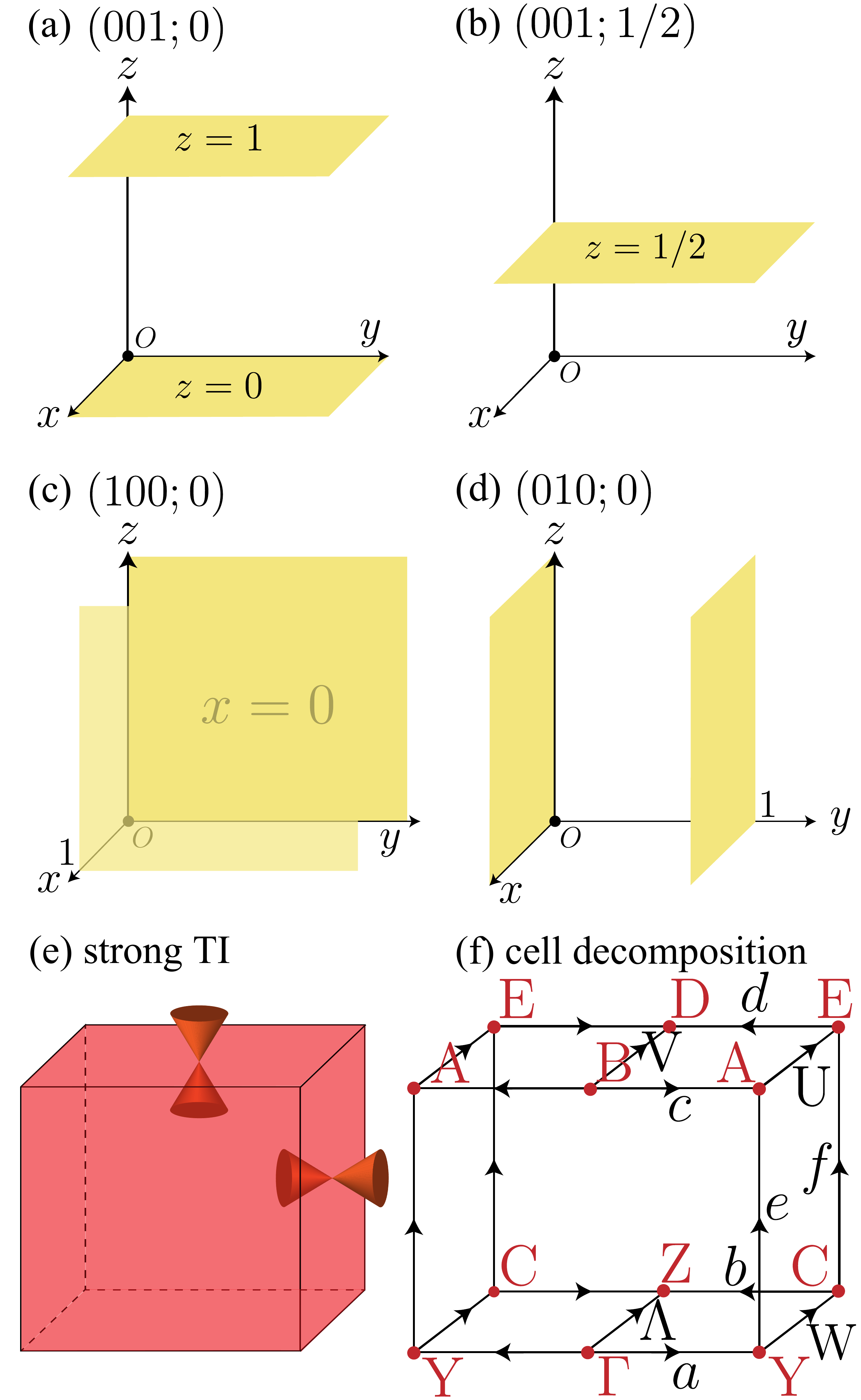}
	\caption{\label{fig:TI_P2}
		(a, b, c, d) Illustrations of topological crystalline insulators introduced in Ref.~\cite{SI_NC_Song}. Yellow planes represent two-dimensional topological insulators that are stacked along $z$-direction.
		(e) Illustration of a strong topological insulator (TI). 
		(f) Cell decomposition of the fundamental domain in $P2$ with TRS. Here, the labels of momenta other than $a, b, c, d, e$, and $f$ follow Ref.~\cite{BZ}.
	}
\end{figure}

\begin{table}[t!]
	\begin{center}
		\caption{\label{tab:irrep_P2}
			Character tables of irreps and their EAZ classes on $0$- and $1$-cells in space group $P2$ with TRS. 
		}
		\begin{tabular}{c|c|cc}
			\hline
			irrep. & EAZ & $e$ & $C_{2}^{y}$\\
			\hline\hline
			$\Gamma_1, \text{Y}_1, \text{C}_1, \text{Z}_1, \text{B}_1, \text{A}_1, \text{E}_1, \text{D}_1, \Lambda_1, \text{W}_1, \text{V}_1, \text{U}_1$ & A & $1$ & $-\ii$\\
			$\Gamma_2, \text{Y}_2, \text{C}_2, \text{Z}_2, \text{B}_2, \text{A}_2, \text{E}_2, \text{D}_2,\Lambda_2, \text{W}_2, \text{V}_2, \text{U}_2$ & A & $1$ & $\ii$\\
			\hline
			irrep. & EAZ & $e$ & \\
			\hline\hline
			$a_1, b_1, c_1, d_1, e_1, f_1$ & AI & $1$ & \\
			\hline
		\end{tabular}
	\end{center}
\end{table}

\subsubsection{Example: spinful systems in space group $P2$ with TRS}
\label{sec:Example: spinful systems in space group P2 with TRS}
We generalize the above discussion to spinful systems in space group $P2$ with TRS $\calT$.
This space group is generated by twofold rotation along $y$-axis, denoted by $C_{2}^{y}$, and translations. 
According to Ref.~\cite{Shiozaki-Ono2023}, the $K$-group is $^{\phi}K_{G/\Pi}^{(z,c)+0}(T^3) = \mZ \times (\mZ_2)^8$%
~\footnote{Here, we use the fact that $\mZ_4$ does not appear in the $K$-group~\cite{Shiozaki-Sato2014}.}, %
whose $\mZ\times (\mZ_2)^3$-part is the classification of atomic insulators and the remaining $(\mZ_2)^5$-part corresponds to the strong topological insulator and topological crystalline insulators, as shown in Fig.~\ref{fig:TI_P2}(a--e).
The $\mZ$ part of $^{\phi}K_{G/\Pi}^{(z,c)+0}(T^3)$ can be understood by $E_{2}^{0,0} = \mZ$.
The topological invariant is the filling $\nu_{\text{e}}/2 \in \mZ$ (the number of electrons per unit cell). 
Here, we discuss the $\mZ_2$ invariants to characterize the remaining part of the $K$-group.

Our cell decomposition of the fundamental domain is shown in Fig.~\ref{fig:TI_P2}(f).
For each of $0$- and $1$-cells, irreducible representations and their EAZ classes are shown in Table~\ref{tab:irrep_P2}.
Although there is no unitary symmetry other than identity on $a,b,c,d,e$, and $f$, they are symmetric under the product of twofold rotation $C_{2}^{y}$ and TRS $\calT$, which results in EAZ class AI.
As a result, $E_{1}^{1,-1} = (\mZ_2)^6$ is given by
\begin{align}
	E_{1}^{1,-1} = \bigoplus_{K = a,b,c,d,e,f}\mZ_2[\bm{b}_{K_1}^{(1)}].
\end{align}
From AHSS, we find that the differentials $d_{1}^{0,-1}$ and $d_{1}^{1,-1}$ are trivial, i.e., $E_{2}^{1,-1} = E_{1}^{1,-1} = (\mZ_2)^6$. 
Therefore, topological invariants are determinants of the gauge fixed transition functions [c.f.~\eqref{eq:z2_inv_1cell}].
The transition function on a $1$-cell connecting $\bk_0$ to $\bk_1$ is defined by
\begin{align}
	t_{\bk_0\bk_1} = \Phi^{\dagger}_{\bk_0, (\bk_0+\bk_1)/2}\Phi_{\bk_1, (\bk_0+\bk_1)/2},
\end{align}
where $\Phi_{\bk_i, \bk}$ is a set of occupied states at $\bk$ under independent gauge choices around $\bk_i$ [c.f.~Eq.~\eqref{eq:transition_function_raw}].

Similar to the above $P\calT$-symmetric example, we aim to represent  the transition function using the Berry phase, i.e., 
\begin{align}
	e^{\mathrm{i}\gamma_{\bk_0 \bk_1}} = \lim_{\mathcal{N}\rightarrow\infty} \det \prod_{j = 0}^{\mathcal{N}-1} \Phi_{\bk_0 + (j+1)\bm{\delta}}^{\dagger}\Phi_{\bk_0 + j\bm{\delta}}
\end{align}
for $\bm{\delta} = (\bk_1 - \bk_0)/\mathcal{N}$.
However, since the $1$-cell is not a loop in this case, the Berry phase is not gauge invariant under gauge transformations at the boundary $0$-cells.
We then use the Pfaffian to compensate for the Berry phase, as discussed in Sec.~\ref{sec:AII_Pfaffian}.
Finally, we arrive at the gauge-independent formulas
\begin{widetext}
	\begin{align}
		&(-1)^{\nu_1} := \det \tilde{t}_{\Gamma\text{Y}} = e^{-\mathrm{i}\gamma_{\Gamma\text{Y}}}
		\frac{\mathrm{Pf}[\Phi_{\text{Y}}^{\dagger}U(\calT)\Phi_{\text{Y}}^*]}{\mathrm{Pf}[\Phi_{\Gamma}^{\dagger}U(\calT)\Phi_{\Gamma}^*]}
		\exp\left[\frac{1}{2}\int_a \mathrm{tr}\Phi_{\bk}^{\dagger}(U_{\bk}(C_{2}^{y}\calT)dU^{\dagger}_{\bk}(C_{2}^{y}\calT))\Phi_{\bk}\right],\\
		&(-1)^{\nu_2} := \det \tilde{t}_{\text{C}\text{Z}}= e^{-\mathrm{i}\gamma_{\text{C}\text{Z}}}
		\frac{\mathrm{Pf}[\Phi_{\text{Z}}^{\dagger}U(\calT)\Phi_{\text{Z}}^*]}{\mathrm{Pf}[\Phi_{\text{C}}^{\dagger}U(\calT)\Phi_{\text{C}}^*]}
		\exp\left[\frac{1}{2}\int_b \mathrm{tr}\Phi_{\bk}^{\dagger}(U_{\bk}(C_{2}^{y}\calT)dU^{\dagger}_{\bk}(C_{2}^{y}\calT))\Phi_{\bk}\right],\\
		&(-1)^{\nu_3} := \det \tilde{t}_{\text{B}\text{A}}= e^{-\mathrm{i}\gamma_{\text{B}\text{A}}}
		\frac{\mathrm{Pf}[\Phi_{\text{A}}^{\dagger}U(\calT)\Phi_{\text{A}}^*]}{\mathrm{Pf}[\Phi_{\text{B}}^{\dagger}U(\calT)\Phi_{\text{B}}^*]}
		\exp\left[\frac{1}{2}\int_c \mathrm{tr}\Phi_{\bk}^{\dagger}(U_{\bk}(C_{2}^{y}\calT)dU^{\dagger}_{\bk}(C_{2}^{y}\calT))\Phi_{\bk}\right],\\
		&(-1)^{\nu_4} := \det \tilde{t}_{\text{E}\text{D}}=e^{-\mathrm{i}\gamma_{\text{E}\text{D}}}
		\frac{\mathrm{Pf}[\Phi_{\text{D}}^{\dagger}U(\calT)\Phi_{\text{D}}^*]}{\mathrm{Pf}[\Phi_{\text{E}}^{\dagger}U(\calT)\Phi_{\text{E}}^*]}
		\exp\left[\frac{1}{2}\int_d \mathrm{tr}\Phi_{\bk}^{\dagger}(U_{\bk}(C_{2}^{y}\calT)dU^{\dagger}_{\bk}(C_{2}^{y}\calT))\Phi_{\bk}\right],\\
		&(-1)^{\nu_5} := \det \tilde{t}_{\text{Y}\text{A}}=e^{-\mathrm{i}\gamma_{\text{Y}\text{A}}}
		\frac{\mathrm{Pf}[\Phi_{\text{A}}^{\dagger}U(\calT)\Phi_{\text{A}}^*]}{\mathrm{Pf}[\Phi_{\text{Y}}^{\dagger}U(\calT)\Phi_{\text{Y}}^*]}
		\exp\left[\frac{1}{2}\int_e \mathrm{tr}\Phi_{\bk}^{\dagger}(U_{\bk}(C_{2}^{y}\calT)dU^{\dagger}_{\bk}(C_{2}^{y}\calT))\Phi_{\bk}\right],\\
		&(-1)^{\nu_6} := \det \tilde{t}_{\text{C}\text{E}}=e^{-\mathrm{i}\gamma_{\text{C}\text{E}}}
		\frac{\mathrm{Pf}[\Phi_{\text{E}}^{\dagger}U(\calT)\Phi_{\text{E}}^*]}{\mathrm{Pf}[\Phi_{\text{C}}^{\dagger}U(\calT)\Phi_{\text{C}}^*]}
		\exp\left[\frac{1}{2}\int_f \mathrm{tr}\Phi_{\bk}^{\dagger}(U_{\bk}(C_{2}^{y}\calT)dU^{\dagger}_{\bk}(C_{2}^{y}\calT))\Phi_{\bk}\right].
	\end{align}
\end{widetext}
It should be emphasized that the right-hand sides are gauge-independent quantities. 
In Table~\ref{tab:P2}, we show the values of these topological invariants for representative models of the torsion subgroup of $^{\phi}K_{G/\Pi}^{(z,c)+0}(T^3)$.
Importantly, these invariants cannot fully characterize topological phases in this space group.
For a complete identification, two $\mZ_2$ invariants on the 2-skeleton corresponding to $E_3^{2,-2}=\mZ_2^2$ are required.

\begin{table}[t]
	\begin{center}
		\caption{\label{tab:P2}
			Results of topological invariants for representative models of all topological phases in space group $P2$ with TRS.
			For topological crystalline phases, $(hkl;d_0)$ represents the position of a two-dimensional topological insulator.
			Here, $(hkl)$ denotes the direction perpendicular to the two-dimensional topological insulator, and $d_0$ is the distance from the unit cell origin. 
			For atomic insulators, there are four inequivalent Wyckoff positions, $1\text{a} = (0, y, 0)$, $1\text{b} = (0, y, 1/2)$, $1\text{c} = (1/2, y, 0)$, $1\text{d} = (1/2, y, 1/2)$.
			The notation ``\textbf{AI}@$1\text{x}$'' denotes an atomic insulator whose electrons are at $1\text{x}\ (\text{x} = \text{a}, \text{b}, \text{c}, \text{d})$.
			Also $\nu_{\text{e}} \in 2\mZ$ represents the number of electrons in unit cell. 
		}
		\begin{tabular}{c|c}
			\hline
			Topo. phases & $(\nu_{\text{e}}/2, \nu_1, \nu_2, \nu_3, \nu_4, \nu_5, \nu_6)$\\
			\hline\hline
			strong TI & $(1,1,0,0,0,0,0)$ \\
			$(001; 0)$ & $(1,1,0,1,0,0,0)$\\
			$(001; 1/2)$ & $(1,1,0,1,0,1,1)$\\
			$(100; 0)$ & $(1,0,0,0,0,1,0)$\\
			$(010; 0)$ & $(1,1,1,0,0,0,0)$\\
			\textbf{AI}@$1a$ - \textbf{AI}@$1c$ & $(0,1,1,1,1,0,0)$\\
			\textbf{AI}@$1a$ - \textbf{AI}@$1b$ & $(0,0,0,0,0,1,1)$\\
			\textbf{AI}@$1c$ - \textbf{AI}@$1d$ & $(0,0,0,0,0,1,1)$\\
			\textbf{AI}@$1a$ & $(1,0,0,0,0,0,0)$\\
			\hline
		\end{tabular}
	\end{center}
\end{table}

\subsubsection{Example: spinful systems in space group $P2_12_12_1$ with TRS}
\label{sec:spinful_P2121211'}

\begin{table}[t]
	\begin{center}
		\caption{\label{tab:irrep_P212121}
			Character tables of irreps and their EAZ classes on $1$-cells in space group $P2_12_12_1$ with TRS. 
			Here, we show irreps on $0$- and $1$-cells relevant to the discussion.
		}
		\begin{tabular}{c|c|cccc}
			\hline
			irrep. & EAZ & $e$ & $S_x$& $S_y$ & $S_z$\\
			\hline\hline
			$\Gamma_1$ & AI & $2$ & $0$& $0$ & $0$\\
			\hline
			$\text{X}_2$ & A & $1$ & $1$& $-\ii$ & $-\ii$\\
			$\text{X}_3$ & A & $1$ & $1$& $\ii$ & $\ii$\\
			$\text{X}_4$ & A & $1$ & $-1$& $\ii$ & $-\ii$\\
			$\text{X}_5$ & A & $1$ & $-1$& $-\ii$ & $\ii$\\
			\hline
			$\text{Y}_2$ & A & $1$ & $-\ii$& $1$ & $-\ii$\\
			$\text{Y}_3$ & A & $1$ & $-\ii$& $-1$ & $\ii$\\
			$\text{Y}_4$ & A & $1$ & $\ii$& $-1$ & $-\ii$\\
			$\text{Y}_5$ & A & $1$ & $\ii$& $1$ & $\ii$\\
			\hline
			$\text{Z}_2$ & A & $1$ & $-\ii$& $-\ii$ & $1$\\
			$\text{Z}_3$ & A & $1$ & $-\ii$& $\ii$ & $-1$\\
			$\text{Z}_4$ & A & $1$ & $\ii$& $\ii$ & $1$\\
			$\text{Z}_5$ & A & $1$ & $\ii$& $-\ii$ & $-1$\\
			\hline
			irrep. & EAZ & $e$ & $S_x$&&\\
			\hline\hline
			$\Sigma_3$ & AI & $1$ & $-\mathrm{i}e^{-i k_x/2}$&&\\
			$\Sigma_4$ & AI &$1$ & $\mathrm{i}e^{-i k_x/2}$&&\\
			\hline
			irrep. & EAZ & $e$ & $S_y$&&\\
			\hline\hline
			$\Delta_3$ & AI & $1$ & $-\mathrm{i}e^{-i k_y/2}$&&\\
			$\Delta_4$ & AI &$1$ & $\mathrm{i}e^{-i k_y/2}$&&\\
			\hline
			irrep. & EAZ & $e$ & $S_z$&&\\
			\hline\hline
			$\Lambda_3$ & AI & $1$ & $-\mathrm{i}e^{-i k_z/2}$&&\\
			$\Lambda_4$ & AI &$1$ & $\mathrm{i}e^{-i k_z/2}$&&\\
			\hline
		\end{tabular}
	\end{center}
\end{table}

\begin{figure}[t]
	\begin{center}
		\includegraphics[width=0.9\columnwidth]{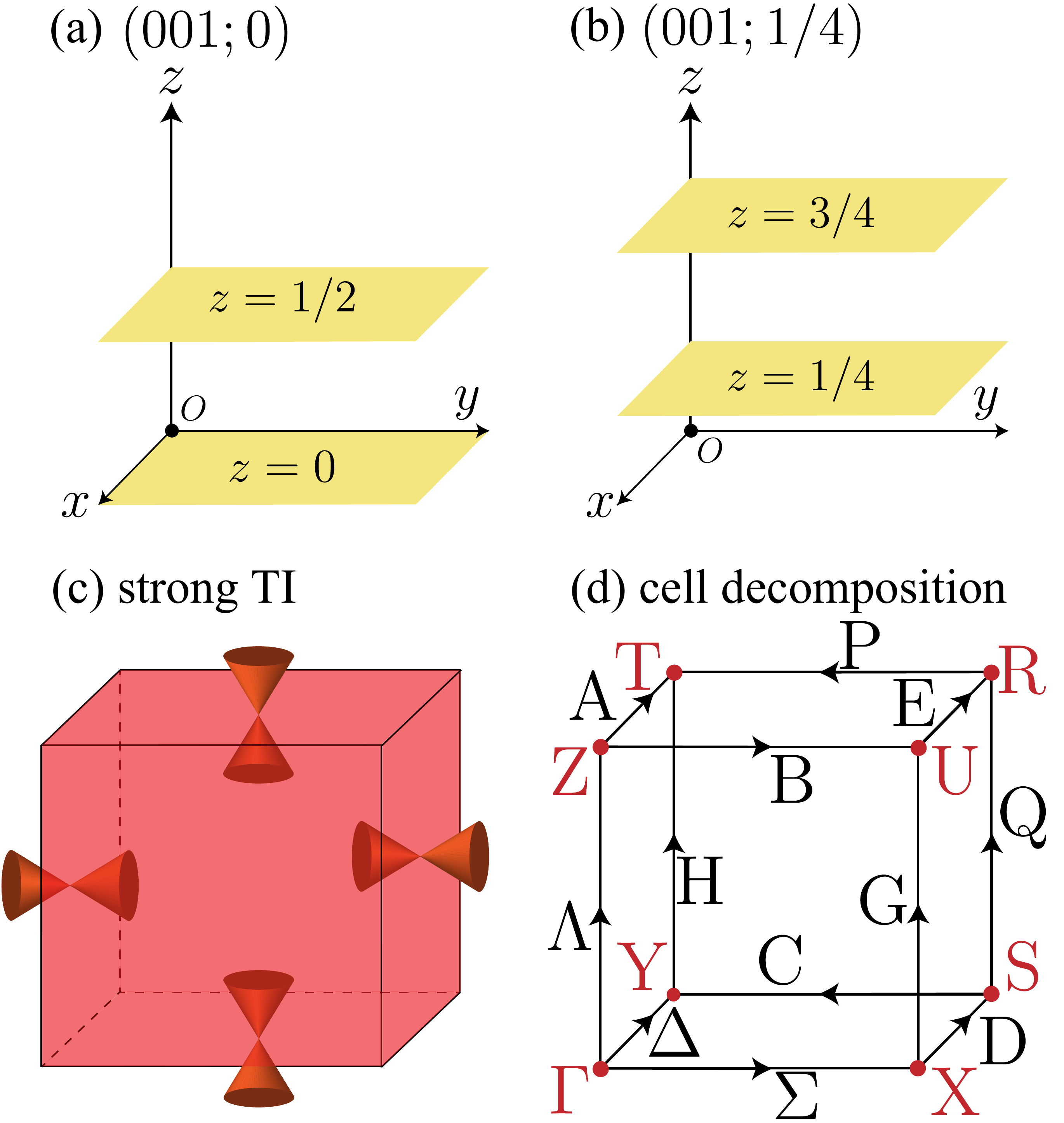}
		\caption{\label{fig:TI_P212121}
			(a, b) Illustrations of topological crystalline insulators introduced in Ref.~\cite{SI_NC_Song}. Yellow planes represent two-dimensional topological insulators that are stacked along $z$-direction.
			(c) Illustration of a strong topological insulator (TI). 
			(d) Cell decomposition of the fundamental domain in $P2_12_12_1$ and $P222$ with TRS. Here, the labels of momenta follow Ref.~\cite{BZ}.
		}
	\end{center}
\end{figure}

The third example is space group $P2_12_12_1$ with TRS $\calT$.
This space is generated by $S_x = \{C_{2}^{x}\vert (1/2, 1/2, 0)^{\top}\}$ and $S_y = \{C_{2}^{y}\vert (0, 1/2, 0)^{\top}\}$.
According to Ref.~\cite{Shiozaki-Ono2023}, the $K$-group is $^{\phi}K_{G/\Pi}^{(z,c)+0}(T^3) = \mZ \times (\mZ_2)^3$, whose $\mZ$-part is the classification of atomic insulators and $(\mZ_2)^3$ corresponds to the strong topological insulator and topological crystalline insulators, as shown in Fig.~\ref{fig:TI_P212121}(a--c). 
The $\mZ$ part of $^{\phi}K_{G/\Pi}^{(z,c)+0}(T^3)$ corresponds to $E_{2}^{0,0} = \mZ$.
The topological invariant to detect $E_{2}^{0,0}$ is the filling $\nu_{\text{e}}/8 \in \mZ$. 
Here, we discuss three $\mZ_2$ invariants to characterize the remaining $(\mZ_2)^3$ part in the $K$-group.

Our cell decomposition of the fundamental domain is shown in Fig.~\ref{fig:TI_P212121}(d).
Only EAZ classes of irreps on $\Sigma, \Delta$, and $\Lambda$ are class AI, and thus $E_{1}^{1,-1} = (\mZ_2)^6$ given by
\begin{align}
	E_{1}^{1,-1} = \bigoplus_{K = \Sigma, \Delta, \Lambda}\mZ_2[\bm{b}_{K_3}^{(1)}] \oplus \mZ_2[\bm{b}_{K_4}^{(1)}],
\end{align}
where the labels of irreps are shown in Table~\ref{tab:irrep_P212121}.
As discussed above, $E_{1}^{1,-1}$ is characterized by transition functions.
In this case, the transition functions are defined for each irrep on $\Sigma, \Delta$, and $\Lambda$.
For example, the transition function for irrep $\Sigma_3$ is defined by
\begin{align}
	t_{\Gamma\text{X}}^{\Sigma_3} = \left(\Phi^{\Sigma_3}_{\Gamma, (\pi/2,0,0)}\right)^{\dagger}\Phi^{\Sigma_3}_{\text{X}, (\pi/2,0,0)},
\end{align}
where $\Phi^{\Sigma_3}_{\Gamma, \bk}$ and $\Phi^{\Sigma_3}_{\text{X}, \bk}$ are occupied states with irrep $\Sigma_3$ under independent gauge choices around $\Gamma$ and $\text{X}$ [c.f.~Eq.~\eqref{eq:transition_function_raw}]. 
For other irreps, the transition functions are defined in the same way.

After computing AHSS and performing the procedures in Appendix~\ref{sec:X_TI}, we find that $E_{2}^{1,-1} = (\mZ_2)^3$ and
\begin{align}
	[X^{(1)}]^{-1} = \left(
	\begin{array}{cccccc}
		\Sigma_3 & \Sigma_4 & \Delta_3 &  \Delta_4 &  \Lambda_3 &  \Lambda_4 \\
		\hline
		1 & 1 & 0 & 0 & 0 & 0\\
		0 & 0 & 1 & 1 & 1 & 1 \\
		0 & 0 & 0 & 0 & 0 & 1 \\
		0 & 0 & 0 & 1 & -1 & 0 \\
		0 & 1 & 0 & 0 & -1 & 0 \\
		0 & 0 & 0 & 0 & -1 & 1 \\
	\end{array}
	\right),
\end{align} 
where the last three rows correspond to three $\mZ_2$-valued topological invariants.
Then, we construct the following three $\mZ_2$-valued topological invariants under certain gauge conditions:
\begin{align}
	\label{eq:nu7}
	&(-1)^{\nu_7} := (\det \tilde{t}_{\Gamma\text{Z}}^{\Lambda_3})^{-1}\det \tilde{t}_{\Gamma \text{Y}}^{\Delta_4}; \\
	\label{eq:nu8}
	&(-1)^{\nu_8}:= (\det \tilde{t}_{\Gamma\text{Z}}^{\Lambda_3})^{-1}\det \tilde{t}_{\Gamma \text{X}}^{\Sigma_4}; \\
	&(-1)^{\nu_9}:= (\det \tilde{t}_{\Gamma\text{Z}}^{\Lambda_3})^{-1}\det \tilde{t}_{\Gamma \text{Z}}^{\Lambda_4},
\end{align}
where $\tilde{t}^{\Lambda_3}_{\Gamma\text{Z}}$ is the gauge fixed transition function [c.f.~Eq.~\eqref{eq:z2_inv_1cell}] and the same applies to the other transition functions.
To use the technique introduced in Sec.~\ref{sec:degeneracy}, we consider
\begin{align}
	\label{eq:nu9}
	(-1)^{\nu'_{9}} := (-1)^{\nu_{9}-\nu_{7}} = (\det \tilde{t}_{\Gamma \text{Y}}^{\Delta_4})^{-1}\det \tilde{t}_{\Gamma \text{Z}}^{\Lambda_4}.
\end{align}

Similar to the above example, we can derive gauge-independent expressions of these invariants.
Again, we use the symmetry enriched Berry phase introduced in Eq.~\eqref{eq:irrep_Berry}, $e^{\mathrm{i}\gamma_{\bk_0 \bk_1}^{\alpha}} := \lim_{\mathcal{N}\rightarrow\infty} \det \prod_{j = 0}^{\mathcal{N}-1} (\Phi^{\alpha}_{\bk_0 + (j+1)\bm{\delta}})^{\dagger}\Phi^{\alpha}_{\bk_0 + j\bm{\delta}}$ for irrep $\alpha$ and $\bm{\delta} = (\bk_1 - \bk_0)/\mathcal{N}$.
Again, the Berry phase is not invariant under gauge transformations at the boundary $0$-cells. 
For $0$-cells $\text{X, Y, and Z}$, the Pfaffian can be used to cancel U(1)-valued quantities originating from the gauge transformations.
On the other hand, at $0$-cell $\Gamma$, there is only a two-dimensional irrep $\Gamma_1$ whose EAZ class is AI. 
According to Eqs.~\eqref{eq:nu7}, \eqref{eq:nu8}, and \eqref{eq:nu9}, $\Gamma$ is an intersection of two $1$-cells in each quantity.
Thus, we can use the technique in Sec.~\ref{sec:degeneracy} to derive gauge invariant quantities.
As a result, we have 
\begin{widetext}
\begin{align}
	&(-1)^{\nu_7} :=  2^{N_{\text{occ}}/2}e^{\mathrm{i}\gamma_{\Gamma \text{Z}}^{\Lambda_3}}e^{-\mathrm{i}\gamma_{\Gamma \text{Y}}^{\Delta_4}}
	\frac{\mathrm{Pf}[(\Phi^{\Delta_4}_{\text{Y}})^{\dagger}U(\calT)(\Phi^{\Delta_4}_{\text{Y}})^*]}{\mathrm{Pf}[(\Phi^{\Lambda_3}_{\text{Z}})^{\dagger}U(\calT)(\Phi^{\Lambda_3}_{\text{Z}})^*]}\det (\Phi^{\Lambda_3}_{\Gamma})^{\dagger}\Phi^{\Delta_4}_{\Gamma}\nonumber \\
	&\quad\quad\quad\times 
	\exp\left[-\frac{1}{2}\int_{\Lambda}\mathrm{tr} (\Phi_{\bk}^{\Lambda_3})^{\dagger}\left(U_{\bk}(S_{y}\calT)d U^{\dagger}_{\bk}(S_{y}\calT)\right)\Phi_{\bk}^{\Lambda_3} \right]\exp\left[\frac{1}{2}\int_{\Delta}\mathrm{tr} (\Phi_{\bk}^{\Delta_4})^{\dagger}\left(U_{\bk}(S_{z}\calT)d U^{\dagger}_{\bk}(S_{z}\calT)\right)\Phi_{\bk}^{\Delta_4} \right]\\
	&(-1)^{\nu_8}:=  2^{N_{\text{occ}}/2}e^{\mathrm{i}\gamma_{\Gamma \text{Z}}^{\Lambda_3}}e^{-\mathrm{i}\gamma_{\Gamma \text{X}}^{\Sigma_4}}
	\frac{\mathrm{Pf}[(\Phi^{\Sigma_4}_{\text{X}})^{\dagger}U(\calT)(\Phi^{\Sigma_4}_{\text{X}})^*]}{\mathrm{Pf}[(\Phi^{\Lambda_3}_{\text{Z}})^{\dagger}U(\calT)(\Phi^{\Lambda_3}_{\text{Z}})^*]}\det (\Phi^{\Lambda_3}_{\Gamma})^{\dagger}\Phi^{\Sigma_4}_{\Gamma}\nonumber \\
	&\quad\quad\quad\times \exp\left[-\frac{1}{2}\int_{\Lambda}\mathrm{tr} (\Phi_{\bk}^{\Lambda_3})^{\dagger}\left(U_{\bk}(S_{y}\calT)d U^{\dagger}_{\bk}(S_{y}\calT)\right)\Phi_{\bk}^{\Lambda_3} \right]\exp\left[\frac{1}{2}\int_{\Sigma}\mathrm{tr} (\Phi_{\bk}^{\Sigma_4})^{\dagger}\left(U_{\bk}(S_{z}\calT)d U^{\dagger}_{\bk}(S_{z}\calT)\right)\Phi_{\bk}^{\Sigma_4} \right]\\
	&(-1)^{\nu'_9}:= 2^{N_{\text{occ}}/2}e^{\mathrm{i}\gamma_{\Gamma \text{Y}}^{\Delta_4}}e^{-\mathrm{i}\gamma_{\Gamma \text{Z}}^{\Lambda_3}}
	\frac{\mathrm{Pf}[(\Phi^{\Lambda_3}_{\text{Z}})^{\dagger}U(\calT)(\Phi^{\Lambda_3}_{\text{Z}})^*]}{\mathrm{Pf}[(\Phi^{\Delta_4}_{\text{Y}})^{\dagger}U(\calT)(\Phi^{\Delta_4}_{\text{Y}})^*]}
	\det (\Phi^{\Delta_4}_{\Gamma})^{\dagger}\Phi^{\Lambda_3}_{\Gamma}\nonumber \\
	&\quad\quad\quad\times 
	\exp\left[-\frac{1}{2}\int_{\Delta}\mathrm{tr} (\Phi_{\bk}^{\Delta_4})^{\dagger}\left(U_{\bk}(S_{x}\calT)d U^{\dagger}_{\bk}(S_{x}\calT)\right)\Phi_{\bk}^{\Delta_4} \right]
	\exp\left[\frac{1}{2}\int_{\Lambda}\mathrm{tr} (\Phi_{\bk}^{\Lambda_3})^{\dagger}\left(U_{\bk}(S_{y}\calT)d U^{\dagger}_{\bk}(S_{y}\calT)\right)\Phi_{\bk}^{\Lambda_3} \right],
\end{align}
\end{widetext}
By explicitly computing these invariants for representative models of a strong topological insulator and topological crystalline insulators, we see that $(\nu_7, \nu_8$, $\nu'_9)$ can diagnose these phases, as shown in Table~\ref{tab:P212121}.

\begin{table}[t]
	\begin{center}
		\caption{\label{tab:P212121}
			Results of topological invariants for representative models of all topological phases in space group $P2_12_12_1$ with TRS.
			For topological crystalline phases, $(hkl;d_0)$ represents the position of a two-dimensional topological insulator.
			Here, $(hkl)$ denotes the direction perpendicular to the two-dimensional topological insulator, and $d_0$ is the distance from the unit cell origin. 
			The notation ``\textbf{AI}@$4\text{a}$'' means an atomic insulator whose electrons are at the maximal Wyckoff position $4\text{a}$.
			Also, $\nu_{\text{e}} \in 8\mZ$ represents the number of electrons in unit cell. 
		}
		\begin{tabular}{c|c}
			\hline
			Topo. phases & $(\nu_{\text{e}}/8, \nu_7, \nu_8, \nu'_9)$\\
			\hline\hline
			strong TI & $(2, 1,0,0)$ \\
			$(001; 0)$ & $(2, 1, 0, 1)$\\
			$(001; 1/4)$ & $(2, 0, 1, 0)$\\
			\textbf{AI}@$4\text{a}$ & $(1, 1,0,1)$\\
			\hline
		\end{tabular}
	\end{center}
\end{table}

\subsubsection{Example: spinless systems with space group $P\bar{\text 4}$ with TRS}
\label{Example: spinless systems with space group P-4 with TRS}

In this symmetry setting, $E_2^{1,-1} = \mZ_2$ and a $\mZ_2$ invariant over the 1-skeleton exists. 
On the one hand, the $K$-group $K \cong \mZ^8$ does not include $\mZ_2$~\cite{Shiozaki-Ono2023}. 
This means that the second differential $d_2^{1,-1}: E_2^{1,-1} \to E_2^{3,-2}$ is nontrivial and the $\mZ_2$ invariant over the 1-skeleton detects a gapless point (Weyl point) in 3-cell. 
We construct the $\mZ_2$ invariant over the 1-skeleton and relate it to the Chern number of Weyl point. 

\begin{figure}[t]
\centering
\includegraphics[width=0.5\linewidth, trim=0cm 0cm 0cm -1cm]{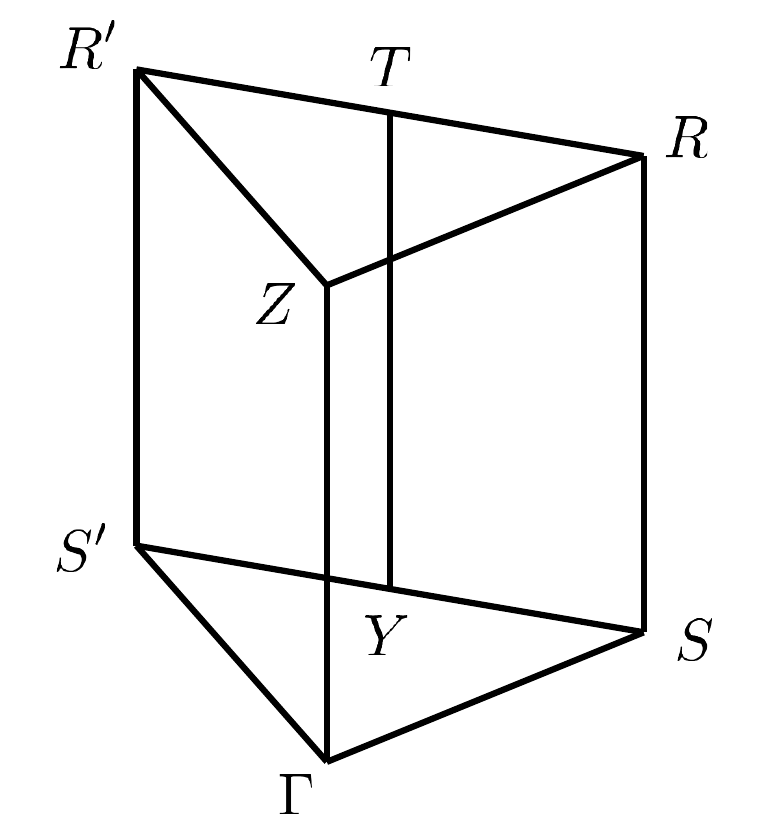}
\caption{Fundamental domain in the momentum space of magnetic space group $P\bar{4}1'$. 
Momenta are set as $\G=(0,0,0), S=(\pi,\pi,0), Y=(0,\pi,0), Z=(0,0,\pi), R=(\pi,\pi,\pi)$ and $T=(0,\pi,\pi)$. }
\label{fig:p-41p}
\end{figure}

Symmetry is summarized as follows. 
\begin{align}
&U_\bk(S_4) H_\bk U_\bk(S_4)^\dag=H_\bk, \\
&U_\bk(\calT) H_\bk^* U_\bk(\calT)^\dag =H_{-\bk}, \\
&U_{S_4^3\bk}(S_4)U_{C_2^z\bk}(S_4)U_{S_4\bk}(S_4)U_\bk(S_4)=1, \\
&U_{-\bk}(\calT)U_\bk(\calT)^*=1, \\
&U_{-\bk}(S_4)U_\bk(\calT)=U_{S_4\bk}(\calT)U_\bk(S_4)^*. 
\end{align}
Here, $S_4\bk = (k_y,-k_x,-k_z)$ is rotoinversion. 
Figure~\ref{fig:p-41p} shows a cell decomposition we use. 
On 1-cells $\G Z$ and $SR$, the stabilizer group is $\{e,S_4 \calT, C_2^z, S_4^3 \calT\}$ and there are two irreps $A$ and $B$ whose characters are $\chi^A(C_2^z) = 1$ and $\chi^B(C_2^z)=-1$. 
The relation $U_{\bk}(S_4\calT)U_{\bk}(S_4\calT)^* = U_\bk(C_2^z)$ over these 1-cells implies that the EAZ class is AI for the irrep $A$ and AII for the irrep $B$, respectively. 
On the 1-cells $\G S$ and $ZR$ the little group is $\{e,C_2^z \calT\}$ and EAZ is AI. 
From the AHSS, we find that the $\mZ_2$ invariant is composed of the transition functions of the $A$-irrep of the 1-cells $\G Z$ and $SR$ and the 1-cells $\G S$ and $ZR$, meaning that the Berry phase formula is composed of these irreps of their 1-cells. 
Since the Berry phase over the irrep whose EAZ class is AII is trivial (see Sec.~\ref{sec:Class AII EAZ and triviality of Berry phase}), multiplying the Berry phase of the $B$-irrep over 1-cells $\G Z$ and $ZR$ does not change the $\mZ_2$ invariant. 
Consequently, the $\mZ_2$ invariant $\nu$ is written as the standard Berry phase $e^{\ii\gamma_{\G S R Z}}$ over the loop $\G \to S \to R \to Z \to \G$ without irreducible decomposition as in 
\begin{widetext}   
\begin{align}
    (-1)^\nu
    &=e^{-\ii\gamma_{\G S R Z}}
    \times \exp \left[\frac{1}{2}\left(\int_{S \to R}+\int_{Z \to \G}\right) \tr[\Phi_\bk^\dag U_\bk(S_4\calT) d U_\bk(S_4\calT)^\dag \Phi_\bk]\right] \nonumber\\
    &\times \exp \left[\frac{1}{2}\left(\int_{\G \to S}+\int_{R \to Z}\right) \tr[\Phi_\bk^\dag U_\bk(C_2^z\calT) d U_\bk(C_2^z\calT)^\dag \Phi_\bk]\right]. 
\end{align}
\end{widetext}

Now we show that the $\mZ_2$ invariant $\nu$ is the parity of Weyl charge inside the fundamental domain of the momentum space. 
Using Stokes's theorem and the symmetry constraints, it should be possible to directly show that $\nu$ is the parity of the Chern number of the boundary of the fundamental domain. 
Instead, here, based on the interpretation of the second differential $d_2^{1,-1}$~\cite{K-AHSS}, we show that the topological transition of $\nu$ results in the emission or absorption of the quartet of Weyl points. 
On the tubular neighborhood of the 1-cell $\G Z$, the topological transition of $\nu$ is modeled by the Hamiltonian 
\begin{align}
    H_\bk = k_z \sigma_x + (m-\bk^2) \sigma_z + k_x k_y \sigma_y 
\end{align}
with $U_\bk(S_4\calT) = 1$. 
When the mass $m$ continuously varies from a positive to a negative value, the $\mZ_2$ invariant $\nu$ changes by 1, and the quartet of Weyl points is created to the fundamental domain, meaning the change of the Chern number by 1.

\subsection{Topological invariants defined on 2-skeleton}
In this work, we have discussed topological invariants defined on $1$-skeletons, which are constructed from $E_{2}^{1,-1}$. 
As shown in the above example of space group $P2$, the introduced topological invariants cannot fully characterize all possible phases. 
In fact, our framework discussed in Sec.~\ref{sec:general_frame} and Appendix~\ref{app:Z2case} is applicable to $E_{2}^{p,-n}$ for arbitrary $p$ and $n$. 
In particular, when the higher-differential $d_{2}^{0,-1}$ is trivial, $E_3^{2,-2} = E_{2}^{2,-2}$ can inform us of topological invariants defined on $2$-skeletons. 
For example, $E_{3}^{2,-2} = E_{2}^{2,-2} = (\mZ_2)^2$ in space group $P2$ with TRS. 
Thus, there should be two $\mZ_2$-valued topological invariants defined on $2$-skeletons. 
It is natural to expect that the topological invariants obtained from $E_{3}^{2,-2}$, together with $\{\nu_i\}_{i=1}^{6}$, enable us to diagnose all possible topological phases. 
Although topological invariants defined on $2$-skeletons are essential for the full characterization of topological phases, we leave a systematic construction of them as a future work.

\section{Conclusions and Outlook}
\label{sec6}
%
In this work, we have established a systematic framework to construct topological invariants based on AHSS in momentum space.
We have applied our scheme to insulators and superconductors in all magnetic space groups and presented necessary information about topological invariants such as $[X^{(1)}]^{-1}$ and $[V^{(0)}]^{-1}$.
Through some illustrative examples, we have shown that the constructed invariants can detect nontrivial topology, for which symmetry indicators do not work at all.

Our work opens new avenues for future research.
First, our scheme is applicable to various kinds of systems.
In recent years, topological phases in non-Hermitian systems have attracted significant attention~\cite{PhysRevLett.120.146402,PhysRevX.8.031079,PhysRevLett.123.066404,PhysRevLett.123.066405,PhysRevX.9.041015,PhysRevLett.124.086801,PhysRevLett.125.126402,RevModPhys.93.015005,Zhang:2022aa,Ding:2022aa,doi:10.1146/annurev-conmatphys-040521-033133}.
Since K-theoretic classification techniques developed for Hermitian Hamiltonians have been successfully adapted to non-Hermitian settings, the AHSS-based construction developed in the present work may also provide a useful framework for constructing topological invariants for non-Hermitian Hamiltonians.
More recently, spin space groups have been comprehensively classified~\cite{SpinSG_Song,SpinSG_Ren,SpinSG_Fang,SpinSG_Watanabe}.
It turns out that there exist a lot of new symmetry classes.
Our general scheme could also be applied to spin space groups.

Second, our invariants can capture nontrivial topology in gapped phases on $2$-skeletons or gapless points on $2$-cells.
This means that our invariants of gapped phases on $2$-skeletons sometimes detect gapless points on $3$-cells.
Since the discovery of gapless points at generic momenta is usually challenging, it would be interesting to differentiate between the invariants for fully gapped phases and those for gapless points on $3$-cells.
This distinction is achieved by computing the invariants for representative models of real-space classifications and examining which ones are for gapless points on $3$-cells.

Third, it is interesting to investigate whether the topological invariants constructed here lead to fundamental relations among topology, quantum geometry, and physical quantities~\cite{PhysRevB.90.165139, Peotta:2015aa, PhysRevB.106.014518, Onishi-Fu-PRX, Onishi-Fu-PRL2024, Z2-Qmetric}.
Recently, it has been shown that the Chern number and the $\mathbb{Z}_2$ topological invariant of time-reversal symmetric insulators give fundamental bounds on the band gap and integrations of quantum metric~\cite{PhysRevB.90.165139, Onishi-Fu-PRX, Z2-Qmetric}.

Last, we expect that first-principles calculations can compute topological invariants constructed based on this work. 
For insulators, our gauge invariant formulas are computable from wave functions obtained by first-principles calculations. 
For superconductors, Fermi surfaces available from first-principles calculations enable us to calculate our invariants under the weak-pairing assumption, as briefly mentioned in Sec.~\ref{sec:Fermi_surfaces}. 
Furthermore, since they are defined on 1-skeletons composed of high-symmetry points and line segments connecting them, the computational cost is relatively low. 
We believe that our work would help to build a more comprehensive database of topological materials than the existing ones based on symmetry indicators and topological quantum chemistry.

\begin{acknowledgments}
	We thank Yohei Fuji, Shuichi Murakami, Kenji Shimomura, and Zhongyi Zhang for fruitful discussions.
	We also thank Isidora Araya Day for informing us about Ref.~\cite{Araya_C4T}.
	We appreciate the YITP workshop YITP-T22-02 on ``Novel Quantum States in Condensed Matter 2022", where a part of this work was carried out.
	SO was supported by KAKENHI Grant No.~JP20J21692 and No.~23K19043 from the Japan Society for the Promotion of Science (JSPS) and RIKEN Special Postdoctoral Researchers Program.
	KS was supported by JST CREST Grant No.~JPMJCR19T2, and JSPS KAKENHI Grant No.~22H05118 and 23H01097. 
\end{acknowledgments}

\section*{Data availability}
All data in this work are available in Ref.~\cite{SM}.

\clearpage
\appendix

\section{Computations for Abelian groups with torsion elements}
\label{app:Z2case}
In this appendix, we show computational procedures for the case where $\{E_{1}^{q,-p}\}_{q = p-1}^{p+1}$ are not a free Abelian group. 
\subsection{General framework}
The following commutative diagram shows the mathematical structure behind our problem:
\begin{equation*}
	\label{eq:Snake}
	\begin{tikzcd}[row sep=small, column sep=scriptsize]
		0 \arrow[r]  &  P^{p-1,-p}_{1} \arrow[r,"i"]  \arrow[d,"\tilde{d}^{p-1,-p}_{1}"]& \tilde{E}^{p-1,-p}_{1} \arrow[d,"\tilde{d}^{p-1,-p}_{1}"] \arrow[r,"\tau"] &   E^{p-1,-p}_{1}  \arrow[r] \arrow[d, "d^{p-1,-p}_{1}"]& 0\\
		0 \arrow[r]  &  P^{p,-p}_{1} \arrow[r,"i"]  \arrow[d,"\tilde{d}^{p,-p}_{1} "]& \tilde{E}^{p,-p}_{1} \arrow[d,"\tilde{d}^{p,-p}_{1}"] \arrow[r,"\tau"] &   E^{p,-p}_{1}  \arrow[r] \arrow[d, "d^{p,-p}_{1} "]& 0\\
		0 \arrow[r]  &  P^{p+1,-p}_{1} \arrow[r,"i"]  & \tilde{E}^{p+1,-p}_{1}  \arrow[r,"\tau"] &   E^{p+1,-p}_{1}  \arrow[r] & 0\\
	\end{tikzcd},
\end{equation*}
where
\begin{align}
	&E^{p,-p}_{1} := \bigoplus_{i=1}^{D_p}\mZ_2[\bm{b}_{i}^{(p)}] \oplus  \bigoplus_{i=D_{p}+1}^{N_p}\mZ[\bm{b}_{i}^{(p)}]; \\
	&\tilde{E}^{p,-p}_{1} := \bigoplus_{i=1}^{D_p}\mZ[\tilde{\bm{b}}_{i}^{(p)}] \oplus  \bigoplus_{i=D_{p}+1}^{N_p}\mZ[\tilde{\bm{b}}_{i}^{(p)}];\\
	&P^{p,-p}_{1} := \bigoplus_{i=1}^{D_p}2\mZ[2\tilde{\bm{b}}_{i}^{(p)}].
\end{align}
In the following, ``$\sim$'' denotes that we forget about the $\mZ_2$-nature of an Abelian group.
By definition, $E^{p,-p}_{1} = \tilde{E}^{p,-p}_{1}/P^{p,-p}_{1}$. 
For simplicity, we use $\tilde{\mathcal{B}}^{(p)} = \left(\tilde{\bm{b}}_{1}^{(p)} \tilde{\bm{b}}_{2}^{(p)} \cdots \tilde{\bm{b}}_{N_p}^{(p)} \right)$ and $\tilde{\mathcal{P}}^{(p)} = \left( 2\tilde{\bm{b}}_{1}^{(p)}\ 2\tilde{\bm{b}}_{2}^{(p)}\ \cdots\  2\tilde{\bm{b}}_{D_p}^{(p)} \right)$. 

To generalize the discussions in Sec.~\ref{sec:topo_invariant}, we introduce a map $f_{1}^{p,-p} = \tilde{d}_{1}^{p, -n} \oplus \mathrm{id}: \tilde{E}^{p,-p}_{1}\oplus P^{p+1,-p}_{1} \rightarrow \tilde{E}^{p+1,-p}_{1}$. 
This map can be represented by a matrix as
\begin{align}
	&f_{1}^{p,-p}\left(\tilde{\mathcal{B}}^{(p)}\ \  \tilde{\mathcal{P}}^{(p+1)}\right)
	= \tilde{\mathcal{B}}^{(p+1)}\begin{pmatrix}
		\tilde{M}_{\tilde{d}_{1}^{p,-p}} & Q^{(p)}
 	\end{pmatrix},
\end{align}
where $Q^{(p)} = \left(\begin{array}{c}
	2 \times \mathds{1}_{D_{p+1}}\\
	\hline
	\multicolumn{1}{c}{O}
\end{array}\right)$.
One can always find the Smith normal form
\begin{align}
	U^{(p)}\begin{pmatrix}\tilde{M}_{\tilde{d}_{1}^{p,-p}} & Q^{(p)} \end{pmatrix}V^{(p)} = \Sigma^{(p)},
\end{align}
where $U^{(p)}$ and $V^{(p)}$ are unimodular matrices. 
As mentioned in Sec.~\ref{sec3}, the Smith normal form can be written as $\Sigma^{(p)}  = \text{diag}\left(s_{1}^{(p)}, s_{2}^{(p)}, \cdots, s_{N_p}^{(p)}\right)\ (s_{i}^{(p)} \in \mZ_{\geq0})$ and $s_{i} \neq 0$ for $i \leq r_p \leq N_p$.
Then, we have
\begin{align}
	\label{eq:f_map}
	&f_{1}^{p,-p}\left(\tilde{\mathcal{B}}^{(p)}\ \  \tilde{\mathcal{P}}^{(p+1)}\right)V^{(p)} \nonumber \\
	&= \tilde{\mathcal{B}}^{(p+1)}[U^{(p)}]^{-1}U^{(p)}\begin{pmatrix}
		\tilde{M}_{\tilde{d}_{1}^{p,-p}} & Q^{(p)}
	\end{pmatrix}V^{(p)} \nonumber \\
&= \tilde{\mathcal{B}}^{(p+1)}[U^{(p)}]^{-1} \Sigma^{(p)},
\end{align}
where $V^{(p)} = (\bm{v}_{1}^{(p)}, \bm{v}_{2}^{(p)}, \cdots, \bm{v}_{N_p+D_{p+1}}^{(p)})$.
Here, we obtain $\ker \tilde{d}_{1}^{p,-p}$ from Eq.~\eqref{eq:f_map}.
First, we restrict $\bm{v}_{i}^{(p)}$ to the first $N_p$ elements, which is denoted by $\bm{v}'^{(p)}_{i}$. 
Then, Eq.~\eqref{eq:f_map} informs us that $\{\bm{v}'^{(p)}_{i}\}_{i = r_p+1}^{N_p+D_{p+1}}$ are elements of $\ker \tilde{d}_{1}^{p,-p}$.
Collecting all independent lists of $\{\bm{v}'^{(p)}_{i}\}_{i = r_p+1}^{N_p+D_{p+1}}$, we have a basis set of $\ker \tilde{d}_{1}^{p,-p}$.

For later convenience, we construct a $N_p$-dimensional invertible matrix $V'^{(p)}$ from $\{\bm{v}'^{(p)}_{i}\}_{i=r_p+1}^{N_p+D_{p+1}}$. 
We obtain a $N_p$-dimensional unimodular matrix $U_{v}$ from the following Smith decomposition
\begin{align}
	\label{eq:smith_ker}
	 &(\bm{v}'^{(p)}_{r_p+1} \cdots \bm{v}'^{(p)}_{N_p+D_{p+1}}) = U_{v}^{-1}\Sigma_{v}V_{v}^{-1},\\
	 &U_{v}^{-1} := \left(\begin{array}{cccc}
	 		\bm{u}_{1} & 
	 		\bm{u}_{2} &
	 		\cdots &
	 		\bm{u}_{N_p}
	 	\end{array}\right)^{\top}.
\end{align}
where $\Sigma_{v}$ is the Smith normal form of $(\bm{v}'^{(p)}_{r_p+1} \cdots \bm{v}'^{(p)}_{N_p+D_{p+1}})$, i.e., a $N_p$-dimensional diagonal matrix whose matrix rank is $N_p - r'_p\ (0  \leq r'_p \leq N_p)$.
$V_{v}$ is a $(N_p + D_{p+1}-r_p)$-dimensional unimodular matrix.
Let us elaborate on the explanation of $\Sigma_{v}$. 
Each diagonal element of $\Sigma_{v}$ has the following features
\begin{itemize}
	\item $[\Sigma_{v}]_{ii} \in \mZ_{>0}$ for $1\leq i \leq (N_{p}-r'_p)$;
	\item $[\Sigma_{v}]_{ii}$ can always divide $[\Sigma_{v}]_{(i+1)(i+1)}$ for $1 \leq i \leq  (N_{p}-r'_p)-1$;
	\item $[\Sigma_{v}]_{ii} = 0$ for $(N_{p}-r'_p)<i$ if $r'_p \neq 0$. 
\end{itemize}
From $U_{v}^{-1}$ and $\Sigma_v$, we have an integer-valued invertible matrix $V'^{(p)}$ as 
\begin{align}
	[V'^{(p)}]_{ij} = \begin{cases}
		[\Sigma_v]_{jj}[\bm{u}_j]_i \quad \text{for $1\leq i \leq (N_{p}-r'_p)$}\\
		[\bm{u}_j]_i \quad \text{for $(N_{p}-r'_p)<j$ if $r'_p \neq 0$}
	\end{cases}.
\end{align}
Note $V'^{(p)}$ is not a unimodular matrix when there exists $j$ such that $[\Sigma_v]_{jj} > 1$.

Since $\im f_{1}^{p-1, -p} \subseteq \ker \tilde{d}_{1}^{p, -p}$,  
\begin{align}
	\label{eq:f_map2}
	f_{1}^{p-1,-p}\left(\tilde{\mathcal{B}}^{(p-1)}\ \  \tilde{\mathcal{P}}^{(p)}\right)&=
	\tilde{\mathcal{B}}^{(p)}\begin{pmatrix}\tilde{M}_{\tilde{d}_{1}^{p-1,-p}} & Q^{(p-1)}\end{pmatrix}\nonumber \\
	&=
	\tilde{\mathcal{B}}^{(p)} V'^{(p)} [V'^{(p)}]^{-1} \begin{pmatrix}\tilde{M}_{\tilde{d}^{1}_{p-1,-p}} & Q^{(p-1)}\end{pmatrix}\nonumber \\
	&= \tilde{\mathcal{B}}'^{(p)}\left(\begin{array}{cc}
		\multicolumn{2}{c}{Y}\\
		\hline
		O_{r'_p\times N_{p-1}} & O_{r'_p\times D_p}\\
	\end{array}\right),
\end{align}
where $\tilde{\mathcal{B}}'^{(p)} = (\bm{b}'^{(p)}_{r_p +1}, \cdots \bm{b}'^{(p)}_{N_p+D_{p+1}}, \bm{b}'^{(p)}_{1}, \cdots, \bm{b}'^{(p)}_{r'_p}) = \tilde{\mathcal{B}}^{(p)} V'^{(p)}$.
The reason why the last $r'_p$ rows are zeros is $\bm{b}_j \notin \ker \tilde{d}_{1}^{p, -p}$. 
These should not be included in $\im f_{1}^{p-1, -p}$.
Again, using the Smith normal form, we rewrite the above equation as
\begin{align}
	\label{eq:map3}
	&f_{1}^{p-1,-p}\left(\tilde{\mathcal{B}}^{(p-1)}\ \  \tilde{\mathcal{P}}^{(p)}\right)V^{(p-1)}\nonumber\\
	&=\tilde{\mathcal{B}}'^{(p)}\left([U^{(p-1)}]^{-1}\oplus \mathds{1}_{r'_p}\right)
	\left(\begin{array}{cc}
		\multicolumn{2}{c}{\Lambda^{(p-1)}}\\
		\hline
		O_{r'_p\times N_{p-1}} & O_{r'_p\times D_p}
	\end{array}\right)\nonumber\\
&=\tilde{\mathcal{B}}''^{(p)}\left(\begin{array}{cc}
	\multicolumn{2}{c}{\Lambda^{(p-1)}}\\
	\hline
	O_{r'_p\times N_{p-1}} & O_{r'_p\times D_p}
\end{array}\right),
\end{align}
where $\Lambda^{(p-1)}$ is the Smith normal form of $Y$, and $\tilde{\mathcal{B}}''^{(p)} =\tilde{\mathcal{B}}^{(p)} X^{(p)}%
=\tilde{\mathcal{B}}^{(p)}V'^{(p)}\left([U^{(p-1)}]^{-1} \oplus \mathds{1}_{r'_p}\right)$.
As with the case where $E_{1}^{p, -p}$ is a free Abelian group, the inverse of $X^{(p)}$ and $V^{(p-1)}$ tell us which $p$- and $(p-1)$-cells are involved in the expression of topological invariants. 
Again $[X^{(1)}]^{-1}$ is denoted by
\begin{align}
    &[X^{(1)}]^{-1} := \left(\begin{array}{cccc}
		\bm{x}_{1} & 
		\bm{x}_{2} &
		\cdots &
		\bm{x}_{N_1}
	\end{array}\right)^{\top}; \nonumber \\
    &[V^{(0)}]^{-1} := \left(\begin{array}{cccc}
		\bm{v}_{1} &
		\bm{v}_{2} &
		\cdots &
		\bm{v}_{N_0}
	\end{array}\right)^{\top} \nonumber
\end{align}

It should be noted that $[X^{(1)}]^{-1}$ sometimes contains fractional numbers since $V^{(1)}$ is not a unimodular matrix in general. 
This happens only when $E_{1}^{p, -n}$ has both $\mZ$- and $\mZ_2$-parts.

\subsection{Example: spinless superconductors in $P3_1211'$}
\begin{figure}[t]
	\begin{center}
		\includegraphics[width=0.99\columnwidth]{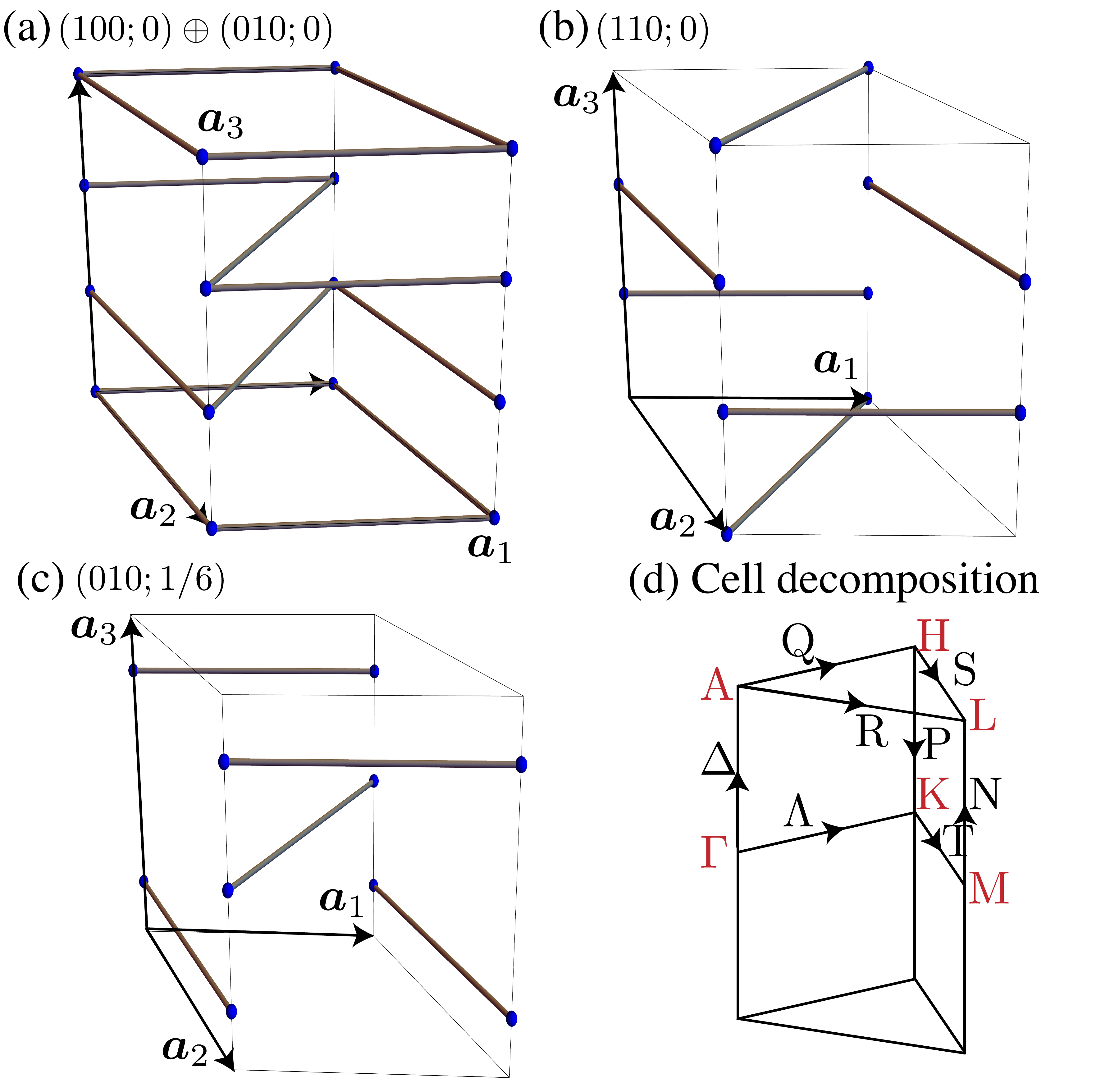}
		\caption{\label{fig:P3121}
			(a, b, c) Real-space pictures of topological crystalline superconductors. Gray objects represent one-dimensional topological superconductors.
			(d) Cell decomposition of the fundamental domain in $P3_121$ with TRS. 
		}
	\end{center}
\end{figure}

As an example, we discuss time-reversal symmetric spinless superconductors in space group $P3_121$ with the conventional pairing symmetry $A_1$.
This space is generated by threefold screw $S_z = \{C_{3}^{z}\vert (0, 0, 1/3)^{\top}\}$, twofold rotation $C_{2}^{[100]}=\{C_{2}^{x}\vert (0, 0, 2/3)^{\top}\}$ along $x$-direction, and translations.
According to Ref.~\cite{Shiozaki-Ono2023}, the $K$-group is $^{\phi}K_{G/\Pi}^{(z,c)+0}(\mathbb{R}^3) = (\mZ_{2})^3 \times \mZ_3 \times \mZ^2$, whose $(\mZ_{2})^3$ classification corresponds to atomic limits and $\mZ_3 \times \mZ^2$ is the classification of topological crystalline superconductors, as shown in Fig.~\ref{fig:P3121} (a--c). 
In addition, we find that $E_{2}^{0,0} = (\mZ_2)^5$ and $E_{2}^{1,-1} = \mZ_3 \times \mZ^2$, which implies that  five $\mZ_2$-, a $\mZ_3$-, and two $\mZ$-valued topological invariants are defined.

Our cell decomposition of the fundamental domain is shown in Fig.~\ref{fig:P3121}(d).
For this cell decomposition, $E_{1}^{0,0}$ and $E_{1}^{1,-1}$ are spanned by
\begin{widetext}
\begin{align}
	\label{eq:E100_P3121}
	&E_{1}^{0,0} = \bigoplus_{K = \Gamma,\text{A}}\bigoplus_{j=1}^{3}\mZ_2[\bm{b}_{K_j}^{(0)}] \oplus \bigoplus_{K = \text{M}, \text{L}}\bigoplus_{j=1}^{2}\mZ_2[\bm{b}_{K_j}^{(0)}] , \\
	\label{eq:E111_P3121}
	&E_{1}^{1,-1} = \bigoplus_{j=1}^{3}\mZ_2[\bm{b}_{\Delta_j}^{(1)}] \oplus \bigoplus_{j=1}^{3}\mZ[\bm{b}_{\text{P}_j}^{(1)}] \oplus \bigoplus_{K =\Lambda, \text{Q}, \text{T}, \text{S}}\bigoplus_{j=1}^{2}\mZ[\bm{b}_{K_j}^{(1)}] \oplus \bigoplus_{K=\text{R}, \text{N}}\mZ_2[\bm{b}_{K_1}^{(1)}]
\end{align}
where the labels of irreps are defined in Table~\ref{tab:irrep_SGP3121}.
After following the procedures discussed above, we obtain
\begin{align}
	&[X^{(1)}]^{-1} = \left(
	\begin{array}{cccccccccccccccc}
		\Delta_1 & \Delta_2 & \Delta_3 & \text{P}_1 & \text{P}_2 & \text{P}_3 & \text{N}_1 & \Lambda_1 & \Lambda_2 & \text{Q}_1 & \text{Q}_2 & \text{R}_1 & \text{T}_1 & \text{T}_2 & \text{S}_1 & \text{S}_2 \\
		\hline
		1 & -1 & 1 & 0 & 0 & 0 & 0 & 0 & 0 & 0 & 0 & -1 & 0 & 0 & 0 & 0 \\
		1 & 0 & 0 & 0 & 0 & 0 & 0 & 0 & 0 & 0 & 0 & -1 & 0 & 0 & 0 & 0 \\
		0 & 0 & 0 & 0 & 0 & 0 & 1 & 0 & 0 & 0 & 0 & 0 & 0 & 0 & 0 & 0 \\
		0 & 0 & 0 & 0 & 1 & 0 & 0 & 0 & 0 & 0 & 0 & 0 & 1 & 0 & 0 & 0 \\
		0 & 0 & 0 & 0 & 1 & 0 & 0 & 0 & -1 & 0 & 0 & 0 & 0 & 0 & 0 & 0 \\
		0 & 0 & 0 & 0 & 1 & 1 & 0 & 0 & -1 & 0 & 0 & 0 & 1 & 0 & 0 & 0 \\
		0 & 0 & 0 & 0 & 0 & 0 & 0 & 0 & 0 & 0 & 0 & 1 & 0 & 0 & 0 & 0 \\
		-1 & 1 & 0 & 0 & 0 & 0 & 0 & 0 & 0 & 0 & 0 & 1 & 0 & 0 & 0 & 0 \\
		0 & 0 & 0 & 0 & 0 & 0 & 0 & 0 & 0 & 0 & 0 & 0 & 0 & 0 & 1 & 0 \\
		0 & 0 & 0 & 0 & 0 & 0 & 0 & 0 & 0 & -1 & -1 & 0 & 0 & 0 & -1 & 0 \\
		0 & 0 & 0 & 0 & 2 & 1 & 0 & 0 & -1 & 0 & 0 & 0 & 1 & 0 & 0 & 0 \\
		0 & 0 & 0 & 0 & 0 & 0 & 0 & -1 & 0 & 0 & 0 & 0 & -1 & 0 & 0 & 0 \\
		0 & 0 & 0 & 0 & 0 & 0 & 0 & 0 & 0 & 0 & 1 & 0 & 0 & 0 & 1 & 0 \\
		0 & 0 & 0 & 0 & 0 & 0 & 0 & 1 & 1 & 0 & 0 & 0 & 1 & 1 & 0 & 0 \\
		0 & 0 & 0 & 1 & 1 & 1 & 0 & -1 & -1 & 1 & 1 & 0 & 0 & 0 & 0 & 0 \\
		0 & 0 & 0 & 0 & 0 & 0 & 0 & 0 & 0 & 1 & 1 & 0 & 0 & 0 & 1 & 1 \\
	\end{array}
	\right);
 \end{align}
 \begin{align}
	&\Sigma^{(0)} = \text{diag}(1, 1, 1, 1, 1, 1, 1, 1, 1, 1, 3, 0, 0);
 \end{align}
 \begin{align} 
	&[V^{(0)}]^{-1} = \left(
	\begin{array}{ccccccccccccccccccccc}
		\Gamma_1 & \Gamma_2 & \Gamma_3 & \text{M}_1 & \text{M}_2 & \text{K}_1 & \text{K}_2 & \text{K}_3 & \text{A}_1 & \text{A}_2 & \text{A}_3  & \text{L}_1 & \text{L}_2 & \text{H}_1 & \text{H}_2 & \text{H}_3 & \Delta_1 & \Delta_2 & \Delta_3 &  \text{R}_1& \text{N}_1\\
		\hline
		1 & 1 & 0 & 0 & 0 & 0 & 0 & 0 & 0 & 0 & 0 & -1 & -1 & 0 & 0 & 0 & 2 & -2 & 2 & 0 & -2 \\
		0 & 0 & 1 & 0 & 0 & 0 & 0 & 0 & 0 & 0 & 0 & -1 & -1 & 0 & 0 & 0 & 2 & 0 & 0 & 0 & -2 \\
		0 & 0 & 0 & 1 & 1 & 0 & 0 & 0 & 0 & 0 & 0 & 1 & 1 & 0 & 0 & 0 & 0 & 0 & 0 & 2 & 0 \\
		0 & 0 & 0 & 0 & 0 & 1 & 0 & 0 & 0 & 0 & 0 & 0 & 0 & 0 & 0 & 1 & 0 & 0 & 0 & 0 & 0 \\
		0 & 0 & 0 & 0 & 0 & 0 & 1 & 0 & 0 & 0 & 0 & 0 & 0 & 0 & 0 & 1 & 0 & 0 & 0 & 0 & 0 \\
		0 & 0 & 0 & 0 & 0 & 0 & 0 & 1 & 0 & 0 & 0 & 0 & 0 & 0 & 0 & 2 & 0 & 0 & 0 & 0 & 0 \\
		0 & 0 & 0 & 0 & 0 & 0 & 0 & 0 & 1 & 1 & 0 & 1 & 1 & 0 & 0 & 0 & 0 & 0 & 0 & 0 & 2 \\
		0 & 0 & 0 & 0 & 0 & 0 & 0 & 0 & 0 & 0 & 1 & 1 & 1 & 0 & 0 & 0 & -2 & 2 & 0 & 0 & 2 \\
		0 & 0 & 0 & 0 & 0 & 0 & 0 & 0 & 0 & 0 & 0 & 0 & 0 & 1 & 0 & 1 & 0 & 0 & 0 & 0 & 0 \\
		0 & 0 & 0 & 0 & 0 & 0 & 0 & 0 & 0 & 0 & 0 & 0 & 0 & 0 & 1 & 1 & 0 & 0 & 0 & 0 & 0 \\
		0 & 0 & 0 & 0 & 0 & 0 & 0 & 0 & 0 & 0 & 0 & 0 & 0 & 0 & 0 & 1 & 0 & 0 & 0 & 0 & 0 \\
		0 & 0 & 0 & 0 & 0 & 0 & 0 & 0 & 0 & 0 & 0 & 1 & 0 & 0 & 0 & 0 & 0 & 0 & 0 & 0 & 0 \\
		0 & 0 & 0 & 0 & 0 & 0 & 0 & 0 & 0 & 0 & 0 & 0 & 1 & 0 & 0 & 0 & 0 & 0 & 0 & 0 & 0 \\
		0 & 0 & 0 & 0 & 1 & 0 & 0 & 0 & 0 & 0 & 0 & 0 & 0 & 0 & 0 & 0 & 0 & 0 & 0 & 0 & 0 \\
		0 & 0 & 0 & 0 & 0 & 0 & 0 & 0 & 0 & 1 & 0 & 0 & 0 & 0 & 0 & 0 & 0 & 0 & 0 & 0 & 0 \\
		0 & 1 & 0 & 0 & 0 & 0 & 0 & 0 & 0 & 0 & 0 & 0 & 0 & 0 & 0 & 0 & 0 & 0 & 0 & 0 & 0 \\
		0 & 0 & 0 & 0 & 0 & 0 & 0 & 0 & 0 & 0 & 0 & 0 & 0 & 0 & 0 & 0 & 1 & 0 & 0 & 0 & 0 \\
		0 & 0 & 0 & 0 & 0 & 0 & 0 & 0 & 0 & 0 & 0 & 0 & 0 & 0 & 0 & 0 & 0 & 1 & 0 & 0 & 0 \\
		0 & 0 & 0 & 0 & 0 & 0 & 0 & 0 & 0 & 0 & 0 & 0 & 0 & 0 & 0 & 0 & 0 & 0 & 1 & 0 & 0 \\
		0 & 0 & 0 & 0 & 0 & 0 & 0 & 0 & 0 & 0 & 0 & 0 & 0 & 0 & 0 & 0 & 0 & 0 & 0 & 1 & 0 \\
		0 & 0 & 0 & 0 & 0 & 0 & 0 & 0 & 0 & 0 & 0 & 0 & 0 & 0 & 0 & 0 & 0 & 0 & 0 & 0 & 1 \\
	\end{array}
	\right).
\end{align}
\end{widetext}
As discussed in Sec.~\ref{sec3}, we can find topological invariants using these data.

Let us begin by constructing topological invariants corresponding to $E_{2}^{1,-1} = \mZ_3 \times \mZ^2$. 
We define the following two quantities:
\begin{align}
	\mathcal{W}^{\text{gapped}}_1 &= \frac{-1}{2\pi \mathrm{i}}\left[\int_{\Lambda} d \log \frac{\det q_{\bk}^{\Lambda_1}}{(\det q_{\bk}^{\Lambda_1})^{\text{vac}}} +  \int_\text{T} d \log \frac{\det q_{\bk}^{\text{T}_1}}{(\det q_{\bk}^{\text{T}_1})^{\text{vac}}}\right];\\
	\mathcal{W}^{\text{gapped}}_2 &= \frac{1}{2\pi \mathrm{i}}\left[\int_\text{Q} d \log \frac{\det q_{\bk}^{\text{Q}_2}}{(\det q_{\bk}^{\text{Q}_2})^{\text{vac}}} +  \int_\text{S} d \log \frac{\det q_{\bk}^{\text{S}_1}}{(\det q_{\bk}^{\text{S}_1})^{\text{vac}}}\right].
\end{align}
Then, we can find the following topological invariant
\begin{widetext}
\begin{align}
	\mathcal{X} = \frac{3}{2\pi} \mathrm{Im}\hspace{-0.5mm}\log\hspace{-0.5mm}\left[\frac{\mathcal{Z}[(q_{\text{H}}^{\text{H}_3})^{\text{vac}}]}{\mathcal{Z}[q_{\text{H}}^{\text{H}_3}]}\exp\left[\frac{1}{3}\left\{-\hspace{-1.5mm}\int_{\text{P}}\left(2d\log \frac{\det q_{\bk}^{\text{P}_2}}{(\det q_{\bk}^{\text{P}_2})^{\text{vac}}} +d\log \frac{\det q_{\bk}^{\text{P}_3}}{(\det q_{\bk}^{\text{P}_3})^{\text{vac}}}\right) +\int_{\Lambda}d\log \frac{\det q_{\bk}^{\Lambda_2}}{(\det q_{\bk}^{\Lambda_2})^{\text{vac}}} -  \int_{\text{T}}d\log \frac{\det q_{\bk}^{\text{T}_1}}{(\det q_{\bk}^{\text{T}_1})^{\text{vac}}}\right\}\right]\right].
\end{align}
\end{widetext}

Next, we construct topological invariants that diagnose the entry in $E_{2}^{0,0} = (\mZ_2)^5$.
As shown in Eq.~\eqref{eq:E100_P3121}, we have $E_{1}^{0,0} = (\mZ_2)^{10}$.
It is well-known that topological phases at $0$-cells are characterized by zero-dimensional topological invariants, as discussed in the context of symmetry indicators~\cite{SI_TSC_Luka, Ono-Po-Shiozaki2021}. 
In this case, the zero-dimensional topological invariants are Pfaffians for each irrep at $\Gamma, \text{M}, \text{A},$ and $\text{L}$; that is, $E_{1}^{0,0}$ is completely characterized by
\begin{align}
		(p_{\Gamma_1}, p_{\Gamma_2}, p_{\Gamma_3}, p_{\text{M}_1}, p_{\text{M}_2}, p_{\text{A}_1}, p_{\text{A}_2}, p_{\text{A}_3}, p_{\text{L}_1}, p_{\text{L}_2}),
\end{align}
where $p_{\rho}$ denotes the Pfaffian invariant for irrep $\rho$. 
Since $E_{2}^{0,0} = \ker d_{1}^{0,0}$, Eq.~\eqref{eq:f_map} informs us about invariants defined on $0$-cells, we find that the following quantities 
\begin{align}
	(p_{\Gamma_2}, p_{\text{M}_2}, p_{\text{A}_2}, p_{\text{L}_1}, p_{\text{L}_2}).
\end{align} 
Although the above set of invariants fully characterize $E_{2}^{0,0}$, it is convenient to change the basis of $E_{2}^{0,0}$ and to have topological invariants to distinguish between atomic limits and other phases. 
This is achieved by symmetry indicators, and we obtain the following five $\mZ_2$-valued invariants:
\begin{align}
	E_{2}^{0,0} = \text{span}\left\{\begin{array}{c}
		\mathfrak{b}_1 = (1, 0, 1, 0, 1, 0, 1, 1, 1, 0) \\
		\mathfrak{b}_2 = (0, 1, 1, 1, 0, 1, 0, 1, 0, 1) \\
		\mathfrak{b}_3 = (1, 0, 1, 0, 1, 1, 0, 1, 0, 1) \\
		\mathfrak{b}_4 = (0, 0, 0, 0, 0, 1, 1, 0, 0, 0) \\
		\mathfrak{b}_5 = (1, 1, 0, 0, 0, 1, 1, 0, 0, 0) \\
	\end{array}\right\}
\end{align}
and
\begin{align}
	\nu_1 &= p_{\text{L}_1} \mod2,\\
	\nu_2 &= p_{\text{M}_2}+ p_{\text{L}_1} + p_{\text{L}_2}\mod2, \\
	\nu_3 &= p_{\text{M}_2}+ p_{\text{L}_1} \mod2, \\
	\nu_4 &= p_{\Gamma_2} + p_{\text{M}_2} + p_{\text{A}_2} + p_{\text{L}_2}\mod2, \\
	\nu_5 &= p_{\Gamma_2} + p_{\text{M}_2} + p_{\text{A}_3} \mod2.
\end{align}
For $\mathfrak{b}_i$, $\nu_i = 1$ and $\nu_{j \neq i} = 0 $.
In fact, $\nu_4$ and $\nu_5$ are symmetry indicators. 

As a result, we have a set of topological invariants
\begin{align}
	\label{eq:topo-P3211}
	\mathcal{V} = (\nu_1, \nu_2, \nu_3, \nu_4, \nu_5, \mathcal{X}, \mathcal{W}^{\text{gapped}}_1, \mathcal{W}^{\text{gapped}}_2). 
\end{align}
We compute all these topological invariants for generators of $^{\phi}K_{G/\Pi}^{(z,c)+0}(\mathbb{R}^3) = (\mZ_{2})^3 \times \mZ_3 \times \mZ^2$, and the computed results are tabulated in Table~\ref{tab:P3_121}.
We make the following two comments. 
First, we find that $\nu_4 = 0$ for all gapped phases.
This implies that $(\mZ_2)^4$-part of $E_{2}^{0,0}$ corresponds to gapped phases.
In other words, the entry of remaining $\mZ_2$-part is gapless, and the second differential $d_{2}^{0,0}:E_{2}^{0,0}\rightarrow E_{2}^{2,-1} = \mZ_2$ is nontrivial. 
To show this, let us consider the following effective low-energy model around $\Gamma$
\begin{align}
	&H_{(k_1, k_2, k_z)} = (\bk^2 - \mu) \tau_z + k_z \tau_y \sigma_x + k_1 k_2 (k_1 + k_2) \tau_y \sigma_x,\\
	&U(\calT) = \tau_0 \sigma_0,\\
	&U(\calC) = \tau_x \sigma_0,\\
	&U(C_{3}^{z}) = \tau_0 \sigma_0,\\
	&U(C_{2}^{[100]}) = \tau_0 \sigma_z,
\end{align}
where $\tau_{i=0,x,y,z}$ and $\sigma_{i=0,x,y,z}$ are Pauli matrices acting on Nambu-spinor space and the eigenvalue space, respectively.
This model corresponds to $\frak{b}_4 + \frak{b}_5 $ and exhibits four gapless points on $(\frac{\sqrt{3\mu}}{2}, 0, 0)$, $(0, \frac{\sqrt{3\mu}}{2}, 0)$, and $(\pm \frac{\sqrt{3\mu}}{2}, \mp\frac{\sqrt{3\mu}}{2}, 0)$, which are the 2-cell $\text{A}$-$\text{A}'$-$\text{L}'$-$\text{L}$ and its symmetry-related 2-cells.
This is exactly the generator of $E_{2}^{2,-1}=\mZ_2$, which implies that $d_{2}^{0,0}$ is nontrivial.

Second, $\mathcal{W}^{\text{gapped}}_1$ and $\mathcal{W}^{\text{gapped}}_2$ are fractional for $(0,0,0,0,1,0), (0,0,0,0,0,1) \in$ $^{\phi}K_{G/\Pi}^{(z,c)+0}(\mathbb{R}^3)$.
This indicates that there is a nontrivial relation among $\mZ_2$ invariants, $\mathcal{W}^{\text{gapped}}_1$, and $\mathcal{W}^{\text{gapped}}_2$.
In fact, $\nu_5$ is nontrivial for $(0,0,0,0,1,0), (0,0,0,0,0,1) \in$ $^{\phi}K_{G/\Pi}^{(z,c)+0}(\mathbb{R}^3)$.
This can generally occur when there is a group extension in the process of deriving $^{\phi}K_{G/\Pi}^{(z,c)+0}(T^3)$ by AHSS.
However, we leave the development of a general theory for the case where nontrivial higher differential and the group extension are involved as future work. 

\begin{table}[t]
	\begin{center}
		\caption{\label{tab:irrep_SGP3121}
			Character table of irreps of little groups and their EAZ classes on $0$- and $1$-cells in space group $P3_121$. 
			When $T_z$ denotes the primitive lattice translation along $z$-direction, the remaining representative elements  are given by $C_{2}^{[110]} = T_{z}^{-1}S_zC_{2}^{[100]}$ and $C_{2}^{[010]} = T_{z}^{-1}S_{z}^{2}C_{2}^{[100]}$.
		}
		\begin{tabular}{c|c|cccccc}
			\hline
			irrep. & EAZ & $e$ & $S_z$ & $S_z^2$ & $C_{2}^{[100]}$ & $C_{2}^{[110]}$ & $C_{2}^{[010]}$ \\
			\hline\hline
			$\Gamma_1$ & BDI & $1$ & $1$ & $1$ & $-1$ & $-1$ & $-1$ \\
			$\Gamma_2$ & BDI &$1$ & $1$ & $1$ & $1$ & $1$ & $1$ \\
			$\Gamma_3$ & BDI & $2$ & $-1$ & $-1$ & $0$ & $0$ & $0$\\
			$\text{A}_1$ & BDI & $1$ & $-1$ & $1$ & $1$ & $1$ & $-1$ \\
			$\text{A}_2$ & BDI &$1$ & $-1$ & $1$ & $-1$ & $-1$ & $1$ \\
			$\text{A}_3$ & BDI & $2$ & $1$ & $-1$ & $0$ & $0$ & $0$\\
			$\text{K}_1$ & AIII & $1$ & $-1$ & $1$ & $1$ & $1$ & $-1$ \\
			$\text{K}_2$ & AIII &$1$ & $-1$ & $1$ & $-1$ & $-1$ & $1$ \\
			$\text{K}_3$ & AIII & $2$ & $1$ & $-1$ & $0$ & $0$ & $0$\\
			$\text{H}_1$ & AIII & $1$ & $-1$ & $1$ & $1$ & $1$ & $-1$ \\
			$\text{H}_2$ & AIII &$1$ & $-1$ & $1$ & $-1$ & $-1$ & $1$ \\
			$\text{H}_3$ & AIII & $2$ & $1$ & $-1$ & $0$ & $0$ & $0$\\
			\hline
			irrep. & EAZ & $e$ &  $C_{2}^{[010]}$ &&&&\\
			\hline\hline
			$\text{M}_1$ & BDI & $1$ & $-1$&&&& \\
			$\text{M}_2$ & BDI &$1$ & $1$ &&&& \\
			$\text{L}_1$ & BDI & $1$ & $-1$ &&&& \\
			$\text{L}_2$ & BDI &$1$ & $1$ &&&& \\
			\hline
			irrep. & EAZ & $e$ &  $S_z$ & $S_z^2$ &&&\\
			\hline\hline
			$\Delta_1$ & BDI & $1$ & $e^{\mathrm{i}(-\tfrac{2\pi}{3} - \tfrac{k_z}{3})}$&$e^{\mathrm{i}(\tfrac{2\pi}{3} - \tfrac{2k_z}{3})}$&&& \\
			$\Delta_2$ & BDI &$1$ & $e^{\mathrm{i}(\tfrac{2\pi}{3} - \tfrac{k_z}{3})}$ &$e^{\mathrm{i}(-\tfrac{2\pi}{3} - \tfrac{2k_z}{3})}$&&& \\
			$\Delta_3$ & BDI & $1$ & $e^{-\mathrm{i} \tfrac{k_z}{3}}$ &$e^{-\mathrm{i} \tfrac{2k_z}{3}}$ &&& \\
			$\text{P}_1$ & AIII & $1$ & $e^{\mathrm{i}(-\tfrac{2\pi}{3} - \tfrac{k_z}{3})}$&$e^{\mathrm{i}(\tfrac{2\pi}{3} - \tfrac{2k_z}{3})}$&&& \\
			$\text{P}_2$ & AIII &$1$ & $e^{\mathrm{i}(\tfrac{2\pi}{3} - \tfrac{k_z}{3})}$ &$e^{\mathrm{i}(-\tfrac{2\pi}{3} - \tfrac{2k_z}{3})}$&&& \\
			$\text{P}_3$ & AIII & $1$ & $e^{-\mathrm{i} \tfrac{k_z}{3}}$ &$e^{-\mathrm{i} \tfrac{2k_z}{3}}$ &&&\\
			\hline
			irrep. & EAZ & $e$ &  $C_{2}^{[110]}$ &&&&\\
			\hline\hline
			$\Lambda_1, \text{Q}_1$ & AIII & $1$ & $-1$&&&& \\
			$\Lambda_2, \text{Q}_2$ & AIII &$1$ & $1$ &&&& \\
			\hline
			irrep. & EAZ & $e$ &  $C_{2}^{[010]}$ &&&&\\
			\hline\hline
			$\text{T}_1, \text{S}_1$ & AIII & $1$ & $-1$&&&& \\
			$\text{T}_2, \text{S}_2$ & AIII &$1$ & $1$ &&&& \\
			\hline
			irrep. & EAZ & $e$ &&&&&\\
			\hline\hline
			$\text{R}_1, \text{W}_1, \text{N}_1$ & BDI & $1$ &&&&& \\
			\hline
		\end{tabular}
	\end{center}
\end{table}

\begin{table}[t]
	\centering
	\caption{\label{tab:P3_121}
		Results of topological invariants for representative models of all topological phases in space group $P3_121$ with TRS.
		The first column represents entry of the K-group $^{\phi}K_{G/\Pi}^{(z,c)}(\mathbb{R}^3) = (\mZ_2)^3 \times \mZ_3 \times \mZ^2$~\cite{Shiozaki-Ono2023}.
		The second column describes the real-space information of topological phases.
		Topological crystalline phases are represented by a symbol $(hkl;d_0)$, where $(hkl)$ denotes the direction along a generating one-dimensional topological superconductor and $d_0$ is the $z$-coordinate of the generating one-dimensional topological superconductor. 
		Atomic limits are denoted by ``$\rho$@$\text{x}$'', where orbital $\rho = A, B$ is at a maximal Wyckoff position $\text{x} = \text{a}, \text{b}$ ($\text{a} = (0, 0, 0)$, $\text{b} = (0, 0, 1/6)$).
		The third column shows the values of topological invariants defined in Eq.~\eqref{eq:topo-P3211}.
	}
	\begin{tabular}{c|c|c}
		\hline
		$^{\phi}K_{G/\Pi}^{(z,c)}(\mathbb{R}^3)$ & Real space & topo. invariants $\mathcal{V}$\\
		\hline\hline
		$(1,0,0,0,0,0)$ & $B$@$\text{a}$ & $(1,0,0,0,0,0,0,0)$\\
		$(0,1,0,0,0,0)$ & $A$@$\text{a}$ & $(0,1,0,0,0,0,0,0)$\\
		$(0,0,1,0,0,0)$ & $B$@$\text{b}$ & $(0,0,1,0,0,0,0,0)$\\
		$(0,0,0,1,0,0)$ & $(100; 0)\oplus(010; 0)$ & $(1,1,0,0,0,1,0,0)$\\
		$(0,0,0,0,1,0)$ & $(110; 0)$ & $(1,0,0,0,1,0,\tfrac{1}{2},\tfrac{1}{2})$\\
		$(0,0,0,0,0,1)$ & $(010; \tfrac{1}{6})$ & $(0,0,1,0,1,0,-\tfrac{1}{2},\tfrac{1}{2})$\\
		\hline
	\end{tabular}
\end{table}

\subsection{Example: $P_c3$}
\label{app:Pc3}
The next example is spinless superconductors in magnetic space group $P_c3$ with the conventional pairing symmetry $A$.
Generators of magnetic space are threefold rotation $C_{3}^{z}$, time-reversal with half translation along $z$-direction, and primitive translations.
Although the Abelian group structure of $K$-group is unknown, we have information about real-space descriptions of topological phases. 
According to Ref.~\cite{Shiozaki-Ono2023}, there are atomic limits and topological crystalline phases classified into $(\mZ_{2})^4$ and $(\mZ_2)^2$, respectively. 

\begin{figure}[t]
	\begin{center}
		\includegraphics[width=0.5\columnwidth]{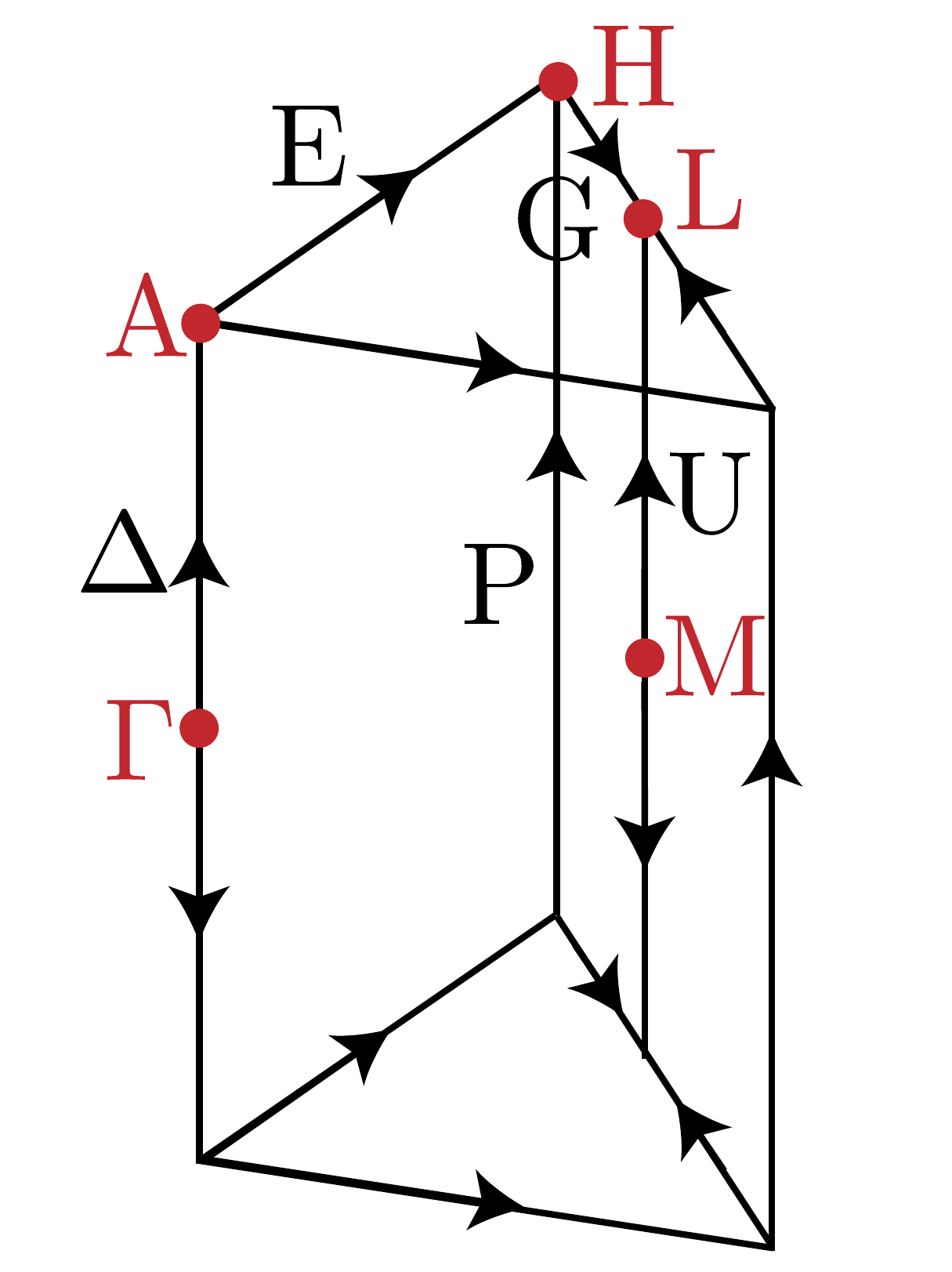}
		\caption{\label{fig:Pc3}
			Cell decomposition of magnetic space group $P_c3$.
		}
	\end{center}
\end{figure}

On the other hand, from AHSS in momentum space, we find that $E_{2}^{0,0} = (\mZ_2)^2$ and $E_{2}^{1,-1} = (\mZ_2)^5$, which implies that seven $\mZ_2$-valued topological invariants can be constructed in our method.  
Our cell decomposition of the fundamental domain is shown in Fig.~\ref{fig:Pc3}.
For this cell decomposition, $E_{1}^{0,0}$ and $E_{1}^{1,-1}$ are spanned by
\begin{align}
	\label{eq:E100_P3c}
	&E_{1}^{0,0} = \mZ_2[\bm{b}_{\Gamma_3}^{(0)}] \oplus \mZ_2[\bm{b}_{\text{M}_1}^{(0)}], \\
	\label{eq:E111_P3c}
	&E_{1}^{1,-1} = \bigoplus_{j=1}^{3}\bigoplus_{K=\Delta, \text{P}}\mZ[\bm{b}_{K_j}^{(1)}] \oplus \bigoplus_{K=\text{U}, \text{E}, \text{G}}\mZ[\bm{b}_{K_1}^{(1)}],
\end{align}
where a threefold rotation symmetric cell $K\ (K = \Gamma, \text{A}, \text{H}, \Delta, \text{P})$ has three irreps $K_1, K_2, K_3$ corresponding to threefold rotation eigenvalues $e^{\ii \frac{4\pi}{3}}, e^{\ii \frac{2\pi}{3}}, 1$; the other cells only have the trivial irrep. 
In fact, $E_{2}^{0,0}=E_{1}^{0,0}$ since $d_{1}^{0,0}$ is trivial. 
Therefore, two Pfaffian invariants $(p_{\Gamma_3}, p_{\text{M}_1})$ are topological invariants defined on $0$-cells.

As for  topological invariants defined on $1$-cells, after performing the same analysis, we have
\begin{align}
	&[X^{(1)}]^{-1}=\left(
	\begin{array}{ccccccccc}
		\Delta_1 & \Delta_2 & \Delta_3 & \text{P}_1 & \text{P}_2 & \text{P}_3 & \text{U}_1 & \text{E}_1 & \text{G}_1\\
		\hline
		0 & -1 & 0 & 0 & 0 & 0 & 0 & 0 & 0 \\
		-1 & -1 & -1 & 0 & 0 & 0 & 0 & -1 & 0 \\
		1 & 1 & -1 & 0 & 0 & 0 & 0 & 0 & 0 \\
		1 & 1 & 0 & 0 & 0 & 0 & -1 & 0 & 0 \\
		-1 & -1 & -2 & 0 & 0 & 0 & 0 & -2 & 0 \\
		1 & 1 & 0 & 0 & 1 & 0 & 0 & 0 & 0 \\
		1 & 1 & 0 & 0 & 0 & 1 & 0 & 0 & 0 \\
		2 & 2 & 2 & 1 & 1 & 1 & 0 & 2 & 0 \\
		1 & 1 & 1 & 0 & 0 & 0 & -1 & 1 & 1 \\
	\end{array}
	\right);\\
	&\Sigma^{(0)} = \text{diag}(1, 1, 2, 2, 2, 2, 2); \\
	&[V^{(0)}]^{-1}=\left(
	\begin{array}{ccccccccc}
		\Gamma_1 &  \Gamma_3 & \text{M}_1 & \text{A}_1 & \text{A}_3 & \text{L}_1 & \text{H}_1 & \text{H}_2 & \text{H}_3\\
		\hline
		1 & 0 & 0 & 1 & 0 & 0 & 0 & 0 & 0 \\
		0 & 0 & 0 & 0 & 0 & 0 & 1 & 1 & 1 \\
		0 & 0 & 0 & -1 & 1 & 0 & 0 & 0 & 0 \\
		0 & 0 & 0 & -1 & 0 & 1 & 0 & 0 & 0 \\
		0 & 0 & 0 & -1 & 0 & 0 & 1 & 1 & 1 \\
		0 & 0 & 0 & -1 & 0 & 0 & 0 & 1 & 0 \\
		0 & 0 & 0 & -1 & 0 & 0 & 0 & 0 & 1 \\
		0 & 1 & 0 & 0 & 0 & 0 & 0 & 0 & 0 \\
		0 & 0 & 1 & 0 & 0 & 0 & 0 & 0 & 0 \\
	\end{array}
	\right).
\end{align}
Based on Sec.~\ref{sec3}, we find the following five $\mZ_2$-valued topological invariants:
\begin{widetext}
	\begin{align}
		&\mathcal{X}_1 = \frac{1}{\pi}\im\hspace{-1mm}\log\hspace{-1mm}\left[ \frac{\det q_{\bk}^{\text{A}_1}}{\det (q_{\bk}^{\text{A}_1})^{\text{vac}}}\left(\frac{\pf [q_{\bk}^{\text{A}_3}]}{\pf[(q_{\bk}^{\text{A}_3})^{\text{vac}}]}\right)^{\hspace{-1mm}-1}\hspace{-3mm}\exp\left[-\frac{1}{2}\int_{\Delta}\left(d\log \frac{\det q_{\bk}^{\Delta_1}}{(\det q_{\bk}^{\Delta_1})^{\text{vac}}}+\hspace{-0.5mm}d\log \frac{\det q_{\bk}^{\Delta_2}}{(\det q_{\bk}^{\Delta_2})^{\text{vac}}}- d\log \frac{\det q_{\bk}^{\Delta_3}}{(\det q_{\bk}^{\Delta_3})^{\text{vac}}}\right)\right]\right];\\
		&\mathcal{X}_2 = \frac{1}{\pi}\im\hspace{-1mm}\log\hspace{-1mm}\left[\frac{\det q_{\bk}^{\text{A}_1}}{\det (q_{\bk}^{\text{A}_1})^{\text{vac}}}\left(\frac{\pf[q_{\bk}^{\text{L}_1}]}{\pf[(q_{\bk}^{\text{L}_1})^{\text{vac}}]}\right)^{\hspace{-1mm}-1}\hspace{-3mm}\exp\left[-\frac{1}{2}\left(\int_{\Delta}\hspace{-0.5mm}d\log \frac{\det q_{\bk}^{\Delta_1}}{(\det q_{\bk}^{\Delta_1})^{\text{vac}}}+\int_{\Delta}\hspace{-0.5mm}d\log \frac{\det q_{\bk}^{\Delta_2}}{(\det q_{\bk}^{\Delta_2})^{\text{vac}}}-\int_{\text{U}}d\log \frac{\det q_{\bk}^{\text{U}_1}}{(\det q_{\bk}^{\text{U}_1})^{\text{vac}}}\right)\right]\right];\\
		&\mathcal{X}_3 = \frac{1}{\pi}\im\hspace{-1mm}\log\hspace{-1mm}\left[\frac{\det q_{\bk}^{\text{A}_1}}{\det (q_{\bk}^{\text{A}_1})^{\text{vac}}}
		\left(\frac{\det q_{\bk}^{\text{H}_1}}{\det (q_{\bk}^{\text{H}_1})^{\text{vac}}}\frac{\det q_{\bk}^{\text{H}_2}}{\det (q_{\bk}^{\text{H}_2})^{\text{vac}}}\frac{\det q_{\bk}^{\text{H}_3}}{\det (q_{\bk}^{\text{H}_3})^{\text{vac}}}\right)^{\hspace{-1mm}-1} \right. \nonumber\\
		&\left. \quad\quad\quad\quad\quad\quad\quad
		\times\exp\left[\frac{1}{2}\int_{\Delta}\left(d\log \frac{\det q_{\bk}^{\Delta_1}}{(\det q_{\bk}^{\Delta_1})^{\text{vac}}}+\hspace{-0.5mm}d\log \frac{\det q_{\bk}^{\Delta_2}}{(\det q_{\bk}^{\Delta_2})^{\text{vac}}}+2d\log \frac{\det q_{\bk}^{\Delta_3}}{(\det q_{\bk}^{\Delta_3})^{\text{vac}}}\right) +\int_{\text{E}}\hspace{-1mm}d\log \frac{\det q_{\bk}^{\text{E}_1}}{(\det q_{\bk}^{\text{E}_1})^{\text{vac}}} \right]\right];\\
		&\mathcal{X}_4 =\frac{1}{\pi}\im\hspace{-1mm}\log\hspace{-1mm}\left[\frac{\det q_{\bk}^{\text{A}_1}}{\det (q_{\bk}^{\text{A}_1})^{\text{vac}}}\left(\frac{\det q_{\bk}^{\text{H}_2}}{\det (q_{\bk}^{\text{H}_2})^{\text{vac}}}\right)^{\hspace{-1mm}-1}\hspace{-3mm}\exp\left[-\frac{1}{2}\left(\int_{\Delta}\hspace{-0.5mm}d\log \frac{\det q_{\bk}^{\Delta_1}}{(\det q_{\bk}^{\Delta_1})^{\text{vac}}}+\int_{\Delta}\hspace{-0.5mm}d\log \frac{\det q_{\bk}^{\Delta_2}}{(\det q_{\bk}^{\Delta_2})^{\text{vac}}}+\int_{\text{P}}d\log \frac{\det q_{\bk}^{\text{P}_2}}{(\det q_{\bk}^{\text{P}_2})^{\text{vac}}}\right)\right]\right];\\
		&\mathcal{X}_5 =\frac{1}{\pi}\im\hspace{-1mm}\log\hspace{-1mm}\left[\frac{\det q_{\bk}^{\text{A}_1}}{\det (q_{\bk}^{\text{A}_1})^{\text{vac}}}\left(\frac{\det q_{\bk}^{\text{H}_3}}{\det (q_{\bk}^{\text{H}_3})^{\text{vac}}}\right)^{\hspace{-1mm}-1}\hspace{-3mm}\exp\left[-\frac{1}{2}\left(\int_{\Delta}\hspace{-0.5mm}d\log \frac{\det q_{\bk}^{\Delta_1}}{(\det q_{\bk}^{\Delta_1})^{\text{vac}}}+\int_{\Delta}\hspace{-0.5mm}d\log \frac{\det q_{\bk}^{\Delta_2}}{(\det q_{\bk}^{\Delta_2})^{\text{vac}}}+\int_{\text{P}}d\log \frac{\det q_{\bk}^{\text{P}_3}}{(\det q_{\bk}^{\text{P}_3})^{\text{vac}}}\right)\right]\right].
	\end{align}
\end{widetext}
As a result, we have the seven $\mZ_2$-valued topological invariants
\begin{align}
	\label{eq:topo-P3c}
	\mathcal{V} = (p_{\Gamma_3}, p_{\text{M}_1}, \mathcal{X}_1, \mathcal{X}_2, \mathcal{X}_3, \mathcal{X}_4, \mathcal{X}_5). 
\end{align}
As shown in Table~\ref{tab:P3c}, we compute all these topological invariants for representative models constructed from the real-space description.

\begin{table}[t]
	\begin{center}
		\caption{\label{tab:P3c}
			Results of topological invariants for representative models of all topological phases in magnetic space group $P3_c$.
			%
			%
			The first column describes the real-space information of topological phases.
			Again, atomic limits are denoted by ``$\rho$@$\text{x}$'', where orbital $\rho = A$ or $E$ is at a maximal Wyckoff position $\text{x} = \text{a}, \text{b}, \text{c}$ ($\text{a} = (0, 0, 0)$, $\text{b} = (1/3, 2/3, 0), \text{c} = (2/3, 1/3, 0)$).
			As for topological crystalline phases, there are two types. 
            One is topological phases constructed one-dimensional topological superconductors along $z$-direction.
            In each unit cell, the one-dimensional topological superconductor is replaced at maximal Wyckoff position $\text{b}$.
            The other one is composed of two-dimensional topological superconductors with Chern numbers $\pm 1$.
            Two-dimensional topological superconductors are stacked by time-reversal with the half-translation.
			The second column shows the values of topological invariants defined in Eq.~\eqref{eq:topo-P3c}.
		}
		\begin{tabular}{c|c}
			\hline
			Real space & invariants \\
			\hline\hline
			$^1E@\text{b}$ & $(0,0,1,0,1,0,1)$ \\
			$A@\text{b}$ & $(0,0,1,1,0,0,0)$ \\
			$A@\text{b}+E@\text{b}+A@\text{d}$ & $(0,0,1,0,1,0,0)$ \\
			$A@\text{b}+E@\text{c}-E@\text{a}-E@\text{b}$ & $(0,0,0,1,1,1,1)$ \\
			1D TSC with $A@\text{b}$  & $(1,1,\frac{1}{2},\frac{1}{2},1,0,0)$ \\
			2D TSC with $E@\text{a}$& $(0,0,1,1,1,1,1)$\\
			\hline
		\end{tabular}
	\end{center}
\end{table}

\section{Detailed information of discussions presented in Sec.~\ref{sec3}}
In this appendix, we provide information about technical details, which are not included to avoid digressing from the main topic of this work.

\subsection{Brief review of definitions of $K$-groups}
\label{app:K-theory}
This subsection reviews definitions of $K$-groups in Refs.~\cite{Shiozaki-Sato-Gomi2017, CREST-Gomi}. 
Before defining $K$-groups, let us consider the triple $(E, H_{\bk}, H^{\text{ref}}_{\bk})$, where $H_{\bk}$ and $H^{\text{ref}}_{\bk}$ are Hamiltonians, and $E$ is a vector bundle on which $H_{\bk}$ and $H^{\text{ref}}_{\bk}$ act.
In the presence of symmetries, symmetry representations $U_{\bk}(g)$ are the same for $H_{\bk}$ and $H^{\text{ref}}_{\bk}$, i.e., 
\begin{align}
	U_{\bk}(g)H_{\bk}^{\phi_g} = c_g H_{\bk}U_{\bk}(g); \\
	U_{\bk}(g)[H_{\bk}^{\text{ref}}]^{\phi_g} = c_g H_{\bk}^{\text{ref}}U_{\bk}(g).
\end{align}

We define relations between two triples $(E, H_{\bk}, H^{\text{ref}}_{\bk})$ and $(E', H'_{\bk}, H'^{\text{ref}}_{\bk})$.
First, we define the sum structure among triples by
\begin{align}
	&(E, H_{\bk}, H^{\text{ref}}_{\bk}) + (E', H'_{\bk}, H'^{\text{ref}}_{\bk}) \nonumber \\
	&\quad\quad\quad\quad\quad\quad:= (E\oplus E', H_{\bk}\oplus H'_{\bk}, H^{\text{ref}}_{\bk}\oplus H'^{\text{ref}}_{\bk}).
\end{align}
Next, let us suppose that $H_{\bk}$ and $H'_{\bk}$, as well as $H^{\text{ref}}_{\bk}$ and $H'^{\text{ref}}_{\bk}$, are homotopy equivalent, i.e., $H_{\bk} \sim H'_{\bk}$ and $H^{\text{ref}}_{\bk} \sim H'^{\text{ref}}_{\bk}$.
Then, we say that these two triples are isomorphic $(E, H_{\bk}, H^{\text{ref}}_{\bk}) \simeq (E', H'_{\bk}, H'^{\text{ref}}_{\bk})$.
As a result, isomorphism classes of $(E, H_{\bk}, H^{\text{ref}}_{\bk})$ form a commutative monoid $\calM(X)$.

We define a submonoid $\mathcal{Z}(X)$ of $\calM(X)$ by $\mathcal{Z}(X) := \{(E, H_{\bk}, H^{\text{ref}}_{\bk})\in \calM(X)\ \vert \ H_{\bk} \sim  H^{\text{ref}}_{\bk}\}$. 
Then, we also define the equivalence relation ``$\approx$'' between $x, x' \in \calM(X)$. 
We say that $x \approx x'$ if there exist $z, z' \in \mathcal{Z}(X)$ such that $x + z \simeq x'+z'$. 
Then, $K$-group $K(X)$ is defined by the equivalence classes $[V, H_{\bk}, H^{\text{ref}}_{\bk}]$, i.e.,
\begin{align}
	K(X) := \calM(X)/\approx.
\end{align}
It should be noted that $z \approx z'$ for $z, z' \in \mathcal{Z}(X)$. 
This immediately follows from $z + z_2 \simeq z_1 + z'$ since there exist $z_1$ and $z_2$ such that $z_1 \simeq z$ and $z_2 \simeq z'$ by definition.
%
%
The $K$-group has the following properties
\begin{itemize}
	\item $[V, H_{\bk}, H^{\text{ref}}_{\bk}] = 0$ when $H_{\bk} \sim H^{\text{ref}}_{\bk}$,
	\item $-[V, H_{\bk}, H^{\text{ref}}_{\bk}] = [V, H^{\text{ref}}_{\bk}, H_{\bk}]$.
\end{itemize}
From the above two properties, one can see that the $K$-group is invariant under adding trivial degrees of freedom 
\begin{align}
	[V, H_{\bk}, H^{\text{ref}}_{\bk}] + [V, H'_{\bk}, H'_{\bk}] \approx [V, H_{\bk}, H^{\text{ref}}_{\bk}].
\end{align}

Similarly, a relative $K$-group $K(X, Y)$ for $Y \subset X$ is defined by triples $[V, H_{\bk}, H^{\text{ref}}_{\bk}]$ such that $H_{\bk} = H^{\text{ref}}_{\bk}$ for $^\forall \bk \in Y$. 
Also the reduced $K$-group $\tilde{K}(X)$ is defined by
\begin{align}
	\tilde{K}(X) := K(X, pt).
\end{align}

\subsection{Stable classification and finite-rank classifications}
\label{app:stable_vs_finite_rank}

We emphasize that the completeness discussed in this work is meant in the stable $K$-theoretic sense. This should be distinguished from finer finite-rank classifications. One may fix both the numbers of occupied and unoccupied bands and classify gapped Bloch or BdG Hamiltonians up to homotopy without adding extra bands. Such a fixed-size Hamiltonian classification can contain unstable information that disappears after increasing the number of bands. A typical example is the Hopf insulator~\cite{Moore_Hopf_Insulator_2008}, which is nontrivial as a two-by-two Hamiltonian but is not detected by the $K$-group.

One may instead fix only the number of occupied bands and classify the occupied-state vector bundle of fixed rank. This classification is also generally finer than stable $K$ theory. A typical example is the Euler insulator~\cite{Ahn_EulerInsulator_2019}, distinguished by the Euler number of a two-band $PT$-symmetric occupied bundle.

Thus, when we say that our invariants completely characterize the classification, we mean that they completely characterize the stable $K$ group, not all fixed-rank Hamiltonians or fixed-rank vector bundles.
This limitation does not mean that the invariants constructed in this paper are irrelevant to fragile or finite-rank topology. Many such phases are diagnosed by concrete topological invariants, such as irrep-counting invariants and quantized Wilson-loop invariants, whose systematic construction is one of the main aims of this work. What lies beyond the $K$-group is the finer finite-rank question of whether these invariants obstruct gap-preserving deformations with the number of occupied bands, or both the numbers of occupied and unoccupied bands, fixed.

\subsection{Choice of a basis for the computation of $q$-matrix}
\label{app:basis_q}

In this subsection, we explain how to choose $\mathcal{U}_{\alpha\pm}$ in Eq.~\eqref{eq:qmatrix} so that $\det q_{\bk}^{\alpha} = \mathcal{Z}[q_{\bk}^{\alpha}]^{\calD_{\alpha}}$ for EAZ class AIII/CI and $\det q_{\bk}^{\alpha} = \mathcal{Z}[q_{\bk}^{\alpha}]^{2\calD_{\alpha}}$ for EAZ class DIII.

Let us recall that $\mathcal{U}^{\alpha}_{\pm}$ satisfies
\begin{align}
	P^{\alpha\pm}\mathcal{U}^{\alpha}_{\pm} = \mathcal{U}^{\alpha}_{\pm},
\end{align}
where $P^{\alpha\pm}$ is a projection matrix defined in Eq.~\eqref{eq:projection}.
Furthermore, $\mathcal{U}^{\alpha}_{\pm}$ is transformed as
\begin{align}
	\mathcal{U}_{\bk}(g)\mathcal{U}^{\alpha}_{\pm} = \mathcal{U}^{\alpha}_{\pm}\mathcal{W}^{\pm}_{\bk}(g)\quad (g\in \calG_{\bk})
\end{align}
for EAZ class AIII and 
\begin{align}
	\mathcal{U}_{\bk}(g)\mathcal{U}^{\alpha}_{\pm} &= \mathcal{U}^{\alpha}_{\pm}\mathcal{W}^{\pm}_{\bk}(g)\quad (g\in \calG_{\bk}) \\
	\mathcal{U}_{\bk}(a)[\mathcal{U}^{\alpha}_{\pm}]^* &= \mathcal{U}^{\alpha}_{\mp}\mathcal{W}^{\mp}_{\bk}(a)\quad (a\in \mathcal{A}_{\bk}) 
\end{align}
for EAZ class CI/DIII.
It should be noted that $\mathcal{W}^{+}_{\bk}(g)$ and $\mathcal{W}^{-}_{\bk}(g)$ are unitary equivalent since they both are composed of irrep $\alpha$. 
Thus, they can be the same matrix by a basis transformation $\mathcal{U}^{\alpha}_{-} \rightarrow \mathcal{U}^{\alpha}_{-}\mathcal{S}$, i.e., there exists a unitary matrix $\mathcal{S}$ such that
\begin{align}
	\label{eq:equiv_Gk}
	\mathcal{W}_{\bk}(g) &:= \mathcal{W}^{+}_{\bk}(g) = \mathcal{S}^{\dagger}\mathcal{W}^{-}_{\bk}(g)\mathcal{S} \quad (g \in \calGk), \\
	\label{eq:equiv_Ak}
	\mathcal{W}_{\bk}(a) &:=\mathcal{W}^{+}_{\bk}(a)\mathcal{S}^* = \mathcal{S}^{\dagger}\mathcal{W}^{-}_{\bk}(a)\quad (a\in\mathcal{A}_{\bk}).
\end{align}
When we define the $q$-matrix by
\begin{align}
	q'^{\alpha}_{\bk} = [\mathcal{U}^{\alpha}_{-}\mathcal{S}]^{\dagger}H_{\bk}\mathcal{U}^{\alpha}_{+},
\end{align}
the $q$-matrix itself is symmetric under $g \in \calGk$ for EAZ class AIII and $g \in (\calGk+\mathcal{A}_{\bk})$ for EAZ class CI/DIII.
In other words,
\begin{align}
\label{eq:w_Gk}
\mathcal{W}_{\bk}(g)q'^{\alpha}_{\bk} &= q'^{\alpha}_{\bk}\mathcal{W}_{\bk}(g) \quad (g \in \calGk),
\end{align}
for EAZ class AIII. For EAZ class CI/DIII, in addition to Eq.~\eqref{eq:w_Gk}, $q'^{\alpha}_{\bk}$ satisfies
\begin{align}
\label{eq:w_Ak}
\mathcal{W}_{\bk}(a)[q'^{\alpha}_{\bk}]^{T} &= q'^{\alpha}_{\bk}\mathcal{W}_{\bk}(a)\quad (a\in\mathcal{A}_{\bk}). 
\end{align}
As a result, the eigenvalues of $q'^{\alpha}_{\bk}$ are $\calD_{\alpha}$-fold degenerate for EAZ class AIII/CI and $2\calD_{\alpha}$-fold degenerate for EAZ class DIII. 
The problem is how to obtain $\mathcal{S}$ numerically. 
In the following, we discuss our implementation to get $\mathcal{S}$.
Equations \eqref{eq:equiv_Gk} and \eqref{eq:equiv_Ak} can be rewritten as
\begin{align}
	\mathcal{W}^{-}_{\bk}(g)\mathcal{S}[\mathcal{W}^{+}_{\bk}(g)]^{\dagger} - \mathcal{S} = O \quad (g \in \calGk),\\
	\mathcal{W}^{-}_{\bk}(a)\mathcal{S}^{\top}[\mathcal{W}^{+}_{\bk}(a)]^{\dagger} - \mathcal{S} = O\quad (a\in\mathcal{A}_{\bk}).
\end{align}
Importantly, we can numerically solve these linear equations.
To achieve this, let us introduce the following tensors
\begin{align}
	[A_g]_{lm;no} = [\mathcal{W}^{-}_{\bk}(g)]_{ln}[\mathcal{W}^{+}_{\bk}(g)]^{*}_{mo} - \delta_{ln}\delta_{mo},\\
	[B_g]_{lm;no} = [\mathcal{W}^{-}_{\bk}(g)]_{lo}[\mathcal{W}^{+}_{\bk}(g)]^{*}_{mn} - \delta_{ln}\delta_{mo}.
\end{align}
By reshaping these tensors, we have matrices $\{A_g\}_{g \in \calGk}$ and $\{B_g\}_{g \in \calA_{\bk}}$.
Then, the solution can be obtained from
\begin{align}
	\bigcap_{g \in \calGk} \ker A_g
\end{align}
for EAZ class AIII/CI and
\begin{align}
	\left(\bigcap_{g \in \calG_{\bk}} \ker A_g\right) \cap \left(\bigcap_{g \in \calA_{\bk}} \ker B_g\right)
\end{align}
for class DIII, as discussed below.
Let $(\bm{a}_1, \bm{a}_2, \cdots \bm{a}_J)$ be a basis set of the intersection of the kernels. 
Then, we consider linear combinations of $\{\bm{a}_i\}_{i=1}^{J}$ with {\it randomly} chosen coefficients $\{c_i\}_{i=1}^{J}$, i.e.,
$\bm{s} = \sum_{i=1}^{J} c_i \bm{a}_i$.
Finally, we have the unitary matrix $\mathcal{S}$ by rearranging the vector $\bm{s}$ and unitarizing the reshaped matrix by singular-value decomposition. 

It is noted that the intersection of the kernels is not one-dimensional in general. 
This happens when ${\cal W}_\bk(g)$ is reducible and composed of $n_\alpha$ degenerate $\alpha$-irreps. 
If this is the case, ${\cal S}$ takes a form ${\cal S} = \bigoplus_\alpha {\cal S}_\alpha \otimes \delta {\cal S}_\alpha$, where ${\cal S}_\alpha$ is a unitary matrix determined by ${\cal W}_\bk^+(g)$ and ${\cal W}_\bk^-(g)$, and $\delta {\cal S}_\alpha \in {\rm GL}(n_\alpha)$ is unfixed. 
A vector $\bm{s}$ gives a set of matrices $\delta {\cal S}_\alpha$, and it is invertible if the coefficients $c_i$ are randomly chosen. 
Unitarizing ${\cal S}$ by singular-value decomposition leaves ${\cal S}_\alpha$ invariant, making that ${\cal S}$ is a desired basis transformation.

\subsection{Check of completeness of topological invariants}
\label{app:check}

Here, we explain how to confirm that our invariants span $E_{2}^{1,-1}$.
After computing $W_l[H_{\bk}], W_{\mathrm{g}}[H_{\bk}], C[H_{\bk}]$ in Eqs.~\eqref{eq:wl_list}, \eqref{eq:wg_list}, and \eqref{eq:c_list} for the $20$ Hamiltonians, we have
\begin{align}
	&L = \left(W_{l}[H^{(1)}_{\bk}], W_{l}[H^{(2)}_{\bk}], \cdots, W_{l}[H^{(20)}_{\bk}]\right),\\
	&F = \left(W_{\text{g}}[H^{(1)}_{\bk}], W_{\text{g}}[H^{(2)}_{\bk}], \cdots, W_{\text{g}}[H^{(20)}_{\bk}]\right), \\
	&T = \left(C[H^{(1)}_{\bk}], C[H^{(2)}_{\bk}], \cdots, C[H^{(20)}_{\bk}]\right).
\end{align}

First, we check if gapless topological invariants can span $\mZ^{r_1}$. 
To see this, we compute the Smith normal form of $L$ as
\begin{align}
	\label{eq:L_smith}
	U_L L V_L = \begin{pmatrix}
		\Sigma_L & O 
	\end{pmatrix}.
\end{align}
If $\mathrm{rank}\Sigma_L = r_1$ and $[\Sigma_L]_{jj} = 1$ for $(j = 1, \cdots r_1)$, our gapless topological invariants span $\mZ^{r_1}$.
Furthermore, Eq.~\eqref{eq:L_smith} implies that the $(r_1+1)$-th through $20$th columns of $L' = LV_L$ are zeros.
Thus, $V_L$ informs us about combinations of Hamiltonians whose gapless topological invariants are trivial.
We note that the above criterion is basis-independent.
The columns of $L$ generate an integer sublattice of $\mathbb{Z}^{r_1}$.
The Smith normal form tests whether the sublattice has full rank and index one, and hence whether it equals the full lattice, independently of the chosen basis.

Next, we confirm that the $\mZ$-valued invariants for gapped phases on $2$-skeletons are sufficient to span $\mZ^{N_\mathrm{f}}$.
From Eq.~\eqref{eq:L_smith}, $(r_1+1)$-th through $20$th columns of $F' = FV_L = (\bm{\mathfrak{f}}'_1, \cdots,\bm{\mathfrak{f}}'_{r_1}, \bm{\mathfrak{f}}'_{r_1+1}, \cdots, \bm{\mathfrak{f}}'_{20})$ give us the computed results of $\mZ$-valued invariants without gapless points on $2$-cells.
Then, we again consider the Smith normal form of $(\bm{\mathfrak{f}}'_{r_1+1}, \cdots, \bm{\mathfrak{f}}'_{20})$
\begin{align}
	U_F (\bm{\mathfrak{f}}'_{r_1+1}, \cdots, \bm{\mathfrak{f}}'_{20}) V_F = \begin{pmatrix}
		\Sigma_F & O 
	\end{pmatrix}.
\end{align}
If $\mathrm{rank}\Sigma_F = N_{\mathrm{f}}$ and $[\Sigma_F]_j = 1$ for $(j = 1, \cdots N_{\mathrm{f}})$, our $\mZ$-valued invariants can fully characterize $\mZ^{N_{\mathrm{f}}}$.

Last, we discuss $\mZ_k$-valued invariants for gapped phases on $2$-skeletons.
In the following, we consider $\mZ_k$-valued topological invariants as $\mZ$-valued quantities, i.e., we forget about the $\mZ_k$-nature of $C[H_{\bk}]$. 
Similar to the case of $\mZ$-valued invariants, for $TV_L = (\bm{\mathfrak{t}}'_{1},\cdots, \bm{\mathfrak{t}}'_{r_1},\bm{\mathfrak{t}}'_{r_1+1}, \cdots, \bm{\mathfrak{t}}'_{20})$, $(\bm{\mathfrak{t}}'_{r_1+1}, \cdots, \bm{\mathfrak{t}}'_{20})$ is a set of results of $\mZ_k$-valued invariants without gapless points on $2$-cells, which is denoted by $\mathfrak{T}$.
Let us suppose that we have $\lambda_{i-r_1}$ such that $\lambda_{i-r_1} \notin \{0, 1\}$ for $R_0+1 \leq i-r_1 \leq R_0+N_{\mathrm{t}} \ (R_0\in\mZ)$.
Then, we define a diagonal matrix by
\begin{align}
	\mathfrak{A} = \text{diag}\left(\lambda_{R_0+1}, \cdots, \lambda_{R_0+N_{\mathrm{t}}}\right).
\end{align}
We construct a basis set of $\begin{pmatrix}\mathfrak{T} & \mathfrak{A}\end{pmatrix}$ from its Smith normal form
\begin{align}
	U_{\mathfrak{T}}\begin{pmatrix}\mathfrak{T} & \mathfrak{A}\end{pmatrix}V_{\mathfrak{T}}  = \begin{pmatrix}\Sigma_{\mathfrak{T}}& O \\
		O & O
	\end{pmatrix},
\end{align}
where $\Sigma_{\mathfrak{T}}$ is a diagonal matrix and its elements are all positive integers. 
When we denote $U_{\mathfrak{T}}^{-1} = \begin{pmatrix} U_{\perp}& U_{\parallel}\end{pmatrix}$, $\mathfrak{B}=U_{\perp}\Sigma_{\mathfrak{T}}$ is a basis set of $\begin{pmatrix}\mathfrak{T} & \mathfrak{A}\end{pmatrix}$. 
By definition, $\mathfrak{A}$ can always be expanded by $\mathfrak{B}$.
In other words, when $\mathfrak{B}^{+}$ denotes the pseudo-inverse matrix of $\mathfrak{B}$, $\mathfrak{B}^{+}\mathfrak{A}$ is always an integer-valued matrix.
Then, we compute the Smith normal form of $\mathfrak{B}^{+}\mathfrak{A}$
\begin{align}
	U_{\mathfrak{B}} (\mathfrak{B}^{+}\mathfrak{A}) V_{\mathfrak{B}} = \begin{pmatrix}
		\Sigma_{\mathfrak{B}} & O 
	\end{pmatrix}.
\end{align}
If $\Sigma_{\mathfrak{B}}=\mathfrak{A}$, our invariants can correctly capture torsion elements of $E_{2}^{1,-1}$.

\section{Compatibility relation of non-Hermitian matrix}
\label{sec:Compatibility relation of non-Hermitian matrix}
Let \( G \) be a finite group and \( \phi: G \to \{\pm 1\} \) be a homomorphism indicating $g\in G$ is unitary or antiunitary. 
The factor system of \( G \) is denoted by \( z_{g,h} \). 
Consider the subgroup \( G_0 = \{g \in G | \phi_g = 1\} \). 
Let the set of irreps of $G_0$ be \( \{\alpha\}_{\alpha} \), with the character of the \( \alpha \)-irrep being \( \chi^\alpha_g \). The EAZ class of \( \alpha \)-irrep is defined by the Wigner criterion
\begin{align}
    W^\alpha = \frac{1}{|G_0|} \sum_{g \in G\backslash G_0} z_{g,g} \chi^\alpha_{g^2} \in \{0,\pm 1\}, 
    \label{eq:app_wigner}
\end{align}
where a representative element is denoted by \( a \in G\backslash G_0 \). 
The the irrep obtained by applying $a$ to the \( \alpha \)-irrep is written as \( a\alpha \), where $\chi^{a\alpha}_g = \frac{z_{g,a}}{z_{a,a^{-1}ga}} (\chi_{a^{-1}ga})^*.$
The EAZ class is classified depending on the presence of anti-unitary elements and the value of \( W_\alpha \) as:
\begin{align*}
    \text{EAZ of } \alpha = \left\{
    \begin{array}{ll}
        \text{A} & \text{if } (G\backslash G_0 = \emptyset), \\
        \text{AI} & \text{if } (G\backslash G_0 \neq \emptyset, W_\alpha=1),  \\
        \text{AII} & \text{if } (G\backslash G_0 \neq \emptyset, W_\alpha=-1), \\
        \text{A}_T & \text{if } (G\backslash G_0 \neq \emptyset, W_\alpha=0). \\
    \end{array}
    \right.
\end{align*}
Furthermore, for the irrep \( \alpha \), the representation including anitunitary symmetries is given as:
\begin{align}
    \tilde \alpha= \left\{
    \begin{array}{ll}
        \alpha & (\text{A, AI}), \\
        \alpha\oplus \alpha & (\text{AII}), \\
        \alpha\oplus a\alpha & (\text{A}_T). \\
    \end{array}
    \right.
    \label{eq:app_tilde_alpha}
\end{align}

Consider a subgroup \( H \subset G \), and its unitary subgroup \( H_0 = \{ g \in H | \phi_g = 1 \} \). An irrep of \( H_0 \) is denoted as \( \beta \), and its representation including antiunitary group elements is denoted similarly to (\ref{eq:app_tilde_alpha}) as \( \tilde \beta \). 
The decomposition of \( \tilde \alpha \) by \( \tilde \beta \) is written as 
\begin{align}
    \tilde \alpha = \bigoplus_{\tilde \beta} \tilde \beta^{\oplus n^{\tilde \alpha}_{\tilde \beta}}, \quad n^{\tilde \alpha}_{\tilde \beta} \in \mathbb{Z}_{\geq 0},
\end{align}
where \( n^{\tilde \alpha}_{\tilde \beta} \) is calculated using the irreducible character, equivalent to the first differential \( d_1^{p,0} \) in insulators~\cite{Shiozaki-Ono2023}.

Consider a \( z \)-projective representation \( \rho \) of \( G \) with dimension \( \mathcal{D}_\rho \). 
Let us denote a representation matrix by $u^\rho_{g\in G}$, which satisfies 
\begin{align}
    \left. 
    \begin{array}{ll}
     u^\rho_g u^\rho_h   & (\phi_g=1)  \\
     u^\rho_g (u^\rho_h)^* & (\phi_g=-1)  \\
    \end{array}\right\}
     = z_{g,h} u^\rho_{gh}, \quad \forall g,h \in G.
     \label{eq:app_z-proj_rep}
\end{align}
The decomposition of \( \rho \) by \( \tilde \alpha \) and \( \tilde \beta \) is written as 
\begin{align}
    \rho = \bigoplus_{\tilde \alpha} \tilde \alpha^{\oplus n^{\rho}_{\tilde \alpha}} = \bigoplus_{\tilde \beta} \tilde \beta^{\oplus n^{\rho}_{\tilde \beta}}. 
\end{align}
Let us set a representation matrix $u^\alpha_g$ of the \( \alpha \)-irrep. 
Then, the representation matrix for the \( \tilde \alpha \) representation reads as
\begin{align}
    u^{\tilde \alpha}_{g \in G_0} =
    \left\{
    \begin{array}{ll}
        u^\alpha_g & (\text{A, AI}), \\
        u^\alpha_g \otimes \mathds{1}_2 & (\text{AII}), \\
        \begin{pmatrix}
        u^\alpha_g \\
        & u^{a\alpha}_g \\
        \end{pmatrix} & (\text{A}_T),
    \end{array}
    \right.
    \label{eq:app_rep_can_uni}
\end{align}
and
\begin{align}
    u^{\tilde \alpha}_{a} =
    \left\{
    \begin{array}{ll}
        u^\alpha_a & (\text{AI}), \\
        u^\alpha_a \otimes (i\sigma_y) & (\text{AII}), \\
        \begin{pmatrix}
        & z_{a,a}u^\alpha_{a^2} \\
        \mathds{1}_{\calD_\alpha} \\
        \end{pmatrix} & (\text{A}_T).
    \end{array}
    \right.
    \label{eq:app_rep_can_antiuni}
\end{align}
(Other matrices for elements $g \in G \backslash G_0$ are given by (\ref{eq:app_z-proj_rep}).)
Here, \( u^\alpha_a \in U(\mathcal{D}_\alpha) \) is a unitary matrix that satisfies \( u^\alpha_a (u^\alpha_a)^* = W_\alpha \times z_{a,a}u^\alpha_{a^2} \). The representation matrix for \( \tilde \beta \) is similarly defined. Using this notation, the representation matrix for \( \rho \) can be chosen as:
\begin{align}
    u^\rho_{g \in G} = \bigoplus_{\tilde \alpha} u^{\tilde \alpha}_g \otimes \mathds{1}_{n^\rho_{\tilde \alpha}}. 
    \label{eq:app_rho_rep}
\end{align}

In the following, we consider symmetries that contain only one of the two types, either transpose or complex conjugate.
More generally, there exist symmetries that contain both transposed and complex conjugate types simultaneously, but this is unnecessary for the construction of invariants in this paper.

\subsection{Transpose}
Given a homomorphism \(\phi_g\), we consider an invertible matrix \(M \in {GL}_n(\mathbb{C})\) with the following transpose-type \(G\)-symmetry:
\begin{align}
    M = \left\{
    \begin{array}{ll}
    u_g M u_g^\dag, & \phi_g=1,  \\
    u_g M^T u_g^\dag, & \phi_g=-1, 
    \end{array}\right. \quad g \in G.
    \label{eq:app_sym_M}
\end{align}
Here, \(M^T\) is the transpose of matrix \(M\).
In the basis where \(u^\rho_g\) is given by (\ref{eq:app_rho_rep}), the matrix \(M\) is block-diagonalized as
\begin{align}
    &M = \bigoplus_{\tilde \alpha} \mathds{1}_{\calD_\alpha} \otimes m^{\tilde \alpha}, \label{eq:app_M_deco}\\
    &m^{\tilde \alpha} \in \left\{
    \begin{array}{ll}
       {GL}_{n^\rho_{\tilde \alpha}}(\mathbb{C}), & {\rm A,AI},  \\
       {GL}_{2n^\rho_{\tilde \alpha}}(\mathbb{C}), & {\rm AII,A}_T,  
    \end{array}\right.\label{eq:app_mtilde_constraint}
\end{align}
Moreover, $a$ symmetry implies that 
\begin{align}
    &{\rm AI}:\quad (m^{\tilde \alpha})^T = m^{\tilde \alpha}, \\
    &{\rm AII}:\quad i\sigma_y (m^{\tilde \alpha})^T (i\sigma_y)^\dag = m^{\tilde \alpha}, \\
    &{\rm A}_T:\quad m^{\tilde \alpha}= \begin{pmatrix}
        m^{\alpha}\\
        &(m^\alpha)^T\\
    \end{pmatrix}, \quad m^{\alpha} \in {GL}_{n^\rho_{\tilde\alpha}}(\mathbb{C}). 
    \label{eq:app_M_AT}
\end{align}
Thus, it was observed that when the EAZ is AII or A\(_T\), the eigenvalues of the matrix \(m^{\tilde \alpha}\) are doubly degenerate.

We define the quantity \({\cal Z}^{\tilde \alpha}_G(M)\), which takes values in \(\mathbb{C}^\times = \{z \in \mathbb{C}|z \neq 0\}\), as the product of independent eigenvalues in the \(\tilde \alpha\)-sector.
Introducing the orthogonal projection to the \(\tilde \alpha\) representation
\begin{align}
    &P^{\tilde \alpha} = 
    \left\{\begin{array}{ll}
        P^\alpha & ({\rm A,AI,AII}),  \\
        P^\alpha+P^{a \alpha} & ({\rm A}_T), 
    \end{array}
    \right.\\
    &P^\alpha=\frac{\calD_\alpha}{|G_0|} \sum_{g\in G_0} (\chi^\alpha_g)^* u^\rho_g,
\end{align}
the projection of $M$ to the \(\tilde \alpha\) representation is given by \(M^{\tilde \alpha} = P^{\tilde \alpha} M (= M P^{\tilde \alpha})\).
The matrix rank of \(M^{\tilde \alpha}\) is \(\calD_{\tilde \alpha} n^\rho_{\tilde \alpha}\), and it has \(\calD_{\tilde \alpha}\) degenerate eigenvalues \(\lambda^{\tilde \alpha}_1,\dots,\lambda^{\tilde \alpha}_{n^\rho_{\tilde \alpha}}\).
Writing the non-zero eigenvalues of the matrix \(M\) as the set \({\rm Spec}^\times(M)\), we have
\begin{align}
    {\rm Spec}^{\times}(M^{\tilde \alpha}) = \bigcup_{\mu=1}^{\calD_{\tilde \alpha}} \{\lambda^{\tilde \alpha}_i\}_{i=1}^{n^\rho_{\tilde \alpha}}.
\end{align}
In other words,
\begin{align}
    &\det (\lambda - M^{\tilde \alpha})
    =\lambda^{\calD_\rho-\calD_{\tilde\alpha}} \prod_{i=1}^{n^\rho_{\tilde \alpha}} (\lambda - \lambda^{\tilde \alpha}_i)^{\calD_{\tilde\alpha}}.\end{align}
We define 
\begin{align}
    {\cal Z}_{G}^{\tilde \alpha}(M) := \prod_{i=1}^{n^\rho_{\tilde \alpha}} \lambda^{\tilde \alpha}_i.
    \label{eq:app_Z_phase_def}
\end{align}
Note that this definition of ${\cal Z}_{G}^{\tilde \alpha}(M)$ does not depend on the choice of representation matrix \( u^\rho_g \).

Similarly, for the $\beta$-irrep of the subgroup \(H_0 \subset G_0\), the value \({\cal Z}_{H}^{\tilde \beta}(M) \) in \(\mathbb{C}^\times\) is defined. The following holds:
\begin{align}
    {\cal Z}^{\tilde \beta}_{H}(M) = \prod_{\tilde \alpha} \left[{\cal Z}^{\tilde \alpha}_{G}(M)\right]^{n^{\tilde \alpha}_{\tilde \beta}}.
    \label{eq:app_Z_compatibility}
\end{align}
\noindent
({\it Proof}) 
In the basis that diagonalizes the matrix \( m^{\tilde \alpha} \), the matrix \( M \) can be expressed as:
\begin{align}
    M 
    &= \bigoplus_{\tilde \alpha} \mathds{1}_{\calD_{\tilde \alpha}} \otimes \begin{pmatrix}
        \lambda^{\tilde \alpha}_1\\
        &\ddots\\
        &&\lambda^{\tilde \alpha}_{n^\rho_{\tilde \alpha}}
    \end{pmatrix} \nonumber\\
    &= \bigoplus_{\tilde \alpha} \bigoplus_{\tilde \beta} \mathds{1}_{\calD_{\tilde \beta}} \otimes \mathds{1}_{n^{\tilde \alpha}_{\tilde \beta}} \otimes \begin{pmatrix}
        \lambda^{\tilde \alpha}_1\\
        &\ddots\\
        &&\lambda^{\tilde \alpha}_{n^\rho_{\tilde \alpha}}
    \end{pmatrix}.    
    \label{eq:app_M_decom_beta}
\end{align}
If we decompose the \(\tilde \alpha\) representation into the \(\tilde \beta\) representation as \( u^{\tilde \alpha}_g = \bigoplus_{\tilde \beta} u^{\tilde \beta}_g \otimes \mathds{1}_{n^{\tilde \alpha}_{\tilde \beta}} \), the representation matrix \( u^\rho_g \) is expressed as:
\begin{align}
    u^\rho_g = \bigoplus_{\tilde \alpha} \bigoplus_{\tilde \beta} u^{\tilde \beta}_g \otimes \mathds{1}_{n^{\tilde \alpha}_{\tilde \beta}} \otimes \mathds{1}_{n^\rho_{\tilde \alpha}}.
    \label{eq:app_u_decom_beta}
\end{align}
If we define the projection of \( M^{\tilde \alpha} \) into the \( \tilde \beta \) sector as \( M^{\tilde \alpha,\tilde \beta} = P^{\tilde \beta} P^{\tilde \alpha} M \), from the representations (\ref{eq:app_M_decom_beta}) and (\ref{eq:app_u_decom_beta}), the eigenvalues of the matrix \( M^{\tilde \alpha,\tilde \beta} \) are given by \( n^\rho_{\tilde \alpha} \) eigenvalues \( \lambda^{\tilde \alpha}_1, \dots, \lambda^{\tilde \alpha}_{n^\rho_{\tilde \alpha}} \), which are degenerated \( (\calD_{\tilde \beta} \times n^{\tilde \alpha}_{\tilde \beta}) \) times.
\begin{align}
    {\rm Spec}^{\times}(M^{\tilde \alpha,\tilde \beta}) 
    = \bigcup_{\mu=1}^{\calD_{\tilde \beta}} \bigcup_{\nu=1}^{n^{\tilde \alpha}_{\tilde \beta}}
    \{\lambda^{\tilde \alpha}_i\}_{i=1}^{n^\rho_{\tilde \alpha}}.
\end{align}
Therefore, the eigenvalues of the matrix \( M^{\tilde \beta} = P^{\tilde \beta} M = \sum_{\tilde \alpha} M^{\tilde \alpha,\tilde \beta} \) are obtained by taking the union over \( \tilde \alpha \):
\begin{align}
    {\rm Spec}^{\times}(M^{\tilde \beta}) 
    = \bigcup_{\tilde \alpha} \bigcup_{\mu=1}^{\calD_{\tilde \beta}} \bigcup_{\nu=1}^{n^{\tilde \alpha}_{\tilde \beta}}
    \{\lambda^{\tilde \alpha}_i\}_{i=1}^{n^\rho_{\tilde \alpha}}.
\end{align}
From which we obtain:
\begin{align}
    {\cal Z}^{\tilde \beta}_H(M) 
    = \prod_{\tilde \alpha} \prod_{\nu=1}^{n^{\tilde \alpha}_{\tilde \beta}} \prod_{i=1}^{n^\rho_{\tilde \alpha}} \lambda^{\tilde \alpha}_i
    = \prod_{\tilde \alpha} \prod_{\nu=1}^{n^{\tilde \alpha}_{\tilde \beta}} {\cal Z}^{\tilde \alpha}_G(M).
\end{align}
(Note that \( n^\rho_{\tilde \beta} = \sum_{\tilde \alpha} n^\rho_{\tilde \alpha} n^{\tilde \alpha}_{\tilde \beta} \).)

\subsection{Complex conjugation}
Depending on the homomorphism \(\phi_g\), we consider an invertible matrix \(M \in {GL}_n(\mathbb{C})\) that possesses \(G\)-symmetry of complex conjugation type:
\begin{align}
    M = \left\{
    \begin{array}{ll}
    u_g M u_g^\dag, & \phi_g=1,  \\
    u_g M^* u_g^\dag, & \phi_g=-1, 
    \end{array}\right. \quad g \in G.
    \label{eq:app_sym_M_cc}
\end{align}
Here, \(M^*\) denotes the complex conjugate of the matrix \(M\). In the basis where \(u^\rho_g\) is given by (\ref{eq:app_rho_rep}), the matrix \(M\) is block-diagonalized as (\ref{eq:app_M_deco}) and (\ref{eq:app_mtilde_constraint}).
Furthermore, depending on the EAZ class, the matrix \(m^{\tilde \alpha}\) has the following symmetry constraint by $a$:
\begin{align}
    &{\rm AI}:\quad (m^{\tilde \alpha})^* = m^{\tilde \alpha}, \\
    &{\rm AII}:\quad i\sigma_y (m^{\tilde \alpha})^* (i\sigma_y)^\dag = m^{\tilde \alpha}, \\
    &{\rm A}_T:\quad m^{\tilde \alpha}= \begin{pmatrix}
        m^{\alpha}\\
        &(m^\alpha)^*\\
    \end{pmatrix}, \quad m^{\alpha} \in {GL}_{n^\rho_{\tilde\alpha}}(\mathbb{C}). 
\end{align}
For any EAZ class, no eigenvalue degeneracy occurs. However, when \(a\)-symmetry is present, the eigenvalues exhibit the following structures depending on the EAZ class:
\begin{itemize}
    \item For AI: the eigenvalues of \(m^{\tilde \alpha}=m^{\alpha}\) are either real numbers \(\lambda^{\alpha} \in \mathbb{R}\) or they appear as complex conjugate pairs \(\lambda^{\alpha},(\lambda^{\alpha})^*\).
    \item For AII: the eigenvalues of \(m^{\tilde \alpha}=m^{\alpha}\) appear as complex conjugate pairs \(\lambda^{\alpha},(\lambda^\alpha)^*\).
    \item For A$_T$: the eigenvalue \(\lambda^{\alpha}\) of \(m^{\alpha}\) and the eigenvalue \(\lambda^{a\alpha}\) of \(m^{a\alpha}\) are related as \(\lambda^{\alpha} = (\lambda^{a\alpha})^*\).
\end{itemize}
Thus, if we define \({\cal Z}^\alpha_{G_0}(M)\) in the same manner as (\ref{eq:app_Z_phase_def}) for the $\alpha$-irrep of the unitary subgroup \(G_0\), the following hold:
\begin{align}
    &{\rm A}: \quad {\cal Z}^\alpha_{G_0}(M) \in \mathbb{C}^\times, \\
    &{\rm AI}: \quad {\cal Z}^\alpha_{G_0}(M) \in \mathbb{R}^\times, \\
    &{\rm AII}: \quad {\cal Z}^\alpha_{G_0}(M) \in \mathbb{R}^\times,\quad {\cal Z}^\alpha_{G_0}(M) >0, \\
    &{\rm A}_T: \quad {\cal Z}^\alpha_{G_0}(M) = [{\cal Z}^{a\alpha}_{G_0}(M)]^* \in \mathbb{C}^\times.
\end{align}
(Here, \(\mathbb{R}^\times = \{x \in \mathbb{R}|x\neq 0\}\).)

Similarly, \({\cal Z}_{H_0}^{\beta}(M)\) is defined in the same manner as (\ref{eq:app_Z_phase_def}) for the $\beta$-irrep of the subgroup \(H_0 \subset G_0\). If the irreducible decomposition of the $\alpha$-irrep of \(G_0\) by the irreps of \(H_0\) is given as \(\alpha = \bigoplus_{\beta} \beta^{\oplus n^\alpha_\beta}\), then as a special case of (\ref{eq:app_Z_compatibility}), the following holds:
\begin{align}
    {\cal Z}^{\beta}_{H_0}(M) = \prod_{\alpha} \left[{\cal Z}^{\alpha}_{G_0}(M)\right]^{n^{\alpha}_{\beta}}.
\end{align}

\section{Existence of a canonical gauge fixing condition over a local patch}
\label{app:existence_canonical_gauge}
Let $G$ be a finite group, $\phi: G \to \{\pm 1\}$ a homomorphism specifying unitary/antiunitary, and $z_{g,h}$ a factor system. 
The group $G$ acts on the space $X \cong \R^d$ through a group homomorphism $G \to O(d)$. 
The action of $G$ is expressed as
\begin{align}
k \mapsto gk, \quad k \in X, \quad g \in G,
\end{align}
where $k_0 = 0 \in \R^d$ is a point. 
Note that there exists a homotopy to a point $k_0$ which is compatible with the $G$-action. 

Consider a ``representation on $X$'' satisfying the following:
\begin{align}
U_g(hk)U_h(k)^{\phi_g} = z_{g,h}U_{gh}(k), \quad k \in X, \quad g, h \in G.
\end{align}
Here, $U_g(k)$ is a unitary matrix dependent on $k \in X$. 
We prove the following:
{\it There exists a continuous unitary matrix $V(k)$ on $X$ such that it satisfies:
\begin{align}
U_g(k)V(k)^{\phi_g} = V(gk) U_g(k_0), \quad k \in X, \quad g \in G.
\label{eq:prop}
\end{align}
}

\noindent
(Proof)
First, note that the representation matrices of the group can be considered as taking values in a sort of flag manifold. 
For simplicity, assume that $G$ is unitary, meaning that $\phi_g=1$ for all $g \in G$. 
The case with antiunitary elements will be commented on later. 
Let $\rho$ be a representation of $G$, and $u^\rho_g$ be its representation matrix. 
The irreducible decomposition of $\rho$ can be written as 
\begin{align}
\rho = \bigoplus_\alpha n_\alpha \alpha. 
\end{align}
For each $\alpha$, fix a set of representation matrices $\{u^\alpha_g\}_g$. 
Then, there exists a unitary matrix $V$ such that 
\begin{align}
u^\rho_g V = V \bigoplus_\alpha u^\alpha_g \otimes \mathds{1}_{n_\alpha}. 
\end{align}
The unitary $V$ is not unique, and there is the following ambiguity:
\begin{align}
V \mapsto V \oplus_\alpha \mathds{1}_{\calD_\alpha} \otimes W^\alpha, \quad 
W^\alpha \in U(n_\alpha).
\label{eq:app_gauge_amb}
\end{align}
($\calD_\alpha$ is the representation dimension of the $\alpha$-irrep.) 
The quotient space divided by this ambiguity is written as 
\begin{align}
[V] \in {\cal M}(G,z,\{n_\alpha\}_\alpha) := \frac{U(\sum_\alpha \calD_\alpha n_\alpha)}{\prod_\alpha U(n_\alpha)}. 
\end{align}
There is a one-to-one correspondence between the equivalence class $[V]$ and the representation matrices $\{u^\rho_g\}_g$, meaning that the representation matrices $\{u^\rho_g\}_g$ take values in the space ${\cal M}(G,z,\{n_\alpha\}_\alpha)$. 
Moreover, the ``gauge fixing'' of $V$ corresponds to a choice of $W^\alpha$. 

We now proceed to the proof of the claim.
Divide $X=\R^d$ into cells symmetrically with respect to $G$.
Namely, 
(i) the action of $G$ either keeps points in a $p$-cell $D^p_i$ invariant, $g k = k, k \in D^p_i$, or maps them to another $p$-cell, $g(D^p_i) = D^p_{g(i)}$, 
(ii) every $p$-cell $D^p_i$ is a boundary of some $(p+1)$-cell $D^{p+1}_j$, $D^p_i \subset \partial D^{p+1}_j$.
The stabilizer group of the $p$-cell $D^p_i$ in $G$ is denoted as $G_{D^p_i} = \{g \in G | gk = k, k \in D^p_i\}$.
Every $p$-cell includes the origin $k_0$ in its boundary.
Denote the set of $p$-cells as $C_p$ and define $X_0=C_0= \{k_0\}$, and $X_p = X_{p-1} \cup C_p$.
$X_p$ is referred to as the $p$-skeleton.

\begin{figure}[t]
	\begin{center}
		\includegraphics[width=0.99\columnwidth]{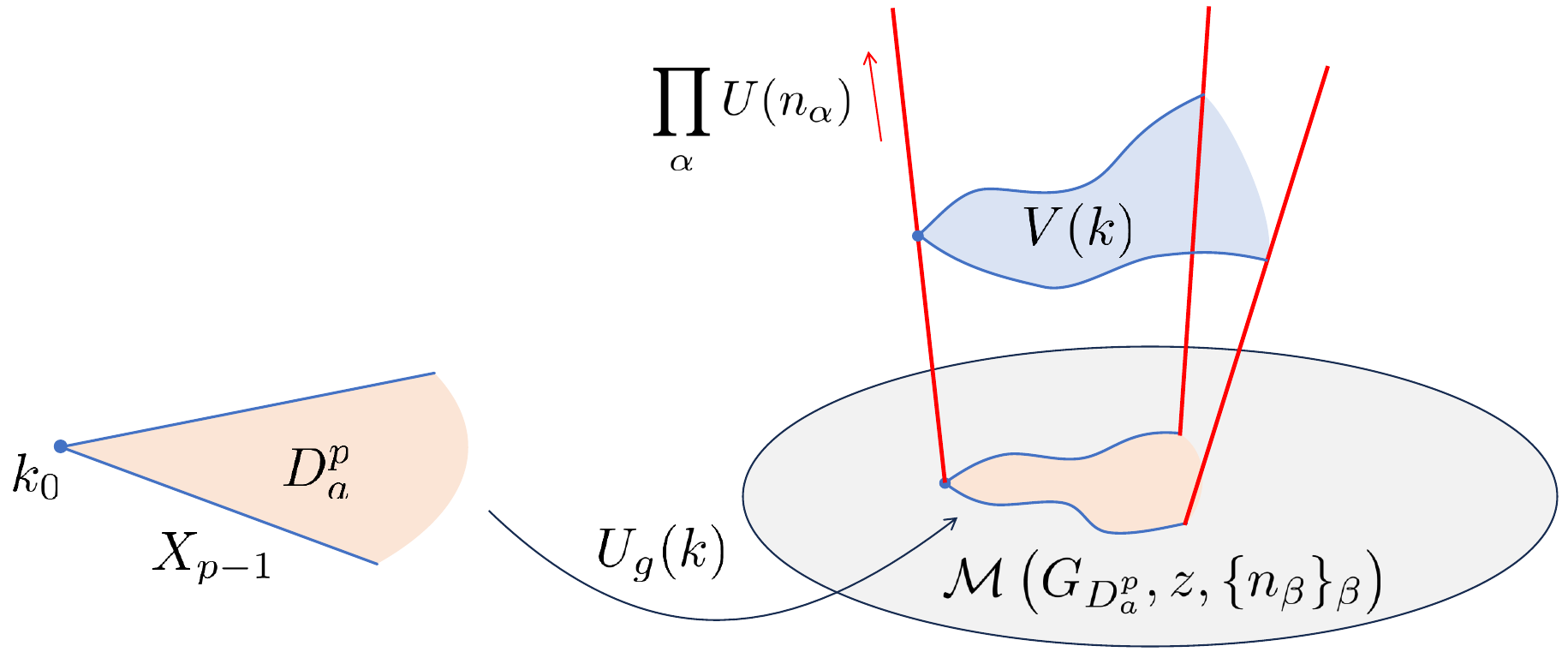}
		\caption{\label{fig:gauge_patch}%
            The representation matrix $U_g(k)$ defines a map from the $p$-cell $D^p_a$ to the quotient space ${\cal M}(G_{D^p_a},z,\{n_\beta\}_\beta)$. 
            The unitary matrix $V(k)$ corresponds to a section in the fiber. 
            The matrix values $V(k)$ at the boundary $\partial D^p_a$ are determined by $V(k)$ on the $(p-1)$-skeleton $X_{p-1}$.
            }
	\end{center}
\end{figure}

We prove the statement by induction.
For the 0-skeleton $X_0=\{k_0\}$, set $V(k_0) = 1$, which satisfies (\ref{eq:prop}).
Assume the existence of $V(k \in X_{p-1})$ satisfying (\ref{eq:prop}) on the $(p-1)$-skeleton $X_{p-1}$. 
We then show that $V(k)$ can be extended to $X_p$.
Consider one of the orbits of the $p$-cell set $C_p$:
\begin{align}
g(D^p_a),\quad g \in G, 
\end{align}
where $D^p_a$ is a representative $p$-cell.
The subgroup keeping $D^p_a$ invariant is written as $G_{D^p_a} = \{g \in G | g (D^p_a) = D^p_a\}$.
On $D^p_a$, restricted to $G_{D^p_a}$, $U_g(k)$ is a group representation:
\begin{align}
U_g(k) U_h(k) = U_{gh}(k), \quad k \in D^p_a, \quad g \in G_{D^p_a}. 
\end{align}
Consider the irreducible decomposition of representation $U$ in the group $G_{D^p_a}$:
\begin{align}
U|_{G_{D^p_a}} = \bigoplus_{\beta} n_\beta \beta. 
\end{align}
Then, the representation matrices $U_{g \in G_{D^p_a}}(k \in D^p_a)$ define a map to the quotient space:
\begin{align}
D^p_a \to {\cal M}(G_{D^p_a},z,\{n_\beta\}_\beta) = \frac{U(\sum_\beta \calD_\beta n_\beta)}{\prod_\beta U(n_\beta)}. 
\end{align}
The unitary matrix $V(k \in \partial D^p_a)$ at the boundary of $D^p_a$ is fixed by $V(k)$ on the $(p-1)$-skeleton $X_{p-1}$.
It remains to be shown whether this can be extended inside $D^p_a$, but this is possible since $D^p_a$ is contractible.
See Fig.\ref{fig:gauge_patch}.
For $h \in G\backslash G_{D^p_a}$, $V(k)$ in the $p$-cell $h (D^p_a)$ should be defined as 
\begin{align}
V(hk) = U_h(k)V(k)U_h(k_0)^\dag,\quad k \in D^p_a. 
\end{align}
This definition is independent of the choice of $h$, as can be confirmed by considering for $g \in G_{D^p_a}$ 
\begin{align}
V(hgk) 
&= U_{hg}(k)V(k)U_{hg}(k_0)\nonumber\\
&= U_h(k)U_g(k)V(k)U_g(k_0)^\dag U_h(k_0^\dag\nonumber\\
&= U_h(k) V(k)U_h(k_0)^\dag. 
\end{align}
This ensures that the definition of $V(k)$ can be consistently extended over the entire $p$-cell $h (D^p_a)$.

When $G$ includes antiunitary symmetries, the ambiguity (\ref{eq:app_gauge_amb}) is changed as follows. 
Depending on the Wigner criterion (\ref{eq:app_wigner}) for the $\alpha$-irrep of the unitary subgroup $G_0 = \{g \in G|\phi_g=1\}$, the representation matrix for $\tilde \alpha$ can be chosen as in (\ref{eq:app_rho_rep}), (\ref{eq:app_rep_can_uni}), and (\ref{eq:app_rep_can_antiuni}). 
In this case, the ambiguity matrix $W: V \mapsto V W$ satisfies the same symmetry as (\ref{eq:app_sym_M}) and can thus be written as in (\ref{eq:app_M_deco} - \ref{eq:app_M_AT}). 
Consequently, the quotient space is obtained as
\begin{align}
    &{\cal M}(G,z,\{n_\alpha\}) \nonumber\\
    &=
    \frac{U(\sum_\alpha \calD_\alpha n_\alpha)}{
    \prod_{\tilde \alpha, W_\alpha=1}O(n_\alpha) \times 
    \prod_{\tilde \alpha, W_\alpha=-1}Sp(\frac{n_\alpha}{2}) \times 
    \prod_{\tilde \alpha, W_\alpha=0}U(n_\alpha)}.
\end{align}
This only slightly changes the structure of the quotient space.
The proof above remains unaffected. 

\section{Proof of (\ref{eq:theorem_canonical_gauge})}
\label{app:proof_canonical_gauge}
On each patch $U_i$, redefine
\begin{align}
    \tilde w_{i,\bk}(g) := e^{ig(\bk-\bk_i)\cdot \bm{a}_g} w_{i,\bk}(g), \quad 
    \bk \in U_i, \quad g \in G, 
\end{align}
so that the factor system of $\tilde w_{i,\bk}(g)$ becomes independent of $\bk$:
\begin{align}
    &\tilde w_{h(i),h\bk}(g) \tilde w_{i,\bk}(h)^{\phi_g}
    =z_{gh\bk_i}(g,h) \tilde w_{i,\bk}(gh), \nonumber\\
    &\bk \in U_i, \quad g, h \in G.
\end{align}
Then, based on the results of the previous section \ref{app:existence_canonical_gauge}, for the stabilizer group $G_i = \{g \in G|g \bk_i = \bk_i\}$, there exists a gauge such that
\begin{align}
    \tilde w_{i,\bk}(g) = \tilde w_{i,\bk_i}(g),\quad \bk \in U_i, \quad g \in G_i,
\end{align}
For $g \in G\backslash G_i$, set the symmetry operator at the 0-cell as
\begin{align}
    \tilde w_{i,\bk}(g) := \tilde w_{i,\bk_i}(g),\quad \bk \in U_i, \quad g \in G \backslash G_i,
\end{align}
Returning the obtained $\tilde w_{i,\bk}(g)$ to $w_{i,\bk}(g)$, it turns out that there exists a gauge
\begin{align}
    w_{i,\bk}(g) = e^{-ig(\bk-\bk_i)\cdot \bm{a}_g} w_{i,\bk_i}(g),\quad \bk \in U_i, \quad g in G. 
\end{align}

\section{Derivation of Eqs.~\eqref{eq:sum_E_F} and \eqref{eq:delta_xi_formula}}
\label{app:sum_E_F}

As stated in Sec.~\ref{sec:Incompatibility of Z2 invariants with band sum}, the $\mZ_2$ invariants $\nu(E)$ and $\nu(F)$ for bands $E$ and $F$ do not maintain an additive structure for the direct sum of bands $E \oplus F$, i.e.,  $\nu(E \oplus F) \neq \nu(E) + \nu(F)$. 
To examine this issue in detail, we first describe a specific construction for the gauge fixing condition $w_{i,\bk}(g)$ and the basis transformation matrix $V_{i\to a,\bk}$ based on the given representation data $\{n_{\beta(\bk_i)}\}_{i,\beta}$ on each 0-cell.

Here, we order the irreps at each 0-cell $i$, which is labeled by $\beta(\bk_i) = 1(\bk_i),\dots,m_i(\bk_i)$.
Let us fix one representation matrix for the irrep $\beta(\bk_i)$ and denote it as $u^{\beta(\bk_i)}_{\bk_i}(g)$.
The irreducible decomposition of band $B\in\{E,F\}$ at 0-cell $i$ can be written as $\bigoplus_{\beta=1}^{m_i} n^B_{\beta(\bk_i)} \beta(\bk_i), n^B_{\beta(\bk_i)} \in \mZ_{\geq 0}$, and thus the gauge fixing condition on patch $U_i$, according to Eq.~\eqref{eq:theorem_canonical_gauge}, becomes
\begin{align}
	w^B_{i,\bk}(g) = e^{-ig(\bk-\bk_i)\cdot \bm{a}_g} \bigoplus_{\beta=1}^{m_i} \mathds{1}_{n^B_{\beta(\bk_i)}} \otimes u^{\beta(\bk_i)}_{\bk_i}(g).
	\label{eq:gauge_fixing_patch_irrep}
\end{align}

Similarly, for each 1-cell $a$, order the irreps as $\alpha(\bk_a)=1(\bk_a),\dots,m_a(\bk_a)$ and denote the representation matrix of irrep $\alpha(\bk_a)$ as $u^{\alpha(\bk_a)}_{\bk_a}(g)$.
For the irreducible decomposition $\bigoplus_{\alpha=1}^{m_a} n^B_{\alpha(\bk_a)} \alpha(\bk_a), n^B_{\alpha(\bk_a)} \in \mZ_{\geq 0}$, of band $B$ on 1-cell $a$, the gauge fixing condition is set as
\begin{align}
	w^B_{a,\bk_a}(g) = \bigoplus_{\alpha=1}^{m_a} \mathds{1}_{n^B_{\alpha(\bk_a)}} \otimes u^{\alpha(\bk_a)}_{\bk_a}(g).
\end{align}

Under the settings described above, we determine the basis transformation matrix $V^B_{i\to a,\bk}$.
For an irrep $\beta(\bk_i)$ of 0-cell $i$, we consider its irreducible decomposition in terms of the irreps $\{\alpha(\bk_a)\}_{q=1,\dots,m_a}$ of the 1-cell $a$ connected to $i$:
\begin{align}
	\beta(\bk_i) 
	= \bigoplus_{\alpha=1}^{m_a} n^{\beta(\bk_i)}_{\alpha(\bk_a)} \alpha(\bk_a), \quad 
	n^{\beta(\bk_i)}_{\alpha(\bk_a)} \in \mZ_{\geq 0}.
\end{align}
Note that
\begin{align}
	\sum_{\beta=1}^{m_i} n^B_{\beta(\bk_i)} n^{\beta(\bk_i)}_{\alpha(\bk_a)} = n^B_{\alpha(\bk_a)}.
\end{align}
Then, a basis transformation matrix $v^{\beta(\bk_i)}_{i\to a,\bk_a}$ from a single irrep $\beta(\bk_i)$ to a representation on the 1-cell $a$ is determined by 
\begin{align}
	&e^{-ig(\bk_a-\bk_i)\cdot \bm{a}_g} u^{\beta(\bk_i)}_{\bk_i}(g)
	v^{\beta(\bk_i)}_{i\to a,\bk_a}\nonumber\\
	&=v^{\beta(\bk_i)}_{i\to a,\bk_a} 
	\bigoplus_{\alpha=1}^{m_a} \mathds{1}_{n^{\beta(\bk_i)}_{\alpha(\bk_a)}} \otimes u^{\alpha(\bk_a)}_{\bk_a}(g), \quad g \in G_a^0.
\end{align}
The transformation matrix $v^{\beta(\bk_i)}_{i\to a,\bk_a}$ does not depend on the band $B \in \{E,F\}$.
Using $v^{\beta(\bk_i)}_{i\to a,\bk_a}$, the representation matrix $w^B_{i,\bk}(g)$ for $g \in G_a^0$ is transformed as follows 
\begin{widetext}
	\begin{align}
		w^B_{i,\bk_a}(g)
		\left(\bigoplus_{\beta=1}^{m_i} \mathds{1}_{n^B_{\beta(\bk_i)}} \otimes v^{\beta(\bk_i)}_{i\to a,\bk_a}\right)
		&= \left(\bigoplus_{\beta=1}^{m_i} \mathds{1}_{n^B_{\beta(\bk_i)}} \otimes v^{\beta(\bk_i)}_{i\to a,\bk_a}\right)
		\bigoplus_{\beta=1}^{m_i} \mathds{1}_{n^B_{\beta(\bk_i)}} \otimes \left(
		\bigoplus_{\alpha=1}^{m_a} \mathds{1}_{n^{\beta(\bk_i)}_{\alpha(\bk_a)}} \otimes u^{\alpha(\bk_a)}_{\bk_a}(g)
		\right).
	\end{align}
	Introduce the permutation matrix $P^B$ that rearranges the sum order from $(\beta,\mu,\alpha) \to (\alpha,\beta,\mu)$:
	\begin{align}
		\left(\bigoplus_{\beta=1}^{m_i} \bigoplus_{\mu=1}^{n^B_{\beta(\bk_i)}} \bigoplus_{\alpha=1}^{m_a} 
		\mathds{1}_{n^{\beta(\bk_i)}_{\alpha(\bk_a)}} \otimes u^{\alpha(\bk_a)}_{\bk_a}(g)\right) P^B 
		&=
		\bigoplus_{\alpha=1}^{m_a} \bigoplus_{\beta=1}^{m_i} \bigoplus_{\mu=1}^{n^B_{\beta(\bk_i)}}
		\mathds{1}_{n^{\beta(\bk_i)}_{\alpha(\bk_a)}} \otimes u^{\alpha(\bk_a)}_{\bk_a}(g) \nonumber \\
		&=
		\bigoplus_{\alpha=1}^{m_a} 
		\mathds{1}_{\sum_{\beta=1}^{m_i} n^B_{\    \beta(\bk_i)} n^{\beta(\bk_i)}_{\alpha(\bk_a)}} \otimes u^{\alpha(\bk_a)}_{\bk_a}(g)
		=w_{a,\bk_a}(g).    
	\end{align}
	Therefore, the basis transformation matrix is given by
	\begin{align}
		V^B_{i\to a,\bk_a} = 
		\left(\bigoplus_{\beta=1}^{m_i} \mathds{1}_{n^B_{\beta(\bk_i)}} \otimes v^{\beta(\bk_i)}_{i\to a,\bk_a} \right)
		P^B.
	\end{align}
	Note that it depends on the representation data $\{n^B_{\beta(\bk_i)}\}_{\beta}$.
	
	For later discussion, we introduce representation bases to explicitly show the order changes in basis.
	Let $\{\phi^{B\beta\mu x}_{i,\bk}\}_{\beta,\mu,x}$ denote the set of Bloch states in band $B$ on patch $U_i$, and write
	\begin{align}
		\Phi^B_{i,\bk}
		&=\bigoplus_{\beta=1}^{m_i} \bigoplus_{\mu=1}^{n^B_{\beta(\bk_i)}} \bigoplus_{x=1}^{\calD_{\beta(\bk_i)}} \phi_{i,\bk}^{B\beta\mu x}.
	\end{align}
	The decomposition of the representation basis $\bigoplus_{x=1}^{\calD_{\beta(\bk_i)}} \phi_{i,\bk_a}^{B\beta\mu x}$ into the representation basis of $G_a^0$ is written as
	\begin{align}
		\left(\bigoplus_{x=1}^{\calD_{\beta(\bk_i)}} \phi_{i,\bk_a}^{B\beta\mu x}\right) v^{\beta(\bk_i)}_{i\to a,\bk_a}
		=
		\bigoplus_{\alpha=1}^{m_a} \bigoplus_{\nu=1}^{n^{\beta(\bk_i)}_{\alpha(\bk_a)}} \bigoplus_{y=1}^{\calD_{\alpha(\bk_a)}} \psi_{i,\bk_a}^{B\beta\mu \alpha\nu y}.
	\end{align}
	Summarizing the changes in basis, we have
	\begin{align}
		\Phi^B_{i,\bk_a}
		= \bigoplus_{\beta=1}^{m_i} \bigoplus_{\mu=1}^{n^B_{\beta(\bk_i)}} \bigoplus_{x=1}^{\calD_{\beta(\bk_i)}} \phi_{i,\bk_a}^{B\beta\mu x}
		&\xrightarrow{\bigoplus_{\beta=1}^{m_i} \mathds{1}_{n^B_{\beta(\bk_i)}} \otimes v^{\beta(\bk_i)}_{i\to a,\bk_a}}
		\bigoplus_{\beta=1}^{m_i} \bigoplus_{\mu=1}^{n^B_{\beta(\bk_i)}} \bigoplus_{\alpha=1}^{m_a} \bigoplus_{\nu=1}^{n^{\beta(\bk_i)}_{\alpha(\bk_a)}} \bigoplus_{y=1}^{\calD_{\alpha(\bk_a)}} \psi_{i,\bk_a}^{B\beta\mu \alpha\nu y} \nonumber \\
		&\xrightarrow{P^B}
		\bigoplus_{\alpha=1}^{m_a} \bigoplus_{\beta=1}^{m_i} \bigoplus_{\mu=1}^{n^B_{\beta(\bk_i)}} \bigoplus_{\nu=1}^{n^{\beta(\bk_i)}_{\alpha(\bk_a)}} \bigoplus_{y=1}^{\calD_{\alpha(\bk_a)}} \psi_{i,\bk_a}^{B\beta\mu \alpha\nu y}.
	\end{align}

	Now, consider
	\begin{align}
		&\xi^{\alpha(\bk_a)}_{\bk_a}\left( (V^E_{s(a)\to a,\bk_a})^\dag (\Phi^E_{s(a),\bk_a})^\dag \Phi^E_{t(a),\bk_a} V^E_{t(a)\to a,\bk_a} \right)
		\times 
		\xi^{\alpha(\bk_a)}_{\bk_a}\left( (V^F_{s(a)\to a,\bk_a})^\dag (\Phi^F_{s(a),\bk_a})^\dag \Phi^F_{t(a),\bk_a} V^F_{t(a)\to a,\bk_a} \right) \nonumber\\
		&=
		\xi^{\alpha(\bk_a)}_{\bk_a}\left( \left(\bigoplus_B V^B_{s(a)\to a,\bk_a}\right)^\dag \left(\bigoplus_B \Phi^B_{s(a),\bk_a}\right)^\dag \left(\bigoplus_B \Phi^B_{t(a),\bk_a}\right) \left(\bigoplus_B V^B_{t(a)\to a,\bk_a}\right) \right) 
	\end{align}
	and
	\begin{align}
		\xi^{\alpha(\bk_a)}_{\bk_a}\left( \left(V^{E \oplus F}_{s(a)\to a,\bk_a}\right)^\dag \left(\Phi^{E\oplus F}_{s(a),\bk_a}\right)^\dag \Phi^{E \oplus F}_{t(a),\bk_a} V^{E \oplus F}_{t(a)\to a,\bk_a}\right) 
		\label{eq:xi_E_oplus_F}
	\end{align}
	for comparison. 
	Recalling that the $\text{U}(1)$-valued quantity $\xi^{\alpha(\bk_a)}_{\bk_a}(\tilde t_{\bk_a})$ is independent of the choice of representation matrices at $\bk_a$, we define another permutation matrix $Q$ for swapping indices $\alpha,B$ as
	\begin{align}
		\left(\bigoplus_B \bigoplus_{\alpha=1}^{m_a} \bigoplus_{\beta=1}^{m_i} \bigoplus_{\mu=1}^{n^B_{\beta(\bk_i)}} \bigoplus_{\nu=1}^{n^{\beta(\bk_i)}_{\alpha(\bk_a)}} \bigoplus_{y=1}^{\calD_{\alpha(\bk_a)}} \psi_{i,\bk_a}^{B\beta\mu \alpha\nu y}\right)Q
		=
		\bigoplus_{\alpha=1}^{m_a} \bigoplus_B \bigoplus_{\beta=1}^{m_i} \bigoplus_{\mu=1}^{n^B_{\beta(\bk_i)}} \bigoplus_{\nu=1}^{n^{\beta(\bk_i)}_{\alpha(\bk_a)}} \bigoplus_{y=1}^{\calD_{\alpha(\bk_a)}} \psi_{i,\bk_a}^{B\beta\mu \alpha\nu y}. 
		\label{eq:psi_alpha_B_beta}
	\end{align}
	Then we have 
	\begin{align}
		&\xi^{\alpha(\bk_a)}_{\bk_a}\left( \left(\bigoplus_B V^B_{s(a)\to a,\bk_a}\right)^\dag \left(\bigoplus_B \Phi^B_{s(a),\bk_a}\right)^\dag \left(\bigoplus_B \Phi^B_{t(a),\bk_a}\right) \left(\bigoplus_B V^B_{t(a)\to a,\bk_a}\right) \right) \nonumber\\
		&=\xi^{\alpha(\bk_a)}_{\bk_a}\left( Q^T \left(\bigoplus_B V^B_{s(a)\to a,\bk_a}\right)^\dag \left(\bigoplus_B \Phi^B_{s(a),\bk_a}\right)^\dag \left(\bigoplus_B \Phi^B_{t(a),\bk_a}\right) \left(\bigoplus_B V^B_{t(a)\to a,\bk_a}\right)Q \right).
	\end{align}
	Note that since the index $\beta$ is not common between the start point $s(a)$ and terminal point $t(a)$, the transformation of basis in $\bk_a$ by matrix $Q$ cannot represent the swapping of direct sum order of $\beta$.
	(Furthermore, it is implicitly used that the compatibility in each $\alpha(\bk_a)$ sector, i.e., $\sum_{\beta=1}^{m_{s(a)}} n^B_{\beta(s(a))} n^{\beta(\bk_{s(a)})}_{\alpha(\bk_a)} = \sum_{\beta=1}^{m_{t(a)}} n^B_{\beta(t(a))} n^{\beta(\bk_{t(a)})}_{\alpha(\bk_a)} = n^B_{\alpha(\bk_a)}$.)
	In Eq.~\eqref{eq:xi_E_oplus_F}, $\Phi^{E \oplus F}_{i,\bk}$ and $V^{E \oplus F}_{i\to a,\bk_a}$ are given by
	\begin{align}
		\Phi^{E \oplus F}_{i,\bk}
		=
		\bigoplus_{\beta=1}^{m_i} \bigoplus_{B} \bigoplus_{\mu=1}^{n^B_{\beta(\bk_i)}} \bigoplus_{x=1}^{\calD_{\beta(\bk_i)}} \phi_{i,\bk}^{B\beta\mu x}, 
	\end{align}
	\begin{align}
		V^{E \oplus F}_{i\to a,\bk_a}
		=
		\left(\bigoplus_{\beta=1}^{m_i} \bigoplus_B \mathds{1}_{n^B_{\beta(\bk_i)}} \otimes v^{\beta(\bk_i)}_{i\to a,\bk_a} \right)
		P^{E \oplus F}, 
	\end{align}
	and we define the change of basis by $P^{E\oplus F}$ as
	\begin{align}
		\Phi^{E \oplus F}_{i,\bk_a}
		= \bigoplus_{\beta=1}^{m_i} \bigoplus_B \bigoplus_{\mu=1}^{n^B_{\beta(\bk_i)}} \bigoplus_{x=1}^{\calD_{\beta(\bk_i)}} \phi_{i,\bk_a}^{B\beta\mu x}
		&\xrightarrow{\bigoplus_{\beta=1}^{m_i} \bigoplus_B \mathds{1}_{n^B_{\beta(\bk_i)}} \otimes v^{\beta(\bk_i)}_{i\to a,\bk_a}}
		\bigoplus_{\beta=1}^{m_i} \bigoplus_B \bigoplus_{\mu=1}^{n^B_{\beta(\bk_i)}} \bigoplus_{\alpha=1}^{m_a} \bigoplus_{\nu=1}^{n^{\beta(\bk_i)}_{\alpha(\bk_a)}} \bigoplus_{y=1}^{\calD_{\alpha(\bk_a)}} \psi_{i,\bk_a}^{B\beta\mu \alpha\nu y} \nonumber \\
		&\xrightarrow{P^{E\oplus F}}
		\bigoplus_{\alpha=1}^{m_a} \bigoplus_{\beta=1}^{m_i} \bigoplus_B \bigoplus_{\mu=1}^{n^B_{\beta(\bk_i)}} \bigoplus_{\nu=1}^{n^{\beta(\bk_i)}_{\alpha(\bk_a)}} \bigoplus_{y=1}^{\calD_{\alpha(\bk_a)}} \psi_{i,\bk_a}^{B\beta\mu \alpha\nu y}. 
		\label{eq:psi_alpha_beta_B}
	\end{align}
	
	Comparing equations (\ref{eq:psi_alpha_B_beta}) and (\ref{eq:psi_alpha_beta_B}), we observe that only the order of the direct sums over indices $B,\beta$ differs.
	Let $\delta V^{(E,F)}_{i\to a,\bk_a}$ denote the permutation matrix for swapping the order of the direct sums over $B,\beta$:
	\begin{align}
		\left(\bigoplus_{\alpha=1}^{m_a} \bigoplus_B \bigoplus_{\beta=1}^{m_i} \bigoplus_{\mu=1}^{n^B_{\beta(\bk_i)}} \bigoplus_{\nu=1}^{n^{\beta(\bk_i)}_{\alpha(\bk_a)}} \bigoplus_{y=1}^{\calD_{\alpha(\bk_a)}} \psi_{i,\bk_a}^{B\beta\mu \alpha\nu y} \right) \delta V^{(E,F)}_{i\to a,\bk_a}
		=
		\bigoplus_{\alpha=1}^{m_a} \bigoplus_{\beta=1}^{m_i} \bigoplus_B \bigoplus_{\mu=1}^{n^B_{\beta(\bk_i)}} \bigoplus_{\nu=1}^{n^{\beta(\bk_i)}_{\alpha(\bk_a)}} \bigoplus_{y=1}^{\calD_{\alpha(\bk_a)}} \psi_{i,\bk_a}^{B\beta\mu \alpha\nu y}.
	\end{align}
	Summarizing the transformation properties, we obtain
	\begin{align}
		\left(\bigoplus_B \Phi^B_{i,\bk_a}\right) \left(\bigoplus_B V^B_{i\to a,\bk_a}\right) Q \delta V^{(E,F)}_{i\to a,\bk_a}
		=
		\Phi^{E \oplus F}_{i,\bk_a} V^{E \oplus F}_{i\to a,\bk_a}.
	\end{align}
	The permutation matrix $\delta V^{(E,F)}_{i\to a,\bk_a}$ is block-diagonal in the $\alpha$-irrep sectors and independent of the index $y$. 
	Thus, it can be represented as
	\begin{align}
		\delta V^{(E,F)}_{i\to a,\bk_a}
		=\bigoplus_{\alpha=1}^{m_a} \delta V^{(E,F),\alpha(\bk_a)}_{i\to a,\bk_a} \otimes \mathds{1}_{\calD_{\alpha(\bk_a)}}
	\end{align}
	and possesses the symmetry under the representation $w^{E\oplus F}_{a,\bk_a}(g)=\bigoplus_{\alpha=1}^{m_a} \mathds{1}_{(n^E_{\alpha(\bk_a)}+n^F_{\alpha(\bk_a)})} \otimes u^{\alpha(\bk_a)}_{\bk_a}(g)$, as noted in Eq.~\eqref{eq:deltaV_sym}.
	The sign of the permutation matrix $\delta V^{(E,F),\alpha(\bk_a)}_{i\to a,\bk_a}$, corresponding to the permutation $(B,\beta,\mu,\nu) \to (\beta,B,\mu,\nu)$, is given by
	\begin{align}
		\delta\xi^{\alpha(\bk_a)}_{\bk_a,i \to a}(E|_{E_2^{0,0}},F|_{E_2^{0,0}})
		&:=\det \left[ \delta V^{(E,F),\alpha(\bk_a)}_{i\to a,\bk_a} \right]\nonumber \\
		&=(-1)^{\sum_{1\leq \beta_F<m_i}\sum_{\beta_F<\beta_E\leq m_i}
			n^F_{\beta_F(\bk_i)}n^{\beta_F(\bk_i)}_{\alpha(\bk_a)}
			n^E_{\beta_E(\bk_i)}n^{\beta_E(\bk_i)}_{\alpha(\bk_a)}     
		}\in \{\pm 1\}. 
	\end{align}
	Here, the notation $E|_{E_2^{0,0}}$ represents the restriction to the 0-cell, and the correction term indeed depends only on elements of $E_2^{0,0}$ in the 0-cell.
	Consequently, the following relation is obtained:
	\begin{align}
		&\xi^{\alpha(\bk_a)}_{\bk_a}\left( \left(V^{E \oplus F}_{s(a)\to a,\bk_a}\right)^\dag \left(\Phi^{E\oplus F}_{s(a),\bk_a}\right)^\dag \Phi^{E \oplus F}_{t(a),\bk_a} V^{E \oplus F}_{t(a)\to a,\bk_a}\right) \nonumber\\
		&=
		\xi^{\alpha(\bk_a)}_{\bk_a}\left( \left(\bigoplus_B V^B_{s(a)\to a,\bk_a}\right)^\dag \left(\bigoplus_B \Phi^B_{s(a),\bk_a}\right)^\dag \left(\bigoplus_B \Phi^B_{t(a),\bk_a}\right) \left(\bigoplus_B V^B_{t(a)\to a,\bk_a}\right) \right) \nonumber\\
		&\times \delta\xi^{\alpha(\bk_a)}_{\bk_a,s(a) \to a}(E|_{E_2^{0,0}},F|_{E_2^{0,0}}) \times \delta\xi^{\alpha(\bk_a)}_{\bk_a,t(a) \to a}(E|_{E_2^{0,0}},F|_{E_2^{0,0}}). 
	\end{align}
	Combining the results, we obtain
	\begin{align}
		\nu_i(E \oplus F)
		\equiv \nu_i(E)+\nu_i(F)+\delta \nu_i(E|_{E_2^{0,0}},F|_{E_2^{0,0}}), 
	\end{align}
	\begin{align}
		(-1)^{\delta \nu_i(E|_{E_2^{0,0}},F|_{E_2^{0,0}})} 
		:=
		&\prod_{(a,\alpha)} \left[\delta\xi^{\alpha(\bk_a)}_{\bk_a,s(a) \to a}(E|_{E_2^{0,0}},F|_{E_2^{0,0}}) \delta\xi^{\alpha(\bk_a)}_{\bk_a,t(a) \to a}(E|_{E_2^{0,0}},F|_{E_2^{0,0}})\right]^{[x_i]_{(a,\alpha)}}. 
	\end{align}
\end{widetext}

\section{Symmetric bilinear forms and quadratic refinement}
\label{sec:quadratic_refinement}
Let $L=\mZ^N$ be a lattice, and 
\begin{align}
b: L \times L \to \mZ_2 = \{0,1\}
\end{align}
be a symmetric bilinear form, meaning for $x,y,z \in L$ and $n,m \in \mZ$, it satisfies
\begin{align}
&b(x+y, z) = b(x,z) + b(y,z), \\
&b(x, y+z) = b(x,y) + b(x,z), \\
&b(nx,my) = nm b(x,y), \\
&b(x,y) = b(y,x).
\end{align}
A function 
\begin{align}
q: L \to \mZ_2
\end{align}
that satisfies that 
\begin{align}
q(x+y) = q(x) + q(y) + b(x,y) \quad \mbox{for all $x,y \in L$} 
\label{eq:qr_def_1}
\end{align}
is called a quadratic refinement of $b$. 

We show the following:

(i) A quadratic refinement $q$ exists.

(ii) $q$ is not unique; the ambiguity of $q$ is given by ${\rm Hom}(L,\mZ_2) \cong \mZ_2^N$.

The proof of the latter is trivial. If $q_1, q_2$ are quadratic refinements, then
\begin{align}
\delta q(x):= q_1(x)-q_2(x)
\end{align}
is linear.
We now prove the former.

Let the basis of $L$ be $e_1,\dots,e_N$, and write
\begin{align}
b_{ij} = b(e_i,e_j),\quad b_{ij}=b_{ji}. 
\end{align}
Introduce the floor function for a real number $x \in \R$, which returns the largest integer not exceeding $x$:
\begin{align}
\lfloor x \rfloor = \max \{ n \in \mZ| n \leq x\}. 
\end{align}
We show that a solution of (\ref{eq:qr_def_1}) is given by
\begin{align}
q\left(\sum_{i=1}x_i e_i\right)
= \sum_{i=1}^N \left\lfloor \frac{x_i}{2} \right\rfloor b_{ii} + \sum_{1\leq i<j \leq N} x_i b_{ij} x_j. 
\end{align}
In fact, 
\begin{align}
&q(x+y)-q(x)-q(y) \nonumber\\
&=\sum_{i=1}^N \left\lfloor \frac{x_i+y_i}{2} \right\rfloor b_{ii} + \sum_{1\leq i<j \leq N} (x_i+y_i) b_{ij} (x_j+y_j)\nonumber \\
&-\left(\sum_{i=1}^N \left\lfloor \frac{x_i}{2} \right\rfloor b_{ii} + \sum_{1\leq i<j \leq N} x_i b_{ij} x_j\right)\nonumber\\
&-\left(\sum_{i=1}^N \left\lfloor \frac{y_i}{2} \right\rfloor b_{ii} + \sum_{1\leq i<j \leq N} y_i b_{ij} y_j\right)\\
&=\sum_{i=1}^N \left( \left\lfloor \frac{x_i+y_i}{2}\right\rfloor 
-\left\lfloor \frac{x_i}{2} \right\rfloor-\left\lfloor \frac{y_i}{2} 
\right)\right\rfloor b_{ii}
+ \sum_{i\neq j} x_i b_{ij} y_j. 
\end{align}
Noticing that when $x_i, y_i$ are integers 
\begin{align}
\left\lfloor \frac{x_i+y_i}{2}\right\rfloor 
=\left\lfloor \frac{x_i}{2} \right\rfloor
+\left\lfloor \frac{y_i}{2} \right\rfloor + x_i y_i \quad \mod 2. 
\end{align}
Thus,
\begin{align}
q(x+y)-q(x)-q(y)
&=\sum_{i,j=1}^N x_i b_{ij} y_j
=b(x,y). 
\end{align}

The ambiguity ${\rm Hom}(L,\mZ_2) \cong \mZ_2^N$ can be represented by $N$ bits $a_1,\dots,a_N \in \mZ_2^N$, and a general form of quadratic refinement is given by
\begin{align}
    q\left(\sum_{i=1}x_i e_i\right)
    = \sum_{i=1}^N \left\lfloor \frac{x_i}{2} \right\rfloor b_{ii} + \sum_{1\leq i<j \leq N} x_i b_{ij} x_j
    +\sum_{i=1}^N a_i x_i.
    \label{eq:qr_formula}
\end{align}

As a comment, quadratic refinement can also be defined as a function $q':L \to \mZ/2$ that satisfies
\begin{align}
b(x,y) = q'(x+y) - q'(x) - q'(y) + q'(0)  
\label{eq:qr_def_2}
\end{align}
for all $x,y \in L$. 
The definition (\ref{eq:qr_def_1}) implies $q(0)=0$. 
If defined as (\ref{eq:qr_def_2}), $q'(0) \in \{0,1\}$ remains indeterminate, and there is a relation $q(x) = q'(x)-q'(0)$ with the quadratic refinement defined by (\ref{eq:qr_def_1}).

\section{Derivation of (\ref{eq:qr_P2221'})}
\label{app:derivation_of_qr_P2221'}
\begin{widetext}

Computing the kernel of the differential $d_1^{0,0} ({\cal B}^{(0)}) = {\cal B}^{(1)} M_{d_1^{0,0}}$ we have the basis transformation ${\cal B}'^{(0)} = {\cal B}^{(0)} V^{(0)} = (\bm{b}'^{(0)}_1,\dots,\bm{b}'^{(0)}_{32})$ with 
\setlength{\arraycolsep}{2pt}
\begin{align}
V^{(0)}=
\left[
\begin{array}{cccc|cccc|cccc|cccc|cccc|cccc|cccc|cccc}
 1 & 0 & 0 & -1 & 0 & 0 & 0 & 0 & 0 & 0 & 0 & 0 & 0 & 0 & 0 & 0 & 0 & 0 & 1 & -1 & 0 & 1 & -1 & 0 & -1 & -1 & 0 & 1 & 1 & 1 & 1 & 0 \\
 0 & 1 & 0 & -1 & 0 & 0 & 0 & 0 & 0 & 0 & 0 & 0 & 0 & 0 & 0 & 0 & 0 & 0 & 1 & -1 & 0 & -1 & 0 & 1 & 1 & 1 & -1 & 0 & 0 & 0 & 1 & 0 \\
 0 & 0 & 1 & -1 & 0 & 0 & 0 & 0 & 0 & 0 & 0 & 0 & 0 & 0 & 0 & 0 & 0 & 0 & 1 & 1 & 1 & -1 & 1 & 0 & 0 & -1 & 1 & 0 & 0 & 0 & -1 & 0 \\
 0 & 0 & 0 & 1 & 0 & 0 & 0 & 0 & 0 & 0 & 0 & 0 & 0 & 0 & 0 & 0 & 0 & 0 & -1 & 1 & -1 & 1 & 0 & -1 & 0 & 1 & 0 & -1 & 0 & 0 & 0 & 1 \\
 \hline
 0 & 0 & 0 & 1 & 1 & 0 & 0 & 0 & 0 & 0 & 0 & 0 & 0 & 0 & 0 & 0 & 0 & 0 & -2 & 0 & -1 & 1 & -1 & -1 & -1 & 0 & 0 & 0 & 1 & 1 & 1 & 1 \\
 0 & 0 & 0 & 1 & 0 & 1 & 0 & 0 & 0 & 0 & 0 & 0 & 0 & 0 & 0 & 0 & 0 & 0 & -2 & 0 & 1 & -1 & 0 & 0 & 1 & 0 & 0 & 0 & 0 & 0 & 0 & 0 \\
 0 & 0 & 0 & 1 & 0 & 0 & 1 & 0 & 0 & 0 & 0 & 0 & 0 & 0 & 0 & 0 & 0 & 0 & -2 & 0 & 0 & -1 & 1 & 1 & 0 & 0 & 0 & 0 & 0 & 0 & 0 & 0 \\
 0 & 0 & 0 & 0 & 0 & 0 & 0 & 0 & 0 & 0 & 0 & 0 & 0 & 0 & 0 & 0 & 0 & 0 & 0 & 0 & 0 & 1 & 0 & 0 & 0 & 0 & 0 & 0 & 0 & 0 & 0 & 0 \\
 \hline
 0 & 0 & 0 & 1 & 0 & 0 & 0 & 1 & 0 & 0 & 0 & 0 & 0 & 0 & 0 & 0 & 0 & 0 & -2 & 0 & 1 & 0 & 0 & 0 & -1 & -1 & 0 & 0 & 1 & 1 & 0 & 0 \\
 0 & 0 & 0 & 1 & 0 & 0 & 0 & 0 & 1 & 0 & 0 & 0 & 0 & 0 & 0 & 0 & 0 & 0 & -2 & 0 & -1 & 0 & 0 & 0 & 1 & 1 & -1 & -1 & 0 & 0 & 1 & 1 \\
 0 & 0 & 0 & 1 & 0 & 0 & 0 & 0 & 0 & 1 & 0 & 0 & 0 & 0 & 0 & 0 & 0 & 0 & -2 & 0 & 0 & 0 & 0 & 0 & 0 & -1 & 1 & 1 & 0 & 0 & 0 & 0 \\
 0 & 0 & 0 & 0 & 0 & 0 & 0 & 0 & 0 & 0 & 0 & 0 & 0 & 0 & 0 & 0 & 0 & 0 & 0 & 0 & 0 & 0 & 0 & 0 & 0 & 1 & 0 & 0 & 0 & 0 & 0 & 0 \\
 \hline
 0 & 0 & 0 & 1 & 0 & 0 & 0 & 0 & 0 & 0 & 1 & 0 & 0 & 0 & 0 & 0 & 0 & 0 & -2 & 0 & 0 & 0 & 0 & 0 & -1 & 0 & 0 & 0 & 1 & 1 & 0 & 0 \\
 0 & 0 & 0 & 0 & 0 & 0 & 0 & 0 & 0 & 0 & 0 & 0 & 0 & 0 & 0 & 0 & 0 & 0 & 0 & 0 & 0 & 0 & 0 & 0 & 1 & 0 & 0 & 0 & 0 & 0 & 0 & 0 \\
 0 & 0 & 0 & 1 & 0 & 0 & 0 & 0 & 0 & 0 & 0 & 1 & 0 & 0 & 0 & 0 & 0 & 0 & -2 & 0 & -1 & 0 & 0 & 0 & 0 & 0 & 0 & 0 & 0 & 0 & 1 & 1 \\
 0 & 0 & 0 & 0 & 0 & 0 & 0 & 0 & 0 & 0 & 0 & 0 & 0 & 0 & 0 & 0 & 0 & 0 & 0 & 0 & 1 & 0 & 0 & 0 & 0 & 0 & 0 & 0 & 0 & 0 & 0 & 0 \\
 \hline
 0 & 0 & 0 & 1 & 0 & 0 & 0 & 0 & 0 & 0 & 0 & 0 & 1 & 0 & 0 & 0 & 0 & 0 & -2 & -1 & 0 & 0 & -1 & 1 & 0 & 0 & 0 & 0 & 1 & 0 & 1 & 0 \\
 0 & 0 & 0 & 1 & 0 & 0 & 0 & 0 & 0 & 0 & 0 & 0 & 0 & 1 & 0 & 0 & 0 & 0 & -2 & -1 & 0 & 0 & 0 & 0 & 0 & 0 & -1 & 1 & 0 & 1 & 1 & 0 \\
 0 & 0 & 0 & 1 & 0 & 0 & 0 & 0 & 0 & 0 & 0 & 0 & 0 & 0 & 1 & 0 & 0 & 0 & -2 & 1 & 0 & 0 & 1 & -1 & 0 & 0 & 1 & -1 & 0 & 0 & -1 & 1 \\
 0 & 0 & 0 & 0 & 0 & 0 & 0 & 0 & 0 & 0 & 0 & 0 & 0 & 0 & 0 & 0 & 0 & 0 & 0 & 1 & 0 & 0 & 0 & 0 & 0 & 0 & 0 & 0 & 0 & 0 & 0 & 0 \\
 \hline
 0 & 0 & 0 & 1 & 0 & 0 & 0 & 0 & 0 & 0 & 0 & 0 & 0 & 0 & 0 & 1 & 0 & 0 & -2 & 0 & 0 & 0 & -1 & 0 & 0 & 0 & 0 & 0 & 1 & 0 & 1 & 0 \\
 0 & 0 & 0 & 1 & 0 & 0 & 0 & 0 & 0 & 0 & 0 & 0 & 0 & 0 & 0 & 0 & 1 & 0 & -2 & 0 & 0 & 0 & 0 & -1 & 0 & 0 & 0 & 0 & 0 & 1 & 0 & 1 \\
 0 & 0 & 0 & 0 & 0 & 0 & 0 & 0 & 0 & 0 & 0 & 0 & 0 & 0 & 0 & 0 & 0 & 0 & 0 & 0 & 0 & 0 & 1 & 0 & 0 & 0 & 0 & 0 & 0 & 0 & 0 & 0 \\
 0 & 0 & 0 & 0 & 0 & 0 & 0 & 0 & 0 & 0 & 0 & 0 & 0 & 0 & 0 & 0 & 0 & 0 & 0 & 0 & 0 & 0 & 0 & 1 & 0 & 0 & 0 & 0 & 0 & 0 & 0 & 0 \\
 \hline
 0 & 0 & 0 & 1 & 0 & 0 & 0 & 0 & 0 & 0 & 0 & 0 & 0 & 0 & 0 & 0 & 0 & 1 & -2 & 0 & 0 & 0 & 0 & 0 & 0 & 0 & 0 & -1 & 1 & 0 & 0 & 1 \\
 0 & 0 & 0 & 1 & 0 & 0 & 0 & 0 & 0 & 0 & 0 & 0 & 0 & 0 & 0 & 0 & 0 & 0 & 0 & 0 & 0 & 0 & 0 & 0 & 0 & 0 & -1 & 0 & 0 & 1 & 1 & 0 \\
 0 & 0 & 0 & 0 & 0 & 0 & 0 & 0 & 0 & 0 & 0 & 0 & 0 & 0 & 0 & 0 & 0 & 0 & 0 & 0 & 0 & 0 & 0 & 0 & 0 & 0 & 1 & 0 & 0 & 0 & 0 & 0 \\
 0 & 0 & 0 & 0 & 0 & 0 & 0 & 0 & 0 & 0 & 0 & 0 & 0 & 0 & 0 & 0 & 0 & 0 & 0 & 0 & 0 & 0 & 0 & 0 & 0 & 0 & 0 & 1 & 0 & 0 & 0 & 0 \\
 \hline
 0 & 0 & 0 & 0 & 0 & 0 & 0 & 0 & 0 & 0 & 0 & 0 & 0 & 0 & 0 & 0 & 0 & 0 & 0 & 0 & 0 & 0 & 0 & 0 & 0 & 0 & 0 & 0 & 1 & 0 & 0 & 0 \\
 0 & 0 & 0 & 0 & 0 & 0 & 0 & 0 & 0 & 0 & 0 & 0 & 0 & 0 & 0 & 0 & 0 & 0 & 0 & 0 & 0 & 0 & 0 & 0 & 0 & 0 & 0 & 0 & 0 & 1 & 0 & 0 \\
 0 & 0 & 0 & 0 & 0 & 0 & 0 & 0 & 0 & 0 & 0 & 0 & 0 & 0 & 0 & 0 & 0 & 0 & 0 & 0 & 0 & 0 & 0 & 0 & 0 & 0 & 0 & 0 & 0 & 0 & 1 & 0 \\
 0 & 0 & 0 & 0 & 0 & 0 & 0 & 0 & 0 & 0 & 0 & 0 & 0 & 0 & 0 & 0 & 0 & 0 & 0 & 0 & 0 & 0 & 0 & 0 & 0 & 0 & 0 & 0 & 0 & 0 & 0 & 1 \\
\end{array}
\right].
\end{align}
Here, the blocks are sequenced as $\G,X,Y,S,Z,U,T,R$, with irreps within each block arranged in the order $A,B_1,B_2,B_3$. 
The basis vectors $\bm{b}'^{(0)}_{20}, \dots, \bm{b}'^{(0)}_{32}$ span $E_2^{0,0} \cong \mZ^{13}$.
The bilinear form $\delta \nu: E_2^{0,0} \times E_2^{0,0} \to \mZ_2$ defined in (\ref{eq:inter}) is 
\setlength{\arraycolsep}{2pt}
\begin{align}
    \left[\delta \nu(\bm{b}'^{(0)}_i,\bm{b}'^{(0)}_j)\right]_{20\leq i,j \leq 32}
=\left[
\begin{array}{ccccccccccccc}
 0 & 0 & 0 & 0 & 1 & 1 & 1 & 1 & 1 & 1 & 0 & 0 & 0 \\
 0 & 0 & 0 & 0 & 1 & 0 & 1 & 0 & 0 & 0 & 0 & 1 & 1 \\
 0 & 0 & 0 & 0 & 0 & 0 & 0 & 0 & 0 & 0 & 0 & 0 & 0 \\
 0 & 0 & 0 & 0 & 0 & 0 & 0 & 0 & 0 & 0 & 0 & 0 & 0 \\
 1 & 1 & 0 & 0 & 0 & 0 & 1 & 1 & 1 & 0 & 0 & 1 & 0 \\
 1 & 0 & 0 & 0 & 0 & 0 & 1 & 0 & 0 & 1 & 0 & 0 & 1 \\
 1 & 1 & 0 & 0 & 1 & 1 & 0 & 0 & 1 & 0 & 0 & 0 & 1 \\
 1 & 0 & 0 & 0 & 1 & 0 & 0 & 0 & 0 & 1 & 0 & 1 & 0 \\
 1 & 0 & 0 & 0 & 1 & 0 & 1 & 0 & 0 & 1 & 0 & 1 & 1 \\
 1 & 0 & 0 & 0 & 0 & 1 & 0 & 1 & 1 & 0 & 0 & 1 & 1 \\
 0 & 0 & 0 & 0 & 0 & 0 & 0 & 0 & 0 & 0 & 0 & 0 & 0 \\
 0 & 1 & 0 & 0 & 1 & 0 & 0 & 1 & 1 & 1 & 0 & 0 & 1 \\
 0 & 1 & 0 & 0 & 0 & 1 & 1 & 0 & 1 & 1 & 0 & 1 & 0 \\
\end{array}
\right]. 
\end{align}
We confirmed that $\delta \nu$ is symmetric. 
A quadratic refinement in the basis of $E_2^{0,0}$ is given by 
\begin{align}
    q(\bm{n}') = \sum_{k,l=20}^{32} n'_i \delta \nu(\bm{b}'^{(0)}_i,\bm{b}'^{(0)}_j) n'_j + \sum_{i=20}^{32} c'_i n'_i \quad 
    \mbox{for} \quad  \bm{n}' = \sum_{i=20}^{32} n'_i \bm{b}'^{(0)}_i \in E_2^{0,0}.
\end{align}
The coefficients $c'_{20},\dots c'_{32}$ are fixed to satisfy (\ref{eq:condition_q_P2221'}). 
There are 8 Wyckoff positions and 4 irreps $\beta \in \{A, B_1, B_2, B_3\}$ for each Wyckoff position, leading to 32 column vectors $\bm{n}'(\bm{a}_{\bx_0}^\beta) = \sum_{i=20}^{32} n'_i(\bm{a}_{\bx_0}^\beta) \bm{b}'^{(0)}_i \in E_2^{0,0}$, 
\begin{align}
\left( \bm{n}'(a_{\bx_0}^\beta) \right)_{\bm{x}_0,\beta}=
\left[
\begin{array}{cccccccccccccccccccccccccccccccc}
 0 & 0 & 0 & 1 & 0 & 0 & 0 & 1 & 0 & 0 & 0 & 1 & 0 & 0 & 0 & 1 & 0 & 0 & 1 & 0 & 0 & 0 & 1 & 0 & 0 & 0 & 1 & 0 & 0 & 0 & 1 & 0 \\
 0 & 0 & 0 & 1 & 1 & 0 & 0 & 0 & 0 & 1 & 0 & 0 & 0 & 0 & 1 & 0 & 0 & 0 & 0 & 1 & 1 & 0 & 0 & 0 & 0 & 1 & 0 & 0 & 0 & 0 & 1 & 0 \\
 0 & 0 & 0 & 1 & 1 & 0 & 0 & 0 & 0 & 0 & 0 & 1 & 1 & 0 & 0 & 0 & 0 & 0 & 0 & 1 & 1 & 0 & 0 & 0 & 0 & 0 & 0 & 1 & 1 & 0 & 0 & 0 \\
 0 & 0 & 1 & 0 & 0 & 1 & 0 & 0 & 0 & 0 & 1 & 0 & 0 & 1 & 0 & 0 & 0 & 0 & 0 & 1 & 1 & 0 & 0 & 0 & 0 & 0 & 0 & 1 & 1 & 0 & 0 & 0 \\
 0 & 0 & 0 & 1 & 1 & 0 & 0 & 0 & 0 & 0 & 0 & 1 & 1 & 0 & 0 & 0 & 0 & 0 & 1 & 0 & 0 & 1 & 0 & 0 & 0 & 0 & 1 & 0 & 0 & 1 & 0 & 0 \\
 0 & 1 & 0 & 0 & 0 & 0 & 1 & 0 & 0 & 0 & 0 & 1 & 1 & 0 & 0 & 0 & 0 & 1 & 0 & 0 & 0 & 0 & 1 & 0 & 0 & 0 & 0 & 1 & 1 & 0 & 0 & 0 \\
 0 & 0 & 0 & 1 & 0 & 0 & 0 & 1 & 0 & 1 & 0 & 0 & 0 & 1 & 0 & 0 & 0 & 0 & 0 & 1 & 0 & 0 & 0 & 1 & 0 & 1 & 0 & 0 & 0 & 1 & 0 & 0 \\
 0 & 0 & 1 & 0 & 0 & 0 & 1 & 0 & 1 & 0 & 0 & 0 & 1 & 0 & 0 & 0 & 0 & 0 & 0 & 1 & 0 & 0 & 0 & 1 & 0 & 1 & 0 & 0 & 0 & 1 & 0 & 0 \\
 0 & 0 & 0 & 1 & 0 & 0 & 0 & 1 & 0 & 1 & 0 & 0 & 0 & 1 & 0 & 0 & 0 & 0 & 1 & 0 & 0 & 0 & 1 & 0 & 1 & 0 & 0 & 0 & 1 & 0 & 0 & 0 \\
 1 & 0 & 0 & 0 & 0 & 0 & 0 & 1 & 0 & 0 & 1 & 0 & 0 & 1 & 0 & 0 & 0 & 1 & 0 & 0 & 0 & 0 & 1 & 0 & 0 & 0 & 0 & 1 & 1 & 0 & 0 & 0 \\
 0 & 1 & 0 & 0 & 0 & 0 & 1 & 0 & 0 & 0 & 0 & 1 & 1 & 0 & 0 & 0 & 1 & 0 & 0 & 0 & 0 & 0 & 0 & 1 & 0 & 0 & 1 & 0 & 0 & 1 & 0 & 0 \\
 0 & 0 & 1 & 0 & 0 & 1 & 0 & 0 & 1 & 0 & 0 & 0 & 0 & 0 & 0 & 1 & 0 & 0 & 0 & 1 & 1 & 0 & 0 & 0 & 0 & 1 & 0 & 0 & 0 & 0 & 1 & 0 \\
 0 & 0 & 0 & 1 & 1 & 0 & 0 & 0 & 0 & 1 & 0 & 0 & 0 & 0 & 1 & 0 & 0 & 0 & 1 & 0 & 0 & 1 & 0 & 0 & 1 & 0 & 0 & 0 & 0 & 0 & 0 & 1 \\
\end{array}
\right]
\end{align}
We find that the equations $q\left(\bm{n}'(\bm{a}^\beta_{\bm{x}_0})\right) \equiv 0$ for all Wyckoff positions $\bm{x}_0$ and all irreps $\beta$ have a solution 
\begin{align}
(c'_i)_{i=20,\dots,32}= 
(0,1,0,0,1,0,0,0,1,0,0,0,0).
\end{align}
This gives the quadratic refinement on the basis of $E_2^{0,0}$. 
The basis transformation $n'_i = \sum_{j=1}^{32} [(V^{(0)})^{-1}]_{ij} n_j$ gives the expression (\ref{eq:qr_P2221'}).

\end{widetext}

\bibliography{ref}

\end{document}